\documentclass[aps,preprint,superscriptaddress,preprintnumbers]{revtex4}
\usepackage{amsmath}
\usepackage{graphicx}
\usepackage{amssymb}

\newcommand{\tmbx}[1]{\mbox{\tiny{$#1$}}}
\newcommand{\gNLt}{\left(g_{\tmbx{L}}^{\tmbx{\nu N}}\right)^2}
\newcommand{\gNRt}{\left(g_{\tmbx{R}}^{\tmbx{\nu N}}\right)^2}
\newcommand{\be}{\begin{equation}}
\newcommand{\ee}{\end{equation}}
\newcommand{\bea}{\begin{eqnarray}}
\newcommand{\eea}{\end{eqnarray}}

\usepackage[normalem]{ulem}
\usepackage[usenames]{color}

\preprint{MSUHEP-110802}

\begin{document}

\title{Discovery and Identification of $W^\prime$ and $Z^\prime$ 
in $SU(2)_1\otimes SU(2)_2\otimes U(1)_X$ Models at the LHC}

\author{Qing-Hong Cao}
\email[]{qinghongcao@pku.edu.cn}
\affiliation{Department of Physics and State Key Laboratory of Nuclear Physics
and Technology, Peking University, Beijing 100871, China}

\author{Zhao Li}
\email[]{zhaoli@pa.msu.edu}
\affiliation{Department of Physics and Astronomy, Michigan State University, East Lansing, MI 48824, U.S.A}

\author{Jiang-Hao Yu}
\email[]{yujiangh@msu.edu}
\affiliation{Department of Physics and Astronomy, Michigan State University, East Lansing, MI 48824, U.S.A}

\author{C.-P. Yuan}
\email[]{yuan@pa.msu.edu}
\affiliation{Department of Physics and Astronomy, Michigan State University, East Lansing, MI 48824, U.S.A}
\affiliation{Center for High Energy Physics, Peking University, Beijing 100871, China}

\begin{abstract}
	We explore the discovery potential of $W'$ and $Z'$ boson searches for various $SU(2)_1\otimes SU(2)_2\otimes U(1)_X$ models at the Large Hadron Collider (LHC), after taking into account the constraints
	from low energy precision measurements and direct searches at both the Tevatron (1.96 TeV) and the LHC (7 TeV).
	In such models, the $W'$ and $Z'$ bosons emerge after the electroweak
	symmetry is spontaneously broken. Two patterns of the symmetry breaking are considered in this work:
	one is 
	$SU(2)_L\otimes SU(2)_2 \otimes U(1)_X \to SU(2)_L\otimes U(1)_Y$ (BP-I), 
	another is $SU(2)_1\otimes SU(2)_2 \otimes U(1)_Y \to SU(2)_L\otimes U(1)_Y$ (BP-II).
	Examining the single production channel
	of $W'$ and $Z'$ with their subsequent leptonic decays, we
	find that the probability of detecting
	$W'$ and $Z'$ bosons in the considered models
	at the LHC (with 14 TeV) is highly limited by the low energy precision data constraints.
	We show that observing $Z'$ alone, without seeing a $W'$, does not rule out new physics models with non-Abelian gauge extension, such as the phobic models in BP-I. Models in BP-II would predict the discovery of
	degenerate $W'$ and $Z'$ bosons at the LHC.

\end{abstract}
\maketitle

%%%%%%%%%%%%%%%%%%%%%%%%%%%%%%%
%                                                                                             %
%        Section I: Introduction                                                  %
%													                              %
%%%%%%%%%%%%%%%%%%%%%%%%%%%%%%%

\section{Introduction}

As remnants of electroweak symmetry breaking, extra gauge bosons exist in many new physics (NP) models,
beyond the Standard Model (SM) of particle physics.  
According to their electromagnetic charges, extra gauge bosons are usually separated into  
two categories:  one is named as $W^\prime$ (charged bosons) and another is $Z^\prime$ 
(neutral bosons). While $Z^\prime$ boson could originate from an additional abelian $U(1)$ group, 
$W^\prime$ boson is often associated with an extra non-Abelian group.  
The minimal extension of the SM, which consists of both $W^\prime$ and $Z^\prime$ 
bosons, exhibits a gauge structure of $SU(2)\times SU(2) \times U(1)$~\cite{Mohapatra:1974gc,
Mohapatra:1974hk,Senjanovic:1975rk,Mohapatra:1980yp,
Chivukula:2006cg,Barger:1980ix,Barger:1980ti,
Georgi:1989ic,Georgi:1989xz,
Li:1981nk,Malkawi:1996fs,He:2002ha,
Hsieh:2010zr}, named as $G(221)$ model~\cite{Hsieh:2010zr}. 
Searching for those new gauge bosons~\cite{Rizzo:2006nw} and determining their quantum numbers~\cite{Berger:2011hn}
would shed light on the gauge structure of NP.

At the Large Hadron Collider (LHC), it is very promising to search for those heavy $Z^\prime$ and $W^\prime$ 
bosons through their single production channel as an $s$-channel resonance with their subsequent 
leptonic decays~\cite{Bauer:2009cc}. 
It yields the simplest event topology to discover  $Z^\prime$ 
and/or $W^\prime$ with a large production rate and clean experiment signature.  
These channels may be one of the most promising early discoveries at the 
LHC~\cite{Khachatryan:2010fa,Chatrchyan:2011wq,Aad:2011yg,Collaboration:2011dca}.
There have been many theoretical studies of searching for the  $Z^\prime$ boson~\cite{Langacker:2008yv,
Carena:2004xs,Salvioni:2009mt,Accomando:2010fz,Lynch:2000md} and the  $W^\prime$ boson~\cite{Schmaltz:2010xr,Maiezza:2010ic,Grojean:2011vu,Nemevsek:2011hz,Torre:2011wy,Jezo:2012rm,Keung:1983uu} at the LHC.
In many NP models with extended gauge groups, the $W^\prime$ boson emerges
together with the $Z^\prime$ boson after symmetry breaking, and usually, the $W^\prime$ boson 
is lighter than, or as heavy as, the $Z^\prime$ boson. It is therefore possible to discover $W^\prime$ 
prior to $Z^\prime$. More often, the masses of the $W^\prime$ and $Z^\prime$ bosons are not 
independent, and so as their couplings to the SM fermions.  Hence, the discovery potential of 
the $W^\prime$ and $Z^\prime$ at the LHC could be highly correlated. In this paper we present a comprehensive 
study of discovery potentials of both the $W^\prime$ and $Z^\prime$ boson searches in the $G(221)$ models at the LHC.  

The $G(221)$ models are the minimal extension of the SM gauge group to include both the $W^\prime$ 
and $Z^\prime$ bosons. The gauge structure is $SU(2)\times SU(2)\times U(1)$.
The model can be viewed as the low energy effective theory of many NP models 
with extended gauge structure when all the heavy particles other than the $W^\prime$ and 
$Z^\prime$ bosons decouple.  
In this paper, based on a linearly-realized effective theory including the $SU(2)\times SU(2) \times U(1)$
gauge group, we present the collider phenomenology related to the simplest event topology
in the resonance $Z^\prime$ and $W^\prime$ processes. % aimed at the early discovery. 
 
In the TeV scale, different electroweak symmetry breaking (EWSB) patterns will induce different
$Z^\prime$ and $W^\prime$ mass relations. In breaking pattern I, which has the
$SU(2)\otimes U(1)$ breaking down to $U(1)_{Y}$, the $W^\prime$ mass
is always smaller than the $Z^\prime$ mass; while in breaking pattern II,
the $SU(2)\otimes SU(2)$ breaking down to $SU(2)_{L}$ requires the
$W^\prime$ and $Z^\prime$ bosons have the same mass at tree level. This feature could assist us
to distinguish these two breaking patterns after the $W^\prime$ and $Z^\prime$ bosons are discovered.

The paper is organized as follows. In Sec.~II we briefly review several typical $G(221)$ models
and present the relevant couplings of $W^\prime$ and $Z^\prime$ to fermions. 
In Sec.~III we discuss the production cross section of the so-called sequential 
$W^\prime$ and $Z^\prime$ bosons in hadron collisions with the next-to-leading (NLO) 
QCD correction included. 
Based on the narrow width approximation, we propose a simple approach to generalize the 
sequential $W^\prime$ and $Z^\prime$ production cross sections to various $G(221)$ models. 
In Sec.~IV we present the allowed theoretical parameter space of various $G(221)$ models after incorporating 
indirect constraints from electroweak precision test observables (EWPTs) and
direct search constraints from Tevatron and 7~TeV LHC (LHC7) data. 
In Sec.~V we explore the potential of the 14~TeV LHC (LHC14). 
Finally, we conclude in Sec.~VI.  

%%%%%%%%%%%%%%%%%%%%%%%%%%%%%%%%%
%                                                                                            		 %
%        Section II: 221 Model                                                   	 %
%													                             		 %
%%%%%%%%%%%%%%%%%%%%%%%%%%%%%%%%%

\section{The model}

In this section we briefly review the $G(221)$ model 
and the masses and couplings of $W^\prime$ and $Z^\prime$ bosons.
In particular we consider various $G(221)$ models categorized as follows: 
left-right (LR)~\cite{Mohapatra:1974gc,Mohapatra:1974hk,Mohapatra:1980yp},
lepto-phobic (LP), hadro-phobic (HP), fermio-phobic (FP)~\cite{
Chivukula:2006cg,Barger:1980ix,Barger:1980ti}, un-unified (UU)~\cite{
Georgi:1989ic,Georgi:1989xz}, and non-universal (NU)~\cite{
Li:1981nk,Malkawi:1996fs,He:2002ha,Berger:2011xk}. 
We also considered a widely-used reference model in the experiment searches: the sequential $W^\prime$ model (SQ).
In the LR model and SQ models, if the gauge couplings  are assigned to be the same for the two $SU(2)$ gauge groups, 
the models are considered as the manifest left-right model (MLR), and manifest sequential model (MSQ).
In the MSQ, the $W^\prime$ couplings to the fermion is the same as the standard model $W$ couplings to fermion, 
which served as the reference model in the experiment searches.
We focus our attention on the couplings of extra gauge boson to SM fermions 
which are involved in extra gauge boson production via the $s$-channel process. 
More details of the $G(221)$ model can be found in our previous paper~\cite{Hsieh:2010zr}. 

The classification of $G(221)$ models is based on the pattern
of symmetry breaking and quantum number assignment of the SM fermions. 
The NP models mentioned above can be categorized into
two symmetry breaking patterns:
\begin{itemize} 
\item[(a)] breaking pattern I (BP-I): \\
$SU(1)_1$ is identified as the $SU(2)_L$ of the SM. The first stage of symmetry breaking 
$SU(2)_2 \times U(1)_{X}\to U(1)_{Y}$ occurs at the TeV scale, while the second stage of 
symmetry breaking $SU(2)_{L}\times U(1)_Y \to U(1)_{em}$ takes place at the electroweak scale;
\item[(b)] breaking pattern II (BP-II):\\
$U(1)_X$ is identified as the $U(1)_Y$ of the SM. The first stage of symmetry breaking 
$SU(2)_{1}\times SU(2)_{2}\to SU(2)_{L}$ occurs at the TeV scale, while the second stage of
symmetry breaking $SU(2)_{L}\times U(1)_Y \to U(1)_{em}$ happens at the electroweak scale.
\end{itemize}
The symmetry breaking is assumed to be induced by fundamental scalar 
fields throughout this paper. 
The quantum number of the scalar fields under the $G(221)$ gauge group 
depends on the breaking pattern. 
In BP-I, the symmetry breaking of $SU(2)_{2}\otimes U(1)_{X}\to U(1)_{Y}$
at the TeV scale could be induced by a scalar doublet field $\Phi\sim(1,2)_{1/2}$,
or a triplet scalar field $(1,3)_{1}$ with a vacuum expectation value
(VEV) $u$, and the subsequent symmetry breaking of 
$SU(2)_{1}\otimes U(1)_{Y}\to U(1)_{Q}$ at the electroweak scale 
is via another scalar field $H\sim(2,\bar{2})_{0}$ with two VEVs $v_1$ and $v_2$, which can be 
redefined as a VEV $v = \sqrt{v_1^2 + v_2^2}$ and a mixing
angle $\beta = \arctan(v_1/v_2)$. 
In BP-II, the symmetry breaking of $SU(2)_{1}\otimes SU(2)_{2}\to SU(2)_{L}$
at the TeV scale is owing to a Higgs bi-doublet $\Phi\sim(2,\bar{2})_{0}$
with only one VEV $u$, and the subsequent breaking of 
$SU(2)_{L}\otimes U(1)_{Y}\to U(1)_{Q}$ at the electroweak scale 
is generated by a Higgs doublet $H\sim(2,1)_{1/2}$ with the
VEV $v$. Since the precision data constraints (including those from CERN LEP
and SLAC SLC experiment data) pushed the TeV symmetry breaking
higher than $1$ TeV, we shall approximate the predictions of physical observables
by taking Taylor expansion in $1/x$ with $x=u^{2}/v^{2}$, which is assumed to be much larger than 1.

Denote $g_1$, $g_2$ and $g_X$ as the coupling of $SU(2)_1$, $SU(2)_2$
and $U(1)_X$, respectively.
Depending on the symmetry breaking pattern, the three couplings are
\begin{eqnarray}
g_{1}=\frac{e}{s_{W}},\quad g_{2}=\frac{e}{c_{W}s_{\phi}},\quad g_{X}=\frac{e}{c_{W}c_{\phi}}, & (\textbf{BP-I})
\end{eqnarray}
\begin{eqnarray}
g_{1}=\frac{e}{s_{W}c_{\phi}},\quad g_{2}=\frac{e}{s_{W}s_{\phi}},\quad g_{X}=\frac{e}{c_{W}}, & (\textbf{BP-II})
\end{eqnarray}
where $s_{W}$ and $c_W$ are sine and cosine of the SM weak mixing angle, 
while $s_{\phi}$ and $c_{\phi}$ are sine and cosine of the new mixing angle 
$\phi$ appearing after the TeV symmetry breaking.

After symmetry breaking both $W^\prime$ and $Z^\prime$ bosons obtain masses
and mix with the SM gauge bosons. 
The masses of the $W^\prime$ and $Z^\prime$ are given as follows: 
\begin{itemize}
\item In BP-I, we find 
\bea
M_{{W^{\prime}}^{\pm}}^{2} & = & \frac{e^{2}v^{2}}{4c_{W}^{2}s_{\phi}^{2}}\left(x+1\right)\,,
\label{mwp_bp1}\\
M_{Z^{\prime}}^{2} & = & \frac{e^{2}v^{2}}{4c_{W}^{2}s_{\phi}^{2}}\left(x+c_{\phi}^{4}\right)\,,
\label{mzp_bp1}
\eea
\item 
In BP-II, we notice that the masses of the $W^\prime$ and $Z^\prime$ bosons are degenerated at the tree level,
and 
\be
M_{{W^{\prime}}^{\pm}}^{2}=M_{Z^{\prime}}^{2}=\frac{e^{2}v^{2}}
{4s_{W}^{2}s_{\phi}^{2}c_{\phi}^{2}}\left(x+s_{\phi}^{4}\right)\,. 
\label{mvp_bp2}
\ee
\end{itemize}

Now consider the gauge interaction of $W^\prime$ and $Z^\prime$ to the SM fermions. 
Note that throughout this work only SM fermions are considered, despited that 
in certain models new heavy fermions are necessary to cancel gauge anomalies. 
Study of $W^\prime$ and $Z^\prime$ bosons in an ultra-violate (UV) completion theory 
is certainly interesting but beyond the scope of this paper. 
Charge assignments of SM fermions in those models of our interest 
are listed in Table~\ref{tab:quantumnumber}. 

\begin{table}
\begin{tabular}{c|c|c|c}
\hline 
Models & $SU(2)_{1}$ ($T_{L},$ $T_{l}$) & $SU(2)_{2}$ ($T_{R}$, $T_{h}$) & $U(1)_{X}$ ($X$, $Y$)\tabularnewline
\hline
\hline 
LRD/LRT
& $\left(\begin{array}{c}
u_{L}\\
d_{L}\end{array}\right),\left(\begin{array}{c}
\nu_{L}\\
e_{L}\end{array}\right)$ & $\left(\begin{array}{c}
u_{R}\\
d_{R}\end{array}\right),\left(\begin{array}{c}
\nu_{R}\\
e_{R}\end{array}\right)$ & $\begin{array}{c}
X_{q}=1/6\\
X_{l}=-1/2\end{array}$$ $\tabularnewline
\hline 
LPD/LPT & $\left(\begin{array}{c}
u_{L}\\
d_{L}\end{array}\right),\left(\begin{array}{c}
\nu_{L}\\
e_{L}\end{array}\right)$ & $\left(\begin{array}{c}
u_{R}\\
d_{R}\end{array}\right)$ & $\begin{array}{c}
X_{q}=1/6\\
X_{l}=Y_{\rm SM}\end{array}$\tabularnewline
\hline 
HPD/HPT & $\left(\begin{array}{c}
u_{L}\\
d_{L}\end{array}\right),\left(\begin{array}{c}
\nu_{L}\\
e_{L}\end{array}\right)$ & $\left(\begin{array}{c}
\nu_{R}\\
e_{R}\end{array}\right)$ & $\begin{array}{c}
X_{q}=Y_{\rm SM}\\
X_{l}=-1/2\end{array}$\tabularnewline
\hline 
FPD/FPT & $\left(\begin{array}{c}
u_{L}\\
d_{L}\end{array}\right),\left(\begin{array}{c}
\nu_{L}\\
e_{L}\end{array}\right)$ &  & $X_{f}=Y_{\rm SM}$\tabularnewline
\hline 
SQD & $\left(\begin{array}{c}
u_{L}\\
d_{L}\end{array}\right),\left(\begin{array}{c}
\nu_{L}\\
e_{L}\end{array}\right)$ &  & $X_{f}=Y_{\rm SM}$\tabularnewline
\hline
TFD & $\left(\begin{array}{c}
u_{L}\\
d_{L}\end{array}\right)_{1st,2nd},\left(\begin{array}{c}
\nu_{L}\\
e_{L}\end{array}\right)_{1st,2nd}$ & $\left(\begin{array}{c}
u_{L}\\
d_{L}\end{array}\right)_{3rd},\left(\begin{array}{c}
\nu_{L}\\
e_{L}\end{array}\right)_{3rd}$ & $X_{f}=Y_{\rm SM}$\tabularnewline
\hline 
UUD & $\left(\begin{array}{c}
u_{L}\\
d_{L}\end{array}\right)$ &  $\left(\begin{array}{c}
\nu_{L}\\
e_{L}\end{array}\right)$ & $X_{f}=Y_{\rm SM}$\tabularnewline
\hline
\end{tabular}
\caption{Assignment of SM fermions under the $G(221)$ symmetry: 
$(T_{L},T_{R})_{X}$ in breaking pattern I while 
$(T_{l},T_{h})_{Y}$ in breaking pattern II. Unless otherwise specified, 
the fermion doublet represents three generations of SM fermions.
LRD (LRT) denotes the left-right doublet (triplet) model, where the $G(221)$ model
is broken by a scalar doublet (triplet). Similarly, LPD (LPT) denotes the lepto-phobic doublet
(triplet) model, HPD (HPT) the hadro-phobic doublet (triplet) model, FPD (FPT) the fermio-phobic
doublet (triplet) model, SQD the sequential $W^\prime$ model with doublet Higgs, TFD the non-universal doublet model, while UUD the un-unified doublet
model. }
\label{tab:quantumnumber}
\end{table}

The most general interaction of the $Z^\prime$ and $W^\prime$ to SM fermions is 
\be
\mathcal{L}_{f}=g_2 Z^\prime_{\mu}\,\bar{f}\,\gamma^{\mu}(g_L P_L+g_R P_R)f
+g_2 W^\prime_{\mu}\,\bar{f}\,\gamma^{\mu}(g_L^\prime P_L + g_R^\prime P_R) f^\prime +h.c.\,,
\ee
where $g_2=e/\sin\theta$ is the weak coupling strength and 
$P_{L,R}=(1\mp\gamma_{5})/2$ are the usual chirality projectors.
For simplicity, we use $g_L$ and $g_R$ for both $Z^\prime$ and $W^\prime$ bosons
from now on. Detailed expressions of $g_L$ and $g_R$ for each individual NP model
are listed in Table~\ref{tab:coupling}. 
According to Table~\ref{tab:quantumnumber} and Table~\ref{tab:coupling}, the couplings of $W'$ to fermions (either leptons or quarks) are suppressed in the FP (either LP or HP) model, while the couplings of $Z'$ to  fermions (either leptons or quarks) are not.

\begin{table}[!htb]
\renewcommand{\arraystretch}{2.0}
 \begin{tabular}{c|c|c}
\hline 
Couplings  & $g_{L}$  & $g_{R}$ \tabularnewline
\hline 
$W^{\prime+\mu}\bar{f}f^{\prime}$ (BP-I) & 
$\displaystyle -\frac{e_{m}}{\sqrt{2}s_{W}^{2}}\gamma_{\rho}T_{L}^{+}\frac{c_{W}s_{2\beta}s_{\phi}}{x}$  
& $\displaystyle \frac{e_{m}}{\sqrt{2}c_{W}s_{\phi}}\gamma_{\rho}T_{R}^{+}$ \tabularnewline 
$Z^{\prime}\bar{f}f$ (BP-I) & 
$\displaystyle \frac{e_{m}}{c_{W}c_{\phi}s_{\phi}}\gamma_{\rho}\left[\left(T_{3L}-Q\right)s_{\phi}^{2}-\frac{c_{\phi}^{4}s_{\phi}^{2}\left(T_{3L}-Qs_{W}^{2}\right)}{xs_{W}^{2}}\right]$  &
$\displaystyle  \frac{e_{m}}{c_{W}c_{\phi}s_{\phi}}\gamma_{\rho}\left[\left(T_{3R}-Qs_{\phi}^{2}\right)+Q\frac{c_{\phi}^{4}s_{\phi}^{2}}{x}\right]$ \tabularnewline 
\hline 
$W^{\prime\pm\mu}\bar{f}f^{\prime}$ (BP-II) & $\displaystyle -\frac{e_{m}s_{\phi}}{\sqrt{2}s_{W}c_{\phi}}\gamma^{\mu}T_{l}^{\pm}\left(1+\frac{s_{\phi}^{2}c_{\phi}^{2}}{x}\right)$  & $0$ \tabularnewline
$W^{\prime\pm\mu}\bar{F}F^{\prime}$ (BP-II) & $\displaystyle \frac{e_{m}c_{\phi}}{\sqrt{2}s_{W}s_{\phi}}\gamma^{\mu}T_{h}^{\pm}\left(1-\frac{s_{\phi}^{4}}{x}\right)$  & $0$ \tabularnewline
$Z^{\prime}\bar{f}f$ (BP-II) & $\displaystyle -\frac{e_{m}}{s_{W}}\gamma^{\mu}\left[\frac{s_{\phi}}{c_{\phi}}T_{3l}\left(1+\frac{s_{\phi}^{2}c_{\phi}^{2}}{xc_{W}^{2}}\right)-\frac{s_{\phi}}{c_{\phi}}\frac{s_{\phi}^{2}c_{\phi}^{2}}{xc_{W}^{2}}s_{W}^{2}Q\right]$  & $\displaystyle \frac{e_{m}}{s_{W}}\gamma^{\mu}\left(\frac{s_{\phi}}{c_{\phi}}\frac{s_{\phi}^{2}c_{\phi}^{2}}{xc_{W}^{2}}s_{W}^{2}Q\right)$ \tabularnewline 
 $Z^{\prime}\bar{F}F$ (BP-II)  & $\displaystyle \frac{e_{m}}{s_{W}}\gamma^{\mu}\left[\frac{c_{\phi}}{s_{\phi}}T_{3h}\left(1-\frac{s_{\phi}^{4}}{xc_{W}^{2}}\right)+\frac{c_{\phi}}{s_{\phi}}\frac{s_{\phi}^{4}}{xc_{W}^{2}}s_{W}^{2}Q\right]$  & $\displaystyle \frac{e_{m}}{s_{W}}\gamma^{\mu}\left(\frac{c_{\phi}}{s_{\phi}}\frac{s_{\phi}^{4}}{xc_{W}^{2}}s_{W}^{2}Q\right)$ \tabularnewline
\hline
\hline 
Couplings & BP-I & BP-II\tabularnewline
\hline 
$H~W_{\nu}~W_{\rho}^{\prime}~$  & $\displaystyle -\frac{i}{2}\frac{e_{m}^{2}s_{2\beta}}{c_{W}s_{W}s_{\phi}}vg_{\nu\rho}\biggl[1+\frac{\left(c_{W}^{2}s_{\phi}^{2}-s_{W}^{2}\right)}{xs_{W}^{2}}\biggr]$ & $\displaystyle -\frac{i}{2}\frac{e_{m}^{2}s_{\phi}}{s_{W}^{2}c_{\phi}}vg_{\nu\rho}\biggl[1+\frac{s_{\phi}^{2}\left(c_{\phi}^{2}-s_{\phi}^{2}\right)}{x}\biggr]$ 
\tabularnewline
$H~Z_{\nu}~Z_{\rho}^{\prime}~$  & $\displaystyle -\frac{i}{2}\frac{e_{m}^{2}c_{\phi}}{c_{W}^{2}s_{W}s_{\phi}}vg_{\nu\rho}\biggl[1-\frac{c_{\phi}^{2}\left(c_{\phi}^{2}s_{W}^{2}-s_{\phi}^{2}\right)}{xs_{W}^{2}}\biggr]$  & $\displaystyle -\frac{i}{2}\frac{e_{m}^{2}s_{\phi}}{c_{W}s_{W}^{2}c_{\phi}}vg_{\nu\rho}\biggl[1-\frac{s_{\phi}^{2}\left(s_{\phi}^{2}c_{W}^{2}-c_{\phi}^{2}\right)}{xc_{W}^{2}}\biggr]$ 
\tabularnewline
$W_{\mu}^{+}~W_{\nu}^{\prime-}~Z_{\rho}~$  & $\displaystyle i\frac{e_{m}s_{2\beta}s_{\phi}}{xs_{W}^{2}}$  & $\displaystyle i\frac{e_{m}c_{\phi}s_{\phi}^{3}}{xs_{W}c_{W}}$ 
\tabularnewline
$W_{\mu}^{+}~W_{\nu}^{-}~Z_{\rho}^{\prime}~$  & $\displaystyle i\frac{e_{m}s_{\phi}c_{W}c_{\phi}^{3}}{xs_{W}^{2}}$  & $\displaystyle i\frac{e_{m}c_{\phi}s_{\phi}^{3}}{xs_{W}}$ \tabularnewline
\hline
\end{tabular}\caption{The fermion couplings and triple boson couplings of the heavy gauge
boson in Breaking Pattern I and II. For the fermion couplings, the
quantum numbers $(T_{L},T_{R})$ in BP-I and $(T_{l},T_{h})$ in BP-II
are implied in Table I, and is given in our previous paper~\cite{Hsieh:2010zr}. In BP-II,
the fermion notation $f$ means the fermions listed in the column
$SU(2)_{1}$, while $F$ means the fermions listed in the column $SU(2)_{2}$
in Table I. For the triple gauge boson couplings, the Lorentz index
$\left[g^{\mu\nu}(k_{1}-k_{2})^{\rho}+g^{\nu\rho}(k_{2}-k_{3})^{\mu}+g^{\rho\mu}(k_{3}-k_{1})^{\nu}\right]$
is implied.}
\label{tab:coupling}
\end{table}

Triple gauge boson couplings as well as the scalar-vector-vector couplings
are also listed as they arise from the symmetry breaking and may contribute 
to the $W^\prime$ and $Z^\prime$ decay.

%%%%%%%%%%%%%%%%%%%%%%%%%%%%%%%%%%%%%%%%%%                                                                                %																										%
%        Section III: W' and Z' production and decay                        					     	%
%																										%
%%%%%%%%%%%%%%%%%%%%%%%%%%%%%%%%%%%%%%%%%%

\section{$W^\prime$ and $Z^\prime$ production and decay}

\subsection{$V^\prime$ production at the LHC}

At the LHC, the cross section of $pp\to V^\prime \to \bar{f}f^\prime$ 
($V^\prime=W^\prime/Z^\prime$) is
\begin{equation}
\sigma_{pp\to V'\to\bar{f}f'}=\sum_{\{ij\}}\int_{\tau_{0}}^{1}
\frac{d\,\tau}{\tau}\cdot\frac{1}{s}\frac{d\,\mathcal{L}_{ij}}{d\,\tau}\cdot[\hat{s}\,\hat{\sigma}_{ij\to V'\to\bar{f}f'}(\hat{s})]\,,
\end{equation}
where $\sqrt{s}$ is the total energy of the incoming proton-proton beam,
$\sqrt{\hat{s}}$ is the partonic center-of-mass (c.m.) energy and $\tau\equiv \hat{s}/s$.
The lower limit of $\tau$ variable is determined by the kinematics threshold of 
the $V^\prime$ production, i.e. $\tau_{0}=M_{V'}^2/s$. 
The parton luminosity $\frac{1}{s}\frac{d\,\mathcal{L}_{ij}}{d\,\tau}$
is defined as
\begin{equation}
\frac{1}{s}\frac{d\,\mathcal{L}_{ij}}{d\,\tau}=\frac{1}{1+\delta_{ij}}
\frac{1}{s}\int_{\tau}^{1}\frac{d\, x}{x}
[f_{i}^{(a)}(x)f_{j}^{(b)}(\tau/x)+f_{j}^{(a)}(x)f_{i}^{(b)}(\tau/x)]\,,
\end{equation}
where $i$ and $j$ denote the initial state partons and $f_i^{(a)}(x)$ is the 
parton distribution of the parton $i$ inside the hadron $a$ with a momentum
fraction of $x=p_i/p_a$. 
Using the narrow width approximation (NWA) one can factorize 
the $pp\to V^\prime \to \bar{f} f^\prime$ process into 
the $V^\prime$ production and the $V^\prime$ decay,
\begin{equation}
\sigma_{pp\to V'\to\bar{f}f'}=\left(\sum_{\{ij\}}\int_{\tau_{0}}^{1}
\frac{d\,\tau}{\tau}\cdot\frac{1}{s}\frac{d\,\mathcal{L}_{ij}}{d\,\tau}\cdot[\hat{s}\,
\hat{\sigma}_{ij\to V'}(\hat{s})]\right)\,\times {\rm Br}(V'\to\bar{f}f'),
\label{eq:nlo_vprime}
\end{equation}
where the branching ratio (Br) is defined as 
$
{\rm Br}(V^\prime\to \bar{f}f^\prime) = \Gamma(V^\prime \to \bar{f}f^\prime)
/\Gamma_{\rm tot}.
$
As to be shown later, the decay widths of $Z^\prime$ 
and $W^\prime$ bosons in most of the allowed parameter space  
are much smaller than their masses, which validates the NWA adapted in this work.

At the next-to-leading-order (NLO) the partonic cross section of the 
$V^\prime$ production is 
\be
\hat{\sigma}_{ij\to V'}(\hat{s})=
\frac{\pi}{6\hat{s}}g_2^2 (g_L^2+g_R^2)\, H_{ij}
\left(\frac{M_{V'}^{2}}{\hat{s}}\right),
\ee
where the functions $H_{ij}(z)$ for different parton flavors 
$ij=(\bar{q}q',qg,\bar{q}g)$ are 
\bea
&&H_{\bar{q}q'}(z)=\delta(1-z) \nonumber \\
&+&\frac{\alpha_{s}}{2\pi}C_{F}
\left[\left(\frac{2\pi^{2}}{3}-8\right)\delta(1-z)-\frac{2(1+z^{2})}{1-z}\log(z)
+4(1+z^{2})\left(\frac{\log(1-z)}{1-z}\right)_{+}\right],
\eea
and
\be
H_{qg}(z)=H_{\bar{q}g}(z)=\frac{\alpha_{s}}{2\pi}T_{F}
\left[\left(z^{2}+(1-z)^{2}\right)\log\frac{(1+z)^{2}}{z}+\frac{1}{2}+3z-\frac{7}{2}z^{2}\right].
\ee
Here, $C_F$ and $T_F$ are the color factor defined as $C_F=4/3$ and $T_F=1/2$.

It is convenient to parametrize 
the $V^\prime$ production cross section into one model-dependent 
piece $C_{q}^{V'}$ and another model-independent piece 
$F_{q}^{V'}(M_{V'},\sqrt{s})$. 
The first piece consists of model couplings, while the second piece,
which includes all the hadronic contributions~\cite{Carena:2004xs},
depends only on $m_{V^\prime}$ and $\sqrt{s}$.  
We separate the up-quark and down-quark contributions in the 
$Z^\prime$ production because $Z^\prime$ couples differently to 
up- and down-quarks in most NP models.  
The NLO cross sections of $Z^\prime$ and $W^\prime$ production 
can then be expressed as
\bea
\sigma_{pp\to Z^\prime\to ff}&=&\frac{\pi}{18\, s}[C_{u}^{Z^\prime}\, 
F_{u}^{Z^\prime}(M_{V'},\sqrt{s})+C_{d}^{Z^\prime}\, F_{d}^{Z^\prime}(M_{V'},\sqrt{s})], \nonumber \\
\sigma_{pp\to W^\prime\to ff'}&=&\frac{\pi}{18\, s}[C_{q}^{W^\prime}\, F_{q}^{W^\prime}(M_{V'},\sqrt{s})]\,,\label{eq:factorization}
\eea
where
 \bea
C_{q}^{V'}&=&g_2^2\left(g_L^2+g_R^2\right)\times {\rm Br}(V'\to ff'),
\label{eq:cq} \\
F_{q}^{V'}(M_{V'},\sqrt{s})&=&\int_{\tau_{0}}^{1}\frac{d\,\tau}{\tau}\cdot\left[\frac{d\,\mathcal{L}_{\bar{q}q'}}{d\,\tau}\cdot H_{\bar{q}q'}(z)+\frac{d\,\mathcal{L}_{\bar{q}g}}{d\,\tau}\cdot H_{\bar{q}g}(z)+(\bar{q}\to q)\right]\,.
\label{eq:fq}
\eea
Note that the decay branching ratio is allocated to the model-dependent piece 
$C_q^{V^\prime}$. After convoluting with PDFs, the model-independent piece
$F_q^{V^\prime}$ is merely a function of $m_{V^\prime}$ and the collider energy $\sqrt{s}$.

\begin{figure}
\includegraphics[width=0.4\textwidth]{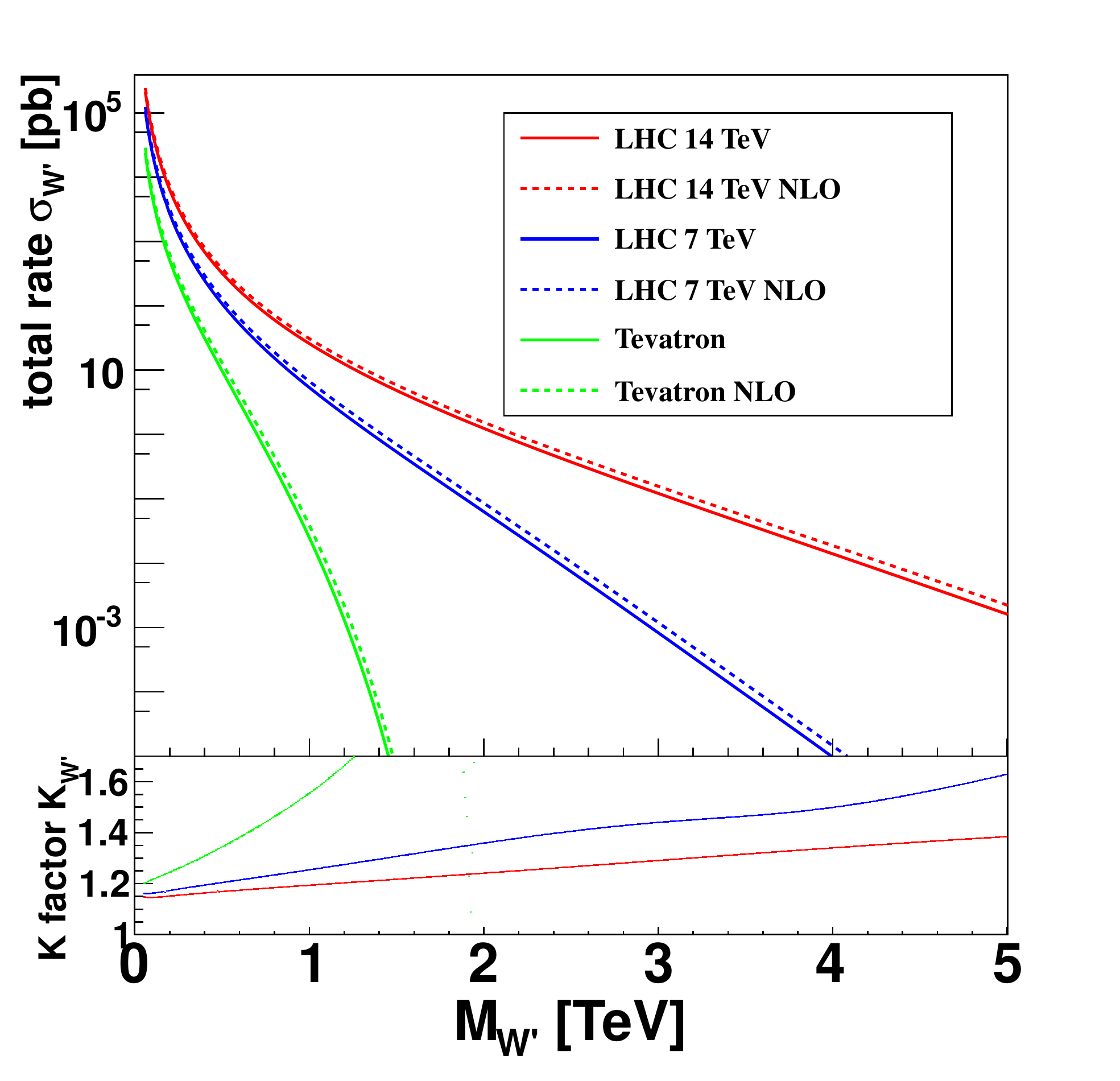}
\includegraphics[width=0.4\textwidth]{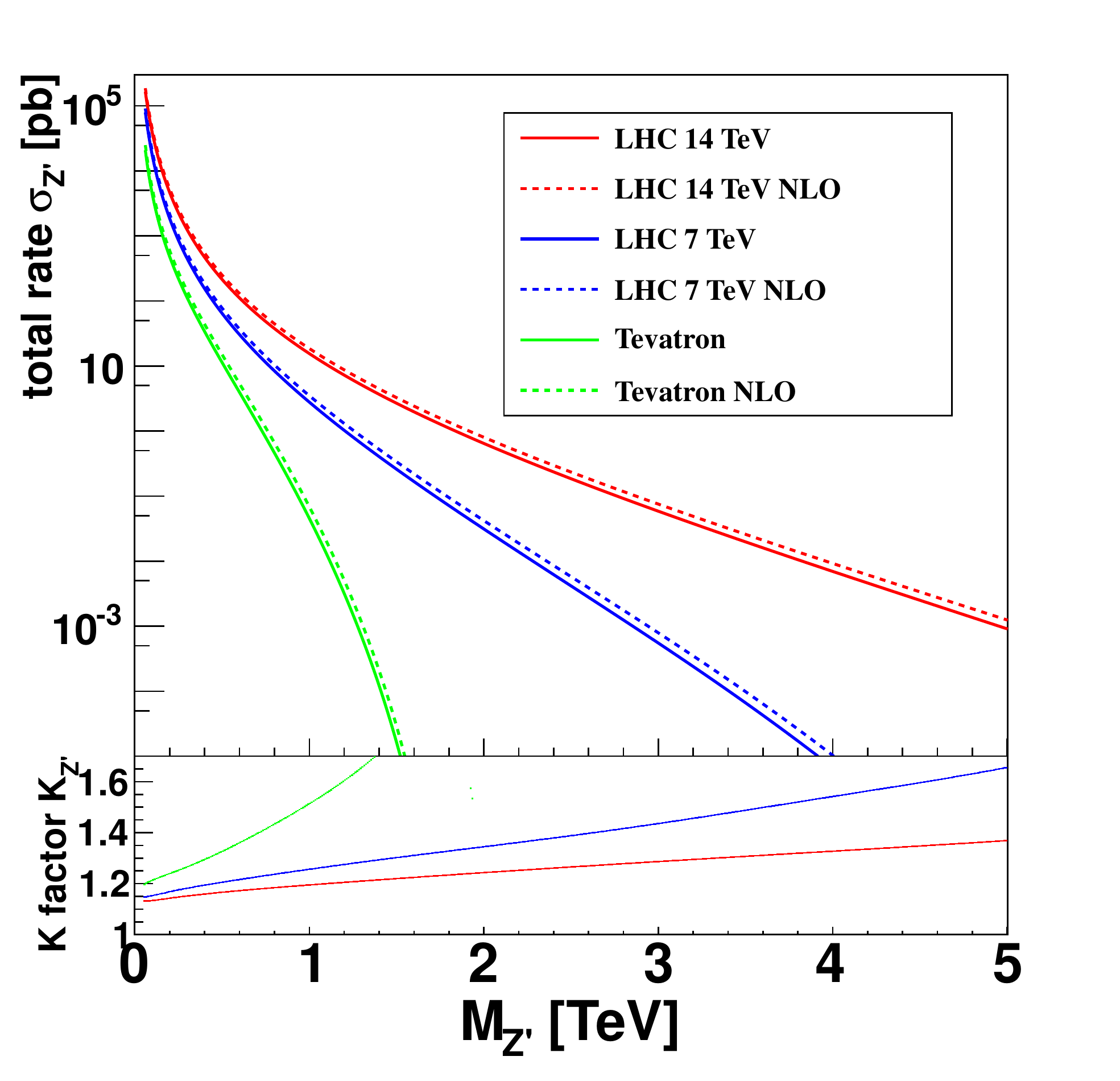}
\caption{Upper panel: the LO and NLO cross sections of $pp\to W^\prime$ (left) and $pp\to Z^\prime$ (right) 
process with a SM like coupling as a function of new heavy gauge boson mass ($m_{V^\prime}$, $V=W,Z$) 
in hadron collisions. Lower panel: the $K$-factor as a function of $m_{V^\prime}$. }
\label{wprimexsec}
\end{figure}

\begin{figure}
\includegraphics[width=0.4\textwidth]{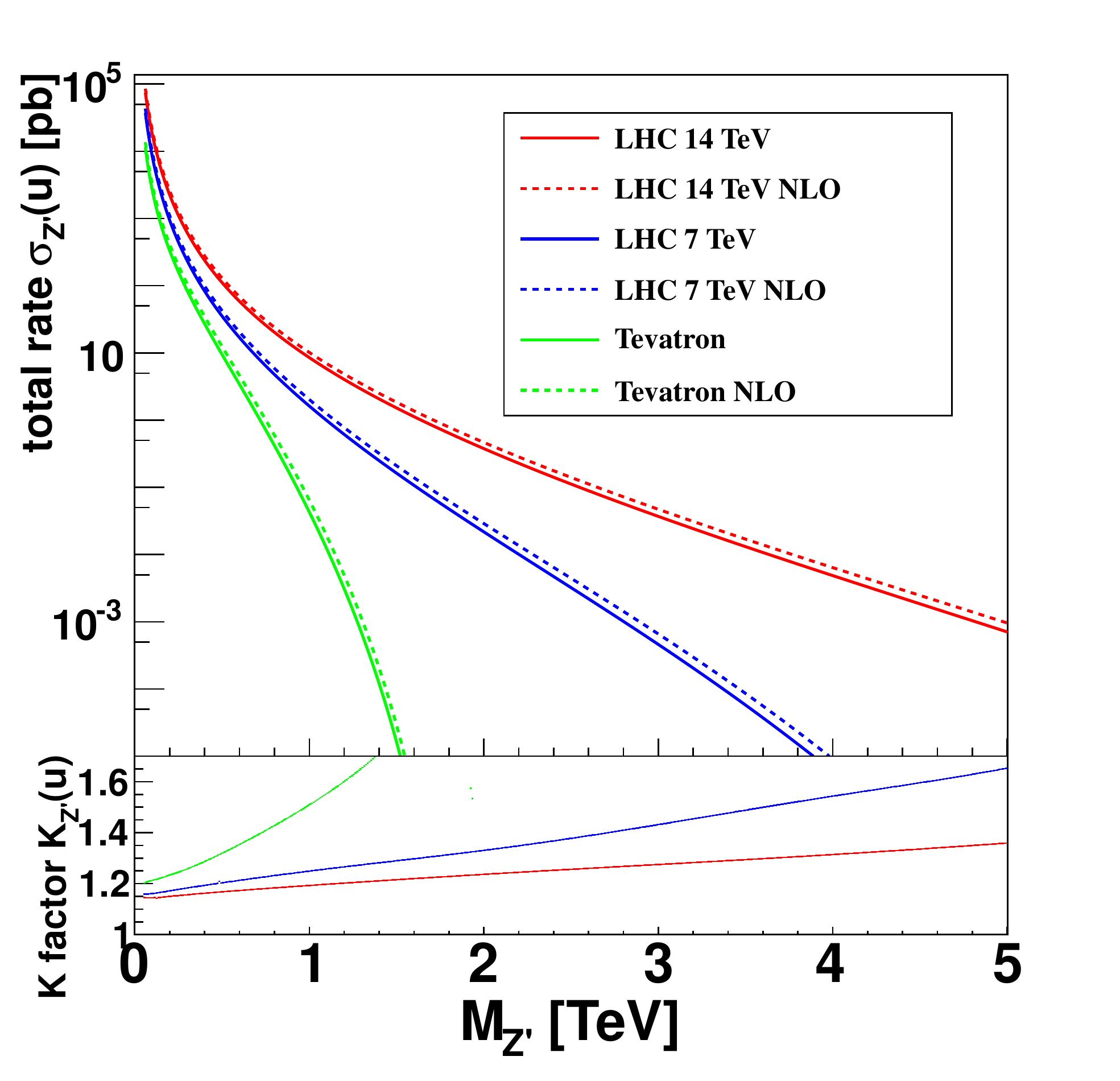}
\includegraphics[width=0.4\textwidth]{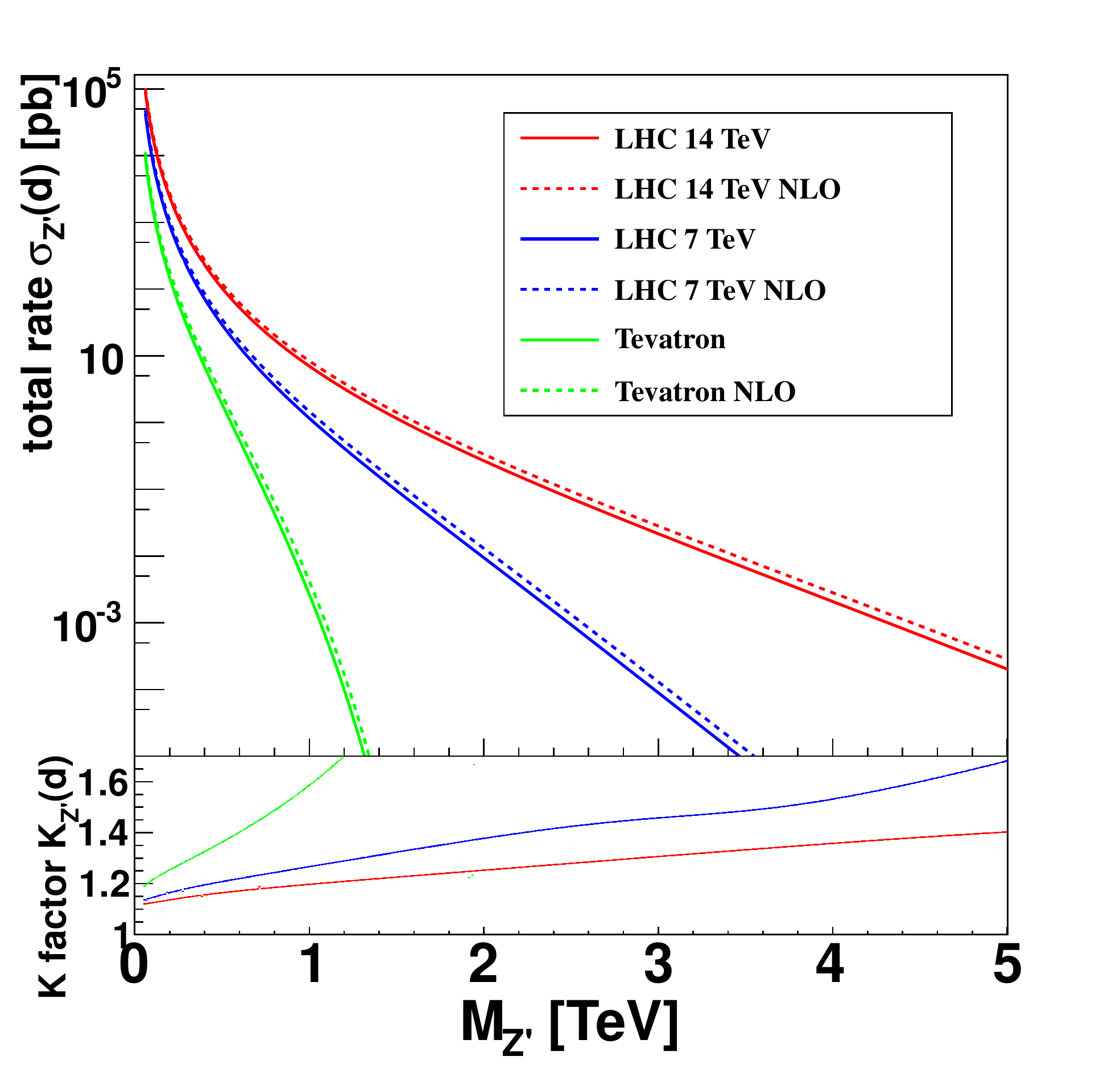}
\caption{Upper panel: the LO and NLO cross sections of $pp\to Z^\prime$ process
with a sequential couplings as a function of $m_{Z^\prime}$ in hadron collision:
(left) induced by up-type quark initial state, (right) induced by down-type quark initial state.
Lower panel: the $K$-factor as a function of $m_{Z^\prime}$. }
\label{zprimexsec}
\end{figure}

Because the model-dependent couplings can be factorized out,
the total cross section in the sequential $W^\prime$ and $Z^\prime$ models can
be used as the reference cross section. The upper panels of Fig.~\ref{wprimexsec} show 
the leading order (LO) and next-to-leading-order (NLO) 
production cross sections of the sequential $W^\prime$ (left) and $Z^\prime$ boson (right)
as a function of the extra gauge boson mass at the Tevatron, the 7~TeV and 14~TeV LHC.
The lower panels display the K-factor, defined as the ratio of the NLO 
to  LO cross sections.  In the upper panels of Fig.~\ref{zprimexsec} we plot the cross section of $Z^\prime$ production induced by $u\bar{u}$ (left) and $d\bar{d}$ (right) initial state, respectively. 
Again, the lower panels show the corresponding $K$-factors. 
Note that the $K$-factors are 
model-independent once one separate the up-quark and down-quark contributions in 
the $Z^\prime$ production. The $K$-factor is defined as
\be
K_q=\frac{\sigma_{NLO}}{\sigma_{LO}}=
\frac{F_q^{V^\prime}(M_{V^\prime},\sqrt{s})_{NLO}}{F_q^{V^\prime_{\rm seq}}(M_{V^\prime},\sqrt{s})_{LO}}.
\ee 
Here we adopt the CTEQ6.6M parton distribution package~\cite{Nadolsky:2008zw}
for both the LO and NLO calculations. Both the factorization and renormalization scales are set to 
be $M_{V^\prime}$.

The NLO cross section of other NP models can be obtained easily from the sequential $W^\prime$ 
and $Z^\prime$ cross sections plotted in Figs.~\ref{wprimexsec} and~\ref{zprimexsec} by:
\begin{itemize}
\item scaling the model-dependent $C^{V^\prime}$-coefficients
($C_{u}^{Z^\prime}/C_{u}^{Z^\prime_{\rm seq}}$, $C_{d}^{Z^\prime}/C_{d}^{Z^\prime_{\rm seq}}$,
$C_{q}^{W^\prime}/C_{q}^{W^\prime_{\rm seq}}$),
\item including the NLO QCD correction with the inclusive $K$-factors ($K_u$, $K_d$ and $K_q$).
\end{itemize}
To be more specific, the NLO cross sections of new gauge boson productions in the $G(221)$ model 
are 
\bea
\sigma_{W^\prime}&=&\frac{C_q^{W^\prime}}{C_q^{W^\prime_{\rm seq}}} \left(F_q^{W^\prime}\right)_{\rm LO}\times K_q, \nonumber \\
\sigma_{Z^\prime}&=&
\frac{C_u^{Z^\prime}}{C_u^{Z^\prime_{\rm seq}}} \left(F_u^{Z^\prime}\right)_{\rm LO}\times K_u
+
\frac{C_d^{Z^\prime}}{C_d^{Z^\prime_{\rm seq}}} \left(F_d^{Z^\prime}\right)_{\rm LO}\times K_d.
\eea

\subsection{$V^\prime$ decay}

In the $G(221)$ model the $W^\prime$ and $Z^\prime$ bosons can decay 
into SM fermions, gauge bosons, or a pair of SM gauge boson and Higgs boson.
In this subsection we give detailed formula of partial decay widths of 
the extra gauge bosons. 

First, consider the fermionic mode. The decay width of $V' \to \bar{f}_{1}f_{2}$ is
\be
\Gamma_{V'\to\bar{f_{1}}f_{2}}=\frac{M_{V'}}{24\pi}\beta_{0}\left[(g_{L}^{2}+g_{R}^{2})\beta_{1}+6g_{L}g_{R}\frac{m_{f_{1}}m_{f_{2}}}{M_{V'}^2}\right]\Theta(M_{V'}-m_{f_{1}}-m_{f_{2}})\,,
\label{eq:v_width}
\ee
where 
\bea
\beta_{0}&=&\sqrt{1-2\frac{m_{f_{1}}^{2}+m_{f_{2}}^{2}}{M_{V'}^{2}}+\frac{(m_{f_{1}}^{2}-m_{f_{2}}^{2})^{2}}{M_{V'}^{4}}}, \nonumber \\
\beta_{1}&=&1-\frac{m_{f_{1}}^{2}+m_{f_{2}}^{2}}{2M_{V'}^{2}}-\frac{(m_{f_{1}}^{2}-m_{f_{2}}^{2})^{2}}{2 M_{V'}^{4}}.
\eea
Note that the color factor is not included in Eq.~(\ref{eq:v_width}) and
the third generation quark decay channel opens only for a heavy $Z^\prime$ and $W^\prime$. 

Second, consider the bosonic decay mode, e.g. $W^\prime$ and $Z^\prime$ decay to gauge 
bosons and Higgs bosons. Such decay modes are induced by gauge interactions
between the extra gauge boson and the SM gauge boson after symmetry breaking. 
Even though the couplings $g_{V'V_{1}V_{2}}$ and $g_{V'V_{1}H}$ 
are suppressed by the gauge boson mixing term $1/x$, 
the bosonic decay channel could be the major decay channel in certain models, e.g.
fermio-phobic model in which the extra gauge boson does not couple to fermions at all.

The decay width of $V'\to V_{1}V_{2}$ is
\be
\Gamma_{V'\to V_{1}V_{2}}=\frac{M_{V'}^{5}}{192\pi M_{V_{1}}^{2}M_{V_{2}}^{2}}g_{V'V_{1}V_{2}}^{2}\beta_{0}^{3}\beta_{1}\Theta(M_{V'}-M_{V_{1}}-M_{V_{2}})\,,
\ee
where 
\bea
\beta_{0}&=&\sqrt{1-2\frac{M_{V_{1}}^{2}+M_{V_{2}}^{2}}{M_{V'}^{2}}+\frac{(M_{V_{1}}^{2}-M_{V_{2}}^{2})^{2}}{M_{V'}^{4}}},\nonumber \\
\beta_{1}&=&1+10\frac{M_{V1}^{2}+M_{V2}^{2}}{2M_{V'}^{2}}+\frac{M_{V_{1}}^{4}+10M_{V2}^{2}M_{V_{2}}^{2}+M_{V_{1}}^{4}}{M_{V'}^{4}}.
\eea
The width of $V'\to V_{1}H$ (where $V_1=W$ or $Z$ boson and $H$ is the lightest Higgs boson) is
\be
\Gamma_{V'\to V_{1}H}=\frac{M_{V'}}{192\pi}\frac{g_{V'V_{1}H}^{2}}{M_{V_{1}}^{2}}\beta_{0}\beta_{1}\Theta(M_{V'}-M_{V_{1}}-M_{V_{2}})\,,
\ee
where 
\bea
\beta_{0}&=&\sqrt{1-2\frac{M_{V_{1}}^{2}+m_{H}^{2}}{M_{V'}^{2}}+\frac{(M_{V_{1}}^{2}-m_{H}^{2})^{2}}{M_{V'}^{4}}}, \nonumber \\
\beta_{1}&=&1+\frac{10M_{V_{1}}^{2}-2m_{H}^{2}}{2M_{V'}^{2}}+\frac{(M_{V_{1}}^{2}-m_{H}^{2})^{2}}{M_{V'}^{4}}.
\eea

The couplings $g_{V'V_{1}V_{2}}$ and $g_{V'V_{1}H}$ for various models are listed in Table II
for reference. 
In this study only left-handed neutrinos are considered while the possible 
right-handed neutrinos are assumed to be very heavy. In addition we also
assume all the heavy Higgs bosons, except the SM-like Higgs boson, decouple
from the TeV scale. As a result, the total decay width of the $W^\prime$ boson is
\be
\Gamma_{W^\prime,\rm tot}=3\Gamma_{W^\prime\to\bar{e}\nu}+2N_{C}\Gamma_{W^\prime\to\bar{u}d}+N_{C}\Gamma_{W^\prime\to\bar{t}b}+\Gamma_{W^\prime\to WZ}++\Gamma_{W^\prime\to WH}\,,
\ee
while the width of the $Z^\prime$ boson is
\be
\Gamma_{Z^\prime,\rm tot}=3\Gamma_{Z^\prime\to\bar{e}e}+3\Gamma_{Z^\prime\to\bar{\nu}\nu}+2N_{C}
\Gamma_{Z^\prime\to\bar{u}u}+3N_{C}\Gamma_{Z^\prime\to\bar{d}d}+N_{C}\Gamma_{Z^\prime\to\bar{t}t}+\Gamma_{Z^\prime\to WW}+\Gamma_{Z^\prime\to ZH}\,,
\ee
where $N_C=3$ originates from summation of all possible color quantum number.

%%%%%%%%%%%%%%%%%%%%%%%%%%%%%%
%                                                                                 		  %
%        Section IV: EWPT and direct search bounds            %
%													                            %
%%%%%%%%%%%%%%%%%%%%%%%%%%%%%%

\section{Indirect and direct constraints}

Even though the $W^\prime$ and $Z^\prime$ bosons are not observed yet, 
they could contribute to a few observables, which can be measured
precisely at the low energy, via quantum effects. In this section 
we perform a global-fit analysis of 37 EWPTs
to derive the allowed model parameter space of 
those NP models of our interest. In addition, we also include 
direct search limits from the Tevatron and the LHC.  

Note that $m_{W^\prime}$ and $m_{Z^\prime}$ are not independent in the $G(221)$ model;
see Eqs.~\ref{mwp_bp1}-\ref{mvp_bp2}.
In this study we choose $M_{W^\prime}$ as an input parameter. In addition, other independent 
parameters are the gauge mixing angle $\phi$, 
and the mixing angle $\beta$ in the EWSB scale between two Higgs VEVs with $s_{2\beta}=\sin(2\beta)$
which only exists in BP-I.
Our parameter scan is not sensitive to the parameter $\beta$ 
as it contributes to physical observables only at the order of $1/x=v/u$.  
We then present our scan results in the plane of $(M_{W^\prime},c_{\phi})$
or $(M_{W^\prime}, M_{Z^\prime})$. 

\subsection{Indirect Search: Electroweak Precision Tests}

Constraints from the EWPTs~\cite{Amsler:2008zzb,Erler:1999ug} on the $G(221)$
model have been presented in our previous study~\cite{Hsieh:2010zr}.
Owing to the tree-level mixing between extra gauge bosons and SM gauge bosons
in the $G(221)$ models, the conventional oblique parameters ($S$, $T$, $U$) 
cannot describe all the EWPT data. 
Therefore, a global fitting is in order. Our global analysis includes 
a set of 37 experiment observables, which is listed as follows:
\begin{itemize}
\item $Z$ pole data (21): $Z$-boson total width $\Gamma_{Z}$, 
cross section $\sigma_{\text{had.}}$, ratios $R\left(f\right)$, 
LR, FB, and charge asymmetries $A_{LR}\left(f\right)$, $A_{FB}\left(f\right)$, and $Q_{FB}$;
\item $W^{\pm}$ and top data (3): $W$-boson mass $M_{W}$  and total width $\Gamma_{W}$, and the top quark  pole mass $m_t$ ;
\item $\nu N$-scattering (5): neutral current (NC) couplings $\gNLt$ and $\gNRt$, 
ratio of neutral current to charged current (NC-CC) $R_{\nu}$ and  $R_{\bar{\nu}}$;
\item $\nu e^{-}$-scattering (2): NC couplings $g_V^{\nu e}$ and $g_A^{\nu e}$;
\item Parity violation (PV) interactions (5): weak charge $Q_{W}\left({}^{133}\text{Cs}\right)$, $Q_{W}\left({}^{205}\text{Tl}\right)$, $Q_{W}\left(e\right)$,
neutral current (NC) couplings ${\cal{C}}_{1},{\cal{C}}_{2}$;
\item $\tau$ lifetime (1).
\end{itemize}
The number inside each parenthesis denotes the number of the low energy precision
observables. In this work we only present the contour of 95\% confidence level
in the plane of $(x,c_{\phi})$ and refer readers to our previous paper for 
all the details.

\subsection{Direct Search at the Tevatron and LHC}

Another important bound on the $G(221)$ models originates from direct
searches at the Tevatron and the LHC. 
Searches for the $W^\prime$ and $Z^\prime$ bosons as a $s$-channel resonance 
have been carried out at the Tevatron and LHC in leptonic decay modes, 
quark decay channels and diboson decays. 
For the constraints from Tevatron, we use the latest Tevatron data:
\begin{itemize}
\item D\O: $p\bar p\to Z^{\prime}\to e^+e^-$ ($\int {\cal L}dt$=5.4 fb$^{-1}$)~\cite{Abazov:2010ti};
\item CDF: $p\bar p\to W^{\prime\pm}\to e \nu$ ($\int {\cal L}dt$=5.3 fb$^{-1}$)~\cite{Aaltonen:2010jj};
\item CDF: $p\bar p\to W^{\prime\pm}\to t\bar b$ ($\int {\cal L}dt$=1.9 fb$^{-1}$)~\cite{Aaltonen:2009qu};
\item CDF: $p\bar p\to Z^{\prime}\to t\bar t$ ($\int {\cal L}dt$=955 pb$^{-1}$)~\cite{:2007dia}.
\end{itemize}
and LHC7 data:
\begin{itemize}
\item ATLAS: $pp\to W^{\prime\pm}\to \ell \nu$ ($\int {\cal L}dt$=1.04 fb$^{-1}$)~\cite{Aad:2011yg};
\item ATLAS: $pp\to Z^{\prime}\to l^+l^-$ ($\int {\cal L}dt$=1.1 fb$^{-1}$)~\cite{Collaboration:2011dca};
\item CMS: $pp\to Z^{\prime}\to t \bar{t}$ in the electron $+$ jets channel 
($\int {\cal L}dt$=4.33  fb$^{-1}$)~\cite{CMS-PAS-EXO-11-092}.
\end{itemize}

\subsection{Parameter constraints}

Using the result of all the indirect and direct searches mentioned above, we scan over the parameter space 
of several typical $G(221)$ models to locate allowed parameter contours at the $95\%$ confidence level (CL).
The NLO QCD correction to new heavy gauge boson production is included using the approach 
described in Sec.~III.  For each individual NP model the total width is calculated with all 
the possible decay channels included, as discussed in Sec.~III.   

\begin{figure}
	\includegraphics[width=0.32\textwidth]{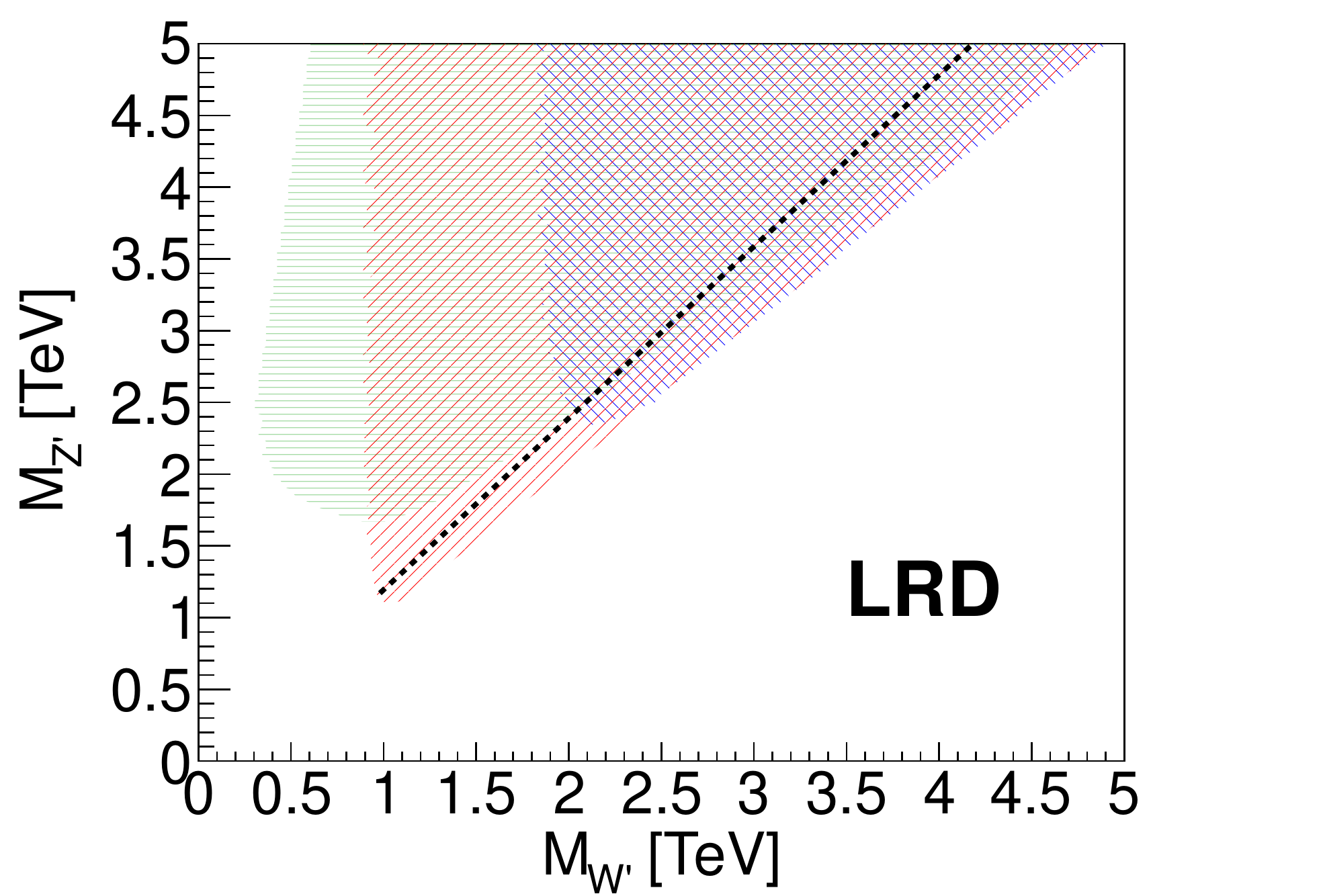}
	\includegraphics[width=0.32\textwidth]{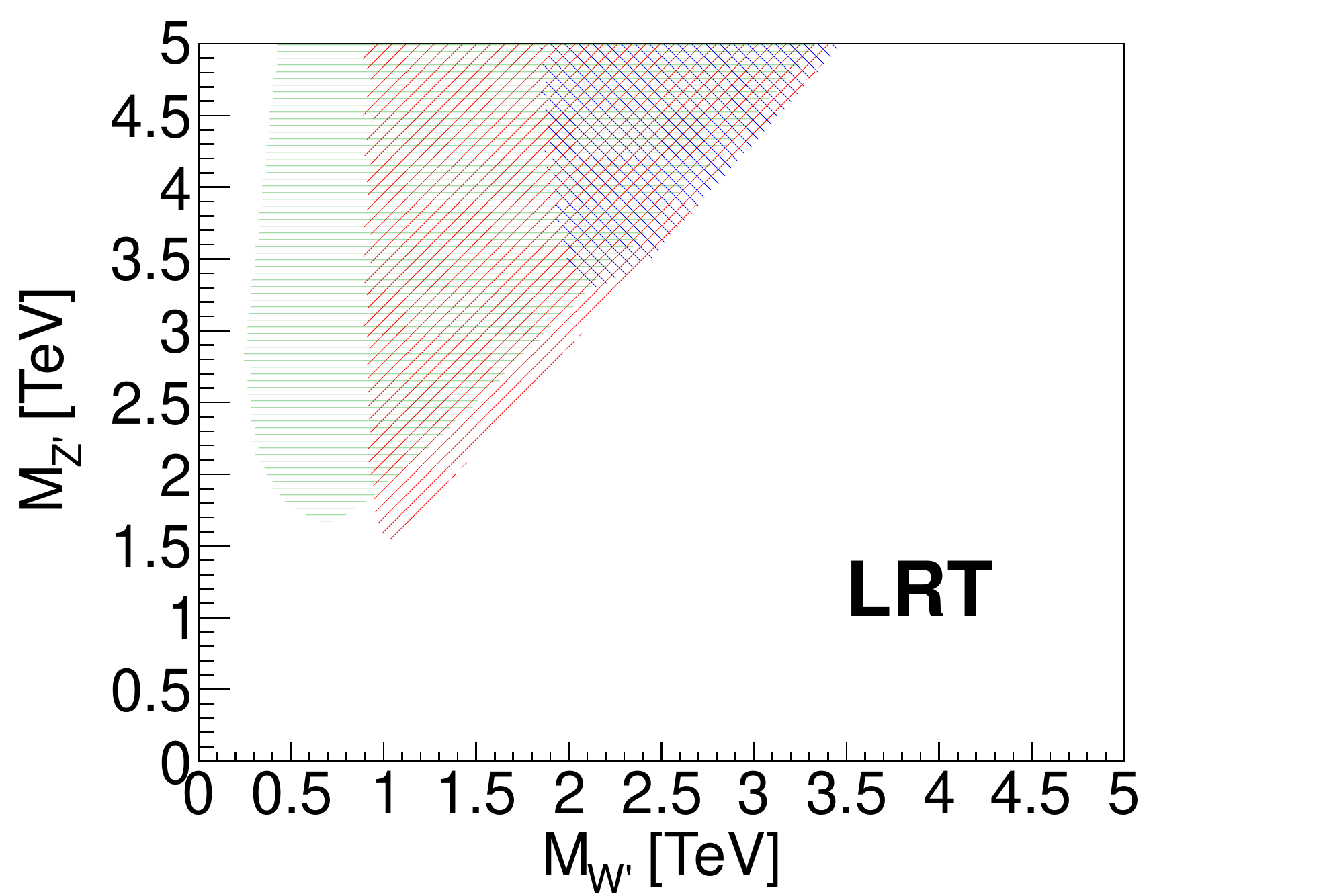}
	\includegraphics[width=0.32\textwidth]{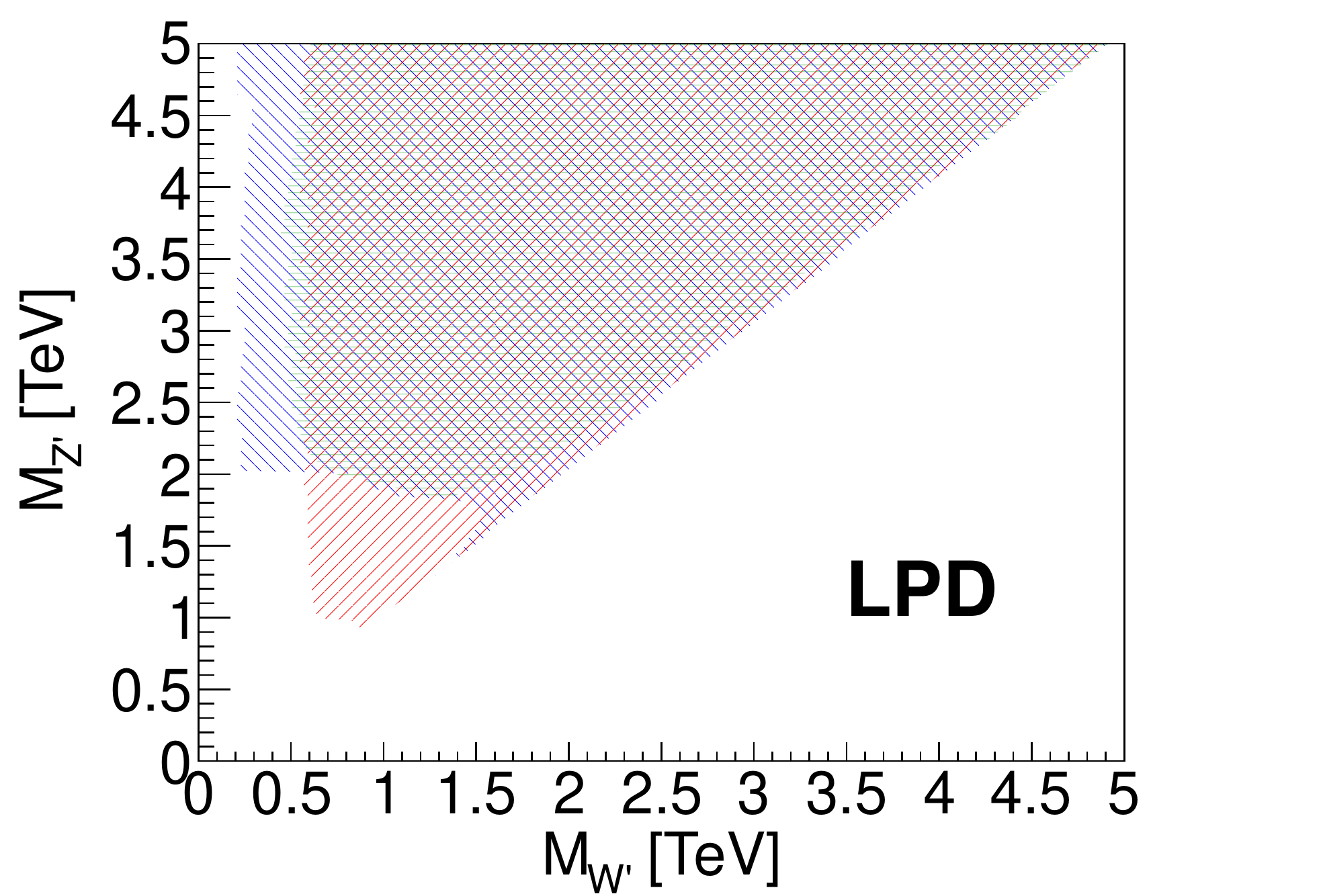}
	\includegraphics[width=0.32\textwidth]{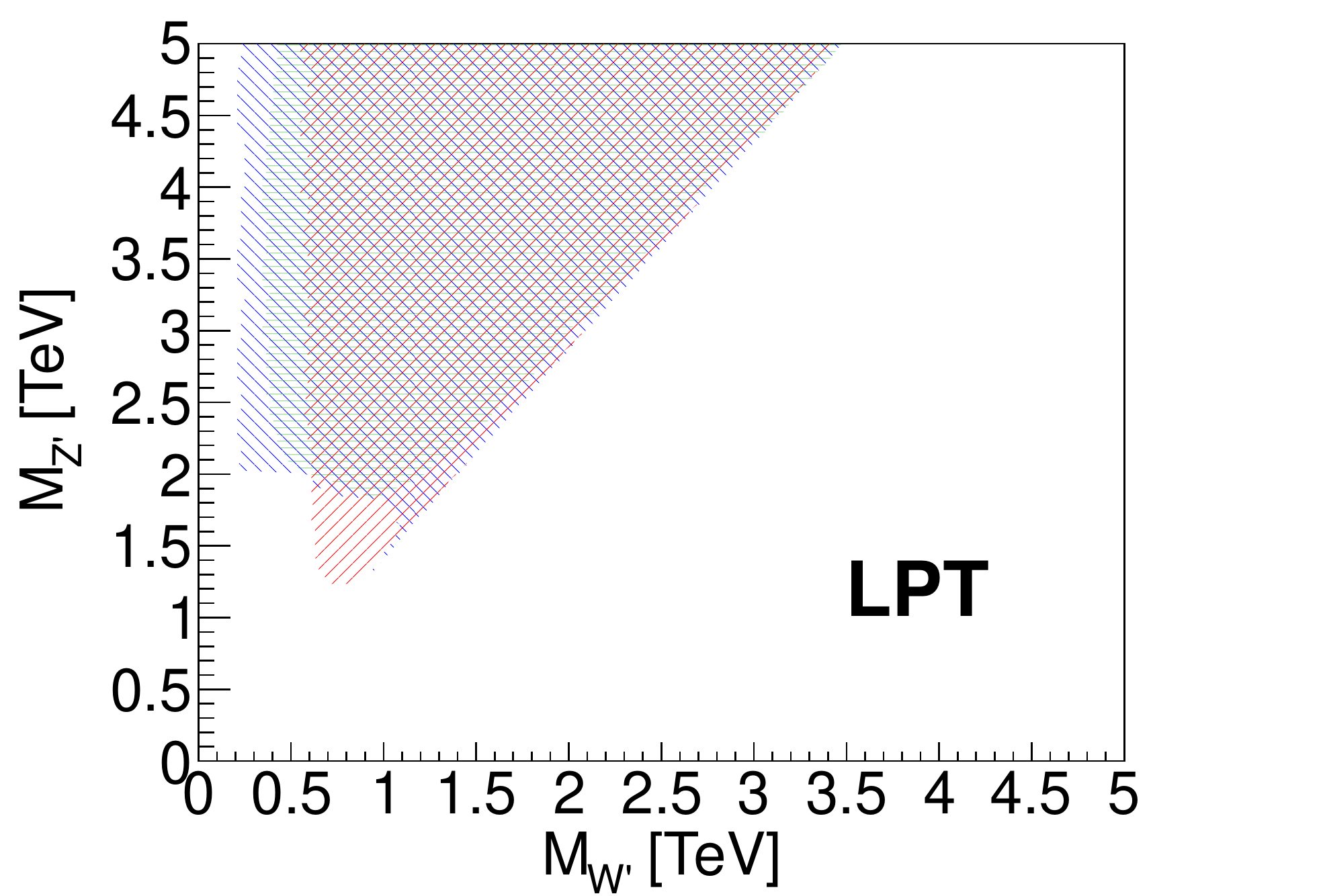}
	\includegraphics[width=0.32\textwidth]{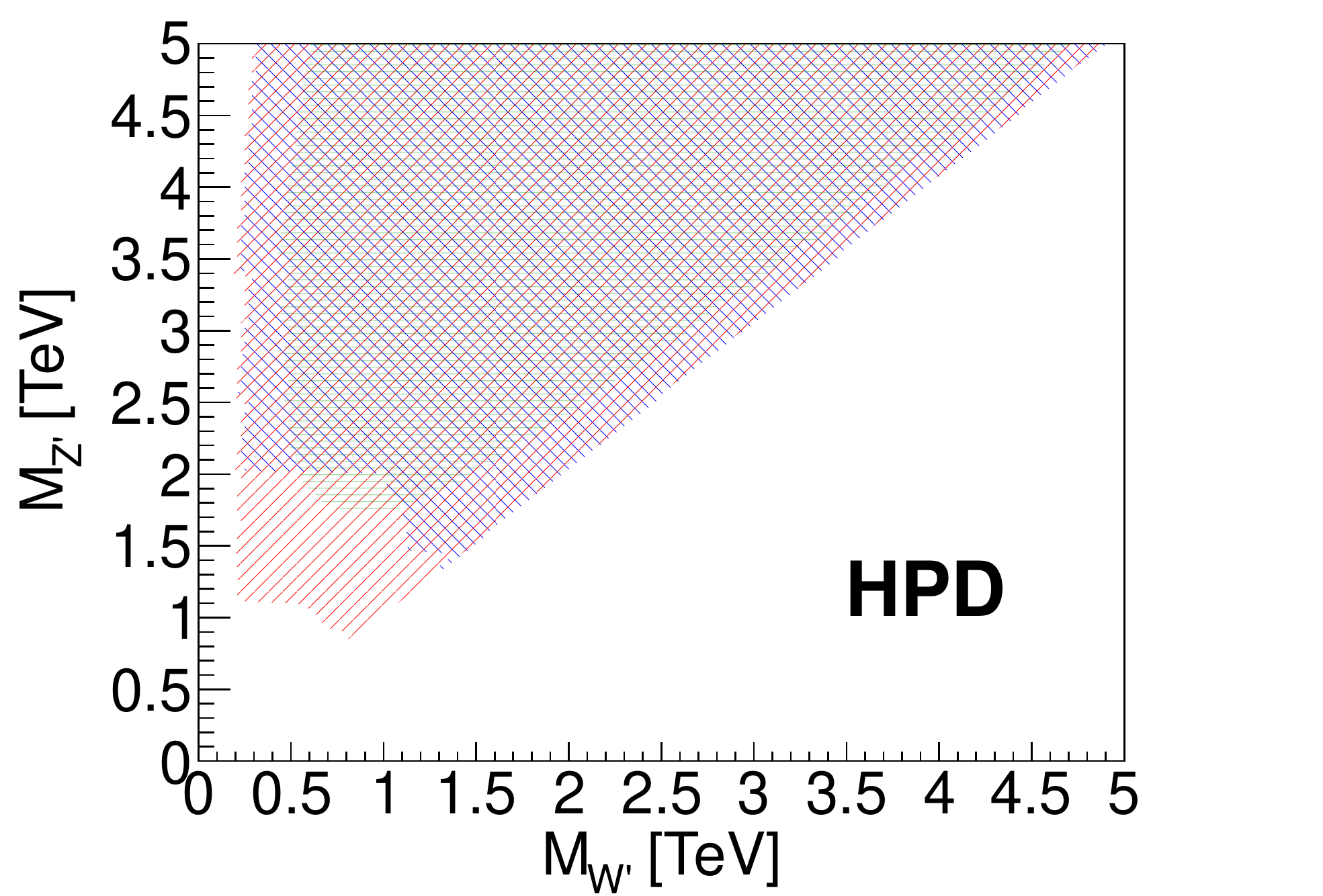}
	\includegraphics[width=0.32\textwidth]{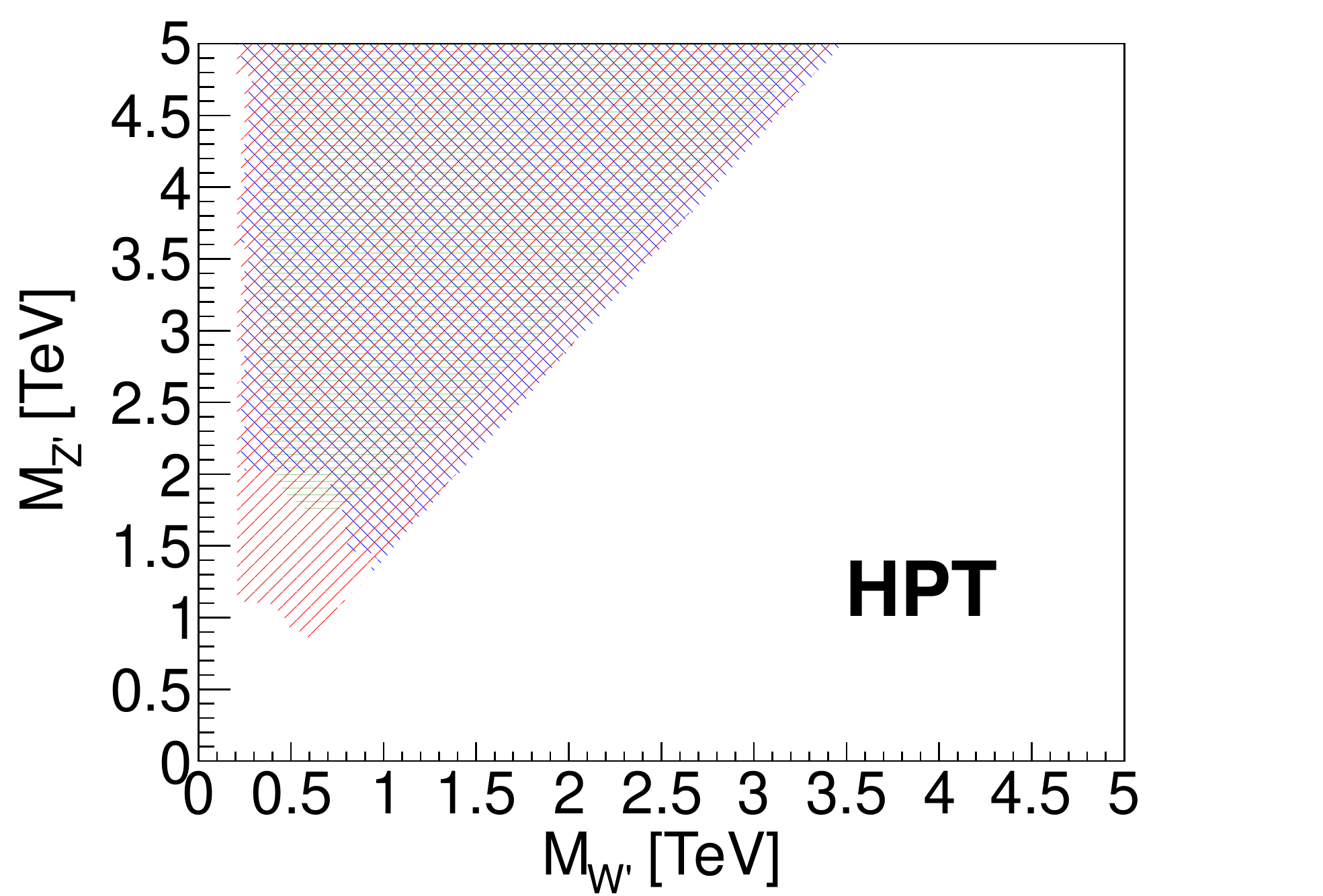}
	\includegraphics[width=0.32\textwidth]{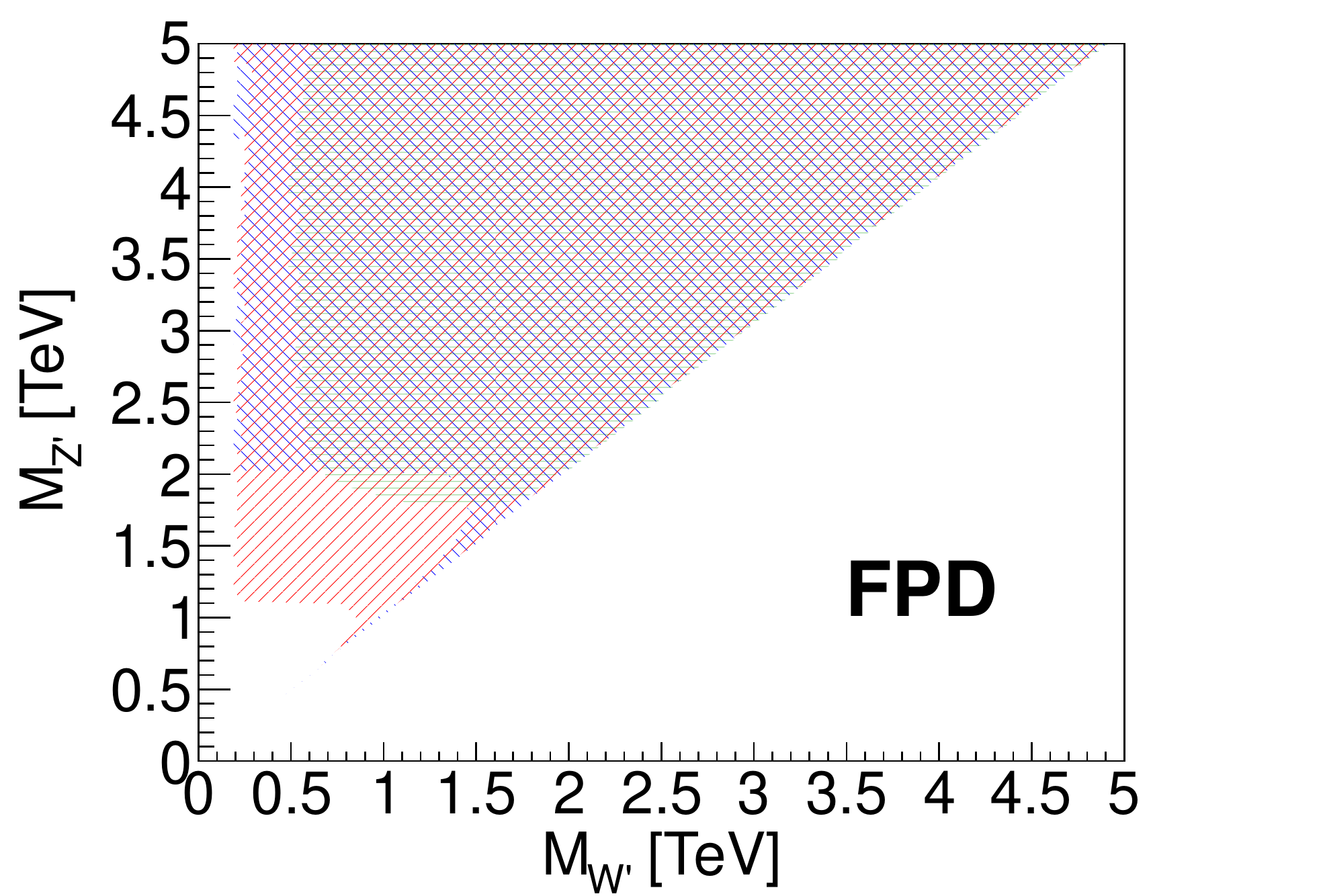}
	\includegraphics[width=0.32\textwidth]{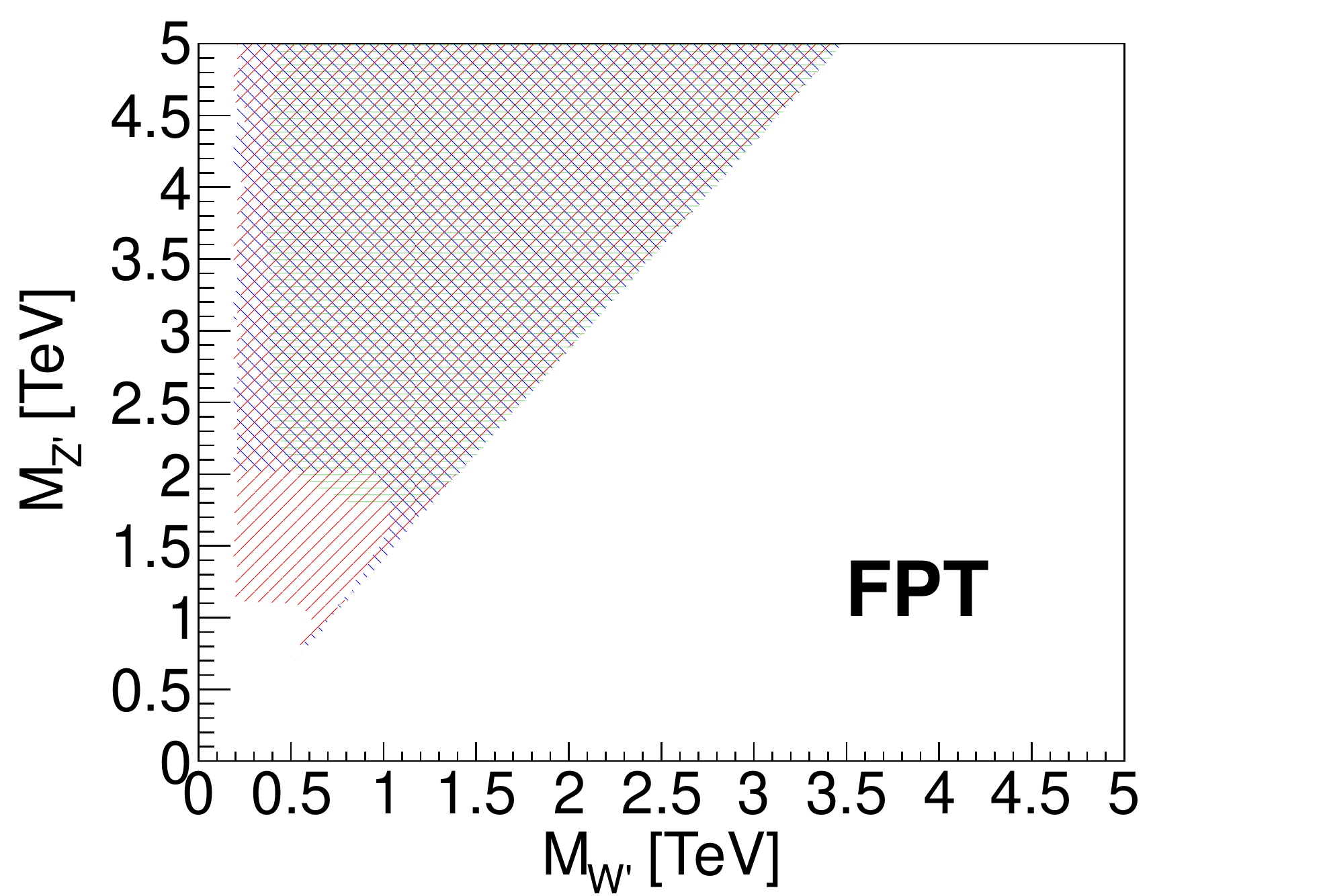}
	\includegraphics[width=0.32\textwidth]{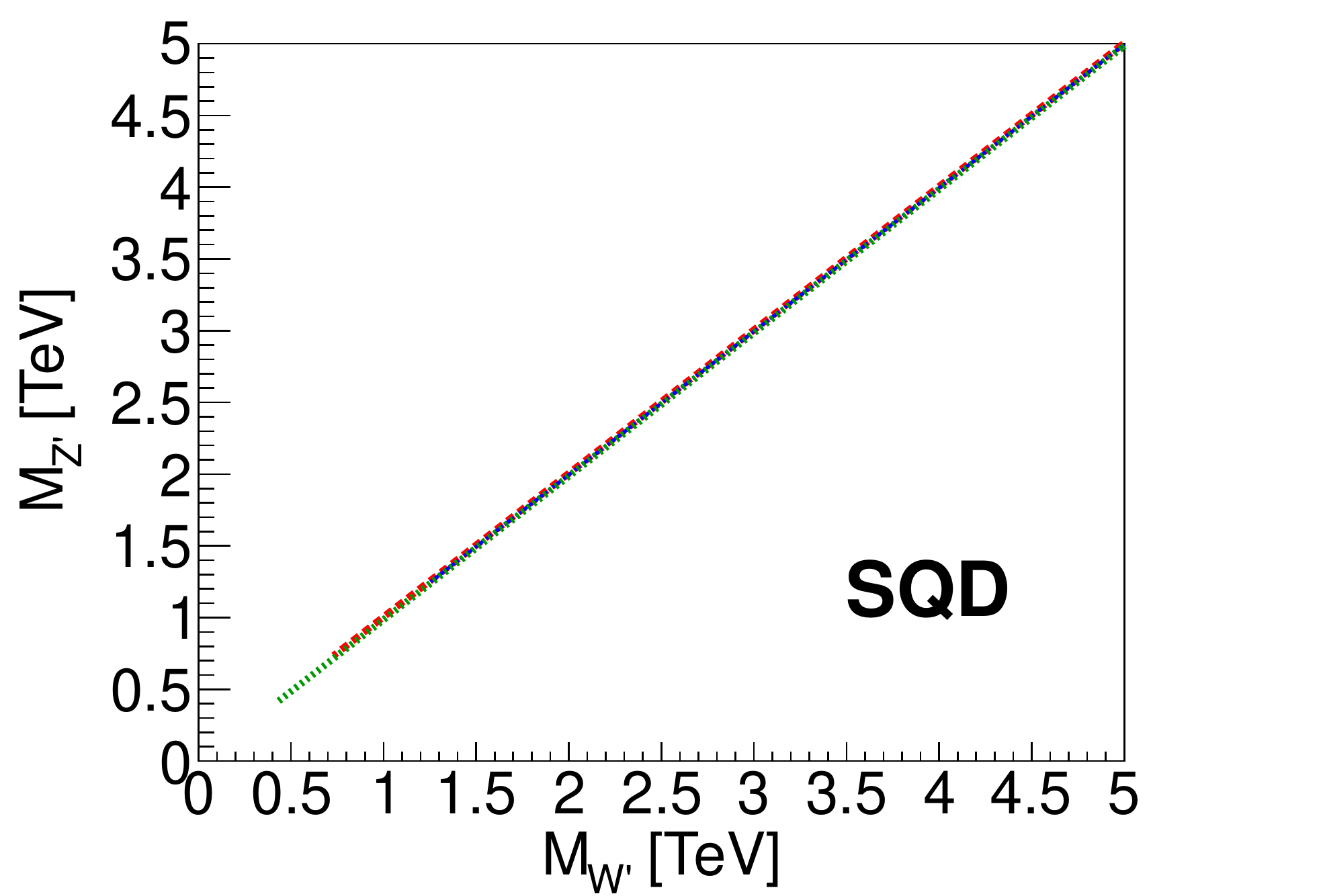}
	\includegraphics[width=0.32\textwidth]{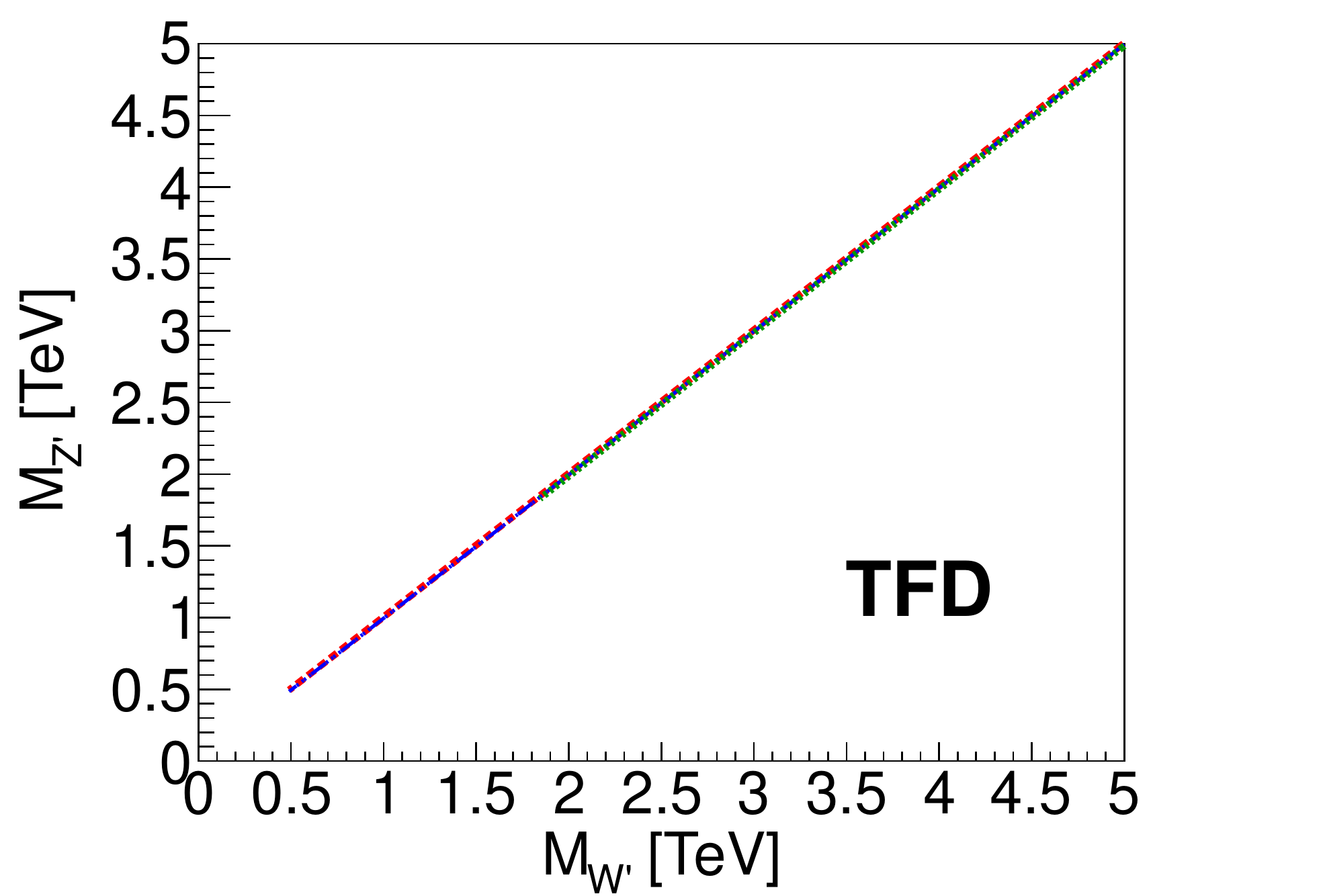}
	\includegraphics[width=0.32\textwidth]{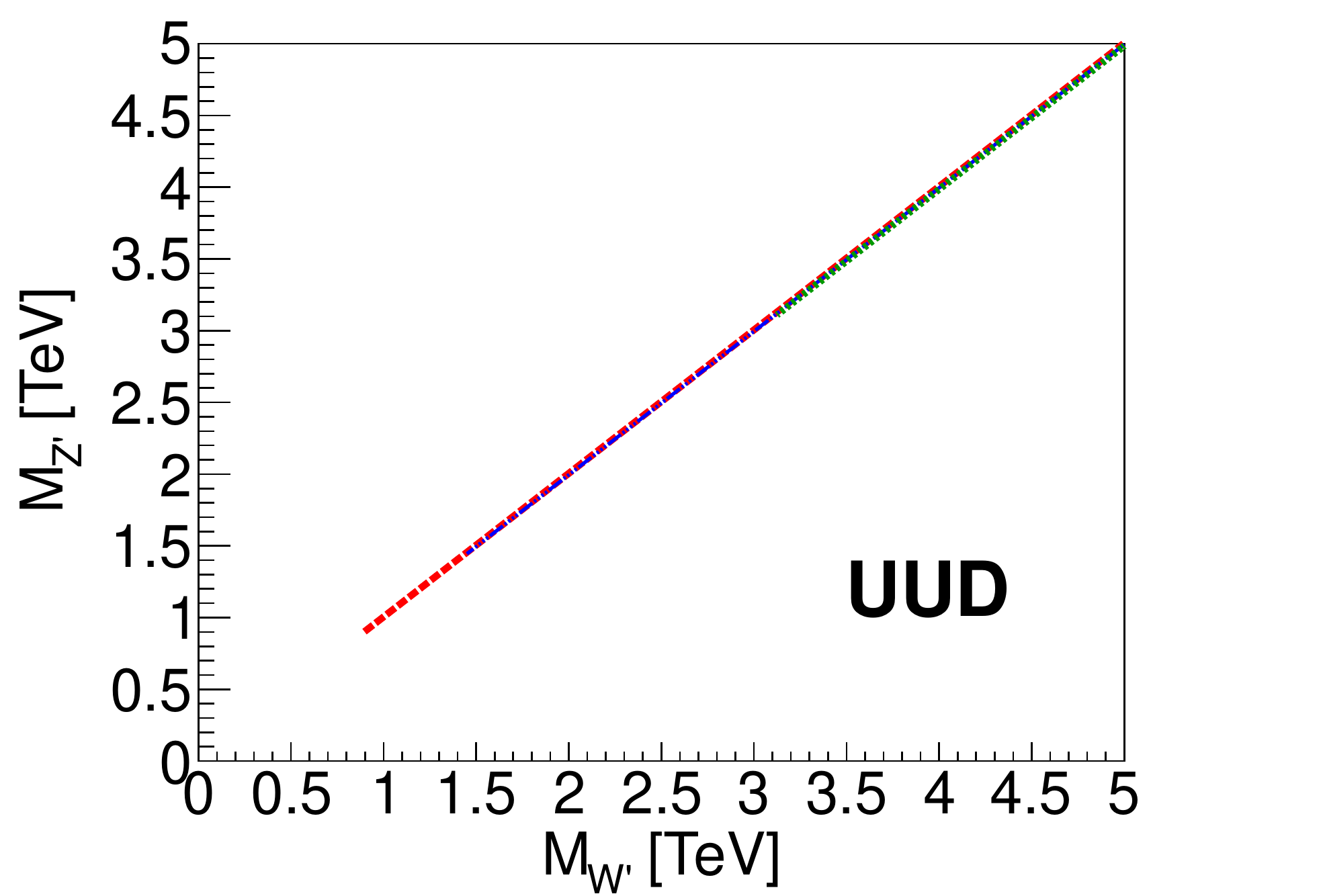}
	\includegraphics[width=0.32\textwidth]{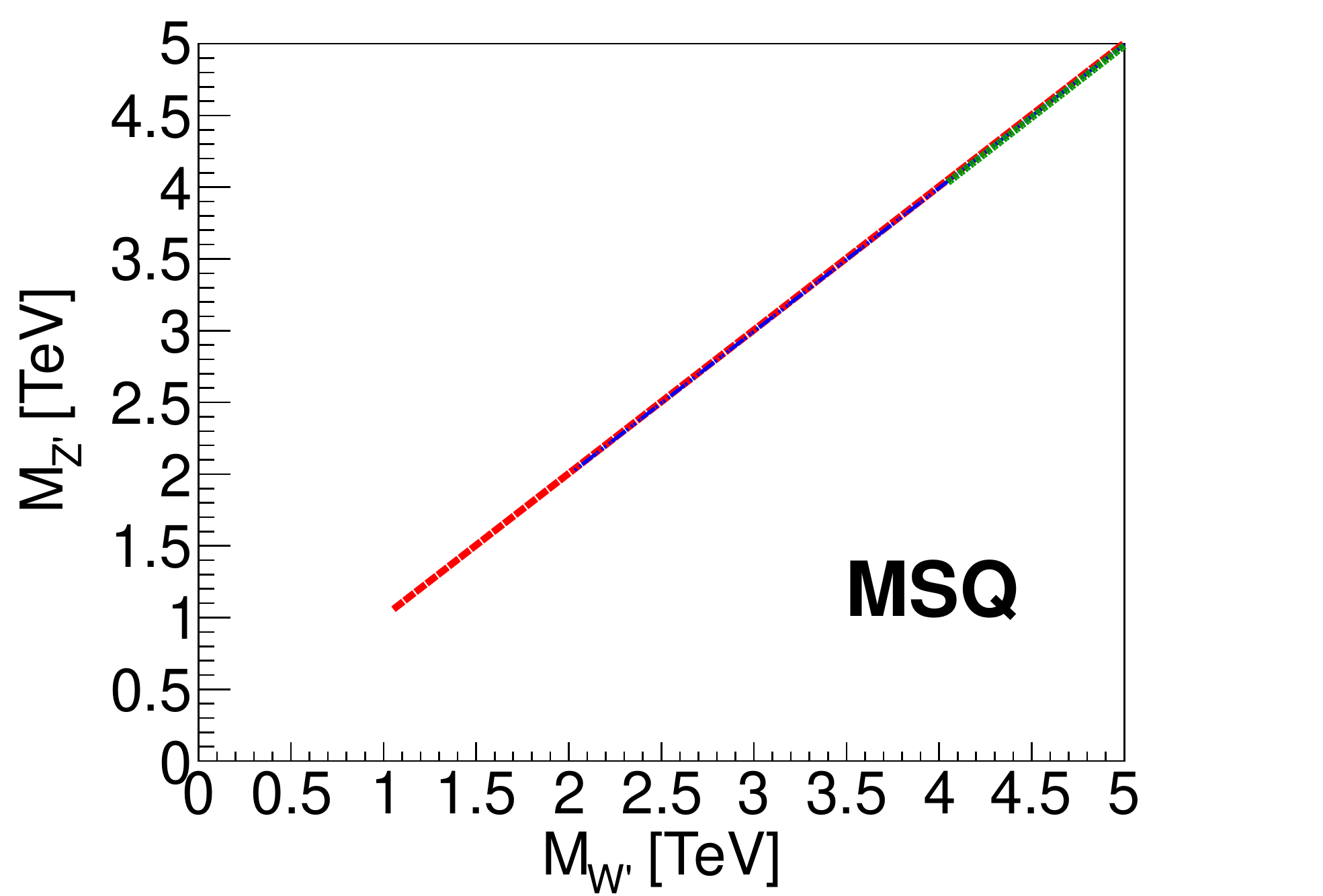}
\caption{Allowed parameter space (colored region) of the $G(221)$ model at $95\%$ CL 
in the $M_{W^\prime}- M_{Z^\prime}$ plane after including
indirect and direct constraints: 
EWPTs (green), Tevatron (red) and LHC7 (blue).}
\label{parameter_constraints}
\end{figure}

\begin{figure}
	\includegraphics[width=0.32\textwidth]{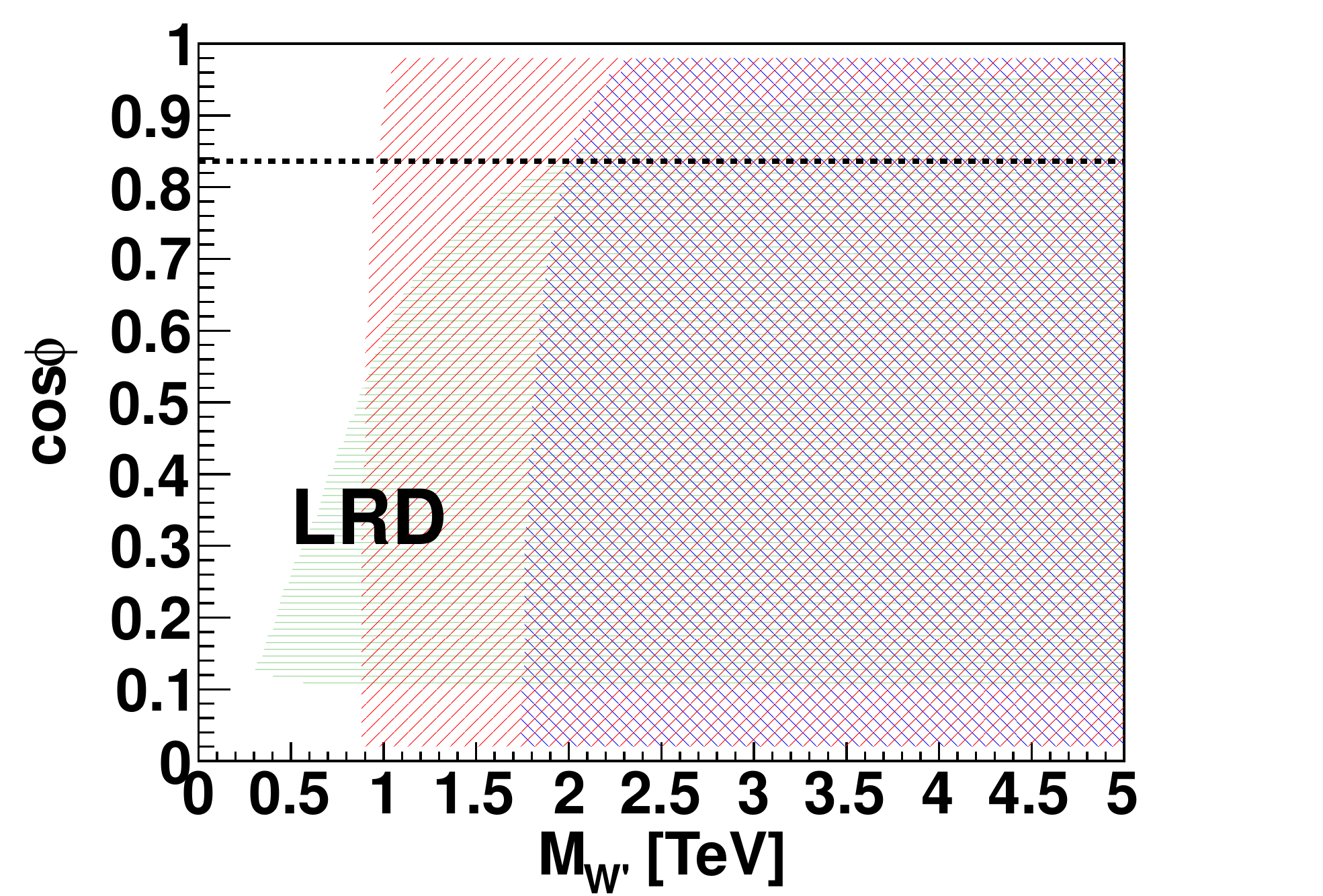}
	\includegraphics[width=0.32\textwidth]{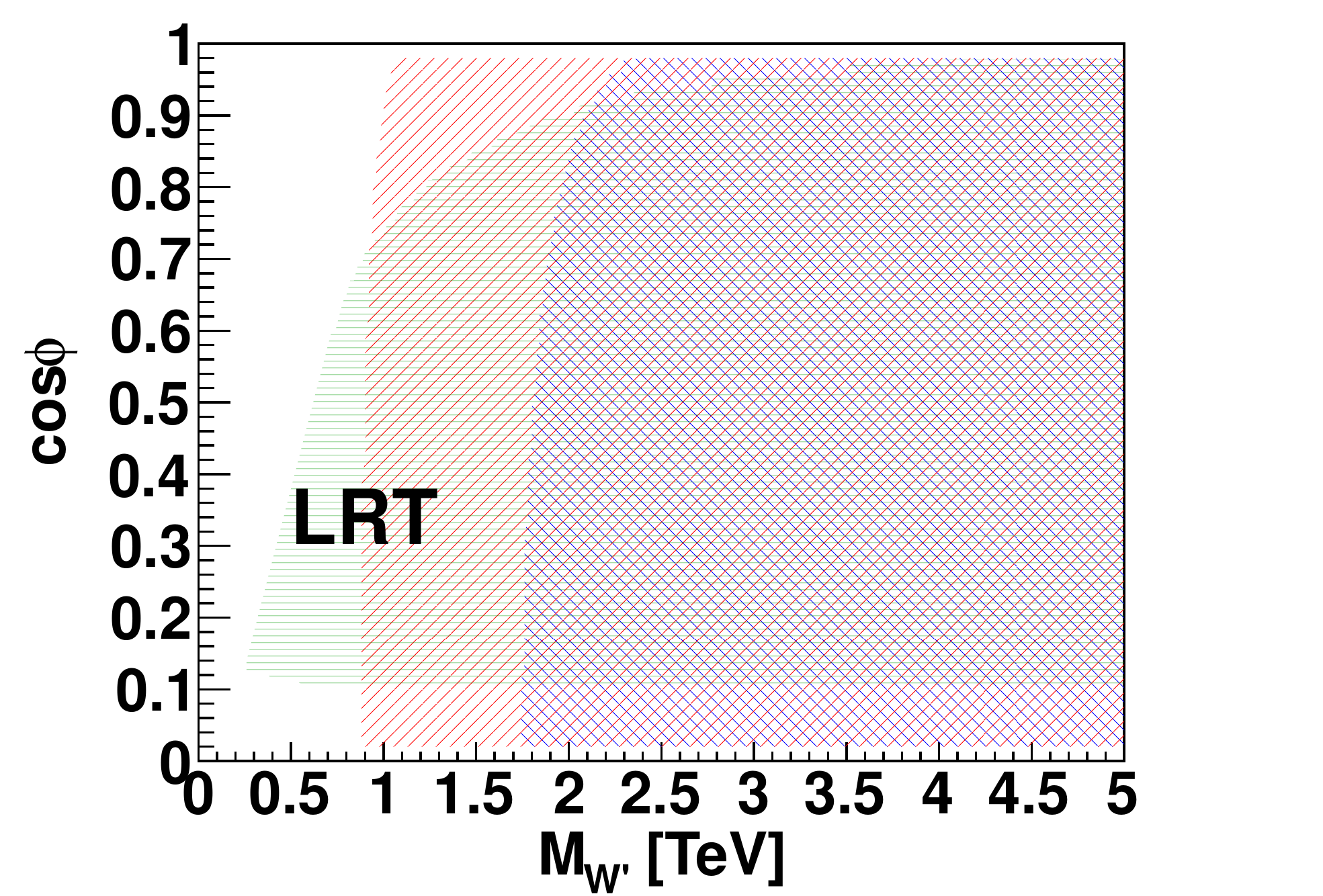}
	\includegraphics[width=0.32\textwidth]{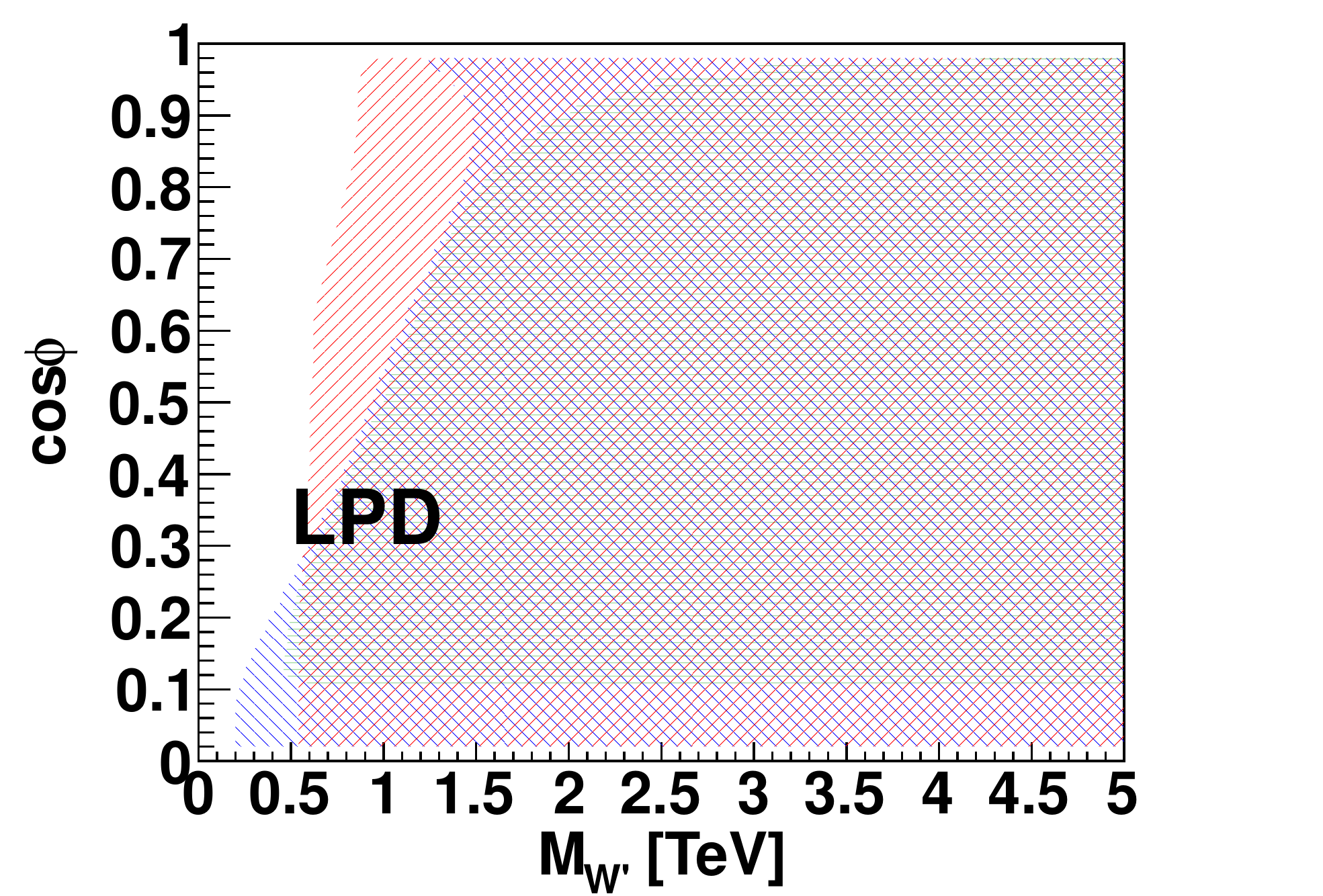}
	\includegraphics[width=0.32\textwidth]{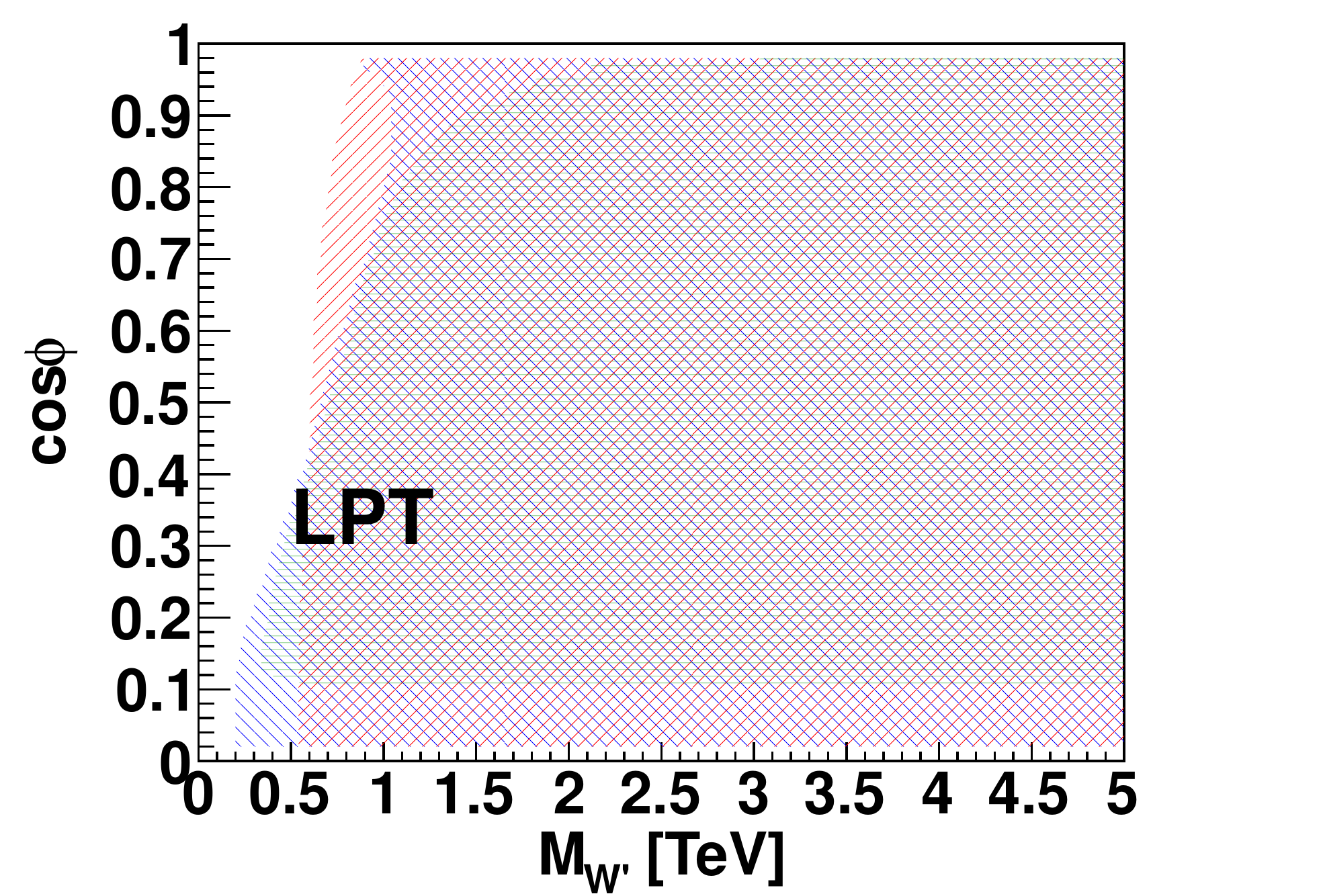}
	\includegraphics[width=0.32\textwidth]{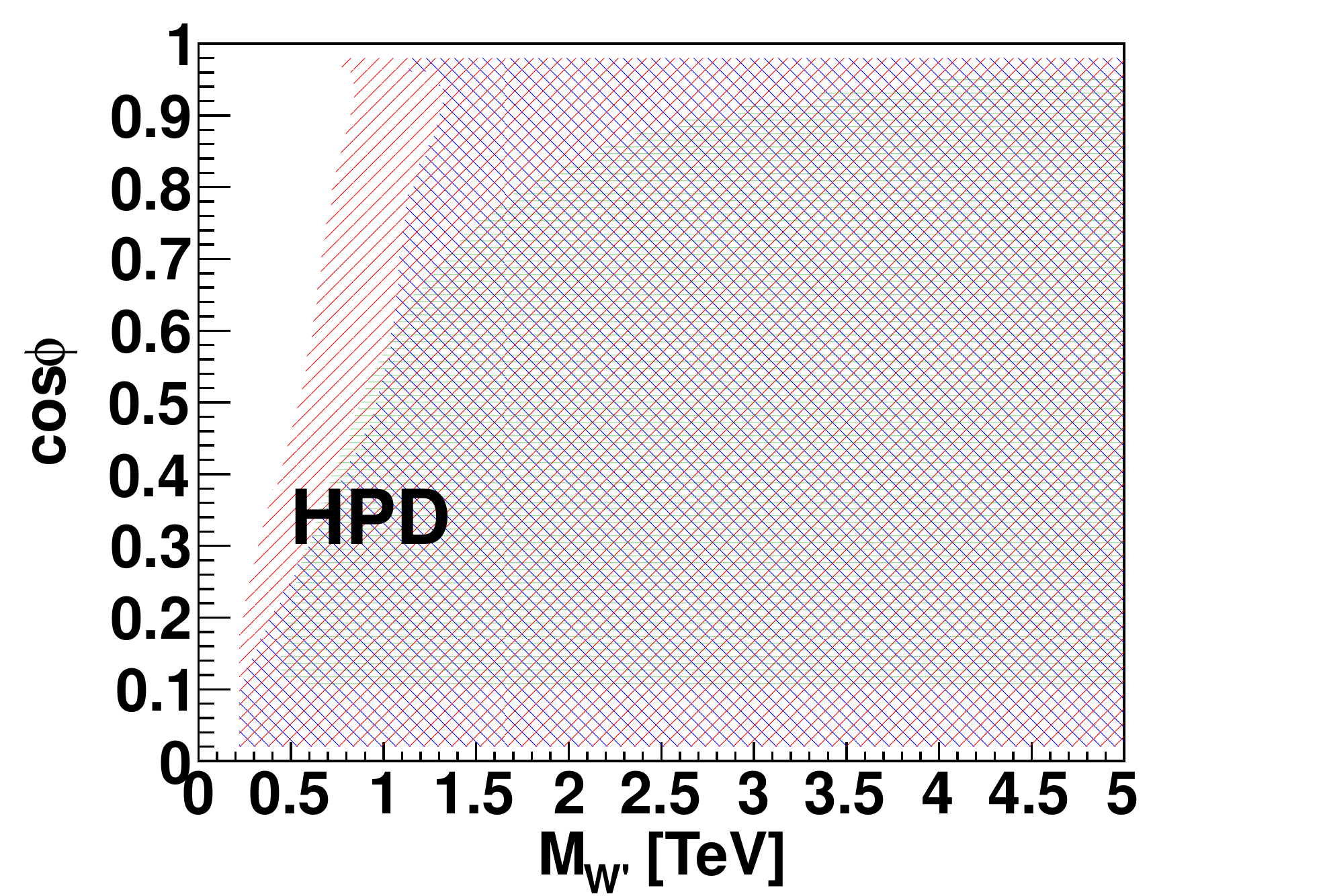}
	\includegraphics[width=0.32\textwidth]{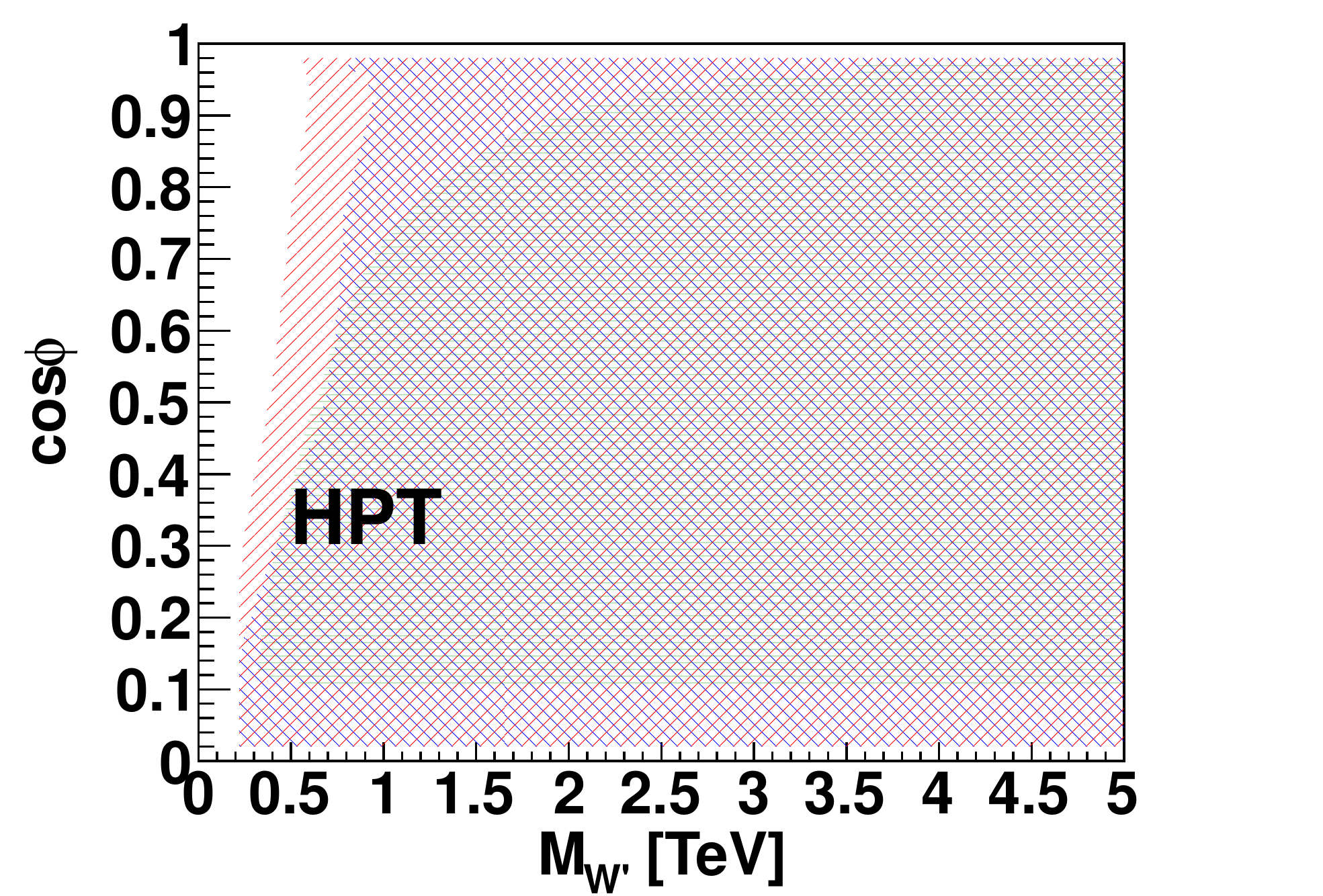}
	\includegraphics[width=0.32\textwidth]{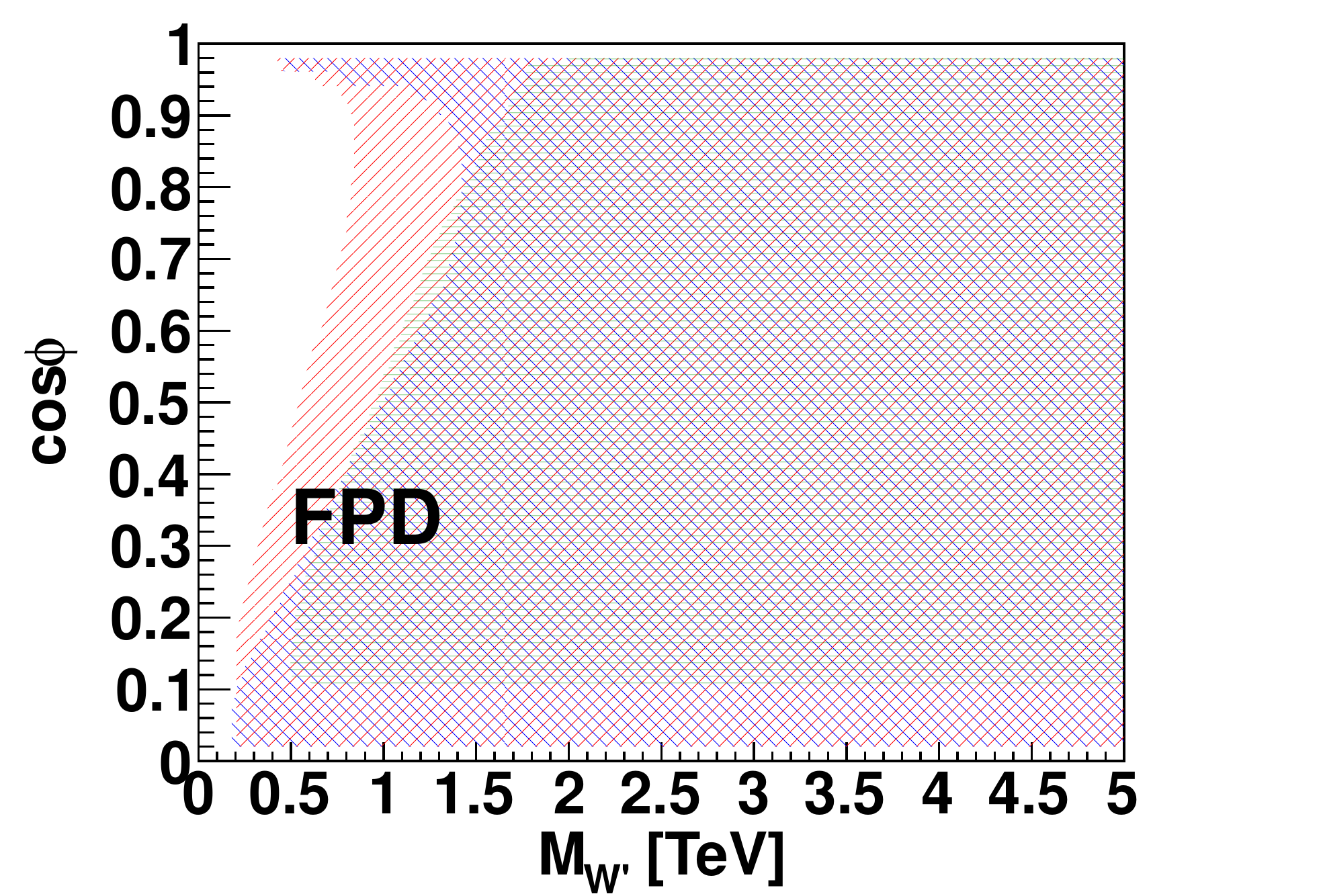}
	\includegraphics[width=0.32\textwidth]{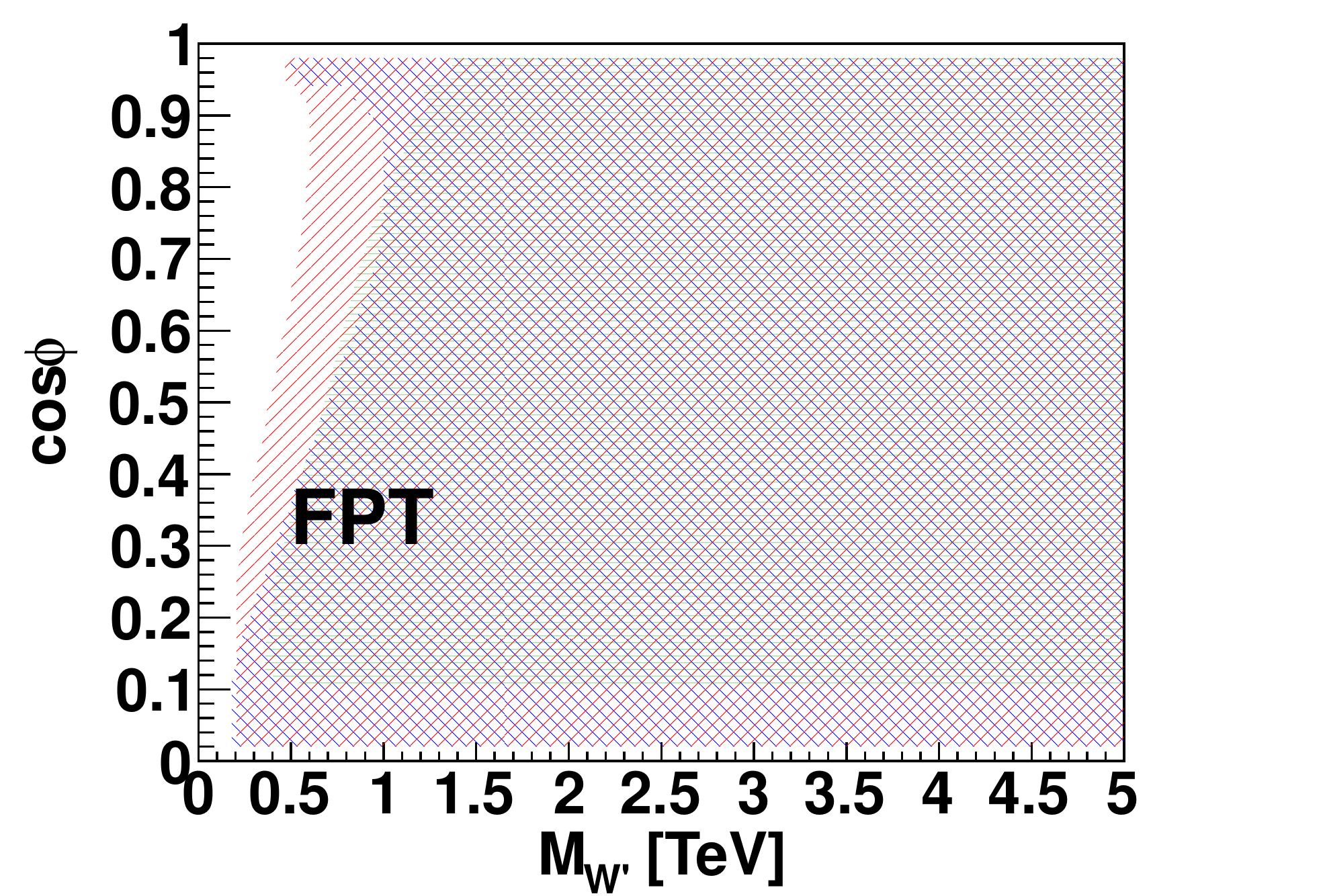}
	\includegraphics[width=0.32\textwidth]{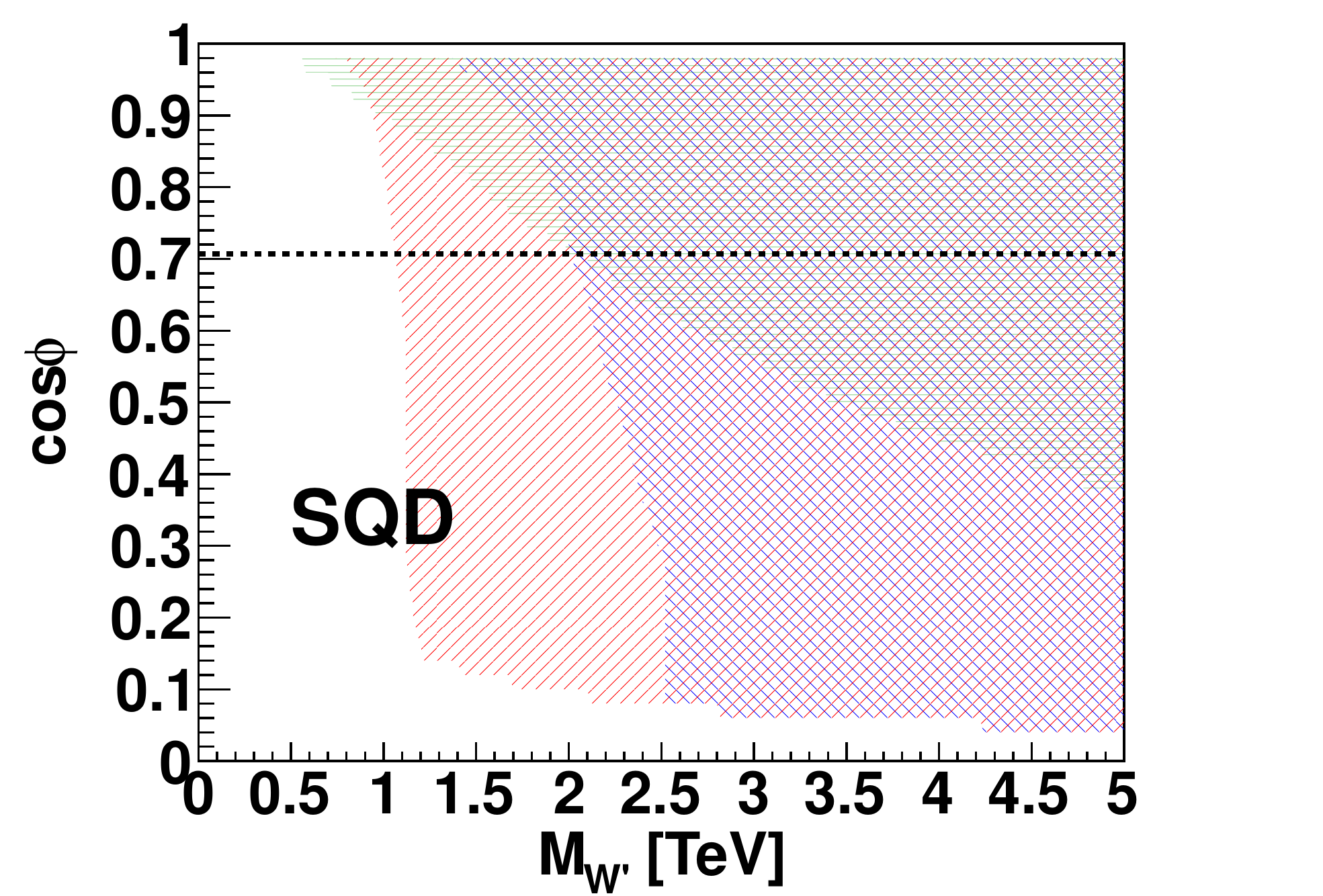}
	\includegraphics[width=0.32\textwidth]{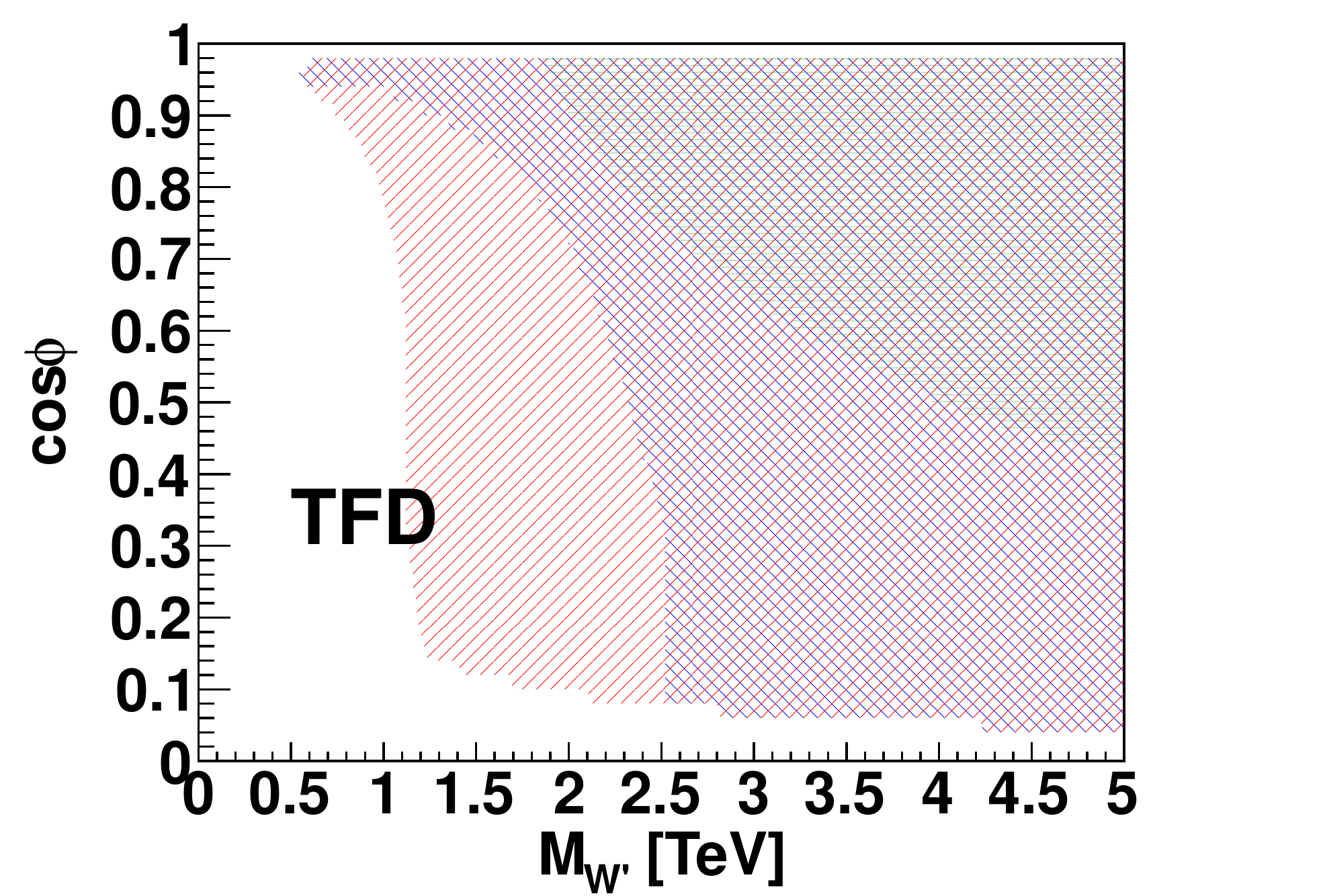}
	\includegraphics[width=0.32\textwidth]{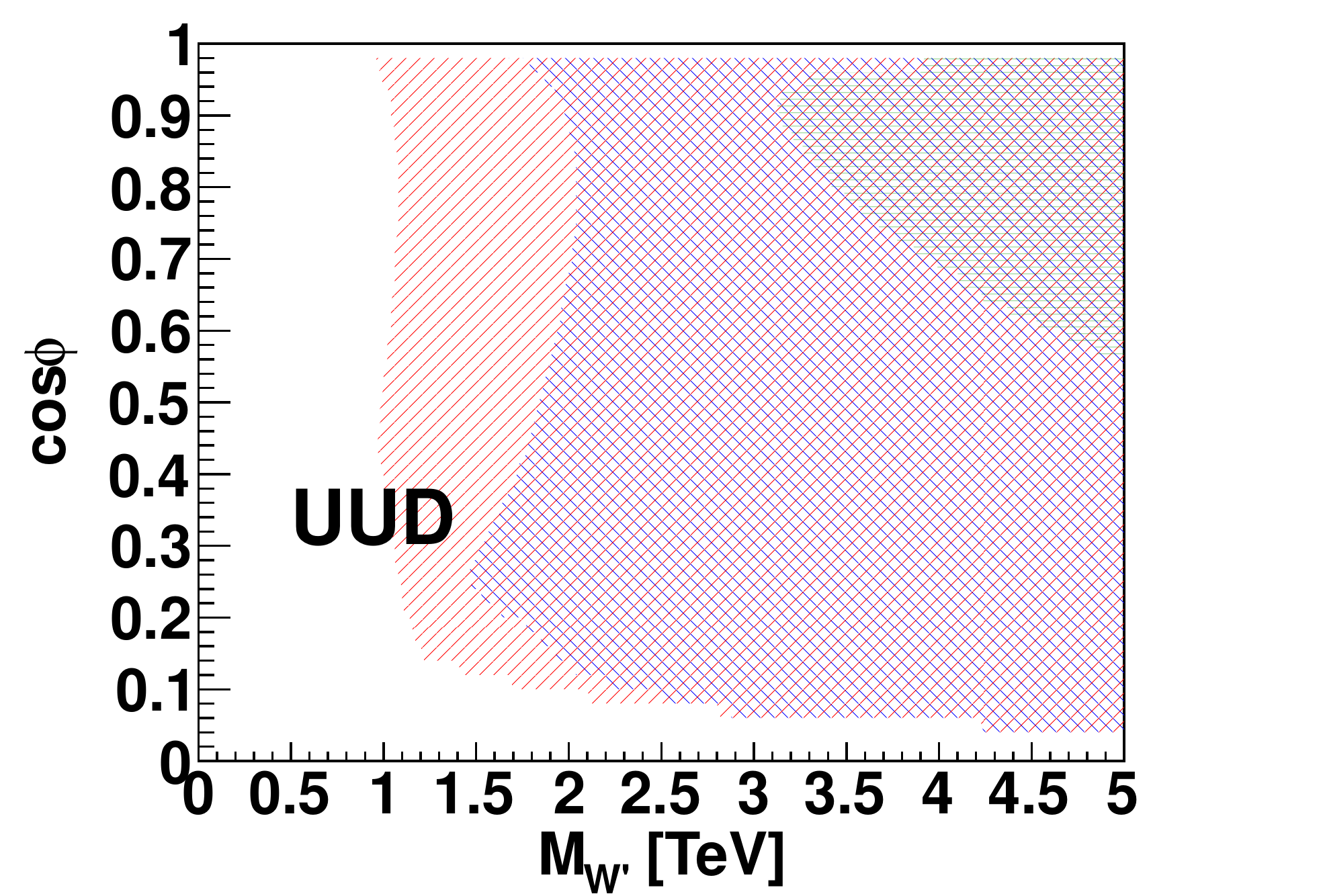}
\caption{Allowed parameter space (colored region) of the $G(221)$ model at the $95\%$ CL 
in the $M_{W^\prime}- c_\phi$ plaine after including indirect
and direct search constraints: EWPTs (green), 
Tevatron (red) and LHC7 (blue). The dashed black lines in LRD and SQD
represent MLR and MSQ models.}
\label{parameter_constraintsWP}
\end{figure}

\begin{figure}
	\includegraphics[width=0.32\textwidth]{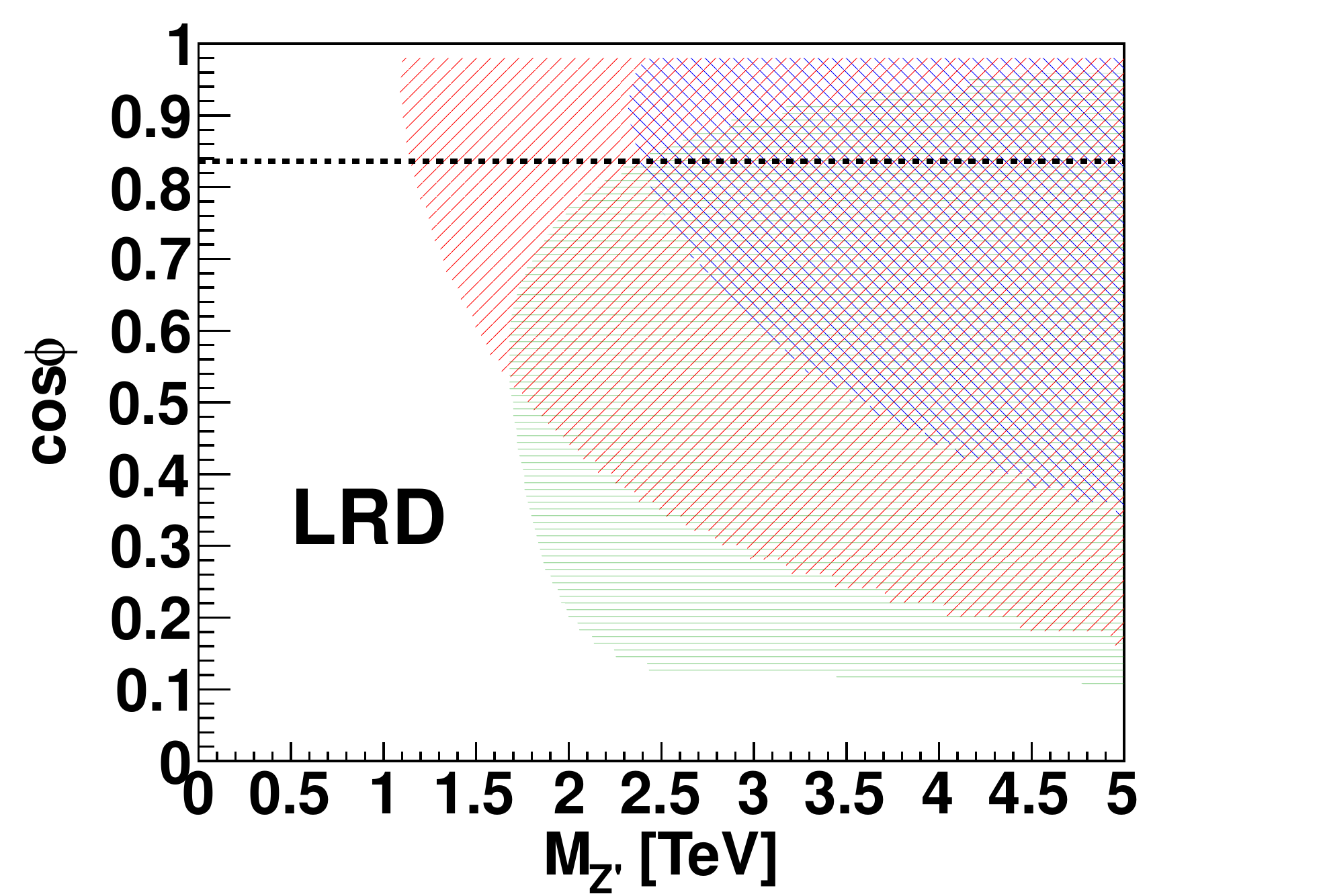}
	\includegraphics[width=0.32\textwidth]{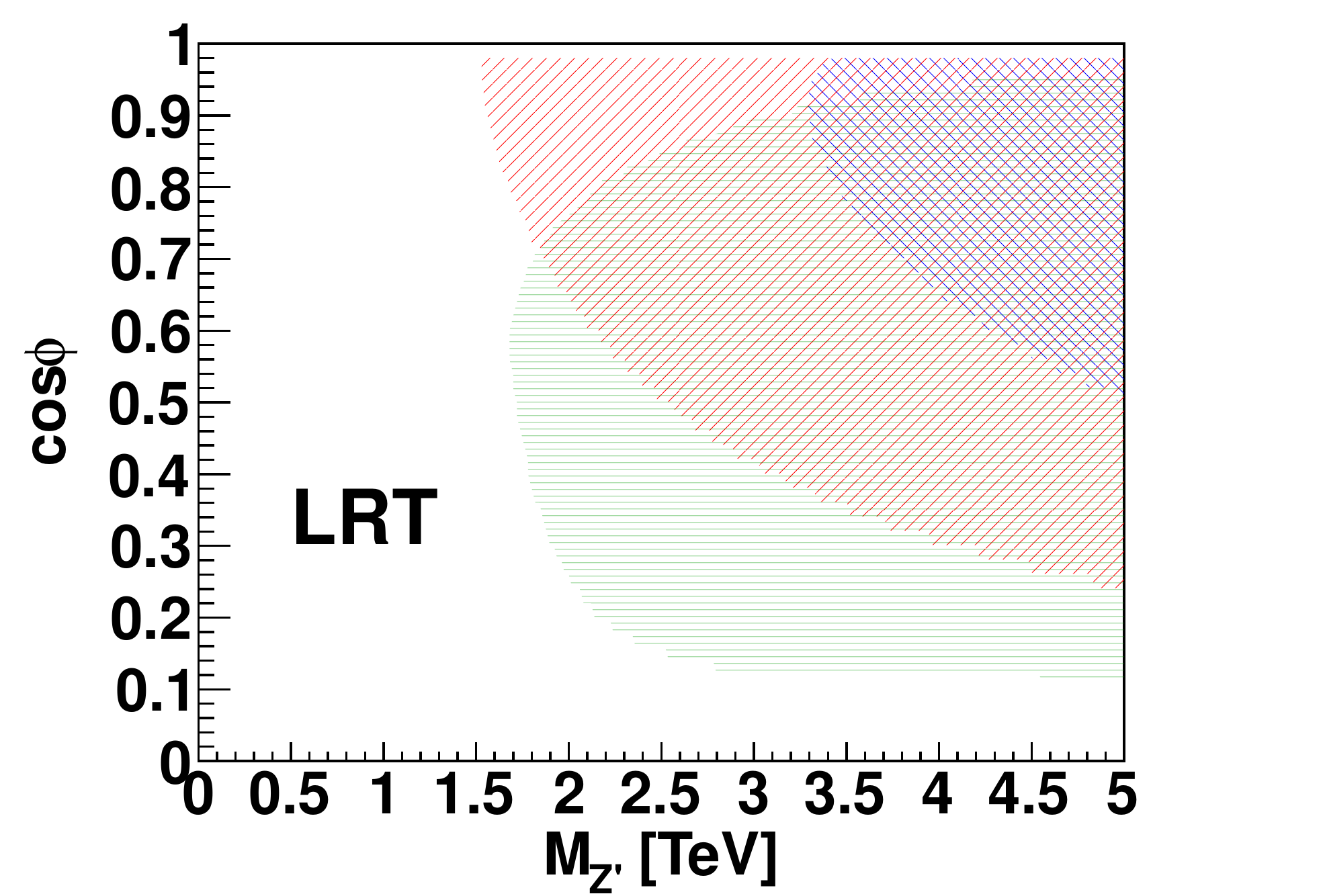}
	\includegraphics[width=0.32\textwidth]{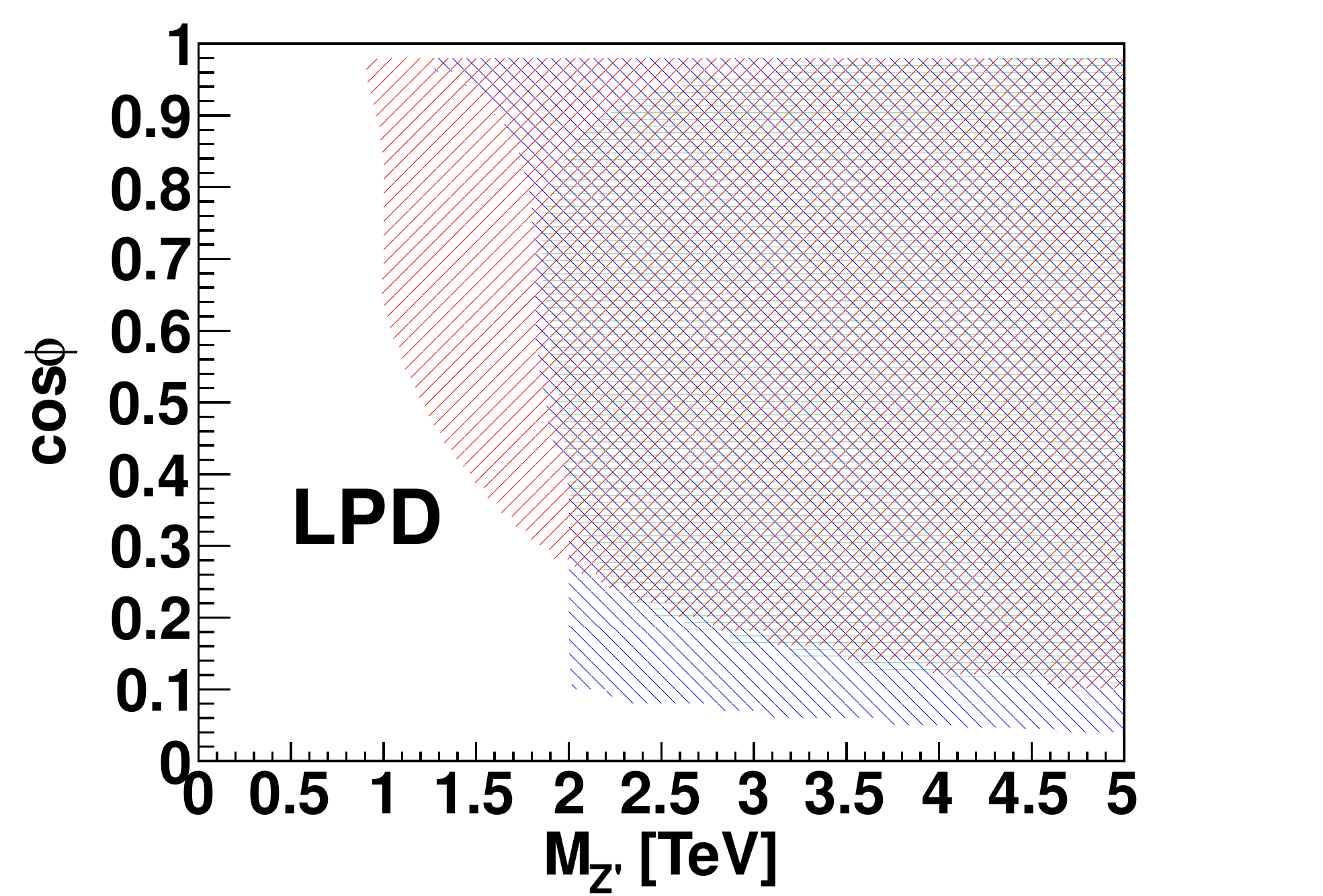}
	\includegraphics[width=0.32\textwidth]{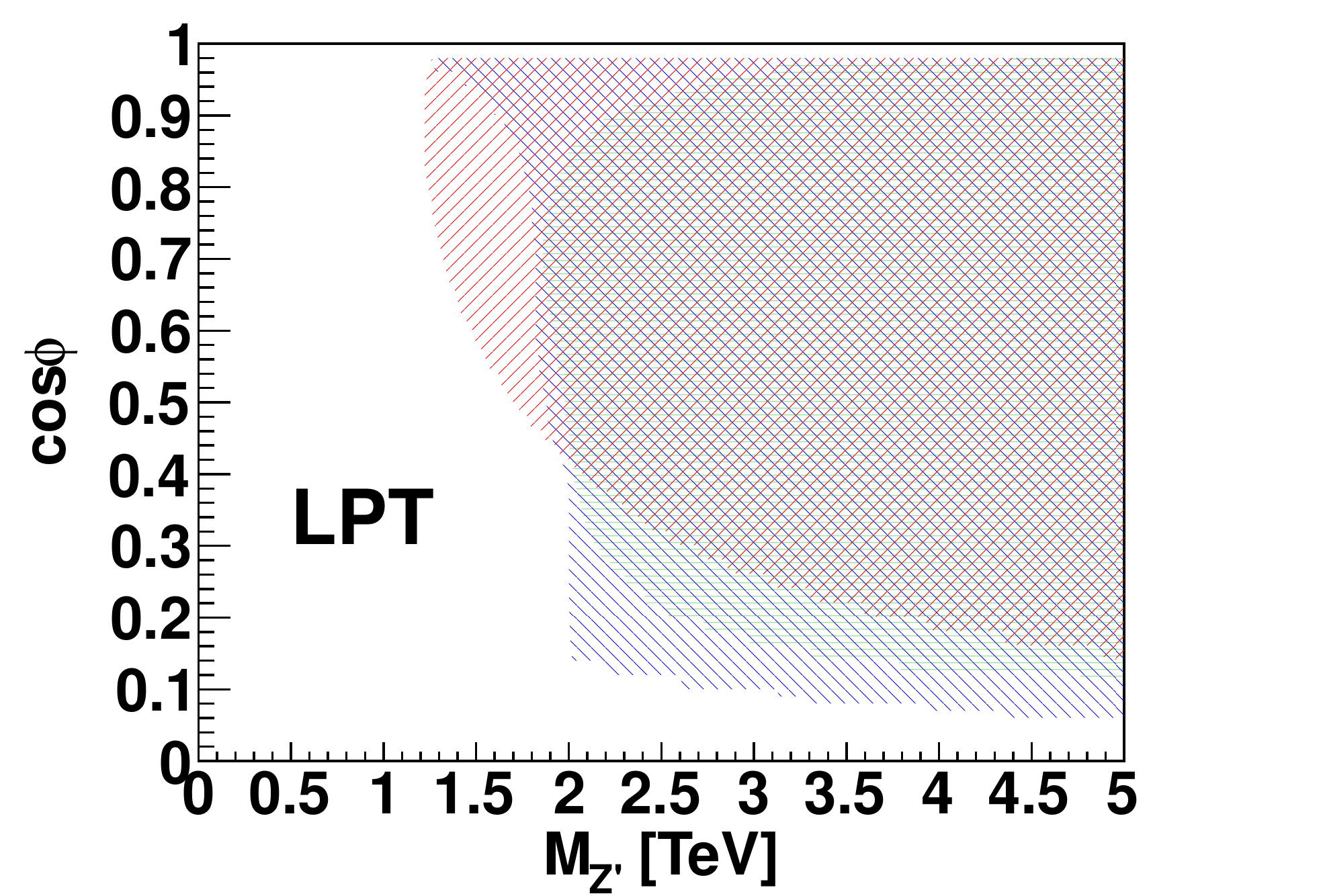}
	\includegraphics[width=0.32\textwidth]{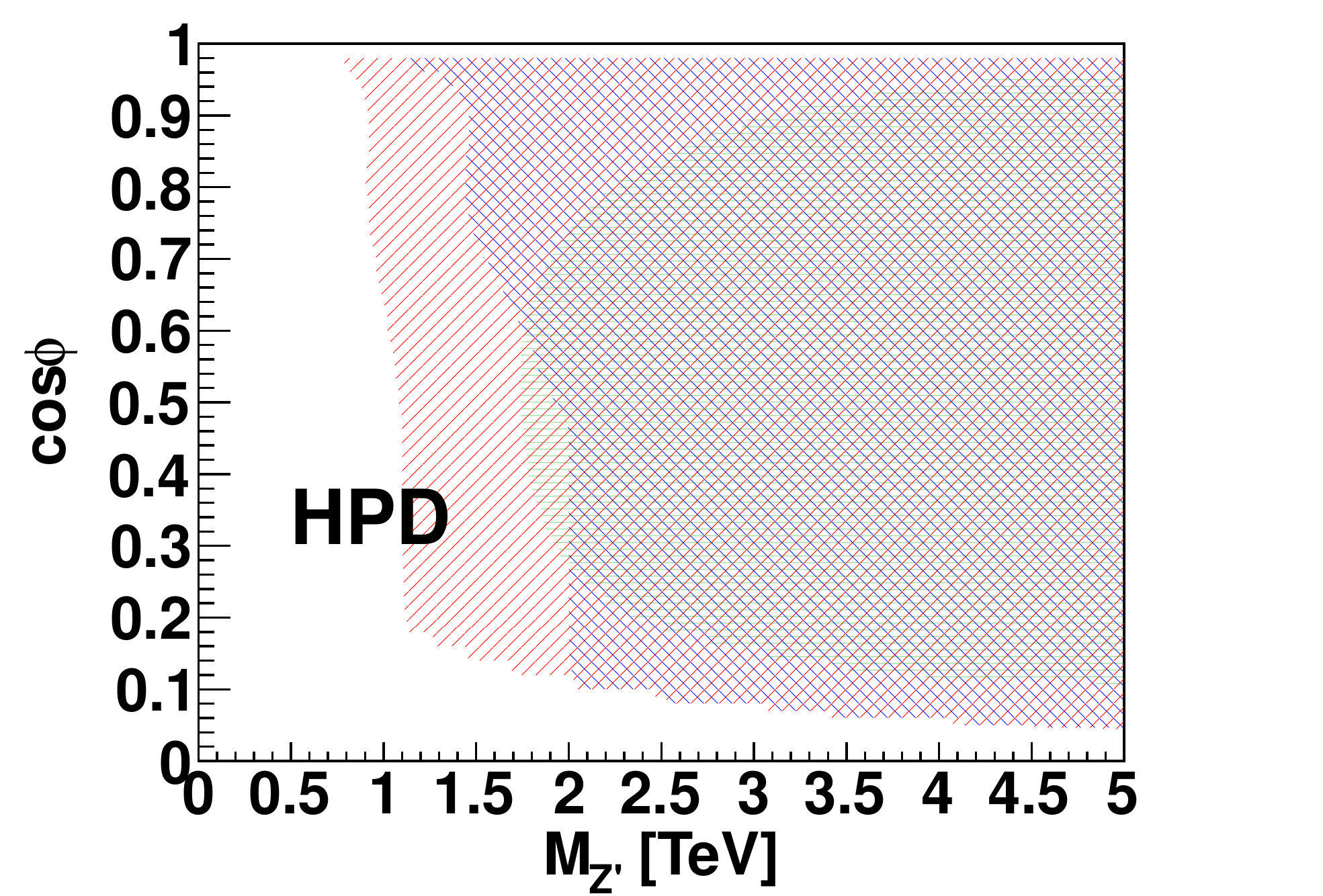}
	\includegraphics[width=0.32\textwidth]{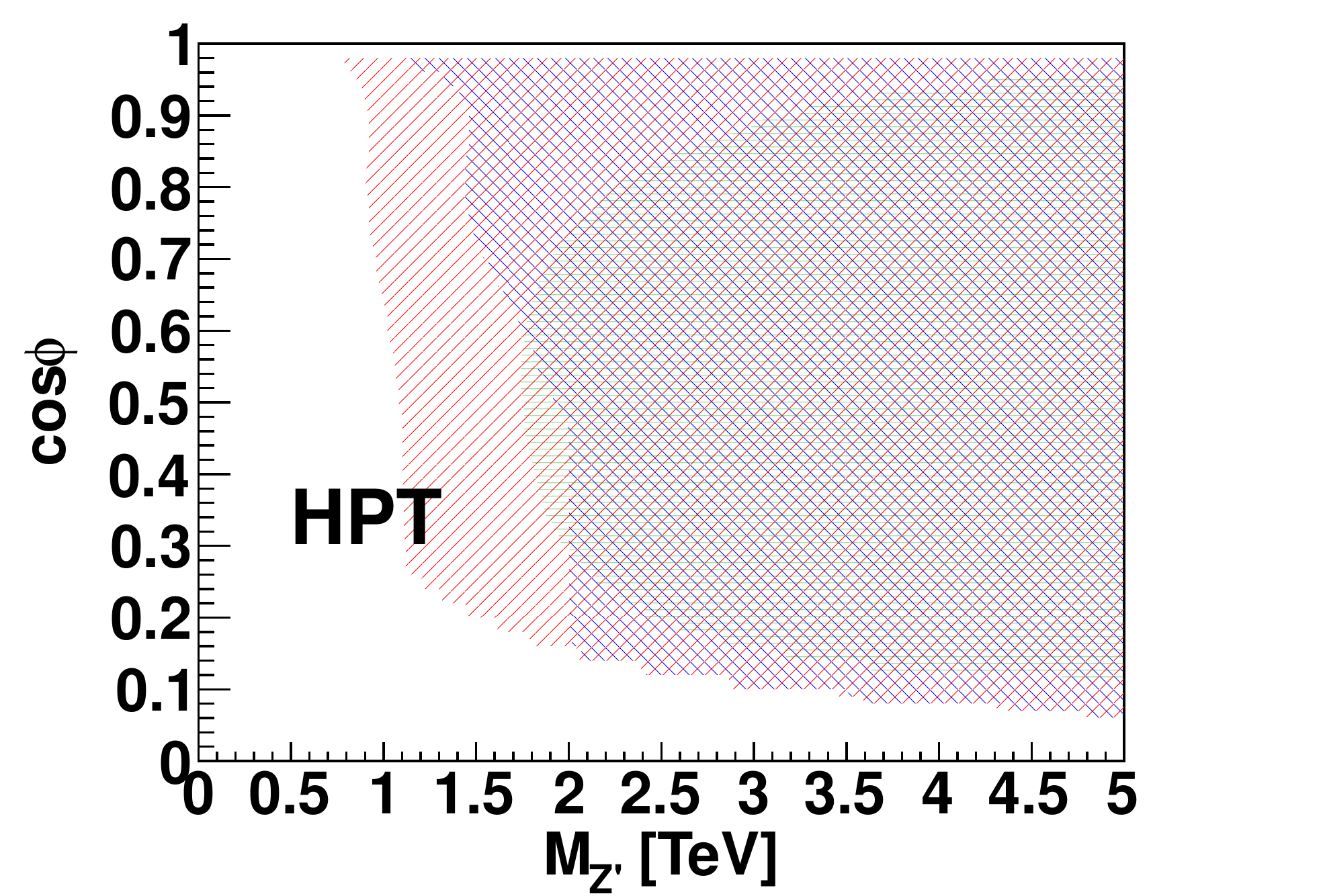}
	\includegraphics[width=0.32\textwidth]{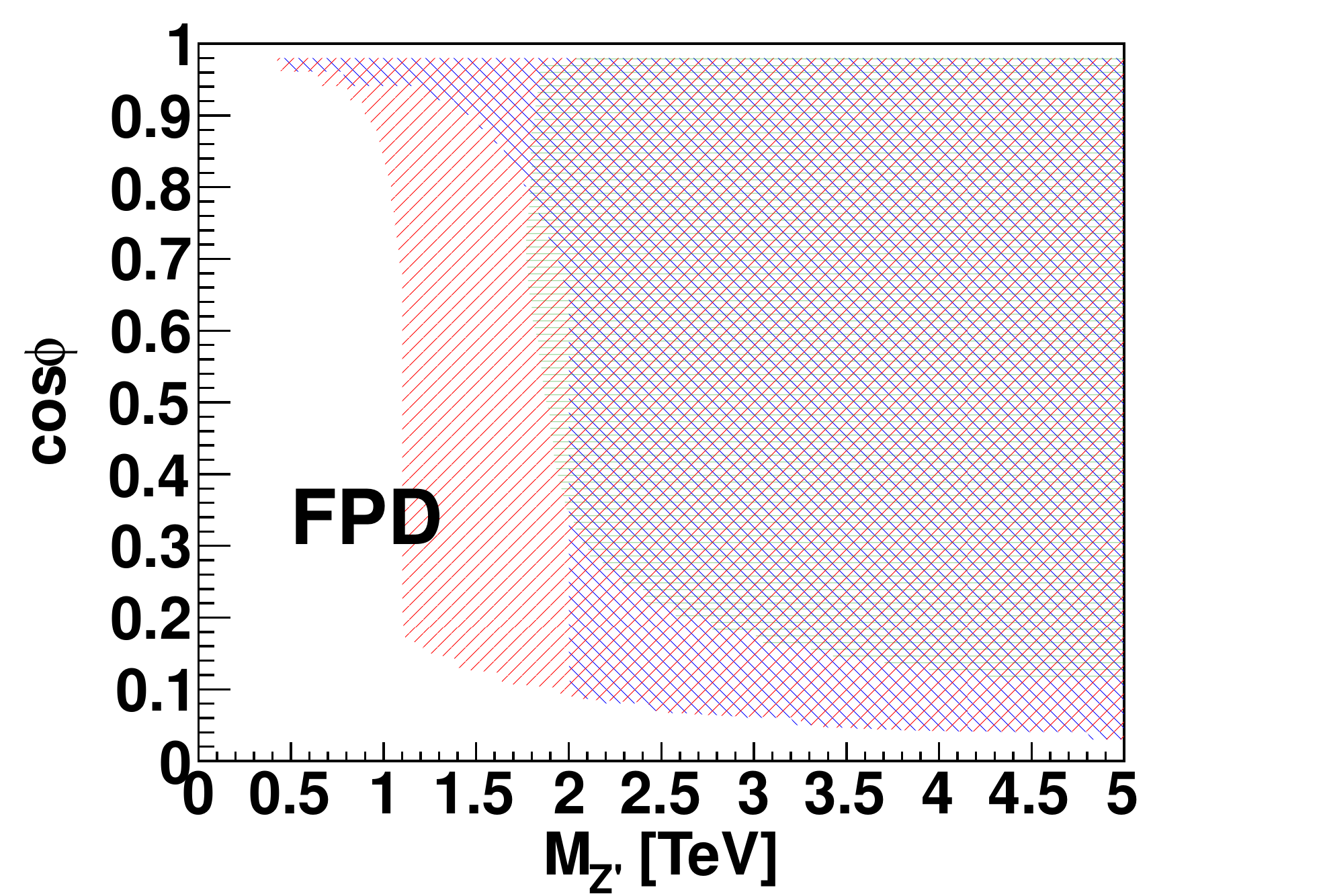}
	\includegraphics[width=0.32\textwidth]{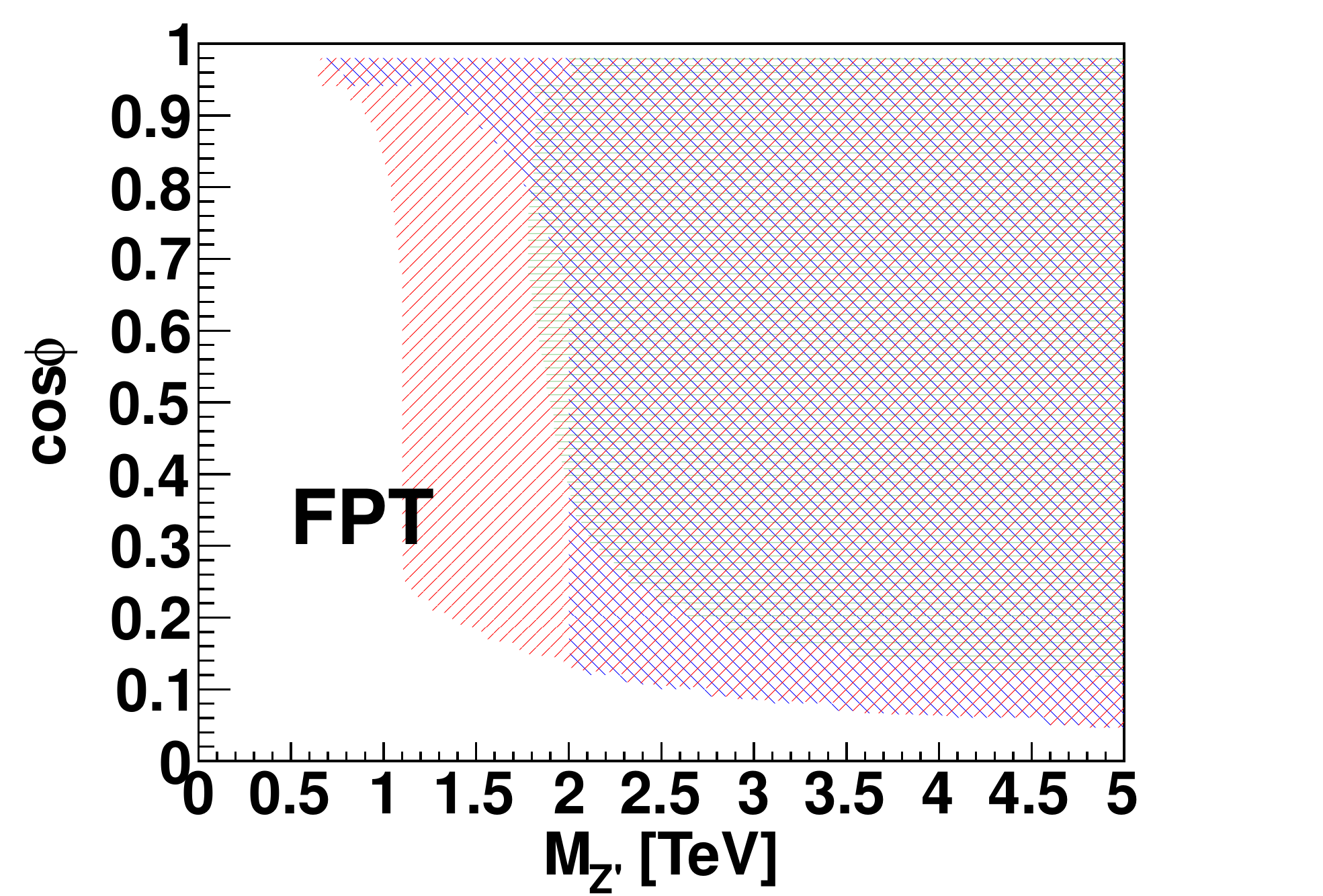}
	\includegraphics[width=0.32\textwidth]{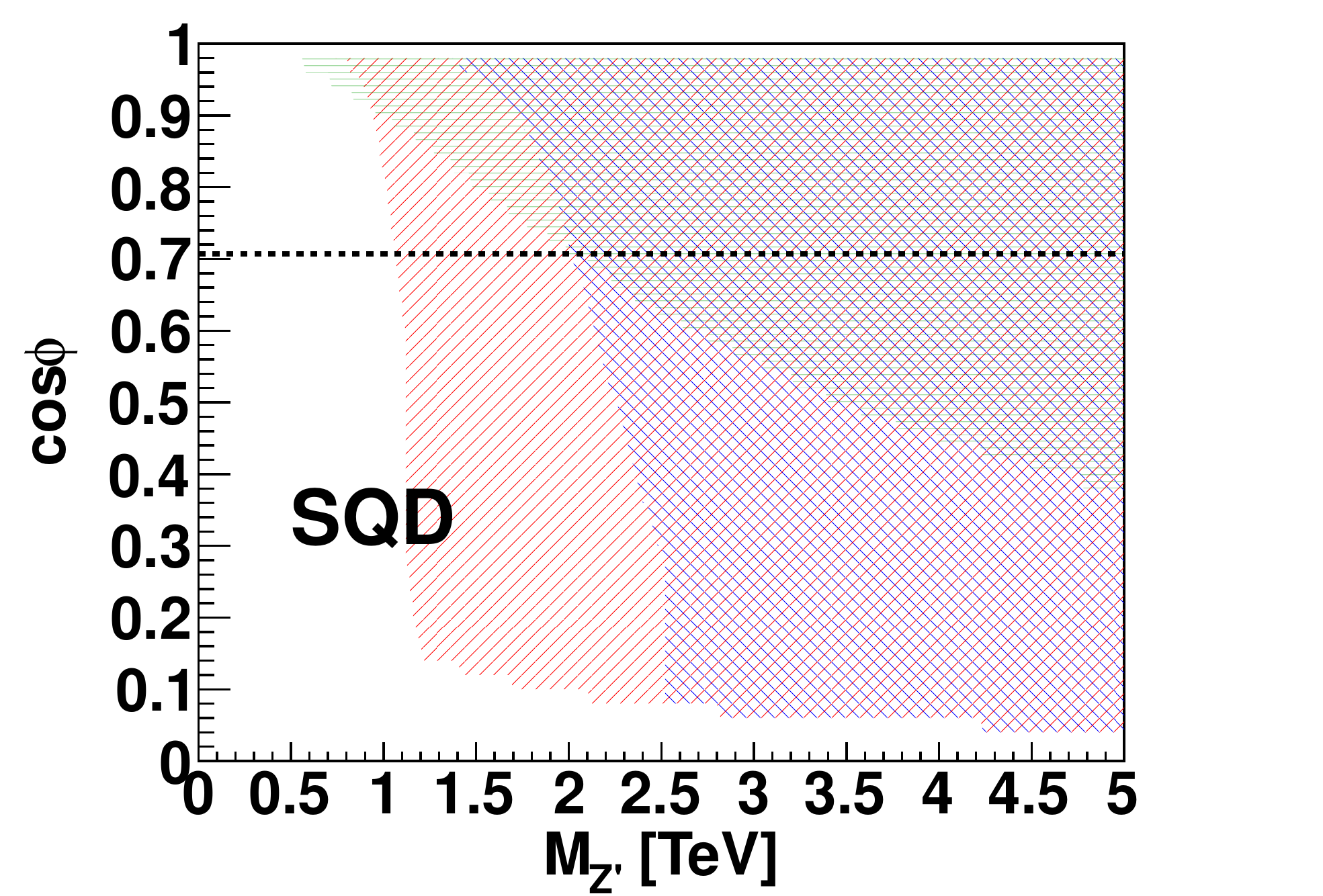}
	\includegraphics[width=0.32\textwidth]{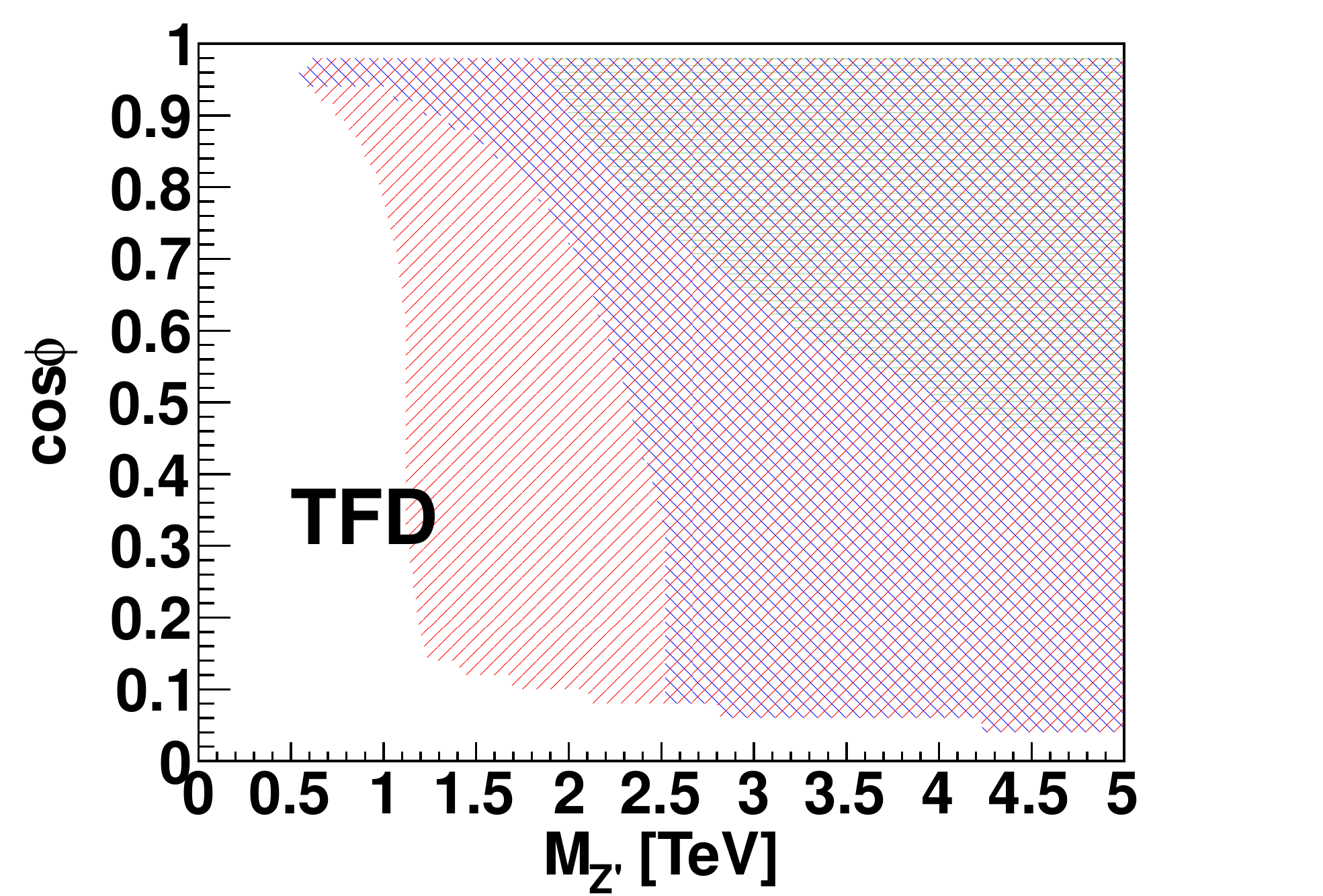}
	\includegraphics[width=0.32\textwidth]{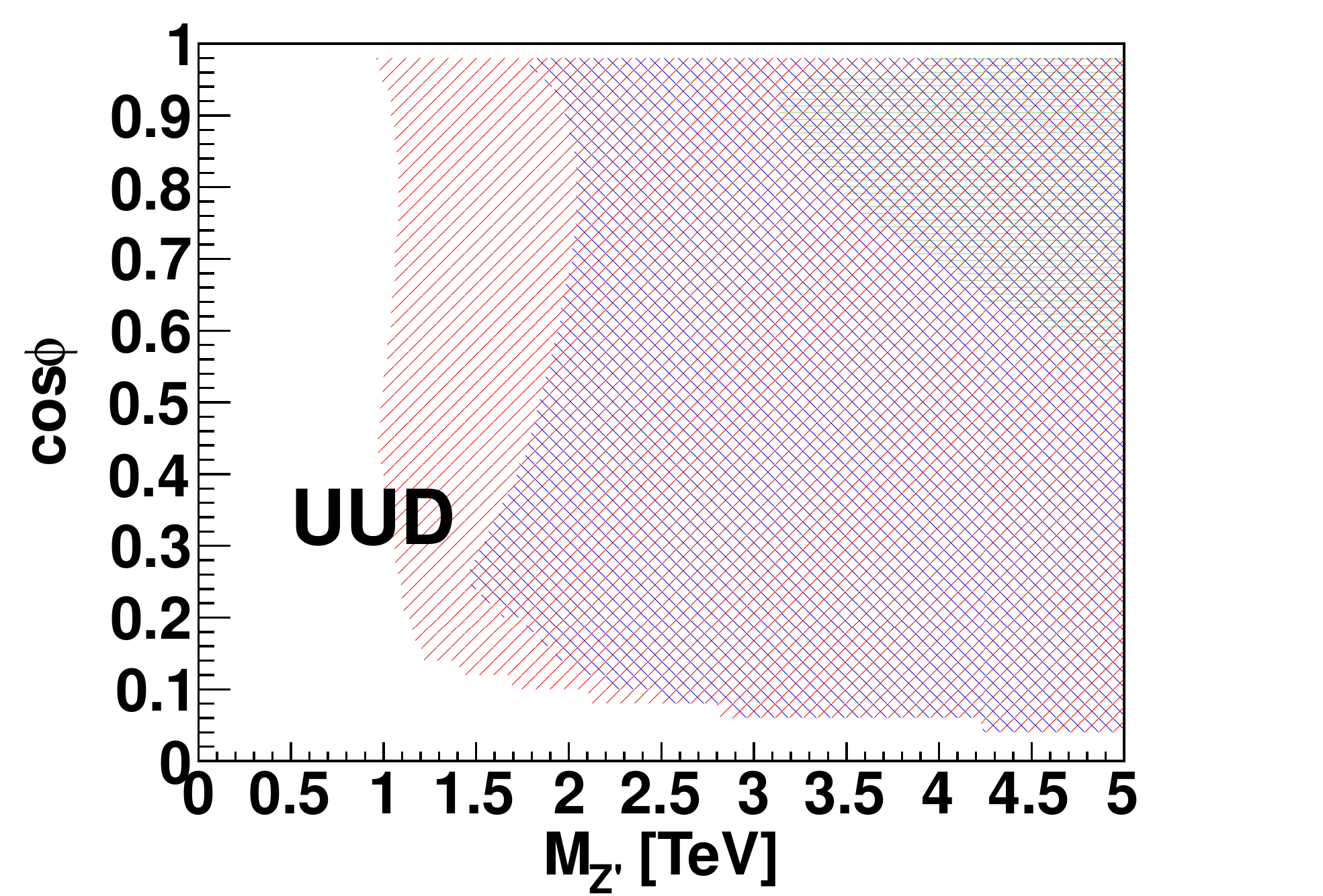}
\caption{Allowed parameter space (colored region) of the $G(221)$ model at $95\%$ CL 
in the $M_{W^\prime}- c_\phi$ plane after including indirect
and direct search constraints: EWPTs (green), 
Tevatron (red) and LHC7 (blue).The dashed black lines in LRD and SQD
represent MLR and MSQ models.}
\label{parameter_constraintsZP}
\end{figure}

The parameter scan results are plotted in 
Figs.~\ref{parameter_constraints},~\ref{parameter_constraintsWP} and~\ref{parameter_constraintsZP}. 
In order to better understand the impact of various experiment data on the parameter
space of the $G(221)$ model, we separate the indirect and direct search constraints 
into three categories:
the electroweak indirect constraints (green region) and the direct search constraints 
from the Tevatron (red region) and the LHC7 (blue region). 
In Fig.~\ref{parameter_constraints}, we note the following points:
\begin{itemize}
	\item For LRD (LRT) model, LHC7 data has stronger constraint on $W^\prime$ and $Z^\prime$ masses
          than both EWPT and Tevatron constraints,
	  and excludes the region where $W^\prime$ mass is smaller than $1.7$~TeV ($1.8$~TeV)
          and $Z^\prime$ mass is smaller than $2.3$~TeV ($3.3$~TeV);
	\item For SQD model, although the $W^\prime$ and $Z^\prime$ with degenerate masses $500$~GeV can be allowed by the EWPTs at large $c_\phi$,
	the limits from Tevatron and  LHC will excludes the region where $W^\prime$ and $Z^\prime$ masses are smaller than $1.5$~TeV.
	\item For all the models except the flavor universal models, such as LRD(T) and SQD, the EWPT data still hold the strongest constraints
	  on the $W^\prime $ and $Z^\prime$ masses, because of the non-universal flavor structure in these models. 
	\item In BP-I, with combined constraints, all the phobic models, in which the couplings of $W'$ to either quarks or leptons are suppressed, can still have relatively light $W^\prime$ around 500~GeV, but heavier $Z^\prime$ (about $1.5$~TeV);
	\item For the non-universal models, such as TFD and UUD, 
          the electroweak indirect constraints are tighter than Tevatron and LHC7 direct search constraints,
	  and push the new gauge boson mass up to more than 2 TeV (TFD) and 3 TeV (UUD), respectively.
\end{itemize}
In Figs.~\ref{parameter_constraintsWP} and~\ref{parameter_constraintsZP}, we also want to point out:
\begin{itemize}
	\item In BP-I, the $M_{W^\prime}- c_\phi$ plane shows that 
          small $c_\phi$ is favored by direct search constraints 
          because the $W^\prime$ coupling is proportional to $1/s_\phi$, 
          which leads to small $W^\prime$ production rate.
	  However, in the $M_{Z^\prime}- c_\phi$ plane, 
          small $c_\phi$ is disfavored by direct search constraints
          because the mass relation $M_{Z^\prime} \simeq M_{W^\prime}/c_\phi$,
          push the exclusion region of small $c_\phi$ to larger $M_{Z^\prime}$.
	\item In BP-II, the shape in small $c_\phi$ region are very similar 
		  because the the production cross section of $W^\prime$ and $Z^\prime$ are proportional 
          to $\tan\phi$ in all models such as SQD, TFD, and UUD. 
		  Because quarks and leptons are un-unified in UUD, the gauge couplings to leptons are proportional to $\cot\phi$,
		  which implies the large $c_\phi$ region is also disfavored. 
	\item Within the direct searches, for LRD(T) the most sensitive constraint
          comes from $W^\prime$ leptonic decay channel, while for phobic models, 
          the tightest constraints comes from $Z^\prime$ leptonic decay channel. 
          This explains that the contours in the phobic models 
          have similar shapes, but different from those in the LRD(T) models. % but not the ones in LRD(T) models. 
\end{itemize}

\subsection{$V^\prime$ decay width}

\begin{figure}
\includegraphics[width=0.32\textwidth]{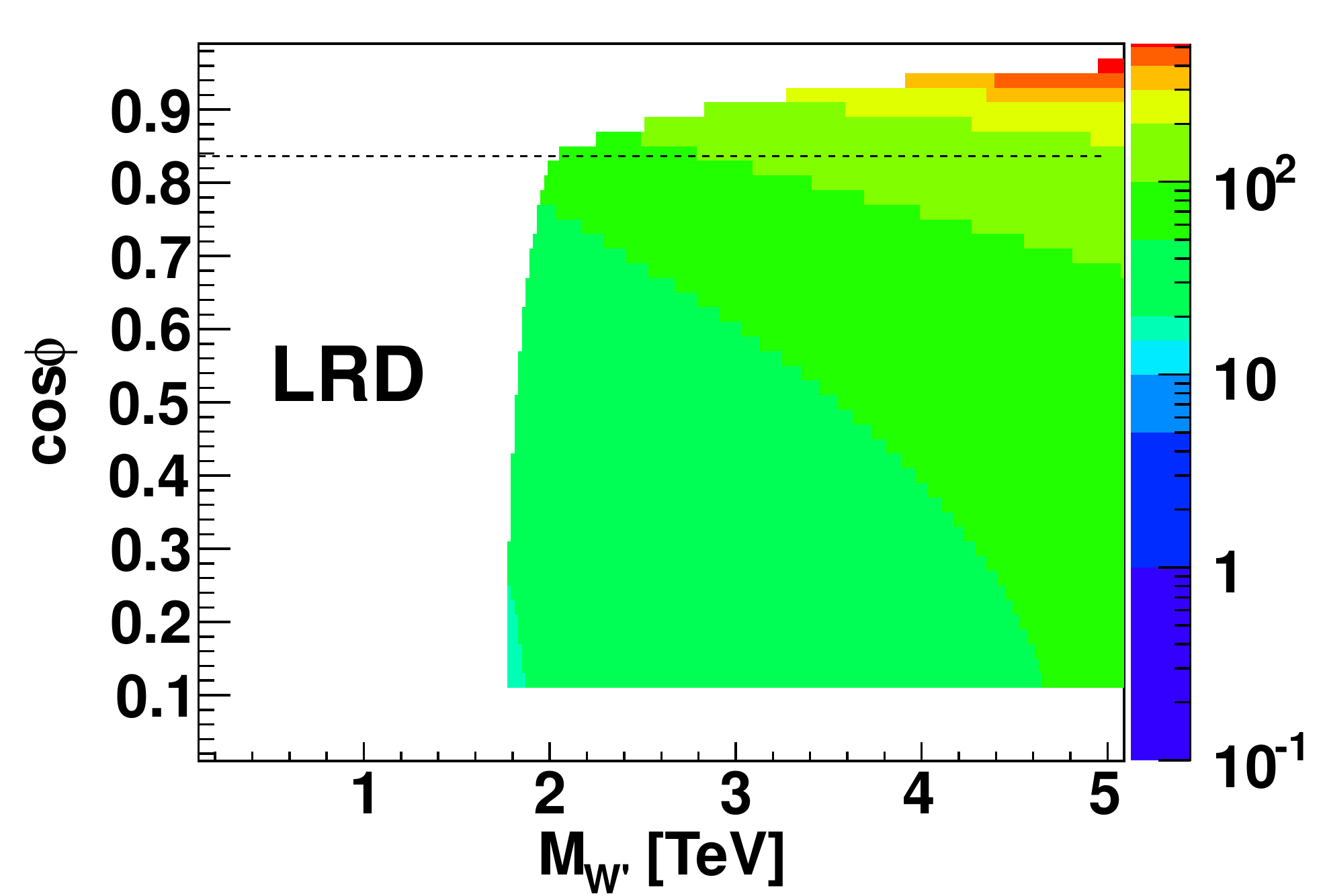}
\includegraphics[width=0.32\textwidth]{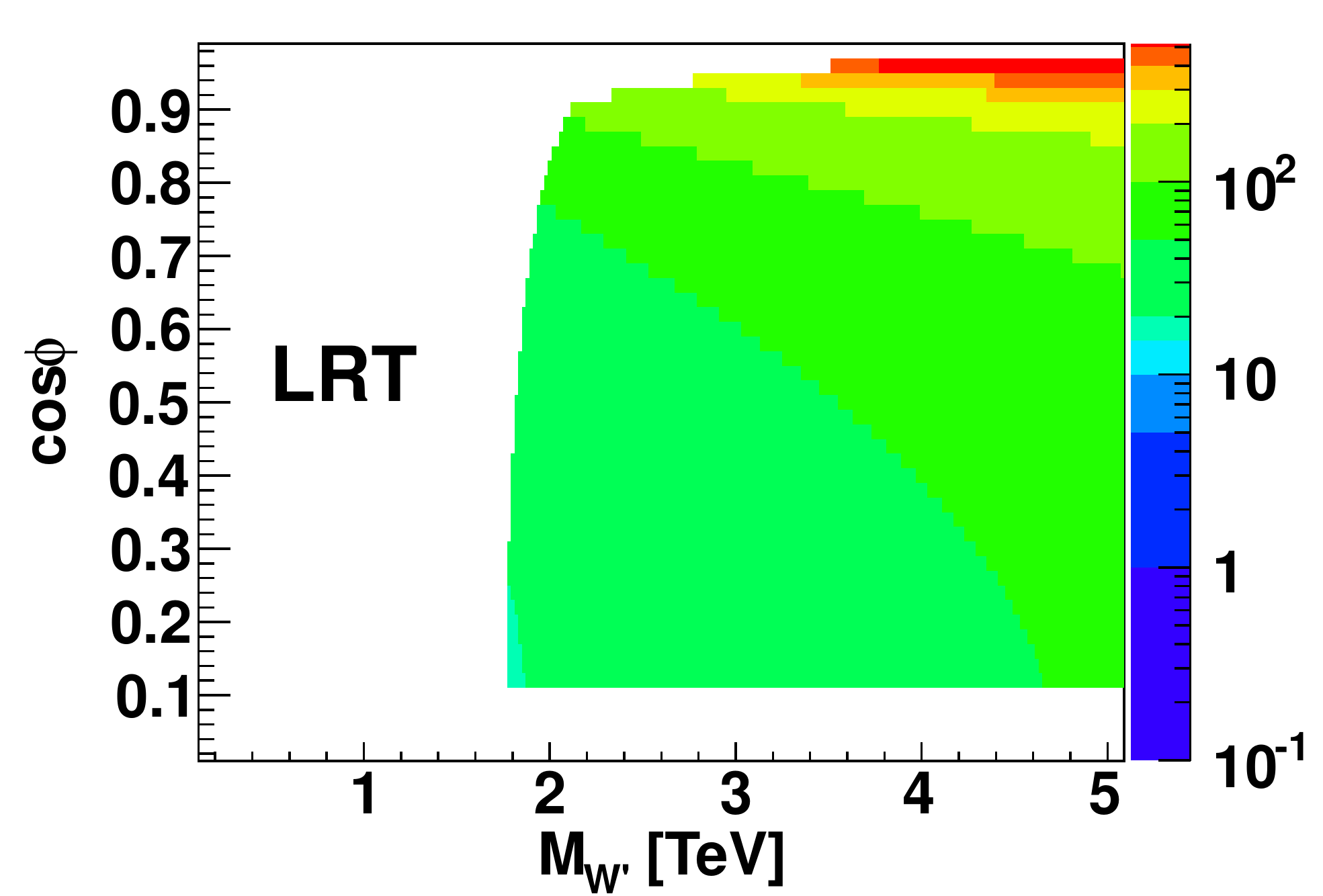}
\includegraphics[width=0.32\textwidth]{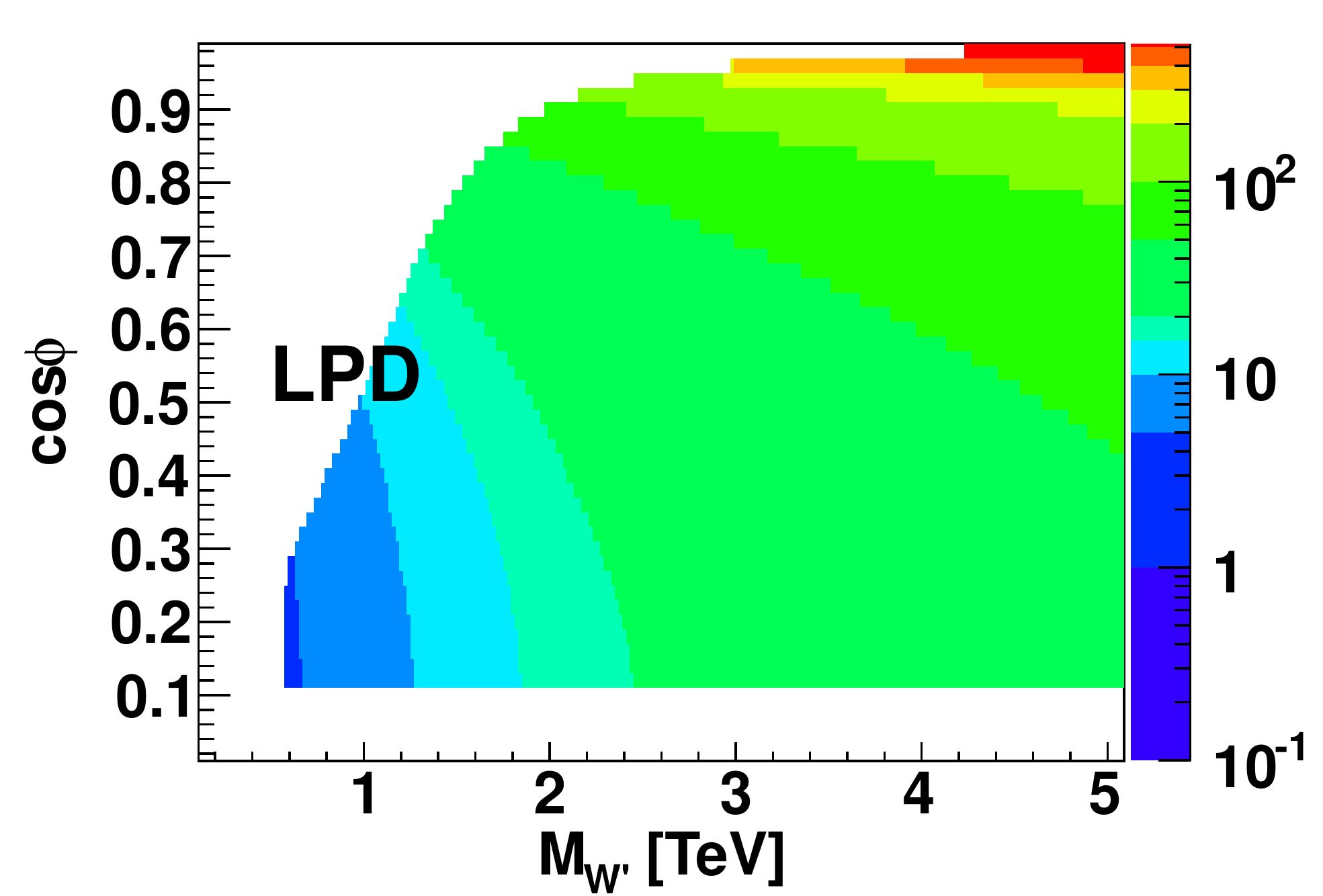}
\includegraphics[width=0.32\textwidth]{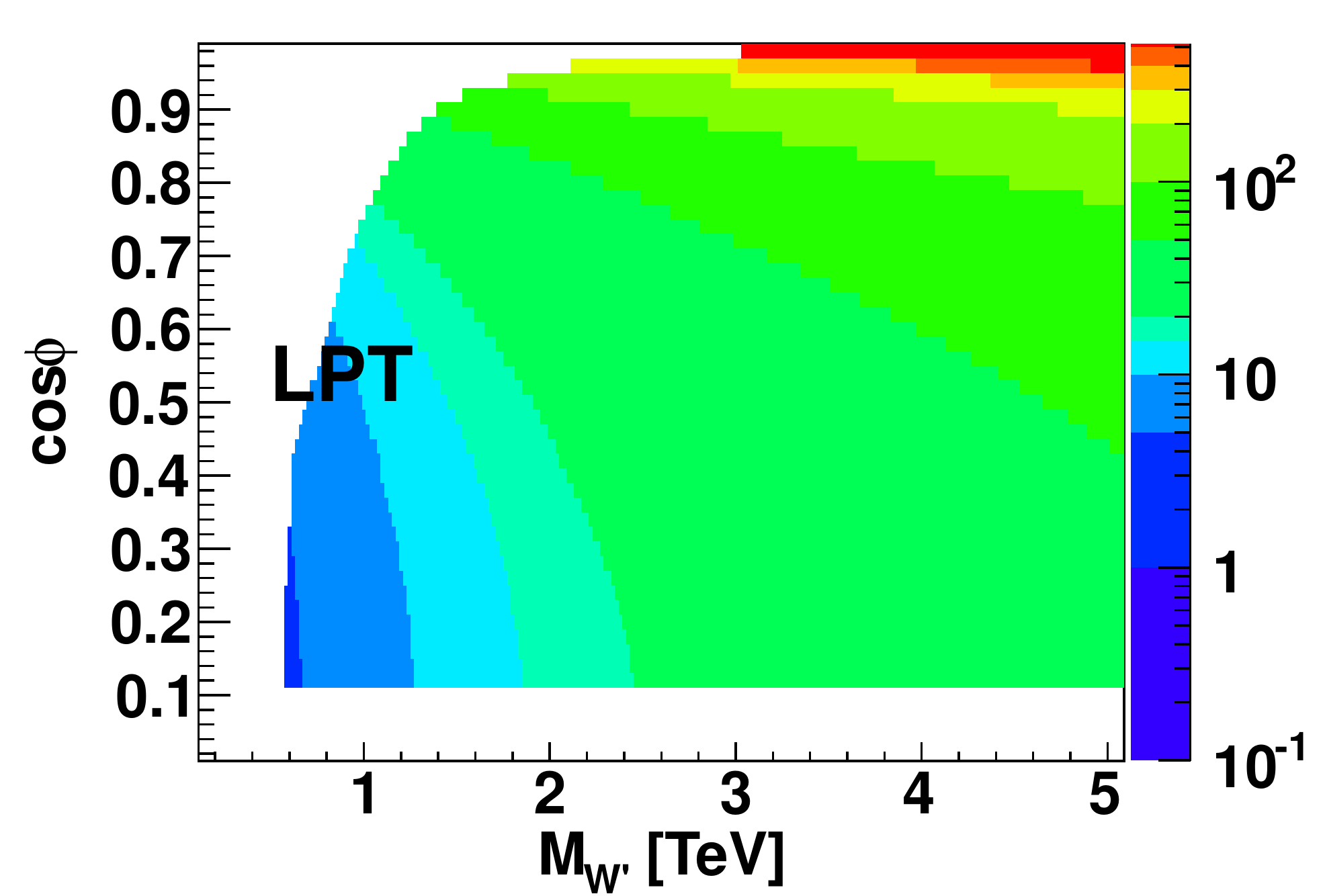}
\includegraphics[width=0.32\textwidth]{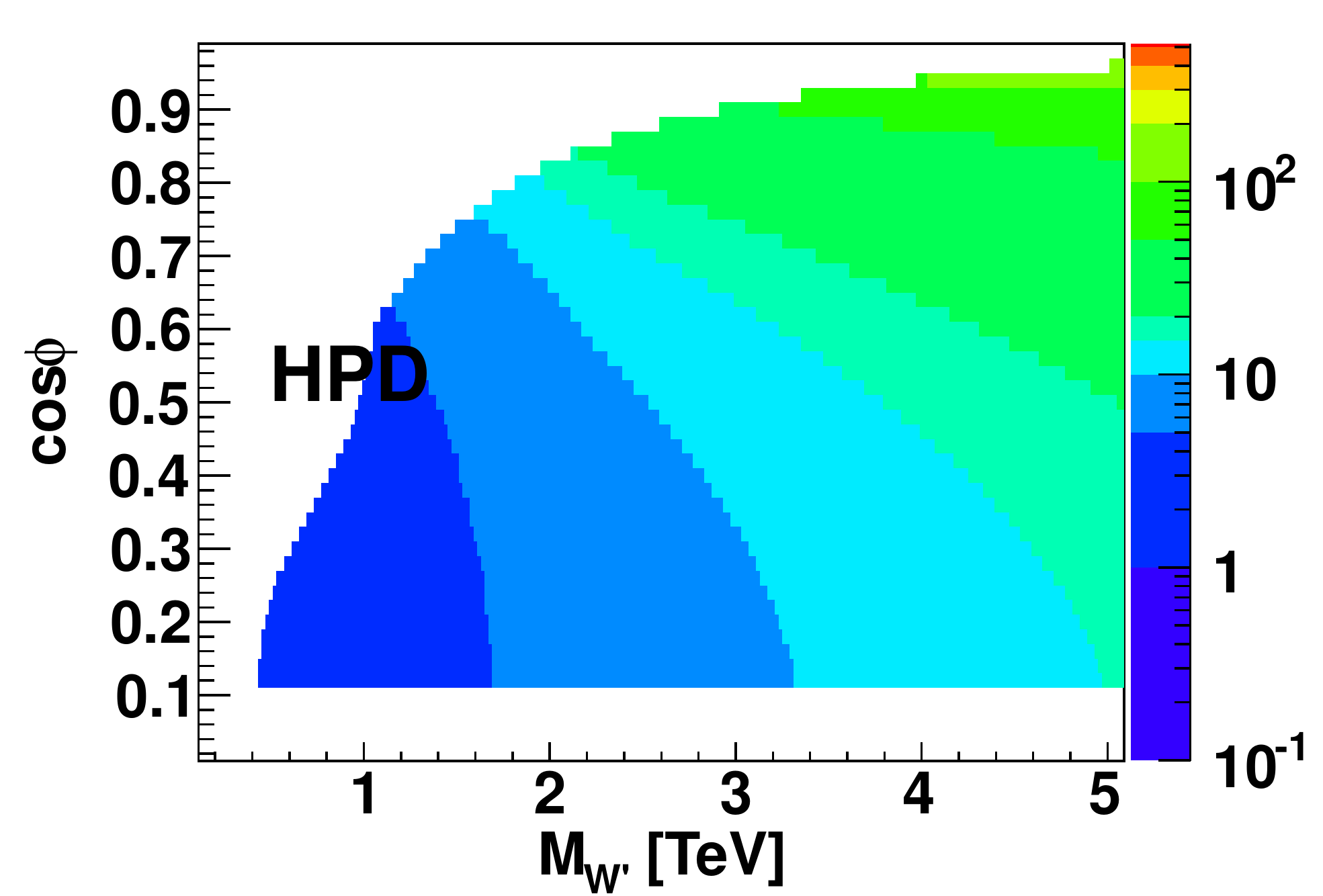}
\includegraphics[width=0.32\textwidth]{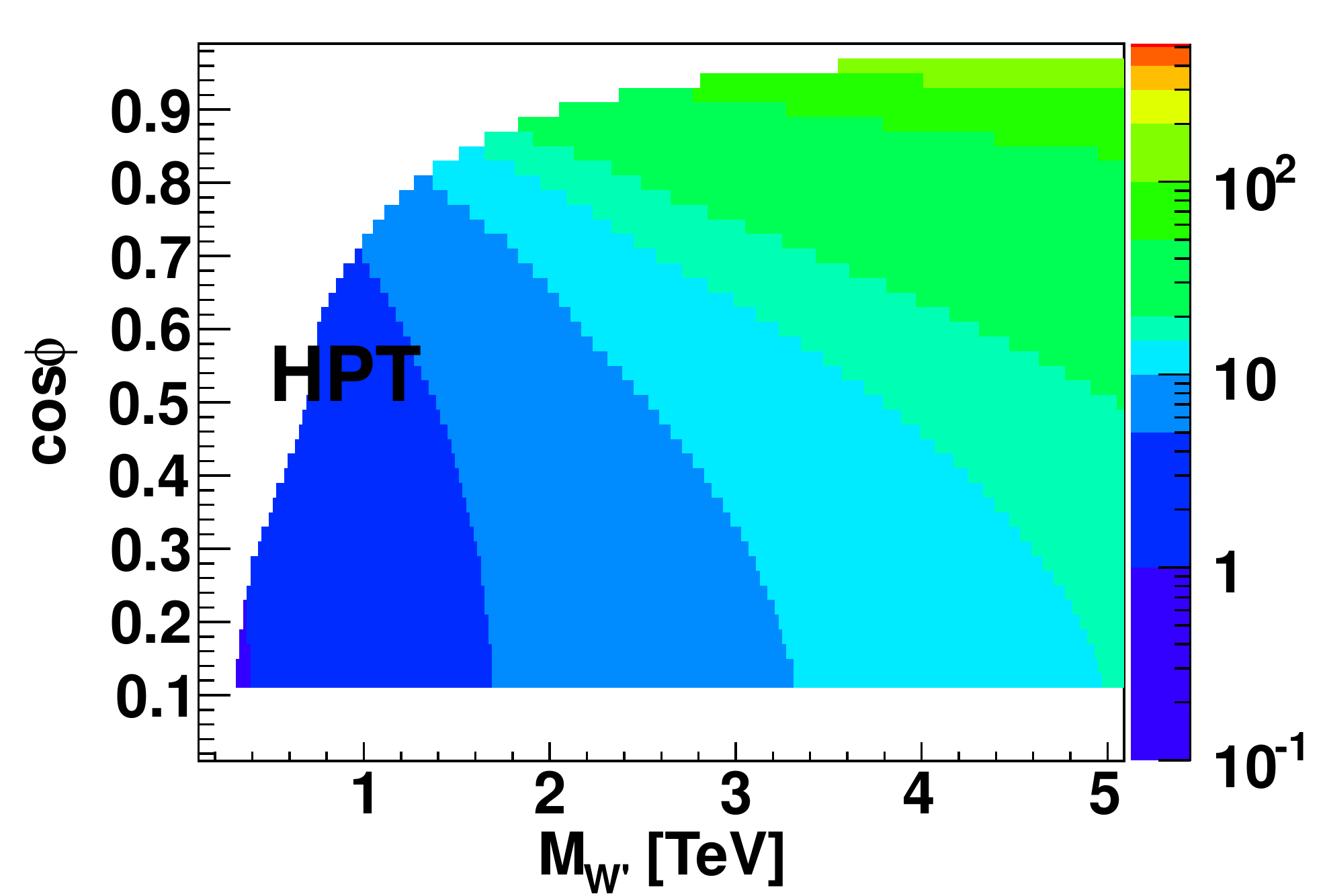}
\includegraphics[width=0.32\textwidth]{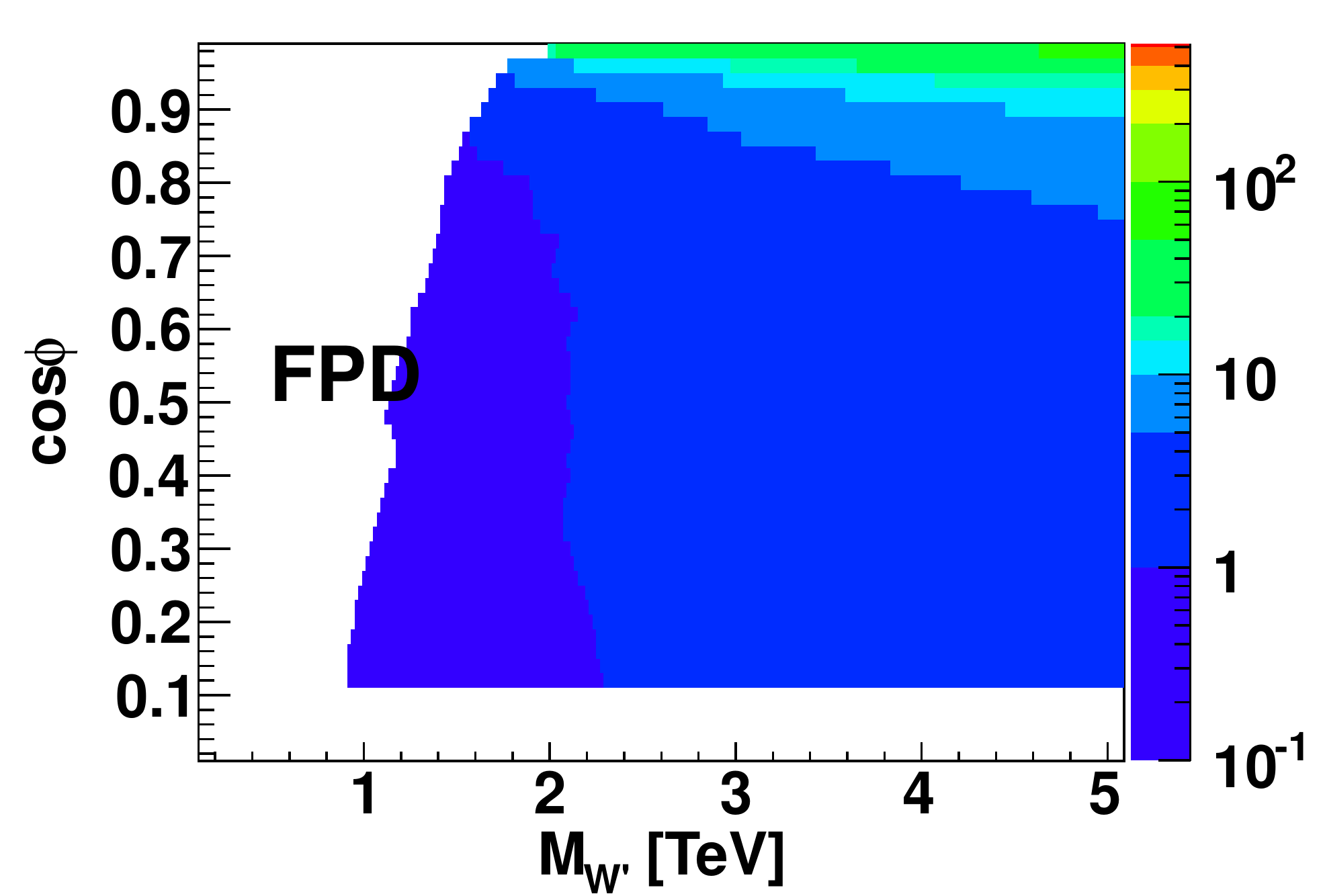}
\includegraphics[width=0.32\textwidth]{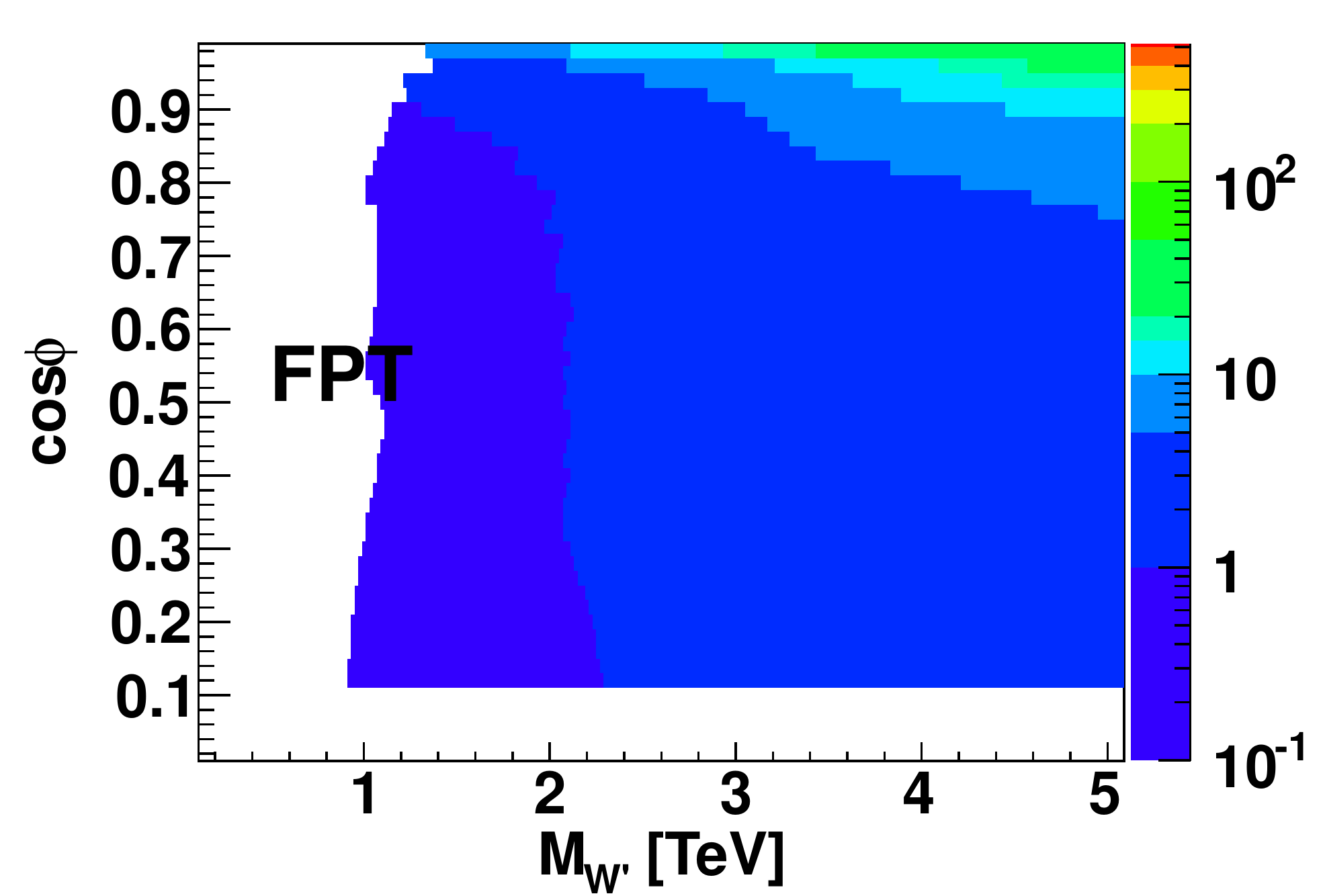}
\includegraphics[width=0.32\textwidth]{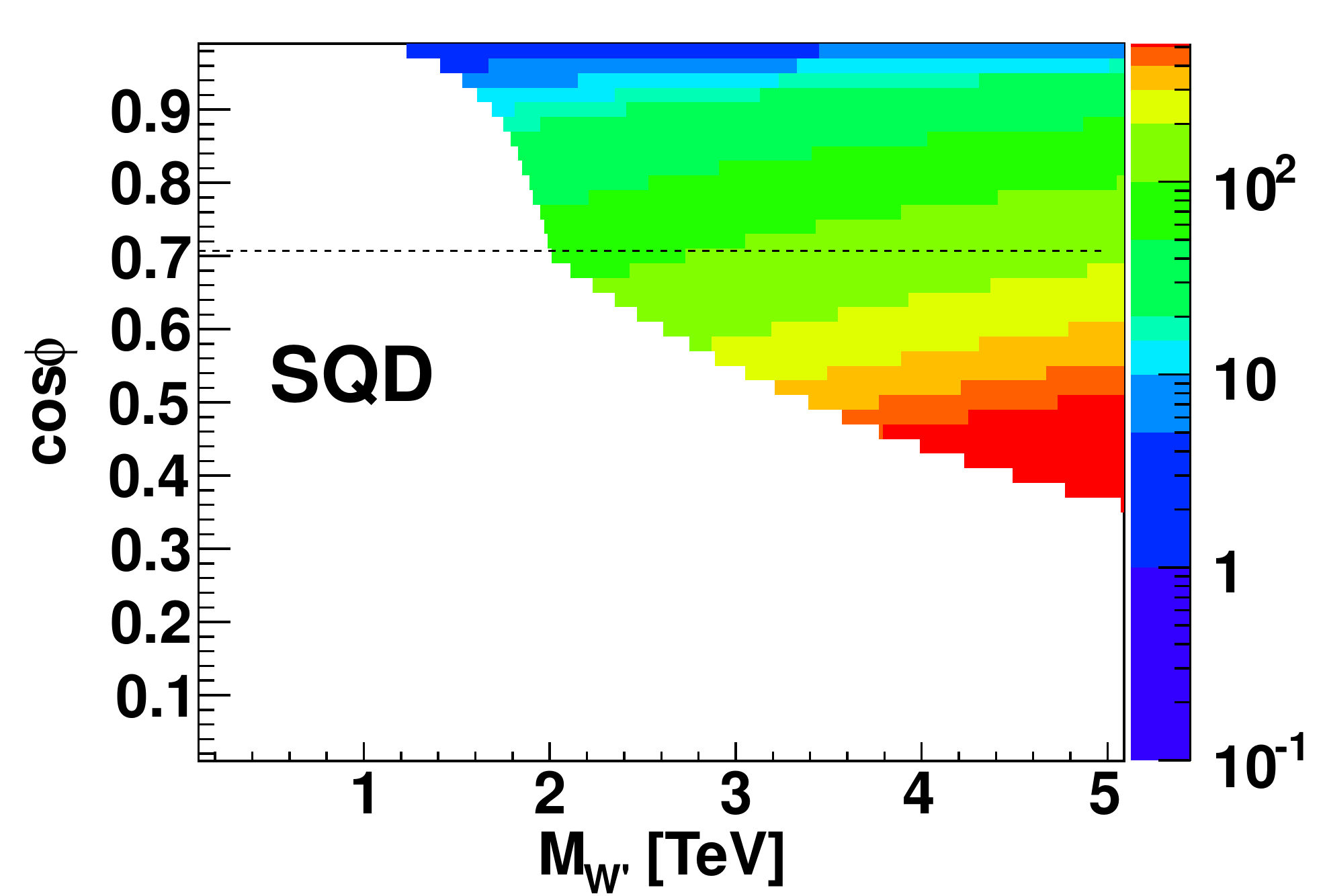}
\includegraphics[width=0.32\textwidth]{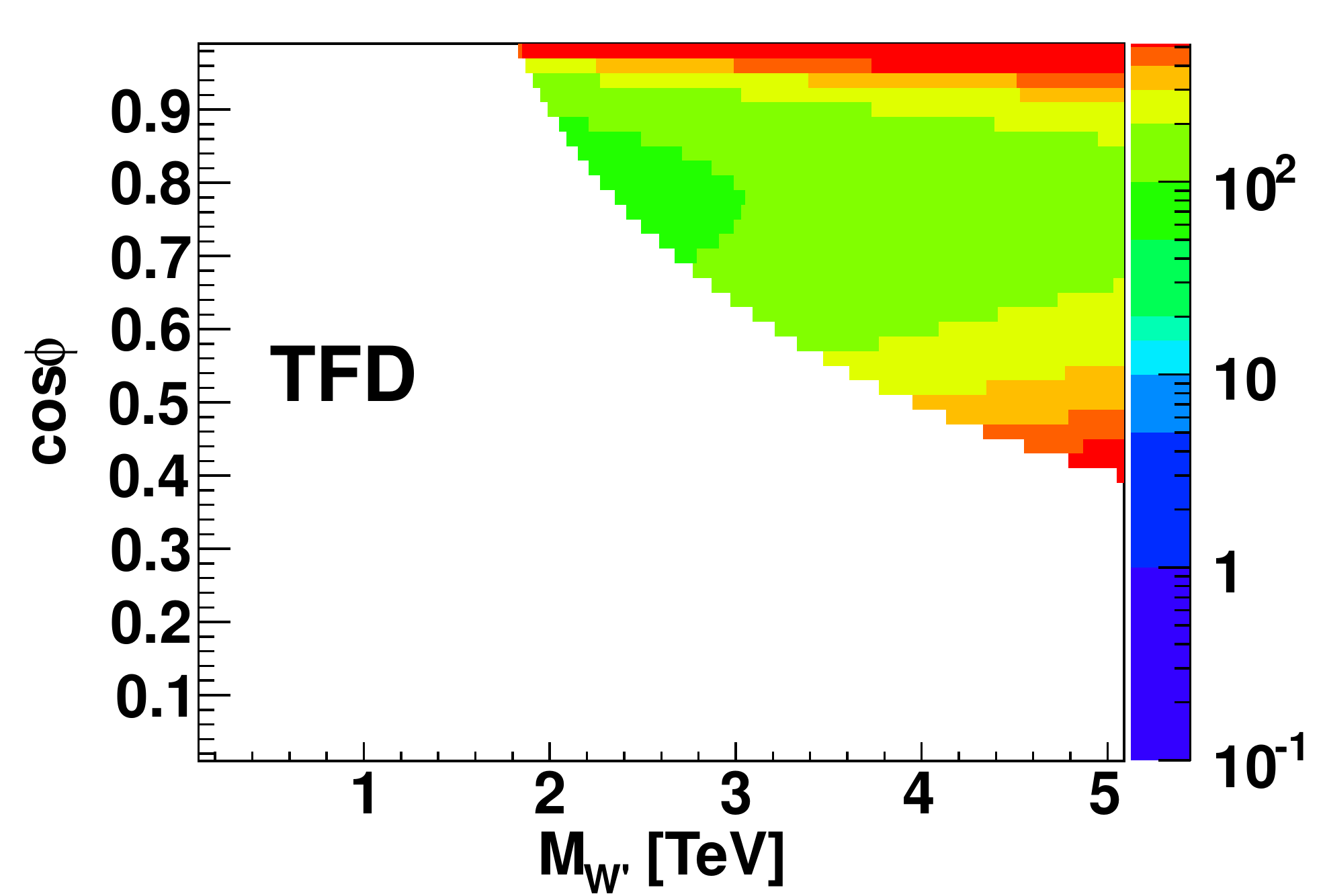}
\includegraphics[width=0.32\textwidth]{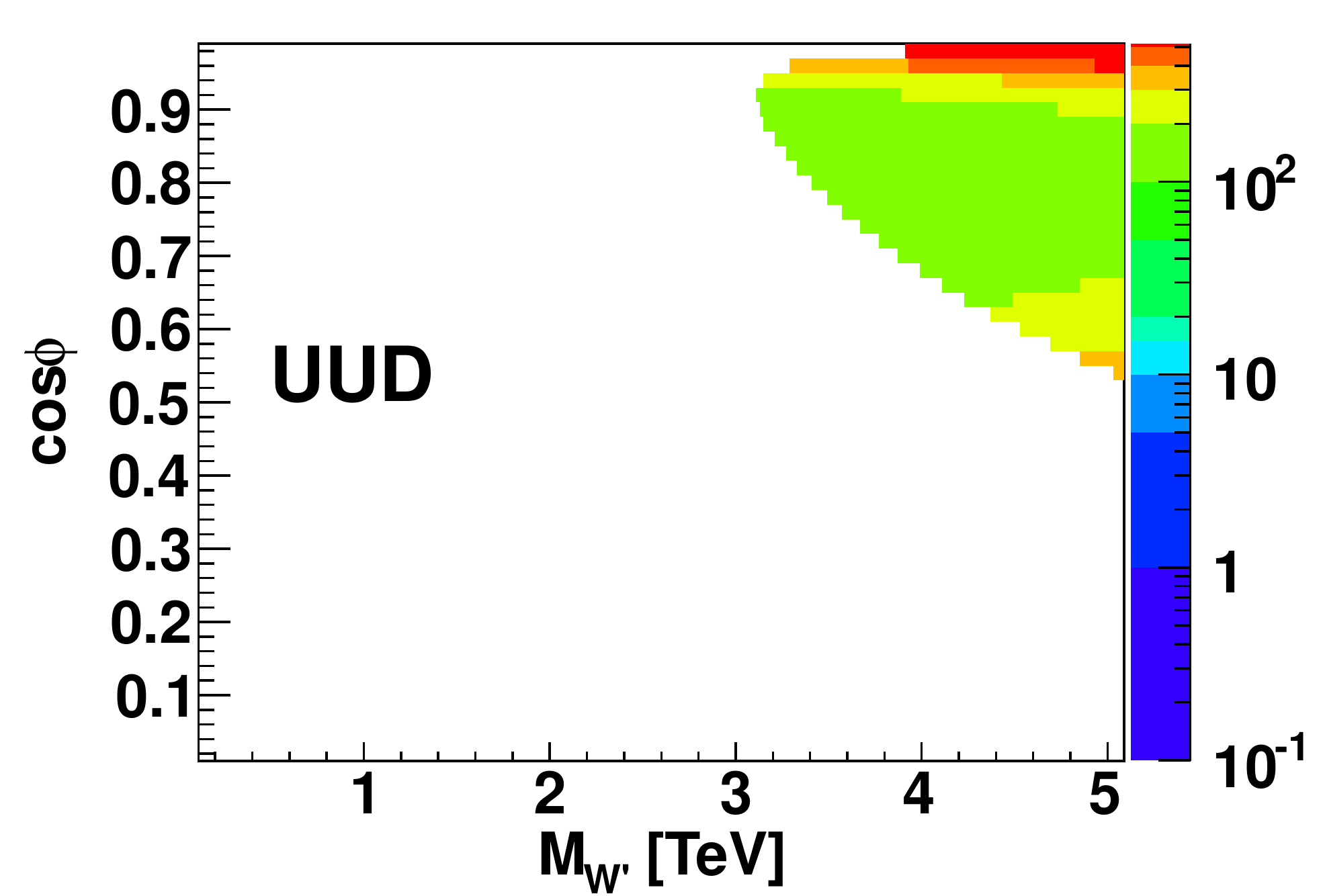}
\caption{The largest total decay widths (GeV) of $W^\prime$ 
in the $M_{W^\prime}- c_\phi$ plane for different $G(221)$ models
within the allowed parameter space constrainted by current experiment data. 
The color palette shows the largest total decay widths in unit GeV. }
\label{Wwidth}
\end{figure}

\begin{figure}
\includegraphics[width=0.32\textwidth]{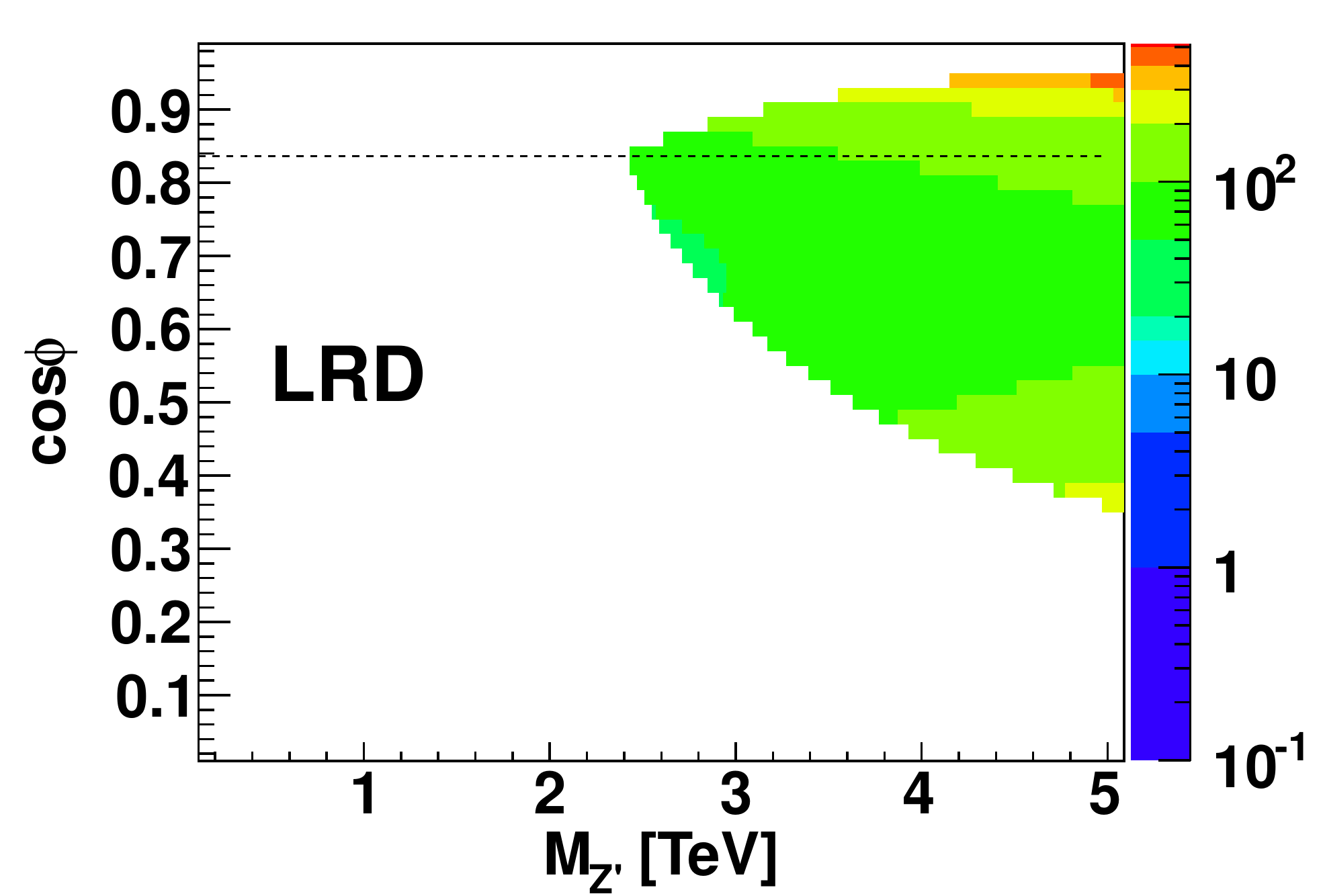}
\includegraphics[width=0.32\textwidth]{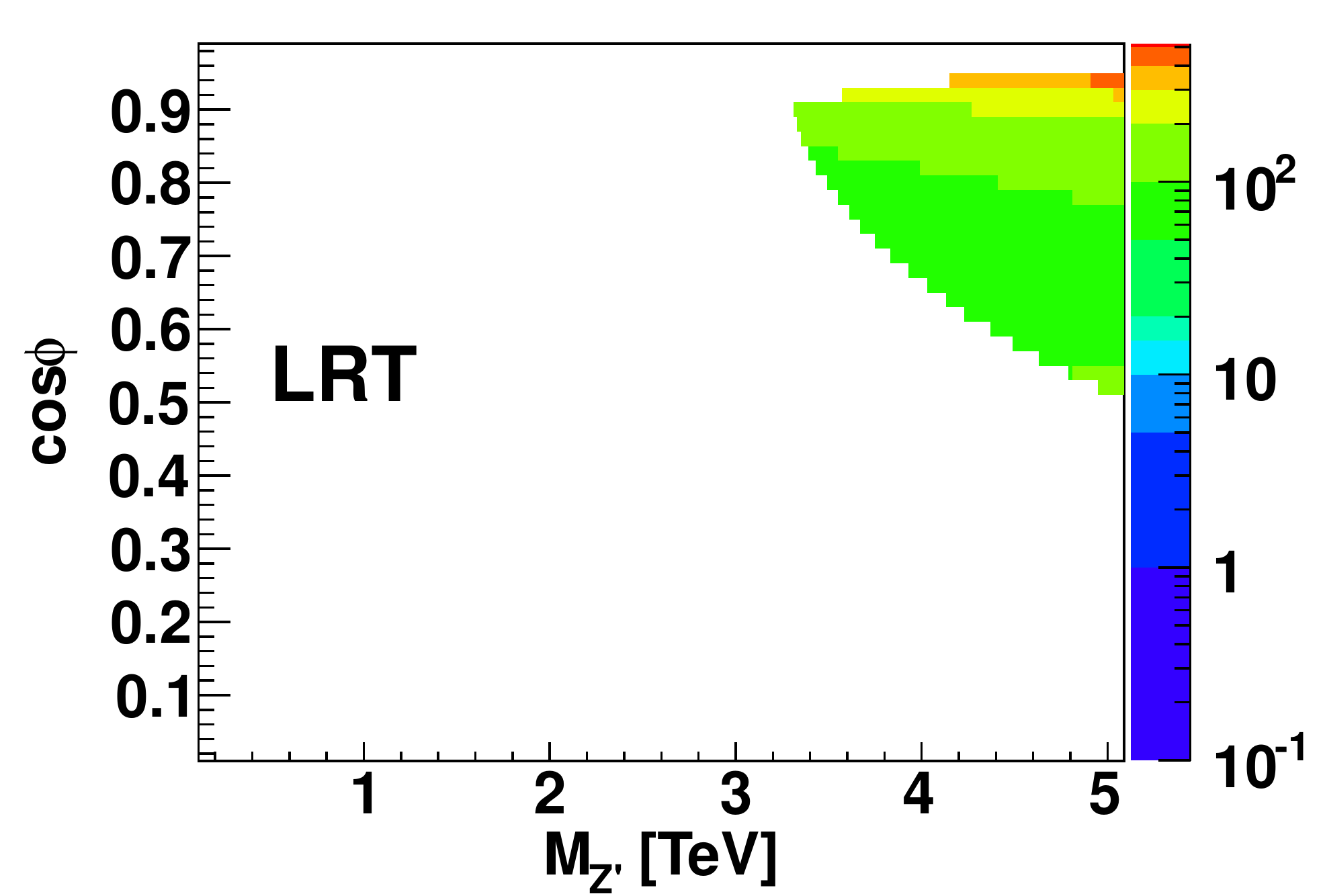}
\includegraphics[width=0.32\textwidth]{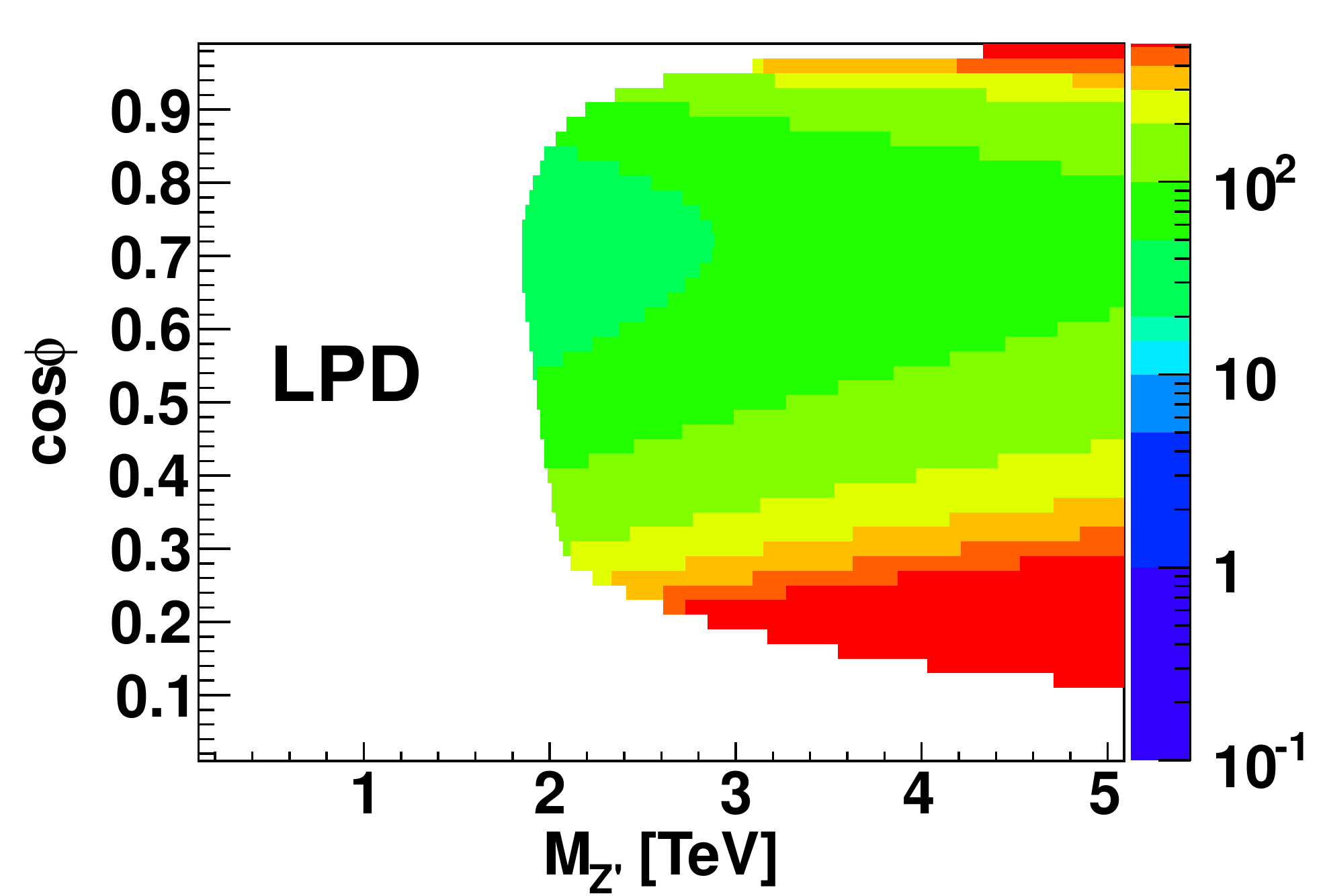}
\includegraphics[width=0.32\textwidth]{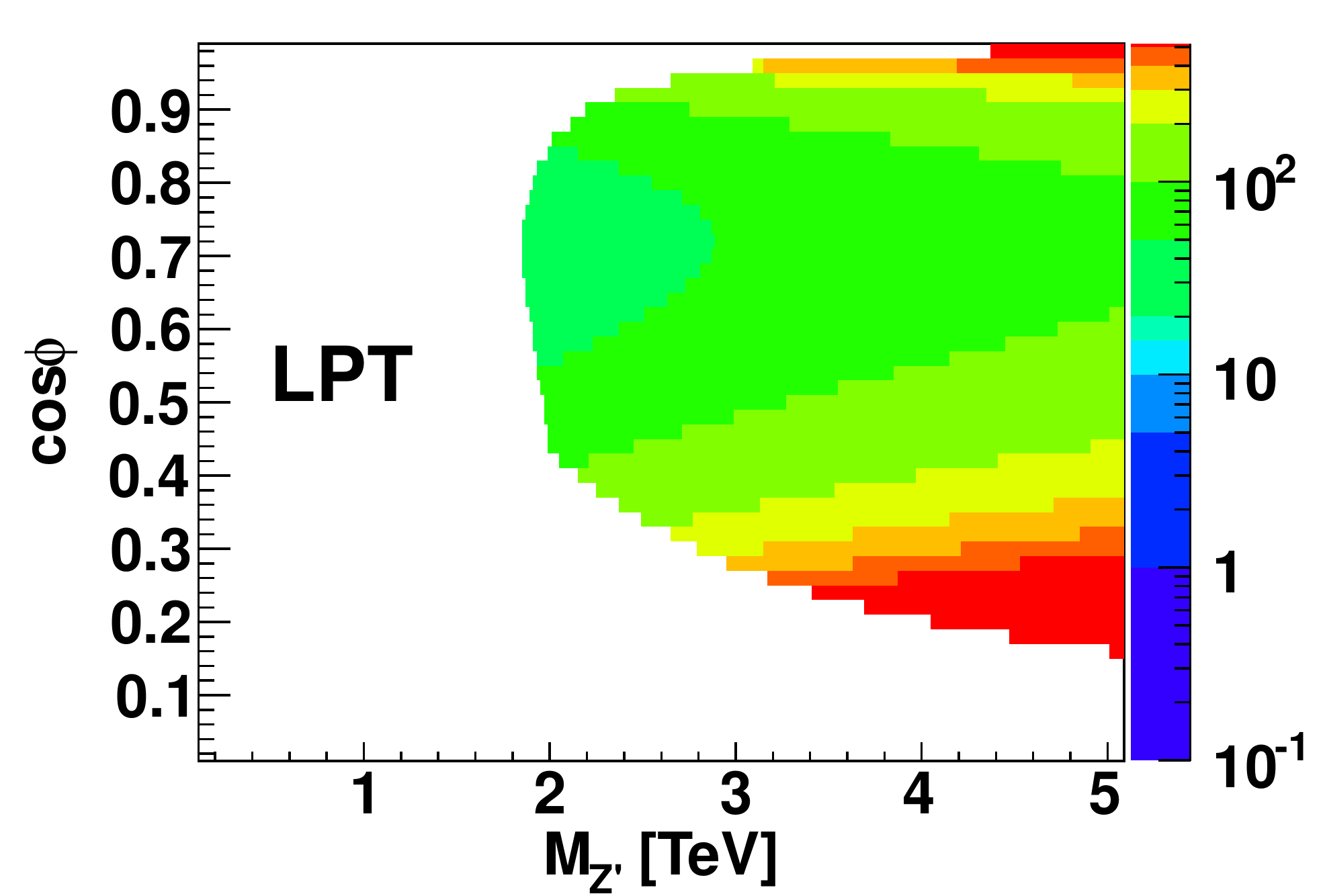}
\includegraphics[width=0.32\textwidth]{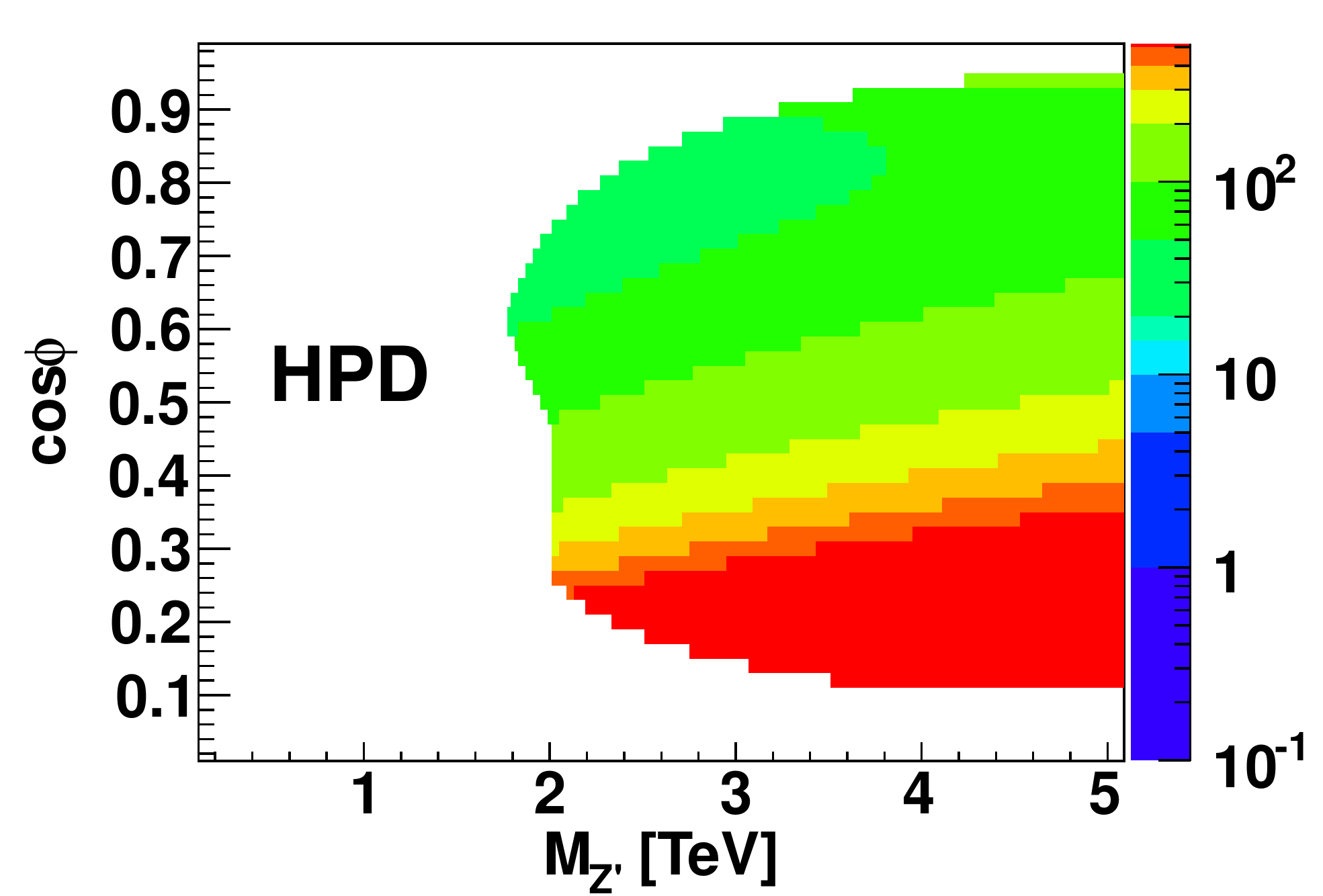}
\includegraphics[width=0.32\textwidth]{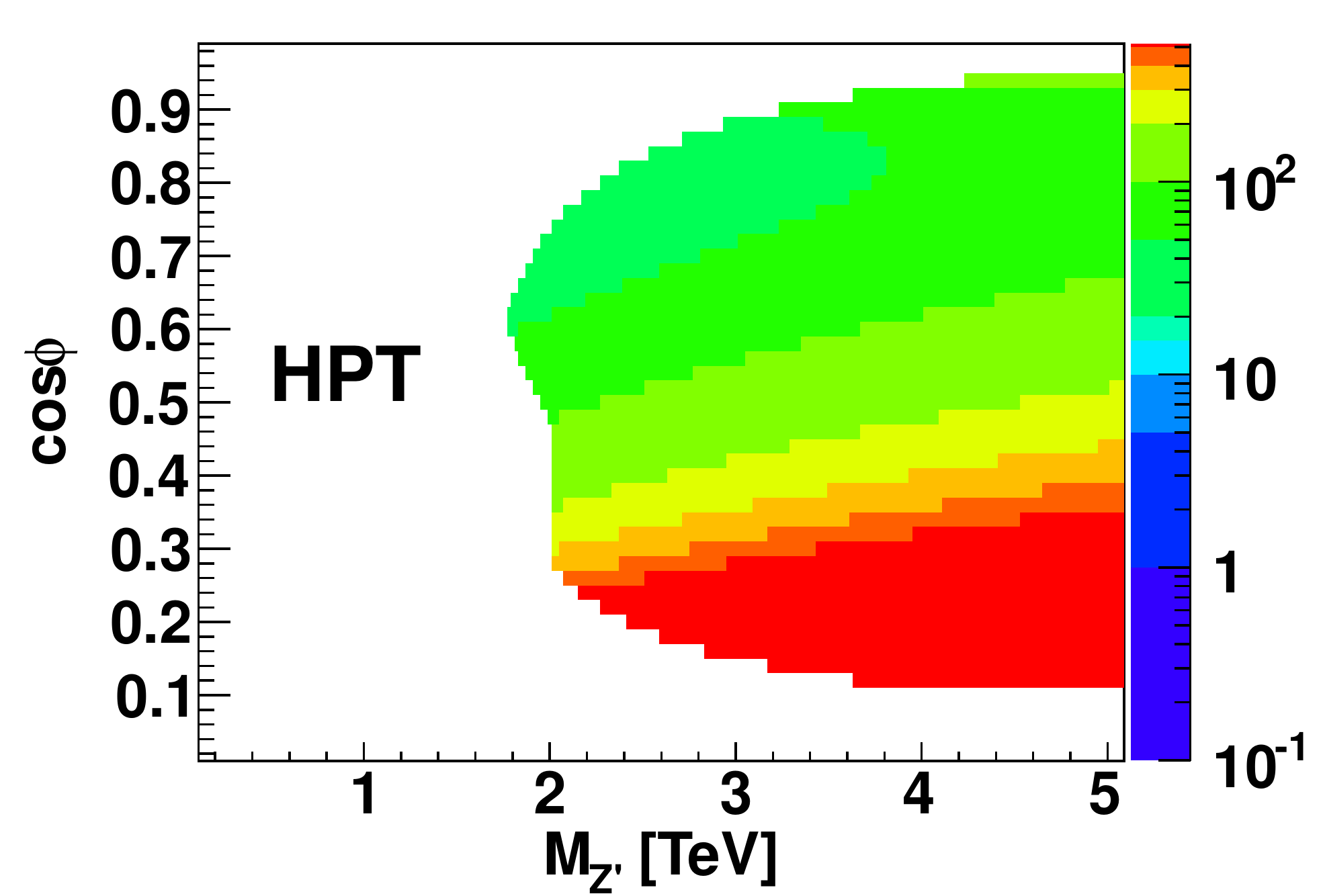}
\includegraphics[width=0.32\textwidth]{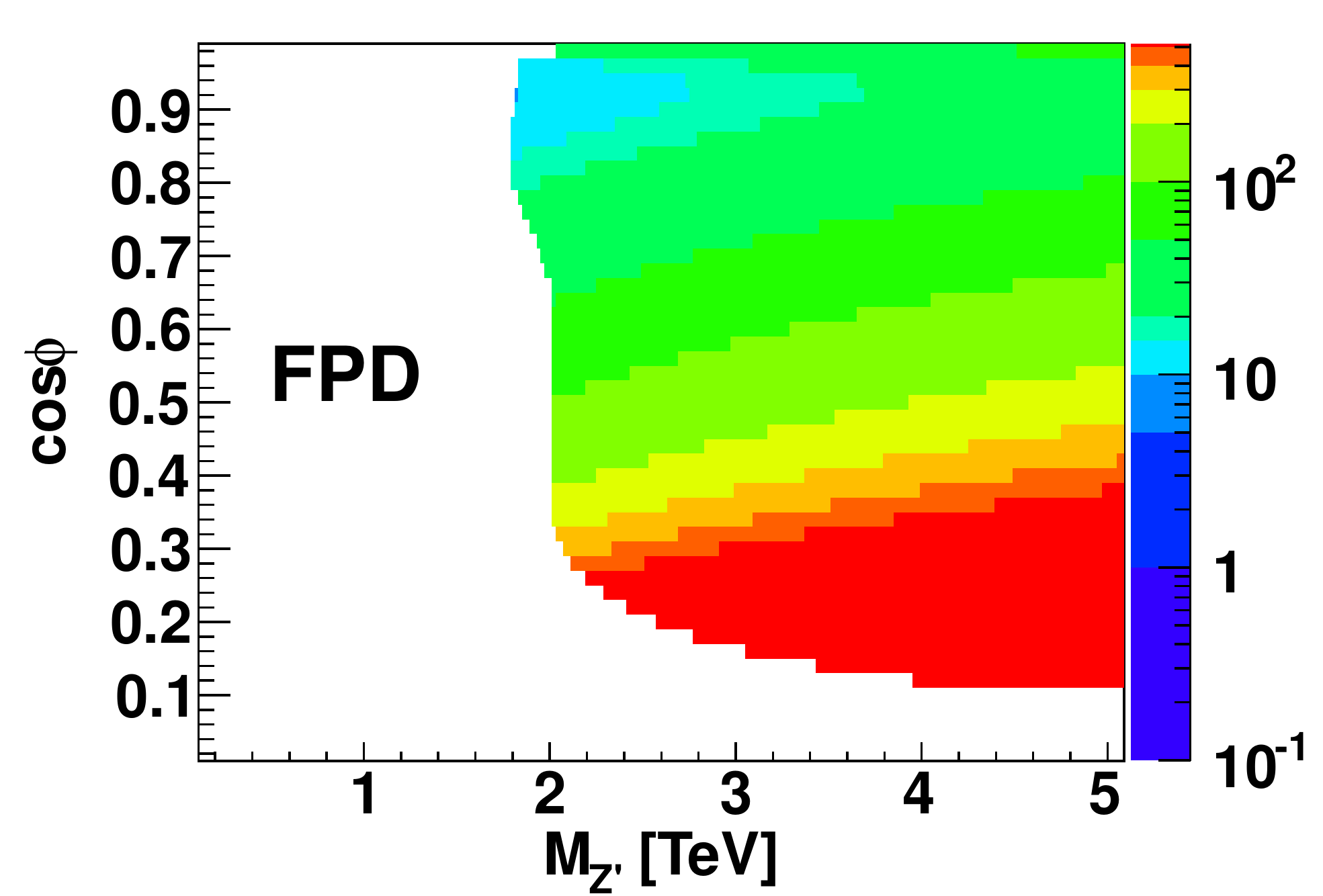}
\includegraphics[width=0.32\textwidth]{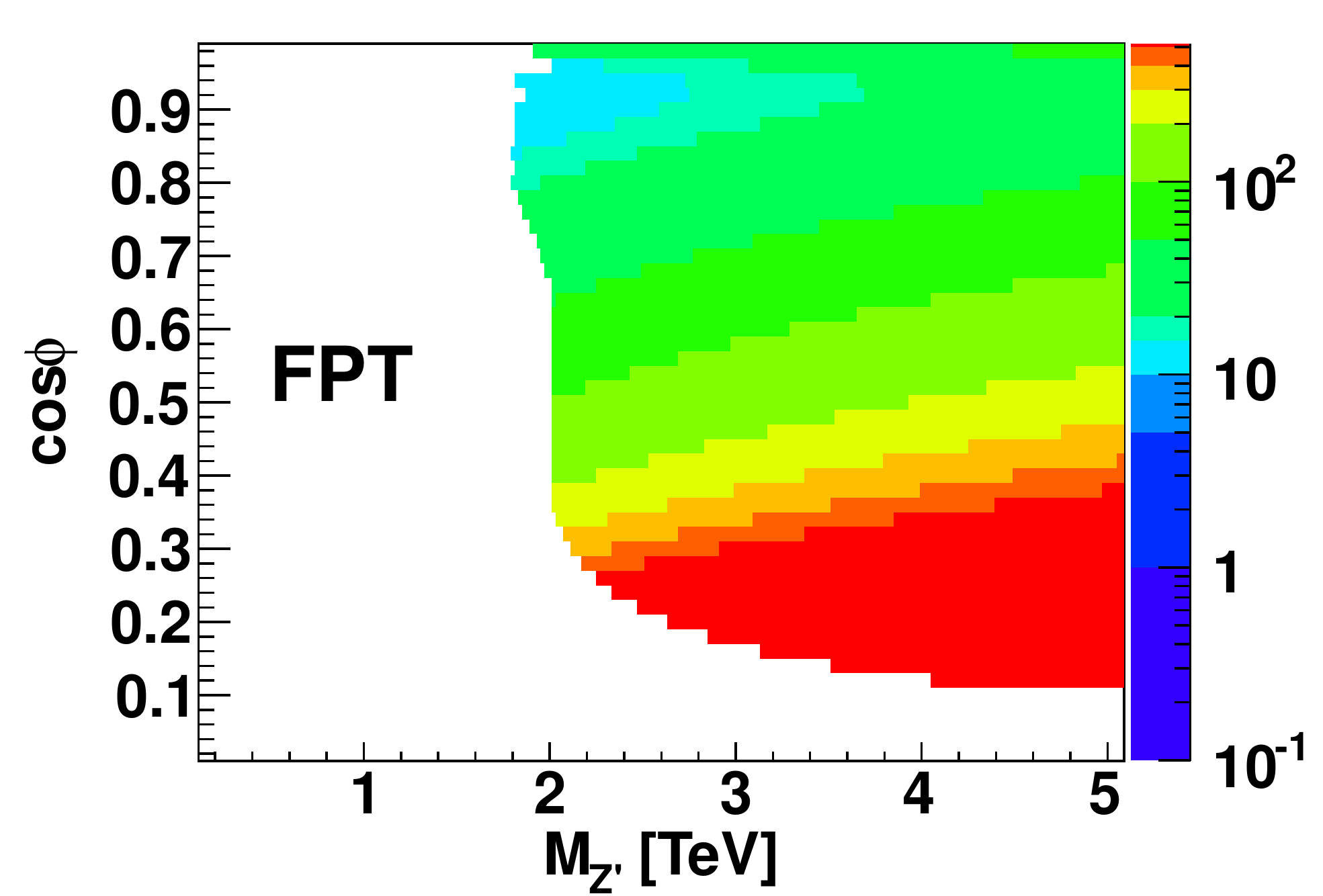}
\includegraphics[width=0.32\textwidth]{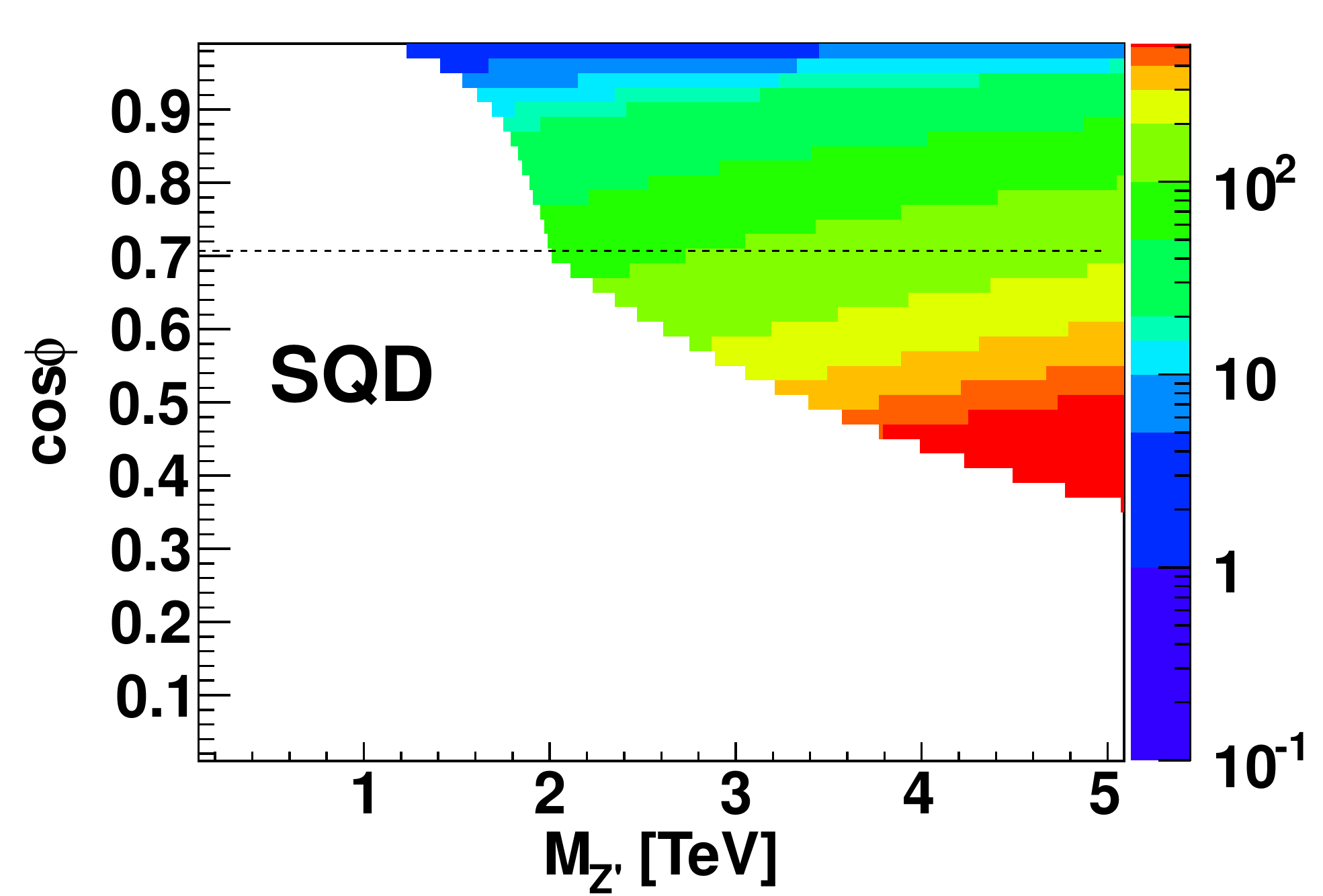}
\includegraphics[width=0.32\textwidth]{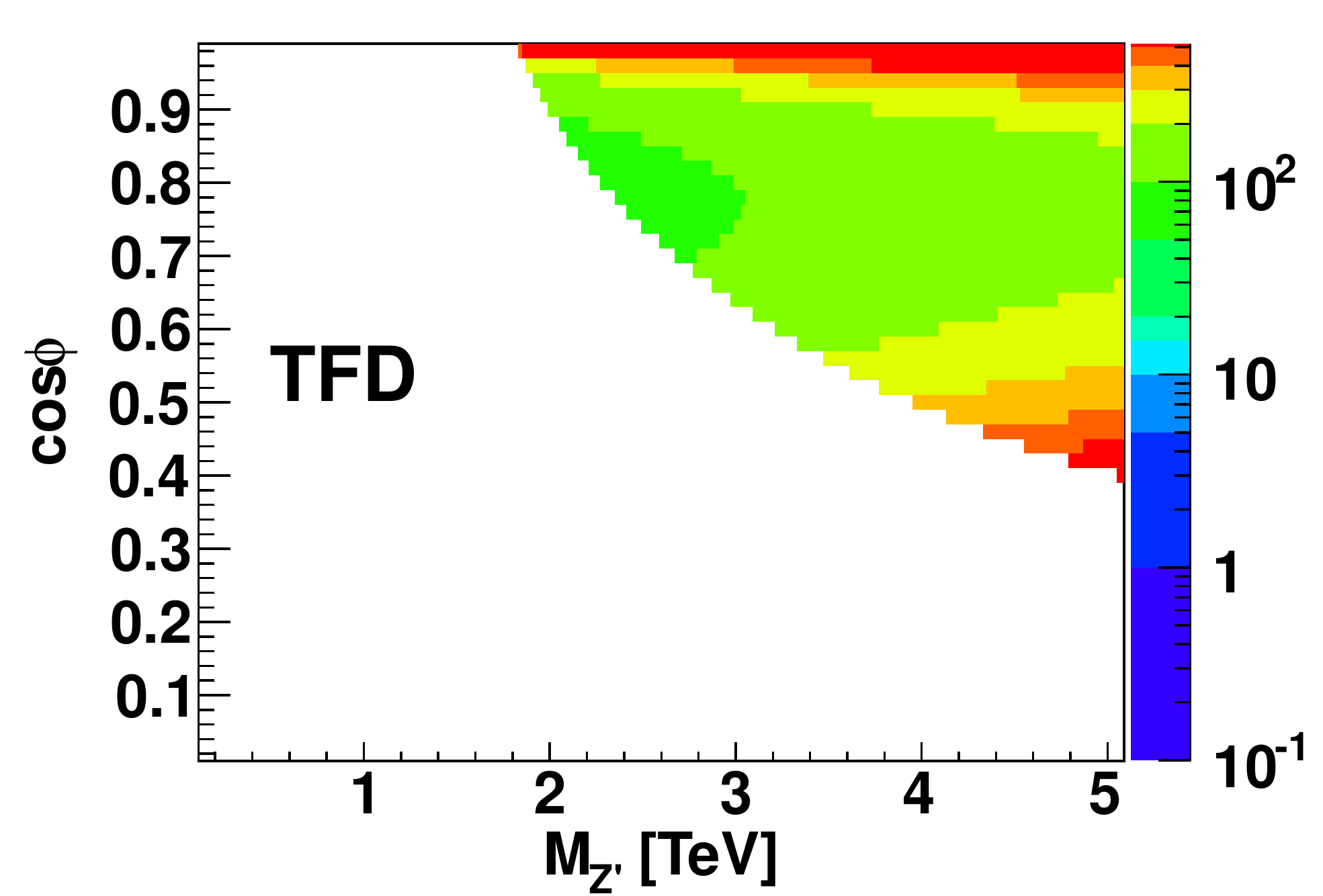}
\includegraphics[width=0.32\textwidth]{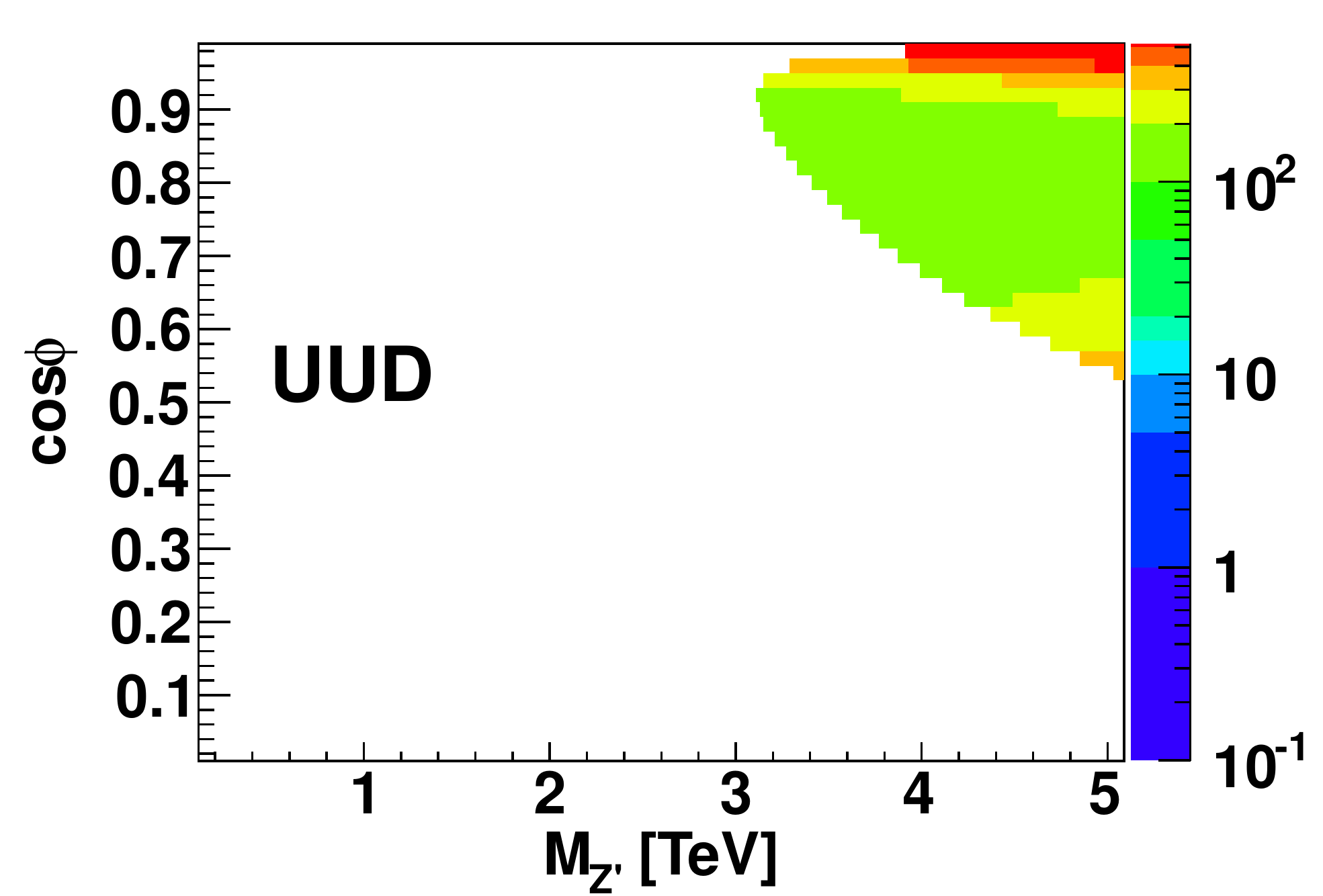}
\caption{The largest total decay widths (GeV) of $Z^\prime$ 
in the $M_{Z^\prime}- c_\phi$ plane for different $G(221)$ models 
within the allowed parameter space constrainted by current experiment data. 
The color palette shows  the largest decay total widths in unit GeV.}
\label{Zwidth}
\end{figure}

Figures~\ref{Wwidth} and~\ref{Zwidth} show 
the largest total decay widths of $W^\prime$ and $Z^\prime$ 
on the parameter space of $G(221)$ models, where we have considered 
the constraint from low energy precision data, LEP, Tevatron and LHC7 data.
We can see that the ratio of total width with respect to the relevant mass
is a few percent in most region of parameter space.
The ratio of total decay width  to mass can reach at most $10\%$ only 
in some edge regions of parameter space.
Therefore, the narrow width approximation in our study is valid. 
Besides, by comparing Figs.~\ref{Wwidth} and~\ref{Zwidth} 
we can see for the phobic models 
the $Z^\prime$ width is much larger than 
the $W^\prime$ width, which is usually below 10 GeV.

%%%%%%%%%%%%%%%%%%%%%%%%%%%%%%
%                                                                               		  %
%        Section ?: LHC potential                                           %
%													                           %
%%%%%%%%%%%%%%%%%%%%%%%%%%%%%%

\section{LHC Discovery Potential and Signature Space }

In early LHC7 results, the combined constraints from current
direct searches and indirect EWPTs play a crucial role in specifying the unexplored
parameter space. Given the allowed parameter space discussed in
the previous sections, we are able to 
provide the following information:
\begin{itemize}
\item  The integrated luminosity, 
with which the LHC can discover the $W^\prime$ and/or $Z^\prime$ 
for certain mass in various $G(221)$ models. 
\item The region of parameter space that could
be accessed for different luminosities and energies in the LHC run.
\item Possibility to identify NP models in our classification,
once the $W^\prime$ and/or $Z^\prime$ are discovered.
\end{itemize}
To be specific, we consider two different scenarios: an early run with $\sqrt{s}=7$ TeV and an
integrated luminosity of $5.61~{\rm fb}^{-1}$ 
(the maximal integrated luminosity reached at the 7~TeV); 
a long run with $\sqrt{s}=14$ TeV and ${\cal O}(10^3)$ fb$^{-1}$ integrated luminosity.  

To get the expected luminosity contour, one has to calculate the signal and background 
cross sections at LHC7 and LHC14 for each point in the parameter space of the models. 
In principle, the complete Monte Carlo simulations for the signal and background
including efficiency analysis in the $G(221)$ models have to be
used to obtain the needed luminosity for the discovery or exclusion
at 7 TeV and 14 TeV. However, in the Drell-Yan production process, 
all the model-independent effects, including the
kinematic cuts, can be factorized out from the model-dependent part, which
only depends on the gauge couplings and branching ratios, as shown
in Sec.~II. Therefore, the simulation on one benchmark model, such as the
sequential $W^\prime$ and $Z^\prime$ model, can provide the needed luminosity information for the other models.
At the LHC7, the complete simulation on the signal and backgrounds including
detection efficiency has been done in Refs.~\cite{Aad:2011yg,Collaboration:2011dca}.
At the LHC14, the ATLAS TDR~\cite{Aad:2009wy} have done the detailed studies 
on the discovery potentials for the sequential $W^\prime$ and $Z^\prime$ model. The luminosity needed for
 other new physics models can be obtained by properly scaling the luminosity obtained for the sequential model.

Here we summarize the event analysis procedures at the current LHC and in the ATLAS TDR.
At the LHC7, we adopt the ATLAS simulation and analysis with integrated luminosity at about $1$ fb$^{-1}$. 
Both electron and muon channels are considered in both $W^\prime$ and $Z^\prime$ searches.
For the $W^\prime$ searches, the missing energy in both channels requires 
to be above the threshold energy of $25$ GeV. 
Furthermore, the cut on the transverse mass of the lepton and 
missing energy system varies as the $W^\prime$ mass increases. 
For more detailed information, please refer to Refs.~\cite{Aad:2011yg,Collaboration:2011dca}. 
In the ATLAS TDR, for the sequential $W^\prime$, the simulation on the lepton 
plus missing transverse energy signal at high mass region
is performed. We list the event selection and cut-based analysis as follows:
\begin{itemize}
	\item Events are required to have exactly one reconstructed lepton with 
          $p_{T} > 50$ GeV within $|\eta| < 0.25$, and isolated from jets with $\Delta R_{\ell j} = 0.5$;
	\item The lepton reconstruction is smeared by $\sigma(1/p_{T}) = 0.011/p_{T}\otimes 0.00017$, 
          while the jet resolution is taken as $\sigma(E_{T}) = 0.45 \times \sqrt{E_T} \otimes 5\%$;
	\item Missing transverse energy $E_{T}^{\rm mis} > 50$ GeV;
	\item To reduce the di-jet and $t\bar{t}$ backgrounds, a lepton fraction 
          is required to be $\sum p_{T}/ \left(\sum p_{T} + \sum E_{T}\right) > 0.5$;
	\item Transverse mass $m_{T} = \sqrt{2 p_{T} E_{T}^{\rm mis} (1 - \cos\Delta \phi)} \,\, > 0.7 \times M_{W^\prime}$,
	  where $\Delta \phi$ is the angle between the momentum of the lepton and the missing momentum.
\end{itemize}
For the sequential $Z^\prime$, we list the event selection and analysis on the di-lepton final states as follows:
\begin{itemize}
	\item Events are required to have exactly two reconstructed same-flavor 
          opposite-charged leptons with at least one lepton $p_{T} > 30$ GeV, within $|\eta| < 0.25$;
	\item Di-lepton invariant mass window $ |m_{\ell \ell}  - M_{Z^\prime} | < 4 \times \Gamma_{Z^\prime}$.
\end{itemize}

Next we explore the LHC sensitivity to $W^\prime$ and $Z^\prime$ bosons.
We can quantify the sensitivity to new physics discovery  or set exclusion limits on it based on statistics. 
Specifically, for the case of discovery we would like to know the statistical significance ($S$) for discovery,
which characterizes the inconsistency of the experiment data with a background-only hypothesis.
If there is no discovery at a given luminosity, we  set exclusion limits on new physics.
In the counting experiments, suppose one has an experiment that counts $n$ events, 
modeled as a Poisson distribution with mean 
$s + b$, where $s$ is the expected signal rate, $b$ is the expected background rate. 
The probability of measuring $n$ events is therefore
\be
	P(n|s,b) = \frac{(s+b)^n}{n!} e^{-(s+b)}\;.
\ee
Using a profile likelihood ratio as the test statistic, the expected significance
is obtained as follows~\cite{AdamBourdarios:2009zza}
\be
S = \sqrt{ 2 \left( (s+b) \ln (1 + s/b) - s \right) }\;.
\ee
For sufficiently large $b$ we can expand the logarithm in $s/b$ and obtain the widely-used significance 
\be
S = \frac{s}{\sqrt{b}} 
\left( 1 + {\cal O}(s/b) \right)\;. 
\ee
In addition to establishing discovery by rejecting the background hypothesis, we can consider the signal hypothesis  as well. 
It is common to use confidence level (CL) $\alpha$ 
and the related $p$-value to quantify the level of incompatibility of data with a signal hypothesis. 
The profile likelihood ratio $q_{\mu}$ is used as the test statistic \cite{AdamBourdarios:2009zza}.
For a sufficiently large data sample the probability density of $q_{\mu}$ 
takes on a well defined $\chi^2$ distribution with mean $\hat{\mu}$ and variance $\hat{\sigma}$ for one degree of freedom. 
Given the $p$-value for each number of signal events $s$,
we can obtain the upper limit $s^{\rm up}$ on the number of signal events,
\begin{equation}
\label{eq:muup}
s^{\rm up} = \hat{\mu} + \hat{\sigma} \Phi^{-1}(1 - \alpha) \;,
\end{equation}
where the mean and variance of the $\chi^2$ distribution are $\hat{\mu}= n-b$, 
and $\hat{\sigma} = \sqrt{b}$ for large data sample,
and $\Phi$ is the cumulative distribution of the standard Gaussian with zero mean and unit variance.
For the expected upper limit, in which 
the data count is taken as the background sum,
the upper limit at confidence level $\alpha = 95\%$ is
\begin{equation}
s^{\rm up} = \Phi^{-1}(0.05) \cdot \sqrt{b} =  1.64 \times \sqrt{b} \;.
\end{equation} 
So for a sufficiently large data sample the equivalent significance $Z$ 
for excluding a signal hypothesis is given by
\be
	Z = \frac{s^{\rm up}}{\sqrt{b}} = 1.64.
\ee
For instance, when expressing the significance for $5\sigma$ discovery 
with the exclusion upper limit at the $95\%$ CL, 
a factor $S/Z = 5/1.64 \simeq 3$ needs to be applied. 

Denoting by
$\sigma_s$ ($\sigma_b$) the inclusive cross section of the signal (background),
$\epsilon_s$ ($\epsilon_b$) the cut acceptance of the signal (background),   
and $\cal L$ the integrated luminosity, the number of signal (background) events can be written as
\bea
s &=& \sigma_{s}\, \epsilon_{s}\,{\cal L}\;,\\
b &=& \sigma_{b}\, \epsilon_{b}\,{\cal L}\;.
\eea
For a sufficiently large data sample, both $S$ and $Z$ have a scaling behavior with the integrated luminosity,
\be
	S \simeq Z \simeq \frac{s}{\sqrt{b}} = \frac{\sigma_{s}\, \epsilon_{s}}{\sqrt{\sigma_{b}\, \epsilon_{b}}}\times \sqrt{\cal L}.
\ee

\begin{figure}
	\includegraphics[width=0.32\textwidth]{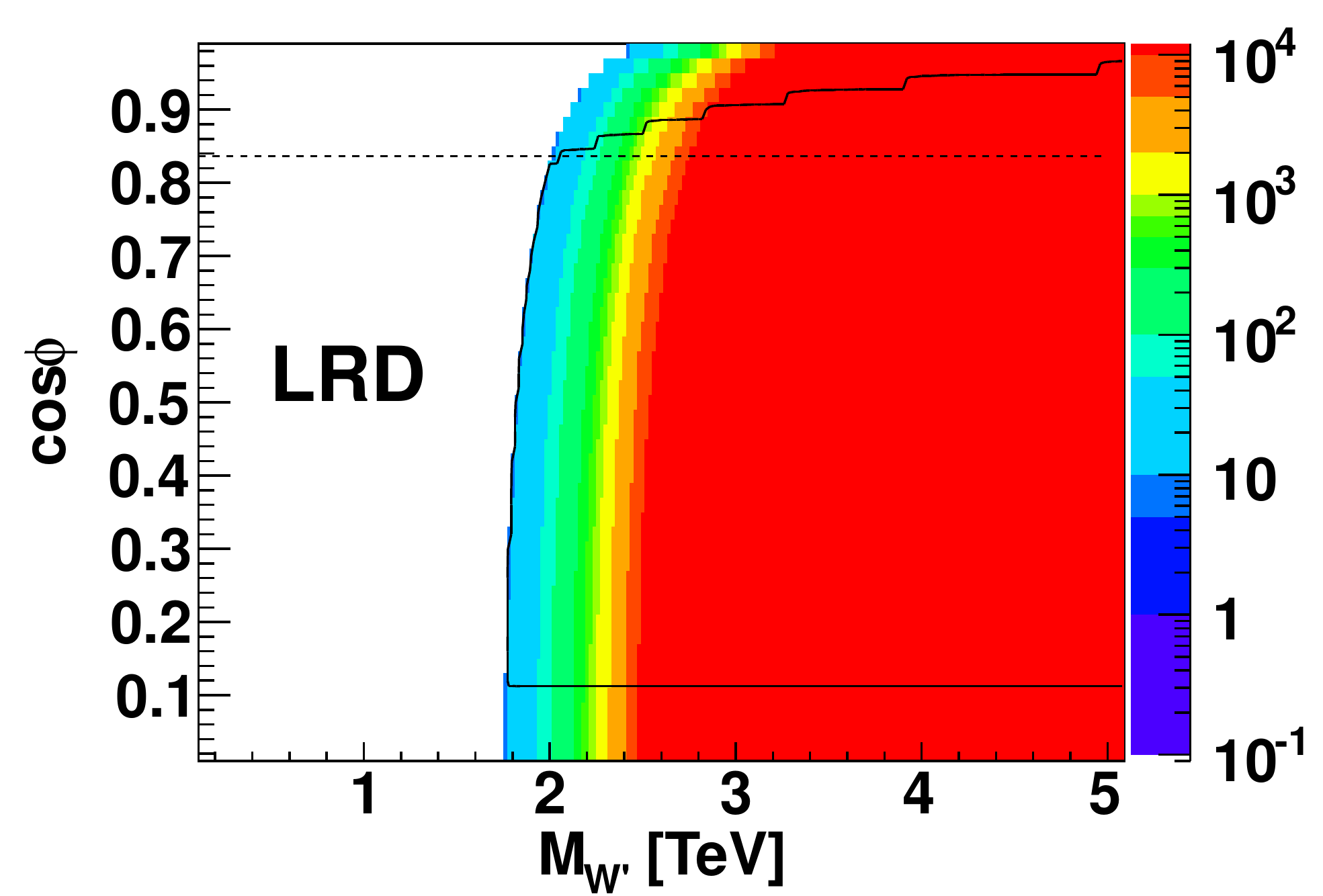}
	\includegraphics[width=0.32\textwidth]{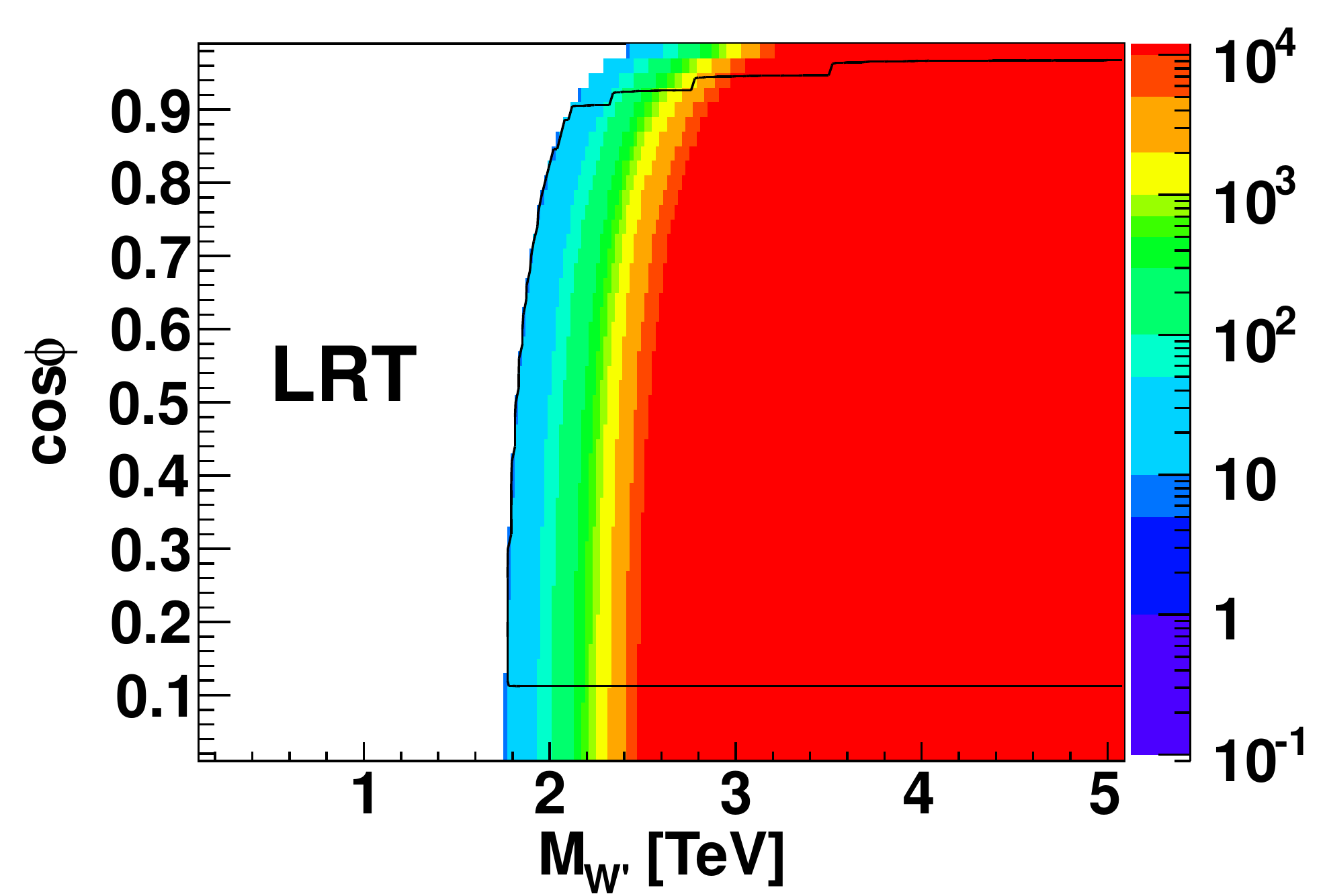}
	\includegraphics[width=0.32\textwidth]{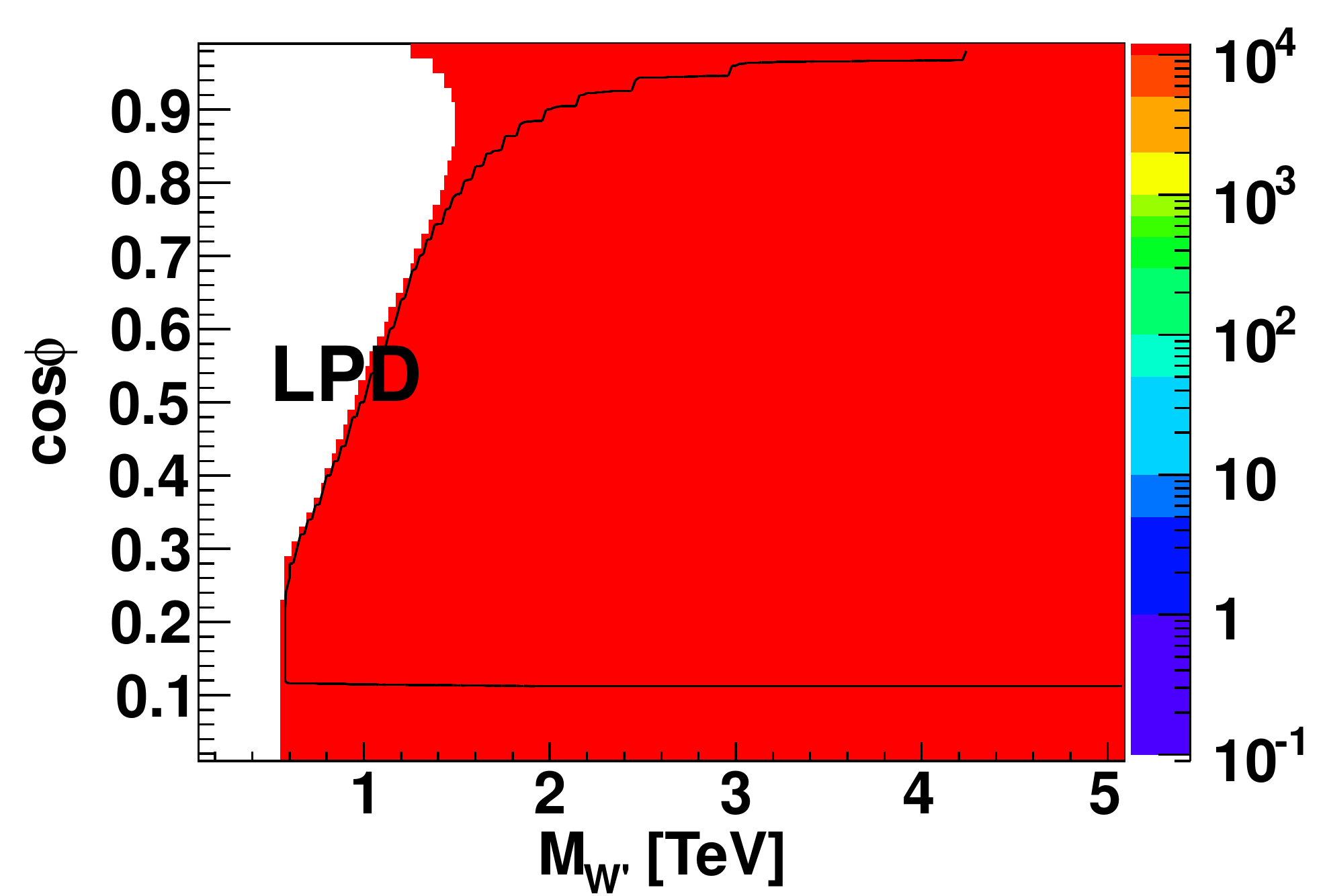}
	\includegraphics[width=0.32\textwidth]{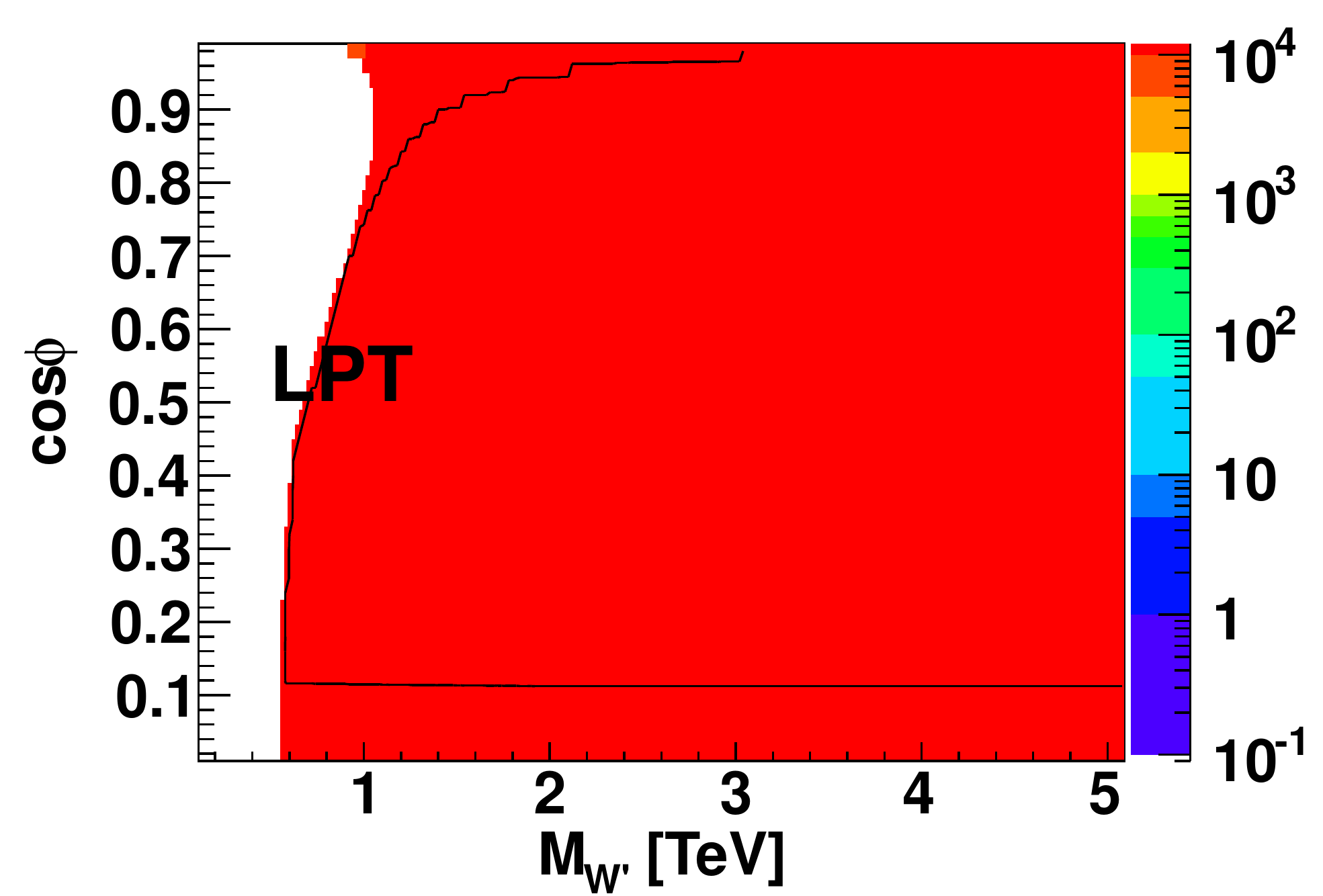}
	\includegraphics[width=0.32\textwidth]{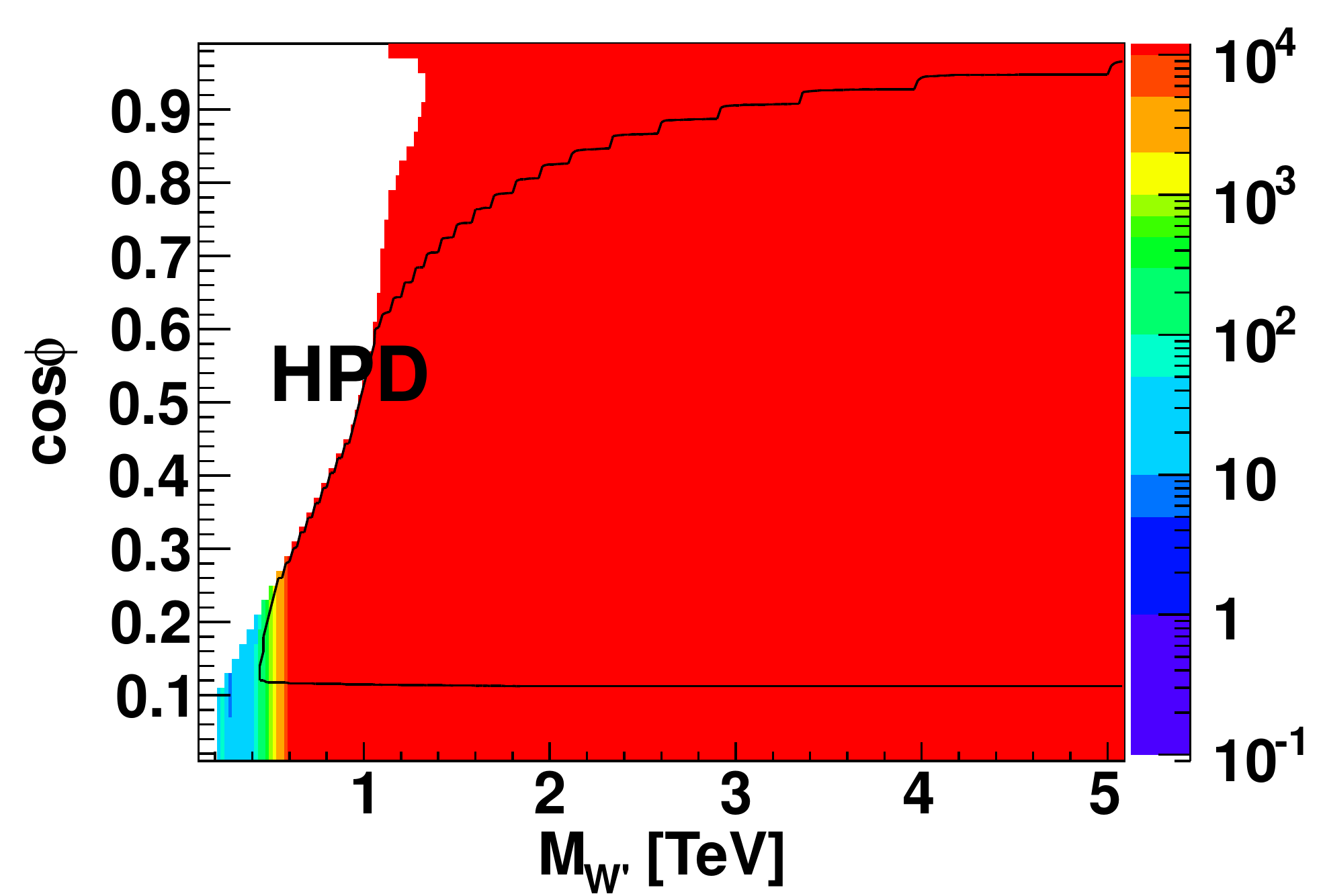}
	\includegraphics[width=0.32\textwidth]{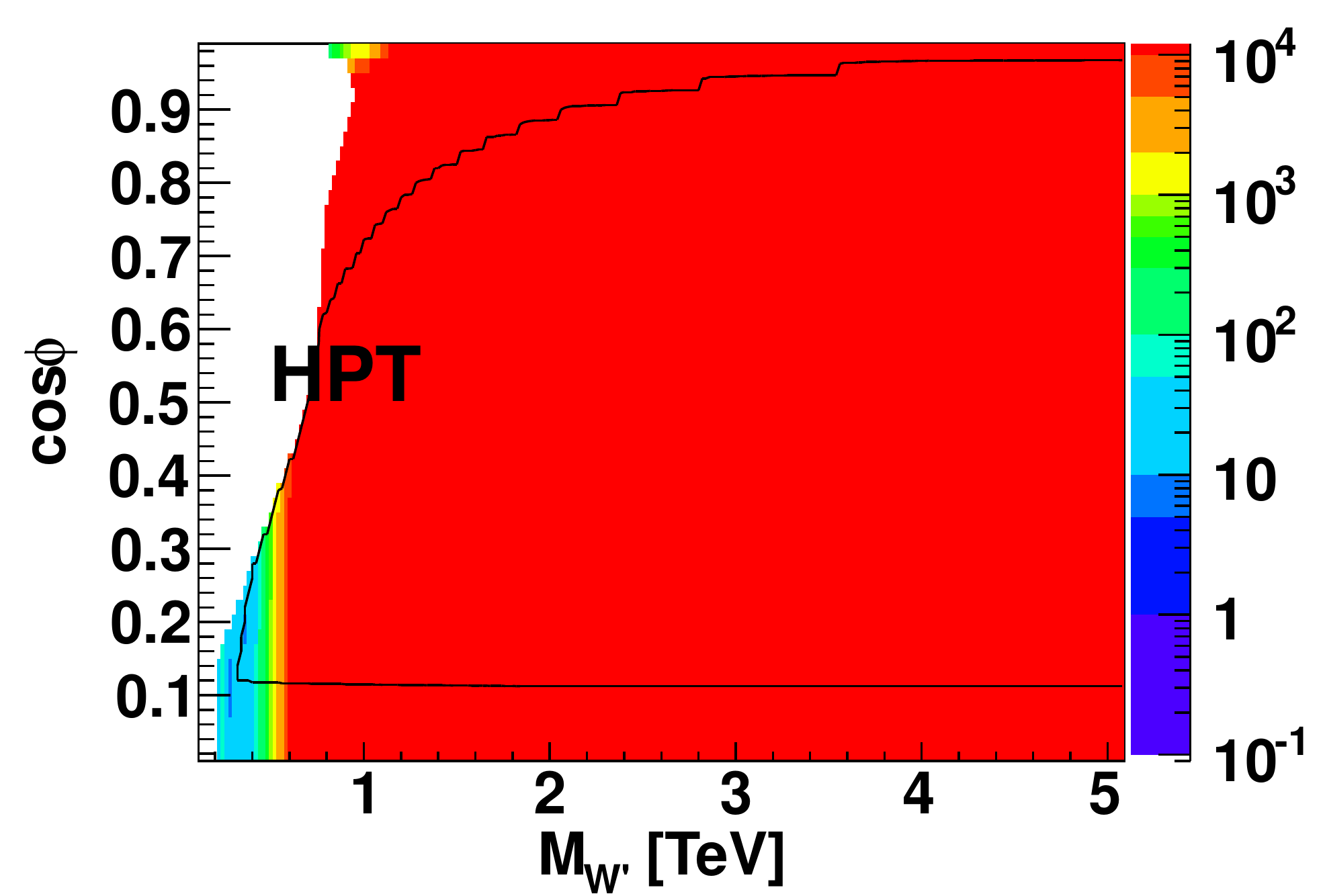}
	\includegraphics[width=0.32\textwidth]{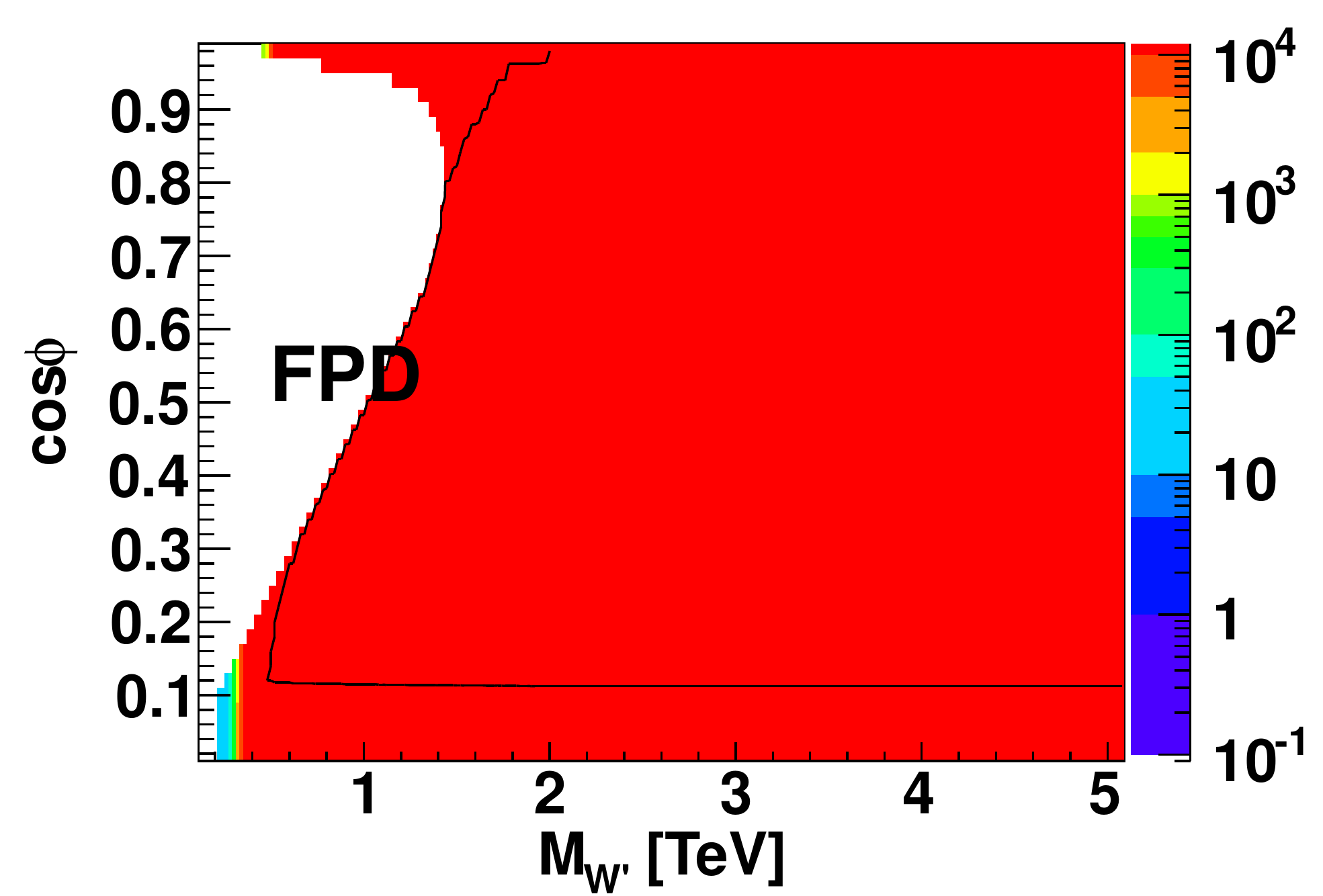}
	\includegraphics[width=0.32\textwidth]{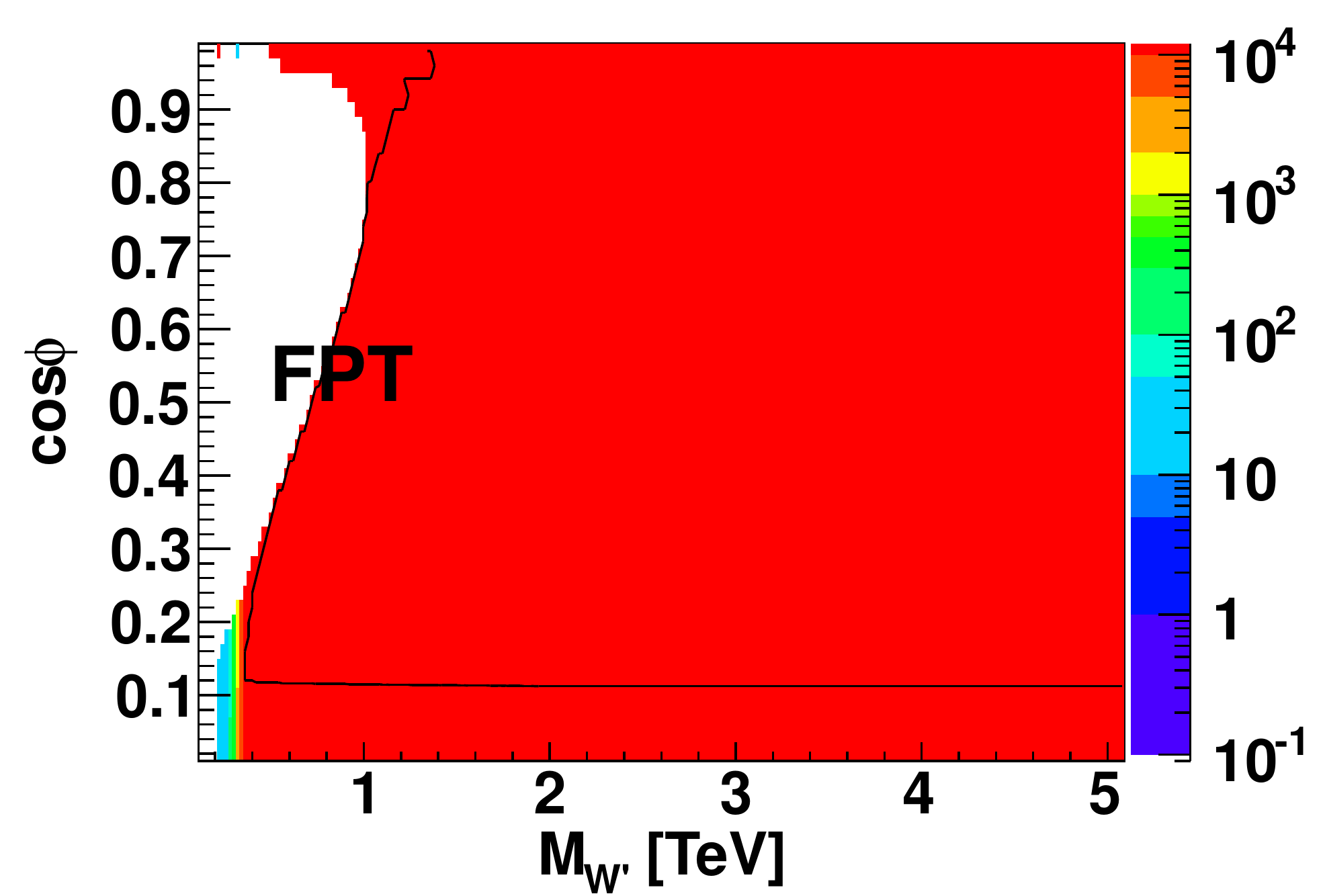}
	\includegraphics[width=0.32\textwidth]{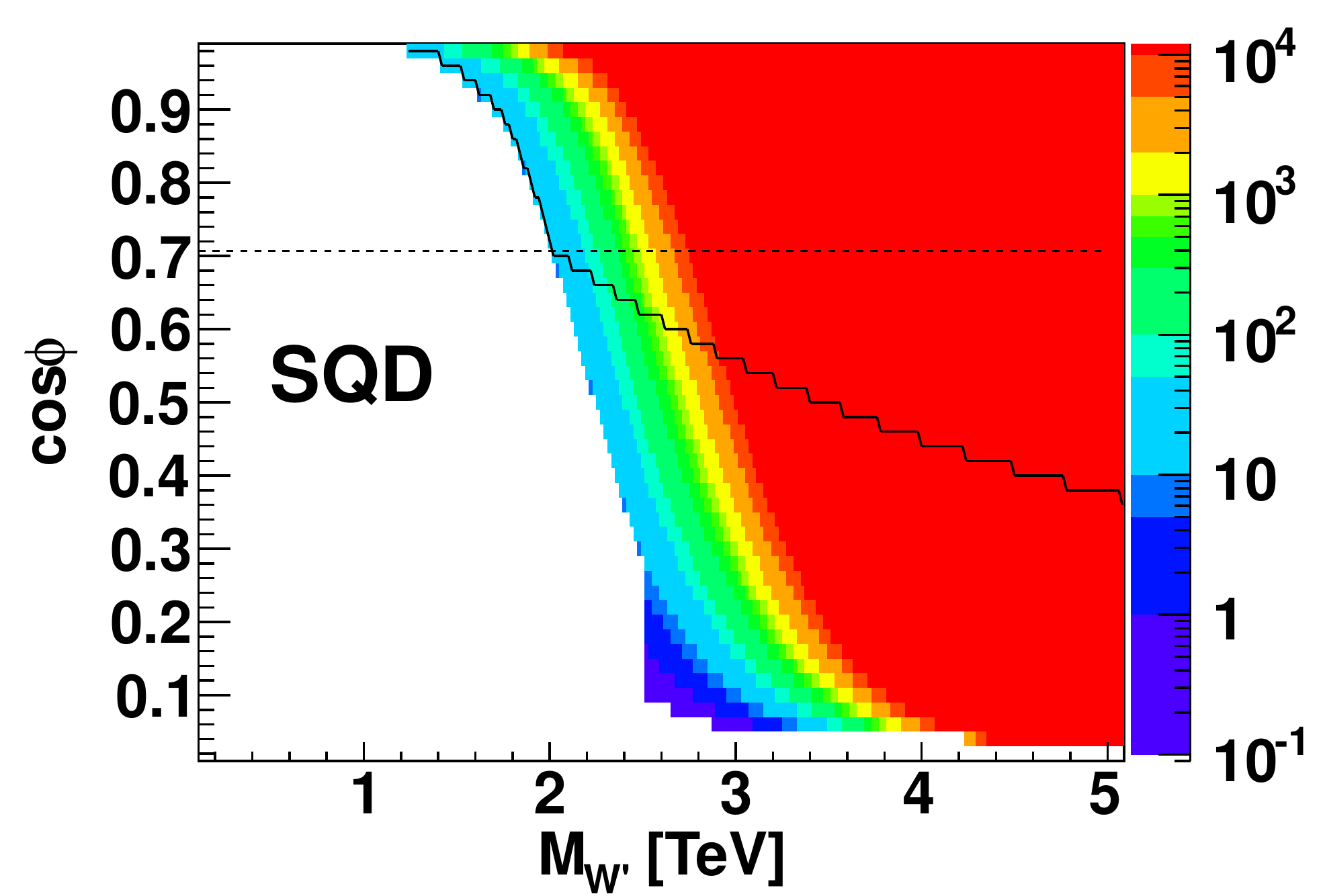}
	\includegraphics[width=0.32\textwidth]{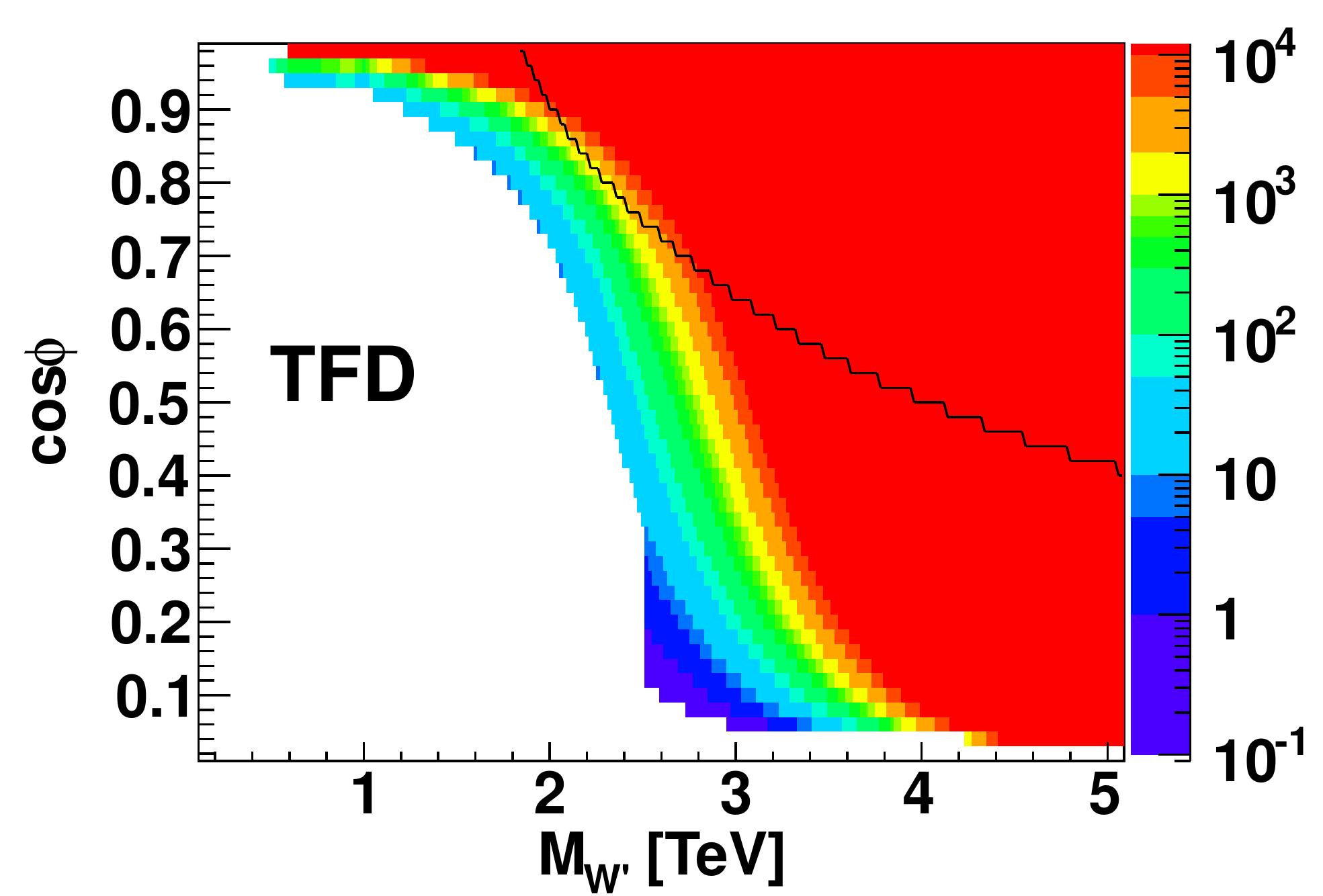}
	\includegraphics[width=0.32\textwidth]{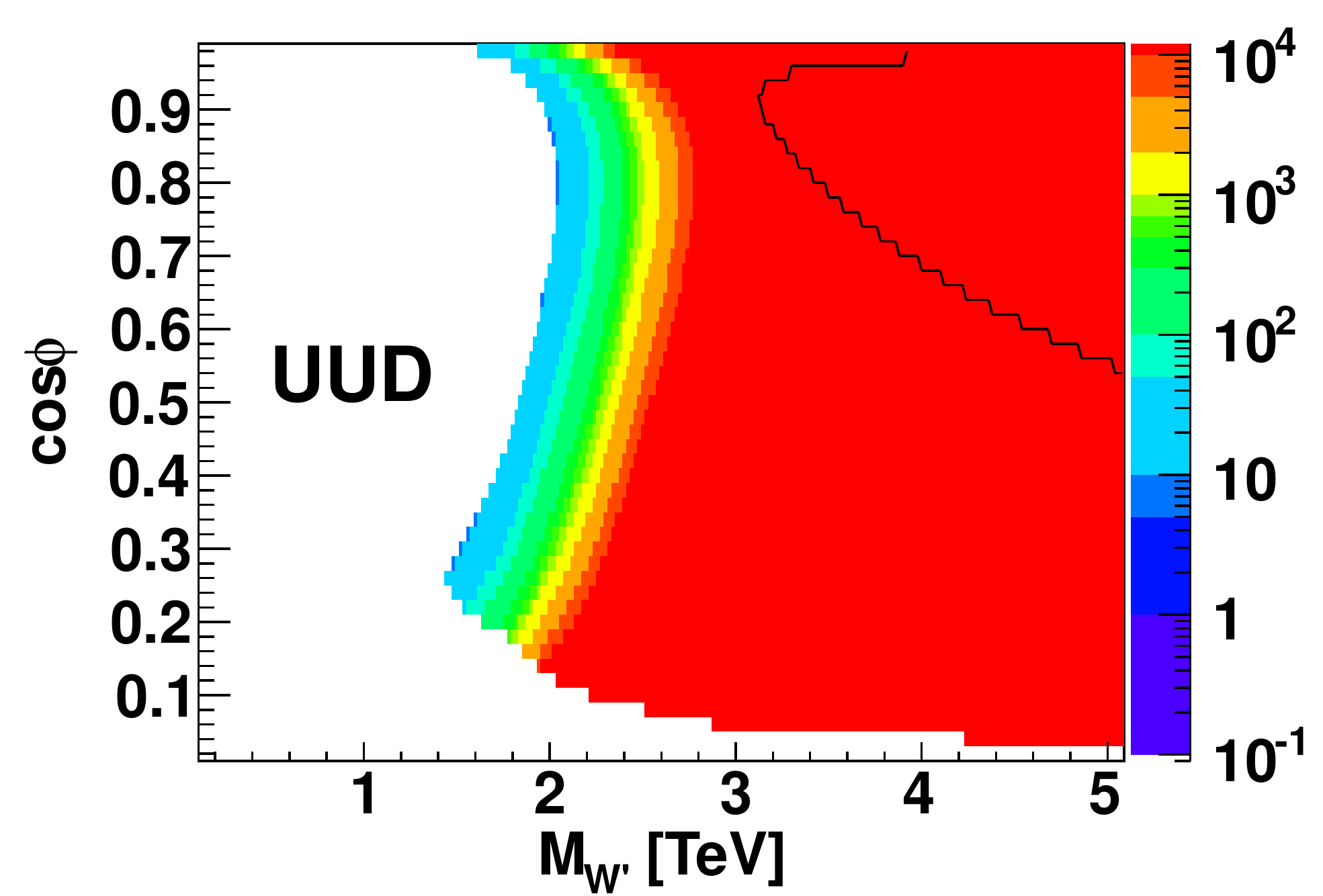}
\caption{$5\sigma$ Discovery potential (fb$^{-1}$) for different 
luminosity at LHC7 via $W^\prime$ leptonic decay channel. 
The color palette shows the integrated luminosity with unit fb$^{-1}$. 
EWPT constraints are within solid black contour.}
\label{lumi7W}
\end{figure}

Figure~\ref{lumi7W} displays the $5 \sigma$ discovery potential (fb$^{-1}$) 
for LHC7 via $W^\prime$ leptonic decay channel,
and current combined constraints are within solid black contour.
The LRD(T) and MLR models can be further constrained 
when the integrated luminosity for LHC7 reaches its maximum 5.6 fb$^{-1}$. 
However, the other models need much more luminosity, 
which even exceeds the total integrated luminosity (5.6 fb$^{-1}$) at LHC7.
Therefore, the $W^\prime$ leptonic decay channel 
cannot make further contributions to discovering these $G(221)$ models,
except for some small region in LRD(T) and MLR.
In Fig.~\ref{parameter_constraints} it shows 
that the EWPTs constraints are stronger than those from the %always have the largest constraint compared to 
Tevatron and the LHC7, except LRD(T) and MLR.
This means that compared to EWPTs,
the LHC7 direct search via the $W^\prime$ leptonic decay channel 
for the new physics models with $G(221)$ gauge group structure
can put further constraint only on LRD(T) and MLR.
For the other models, the direct search at LHC7 for s-channel $W^\prime$ production with leptonic decay 
cannot compete with seeking for deviation from SM predictions via EWPTs.

\begin{figure}
	\includegraphics[width=0.32\textwidth]{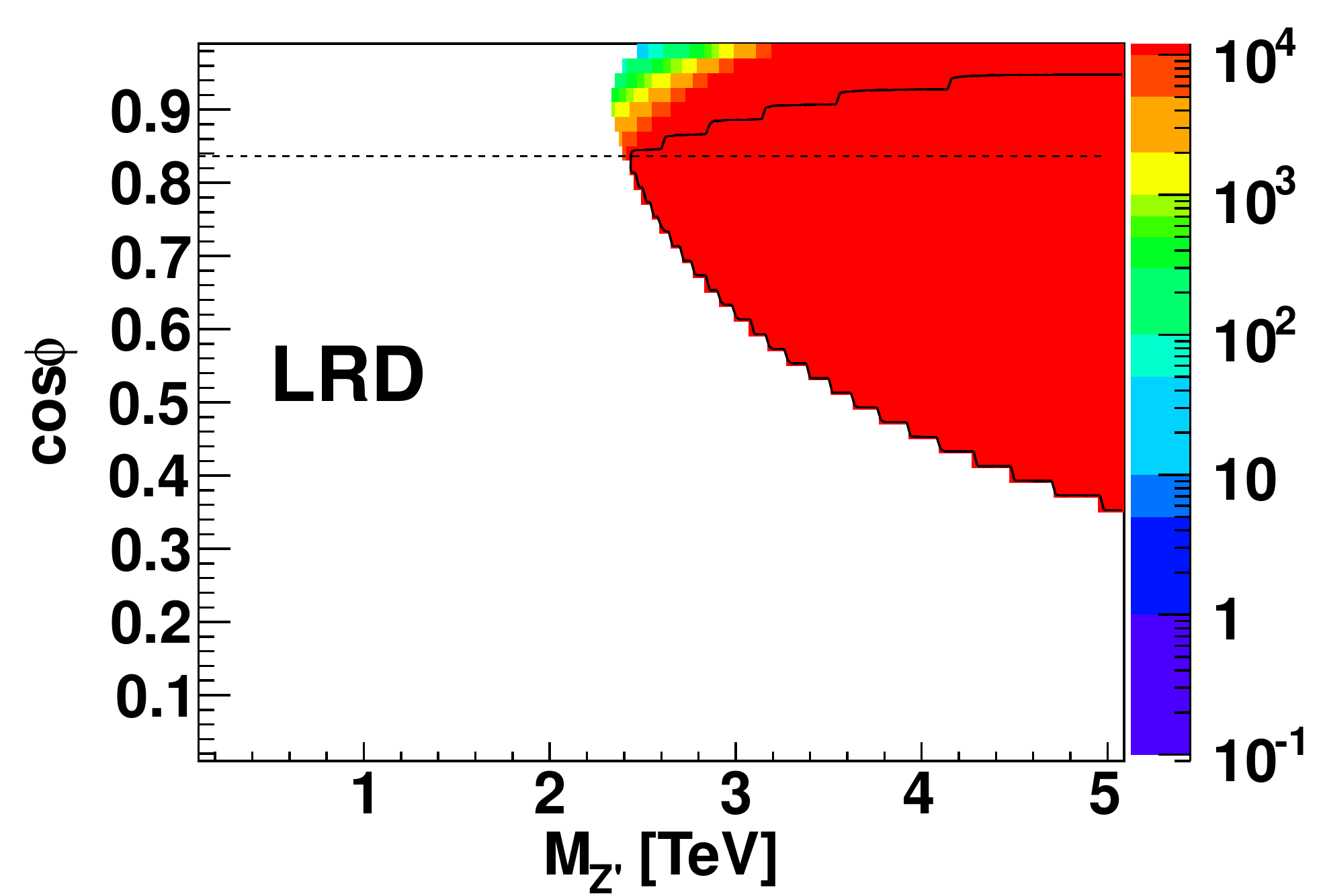}
	\includegraphics[width=0.32\textwidth]{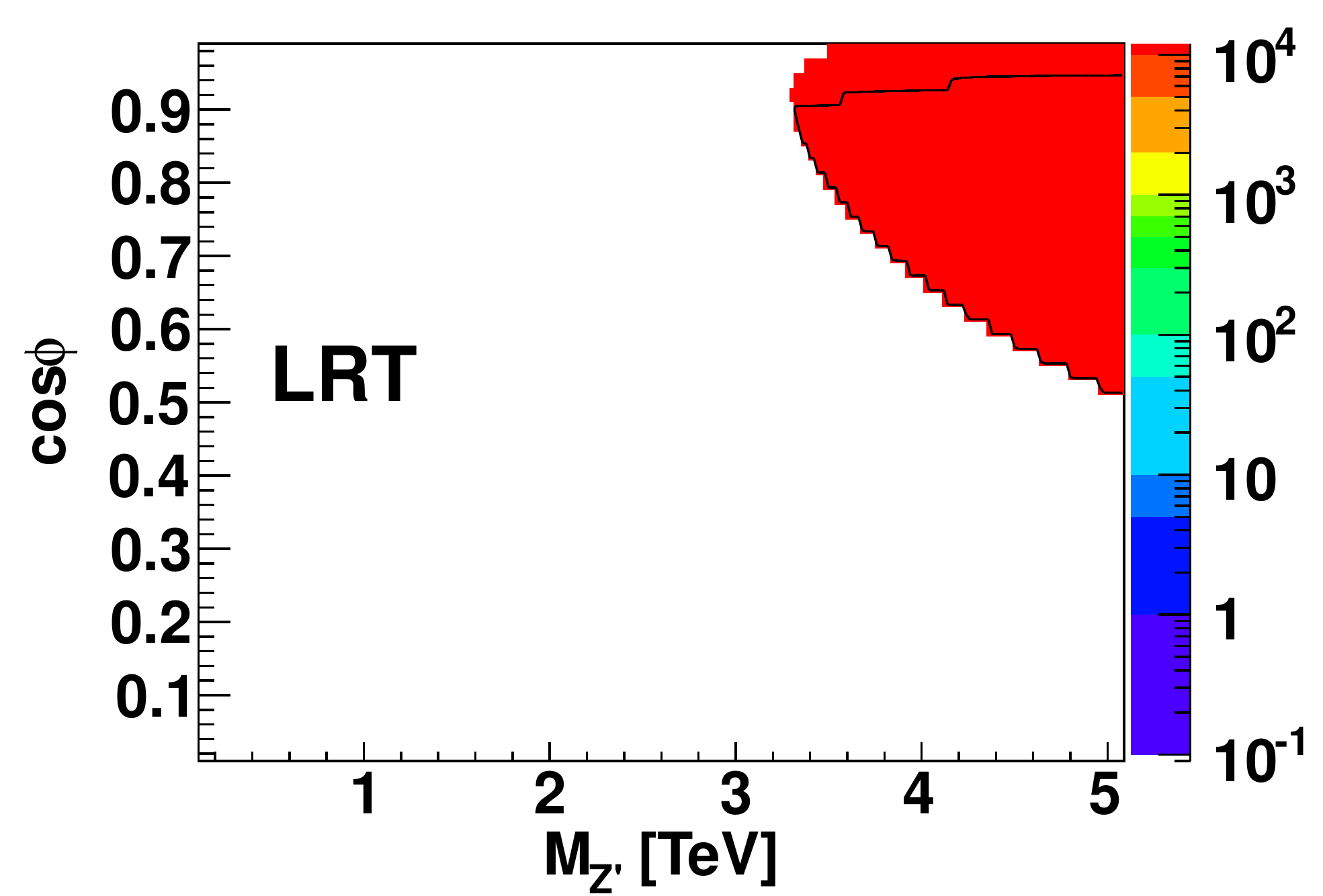}
	\includegraphics[width=0.32\textwidth]{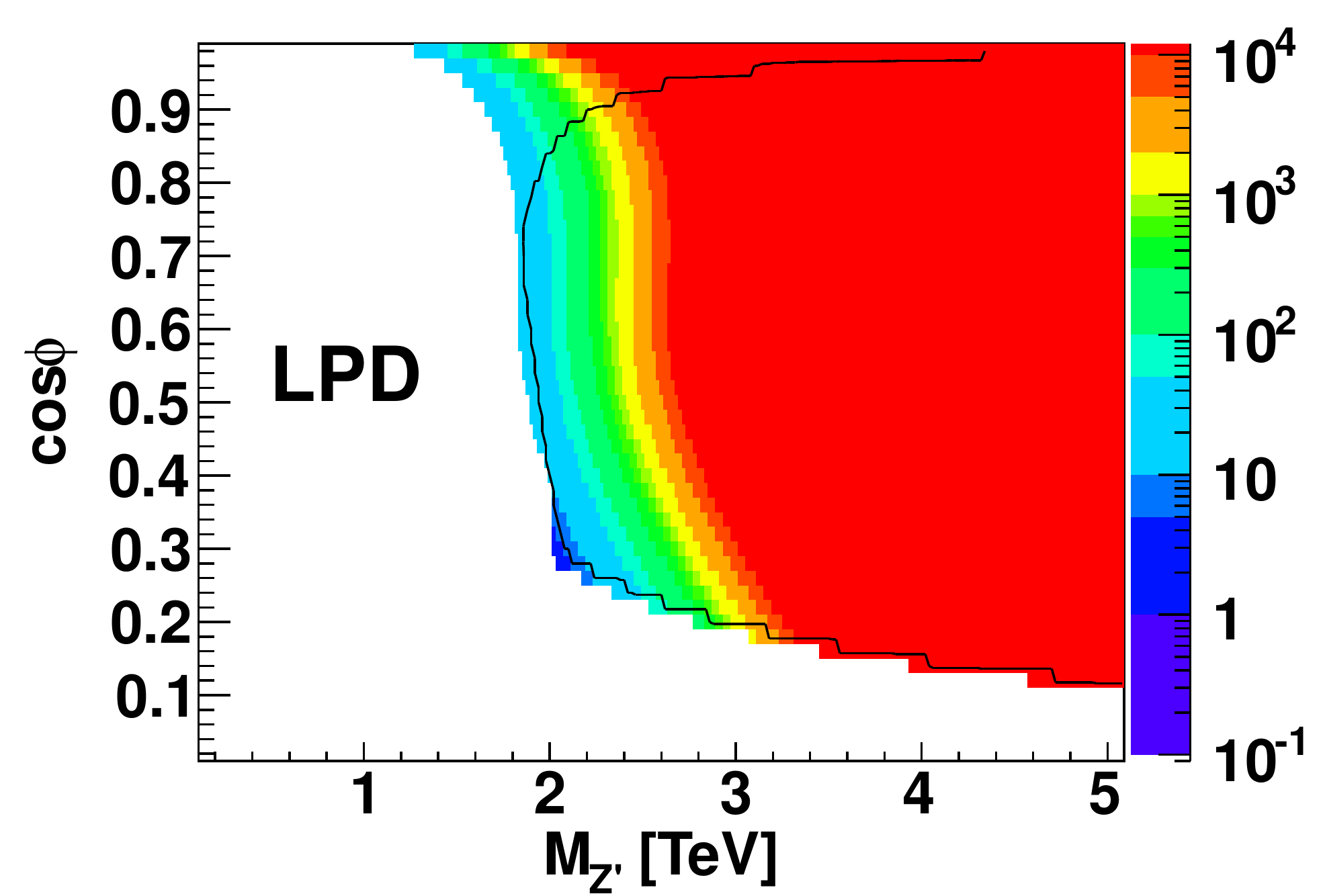}
	\includegraphics[width=0.32\textwidth]{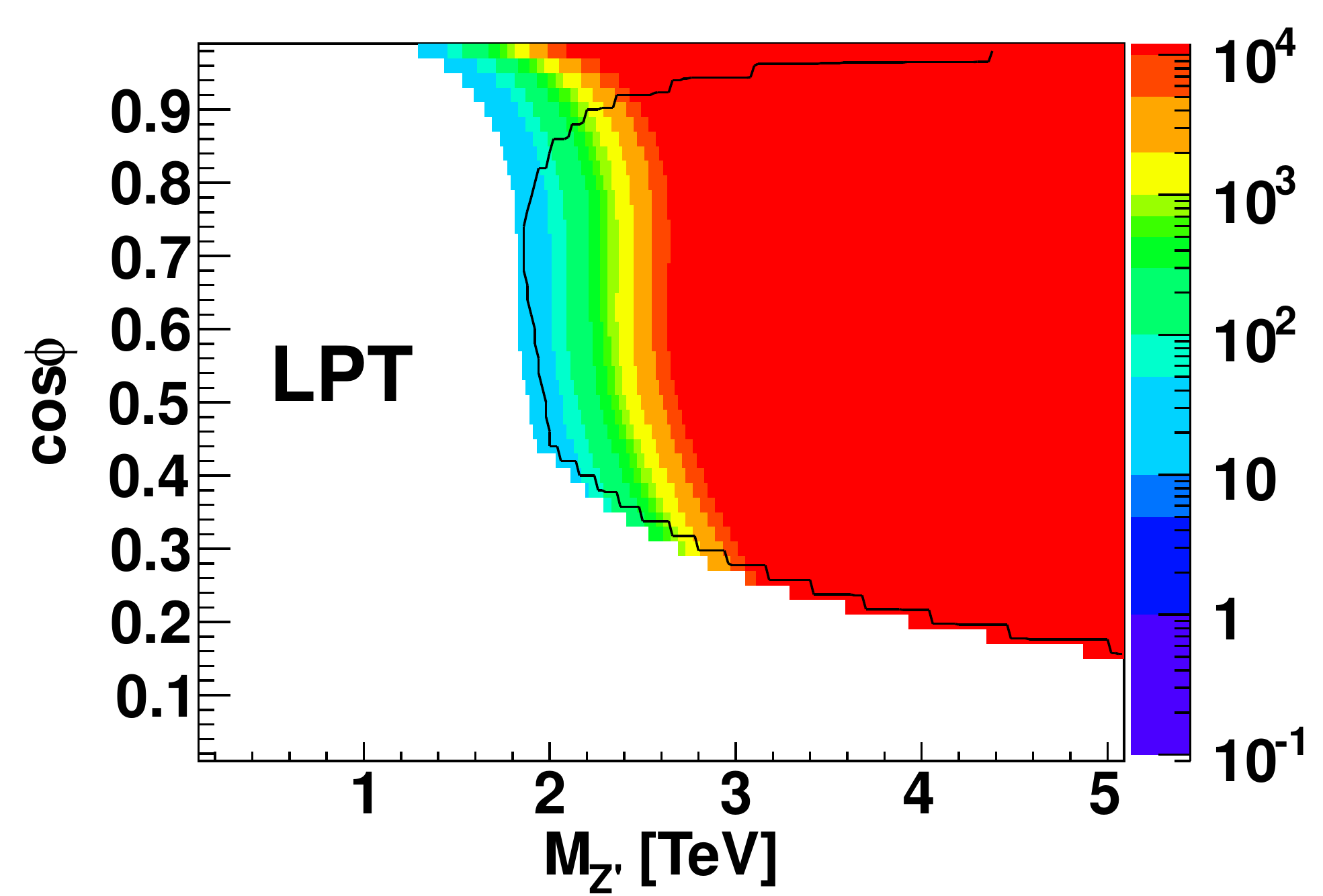}
	\includegraphics[width=0.32\textwidth]{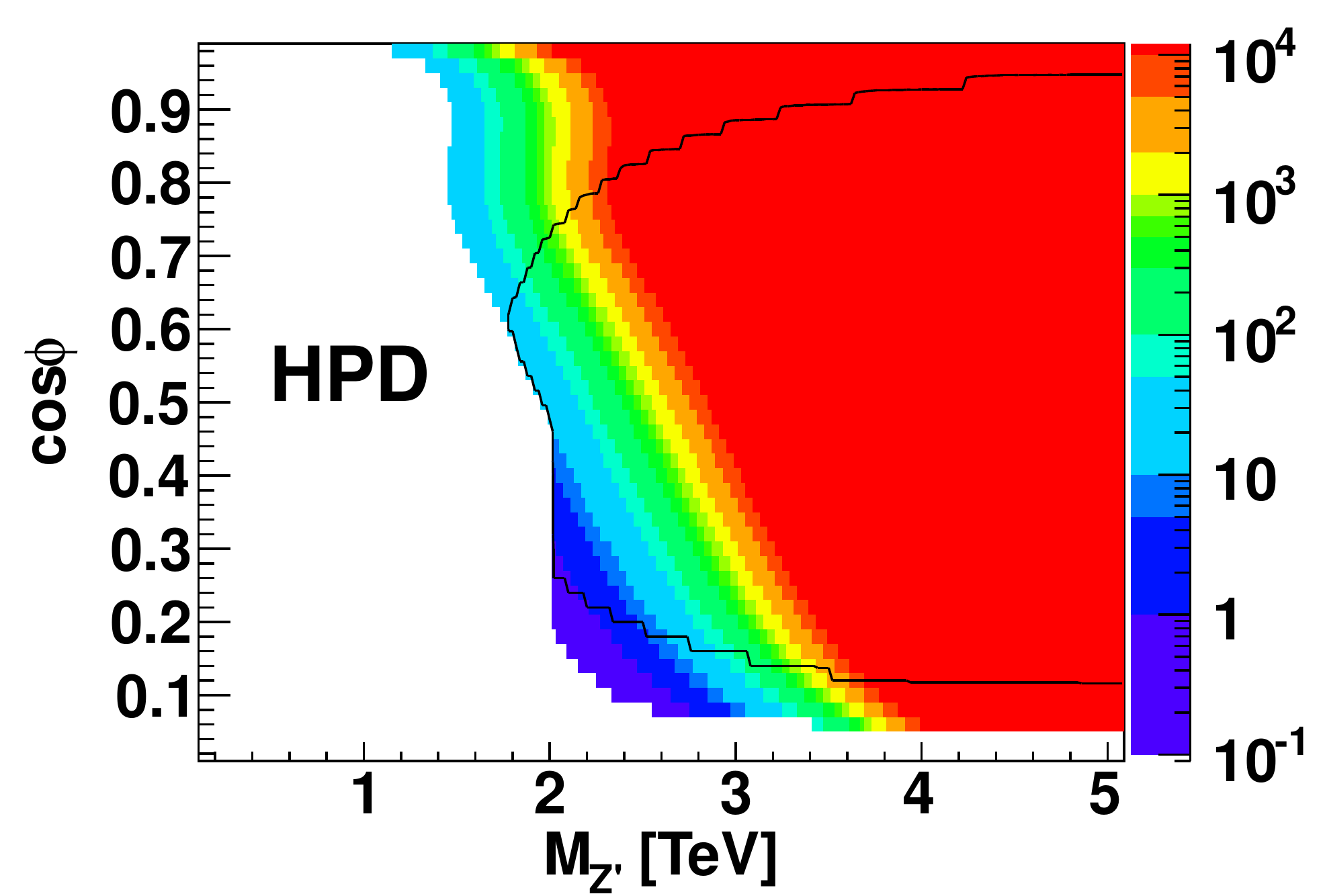}
	\includegraphics[width=0.32\textwidth]{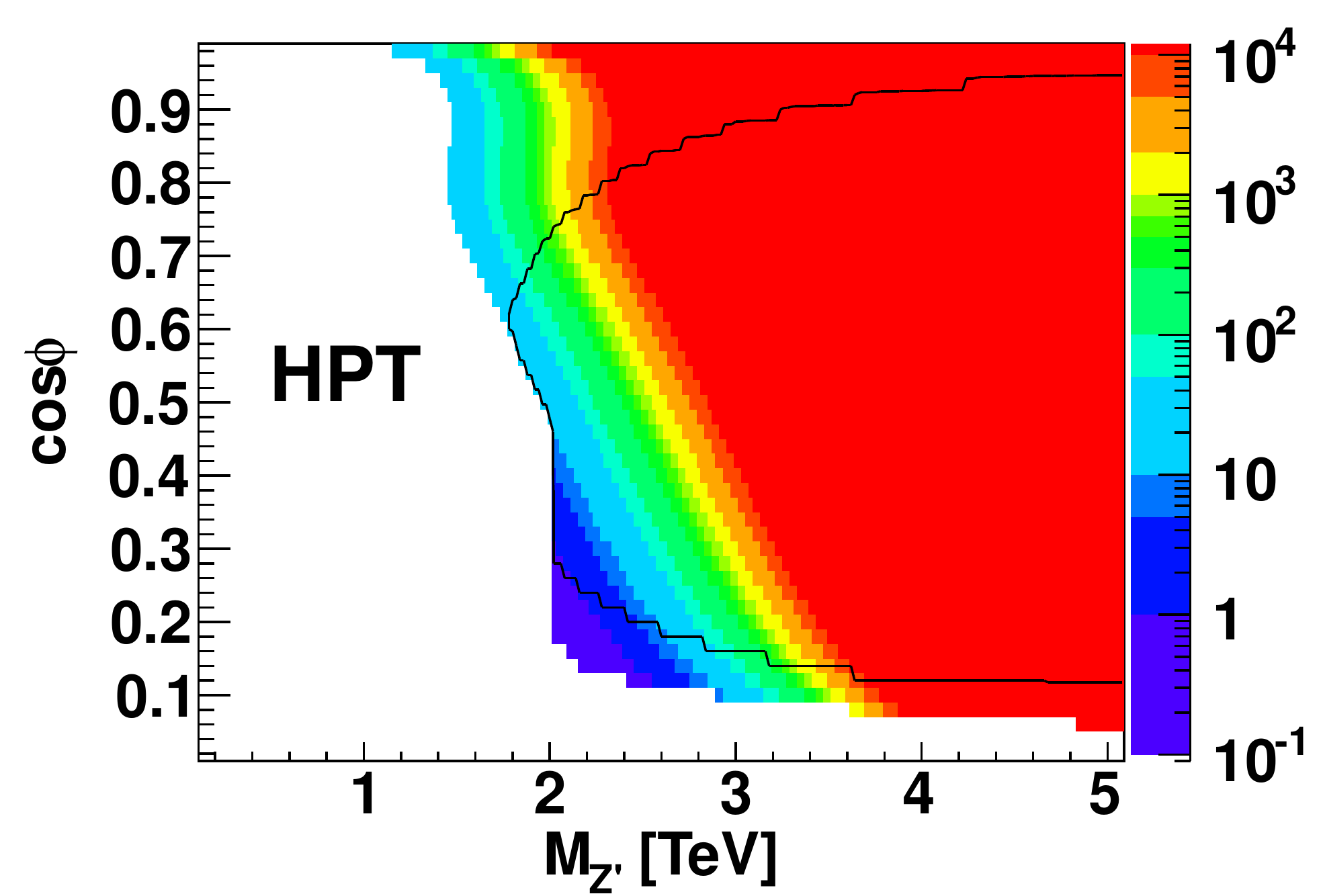}
	\includegraphics[width=0.32\textwidth]{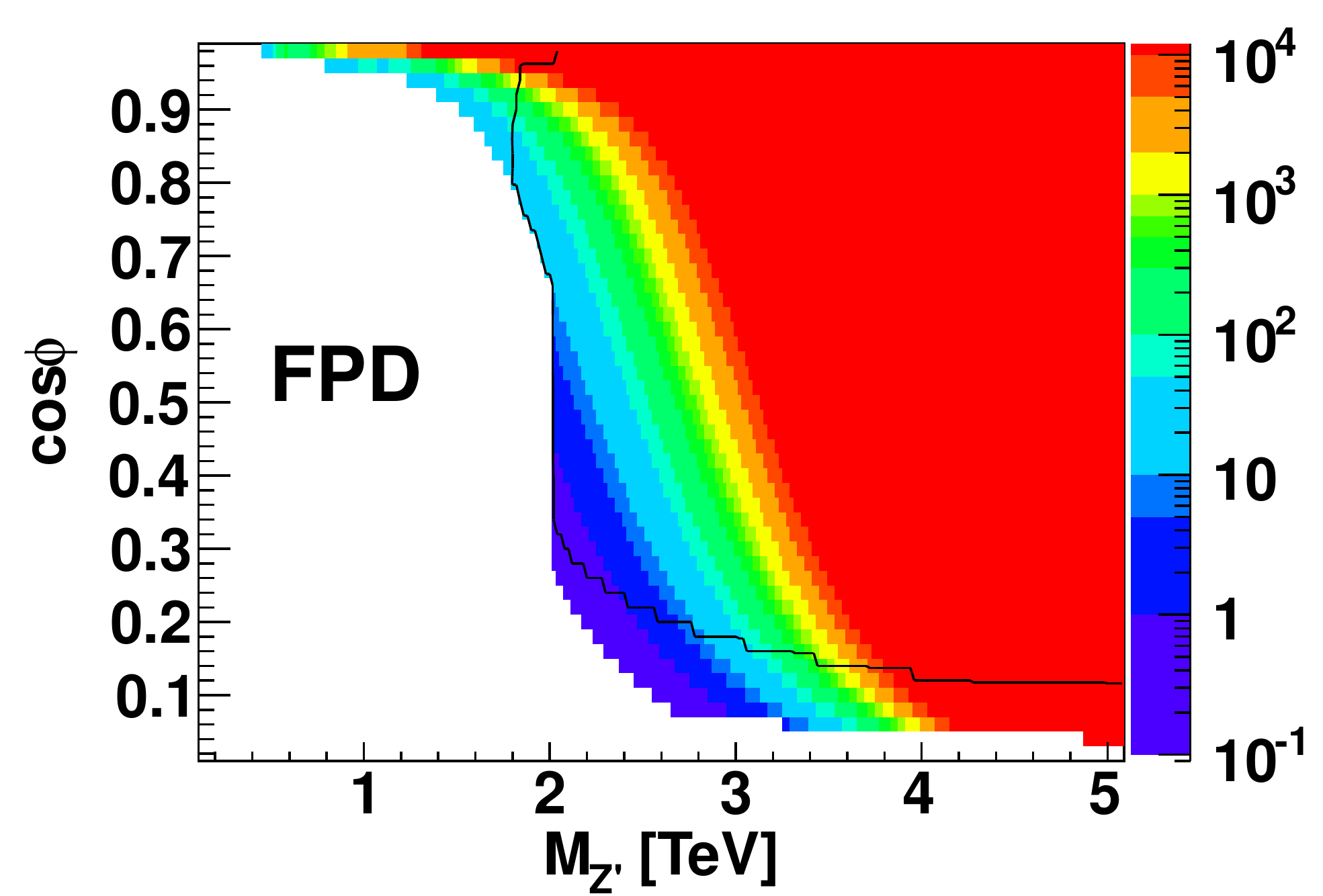}
	\includegraphics[width=0.32\textwidth]{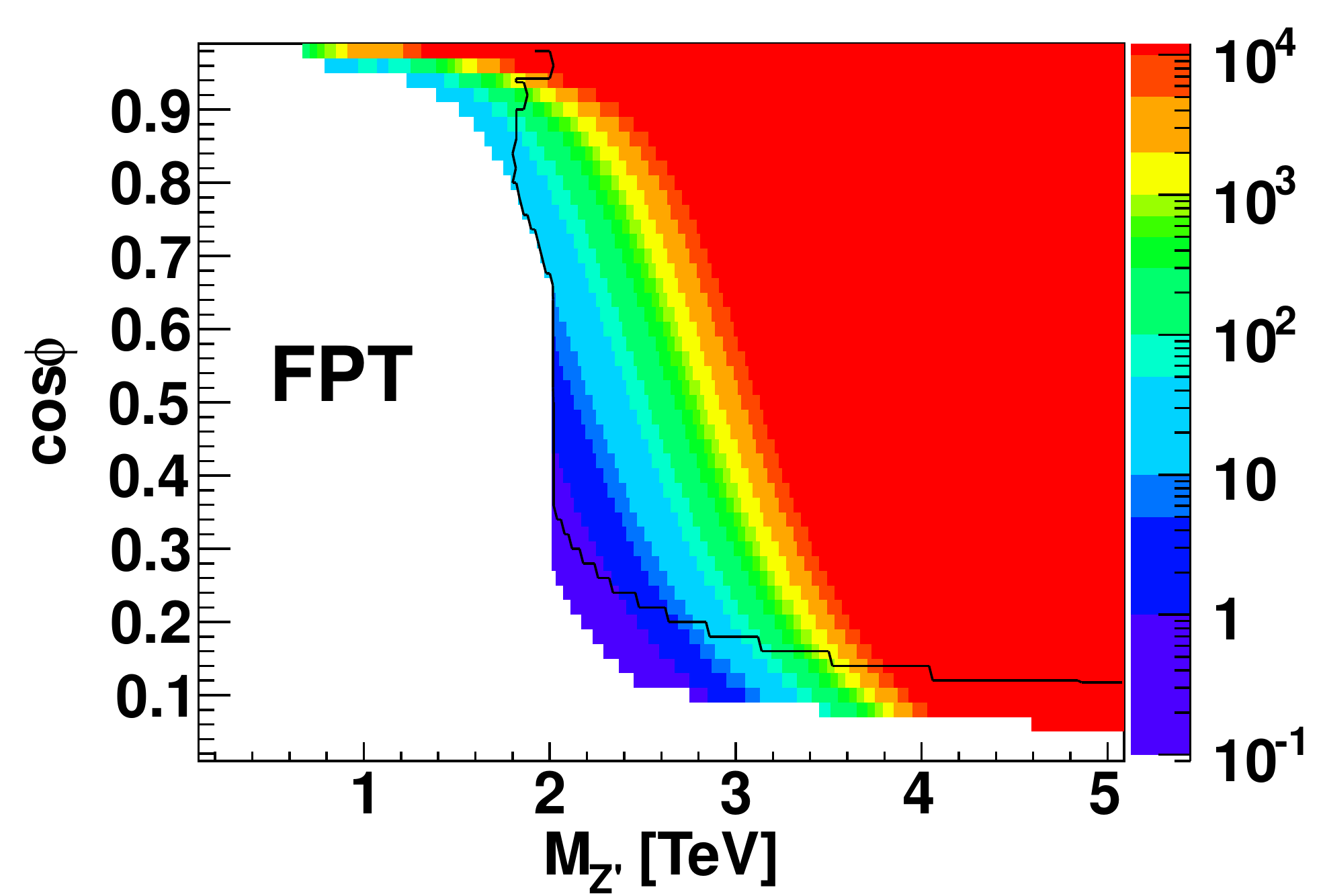}
	\includegraphics[width=0.32\textwidth]{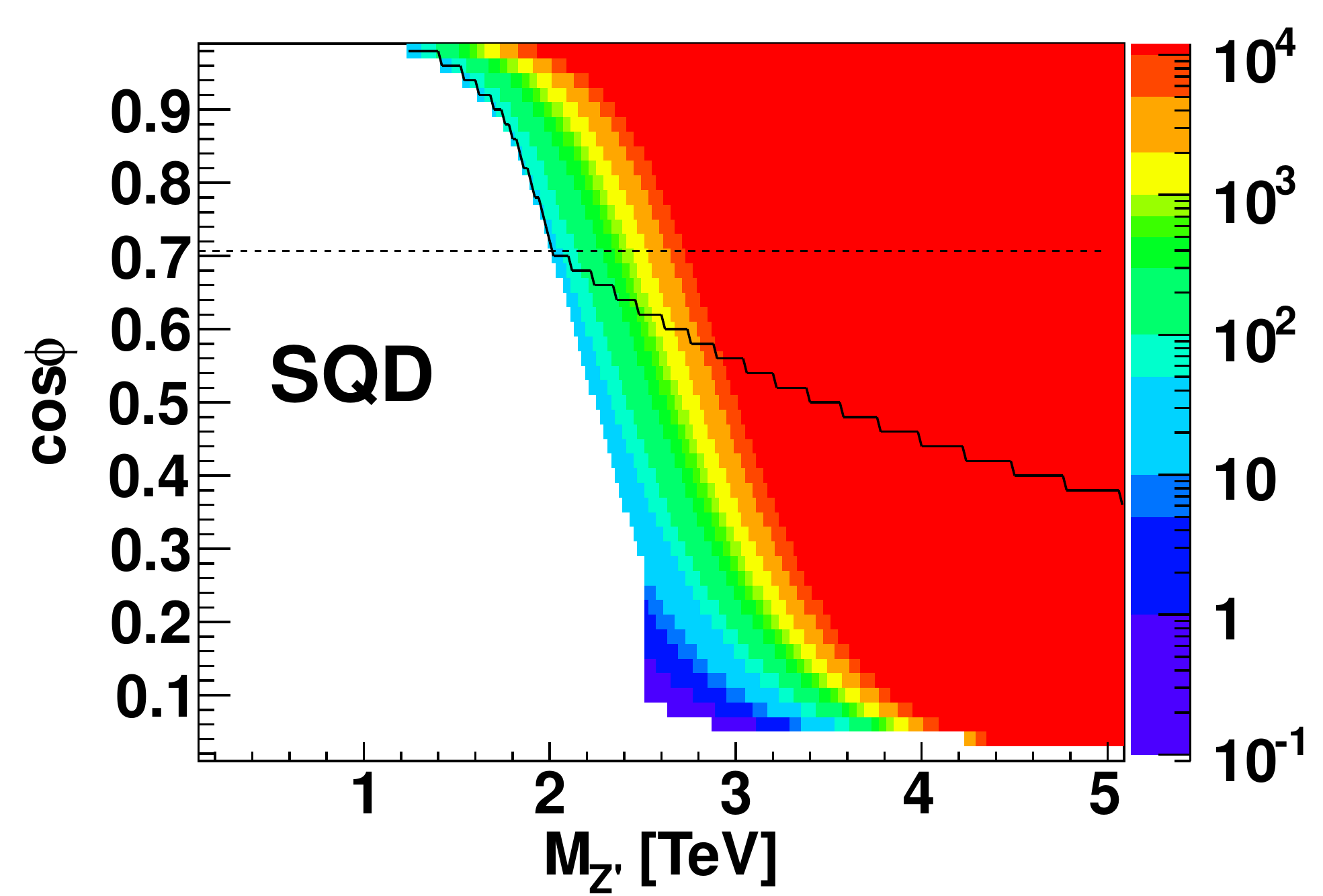}
	\includegraphics[width=0.32\textwidth]{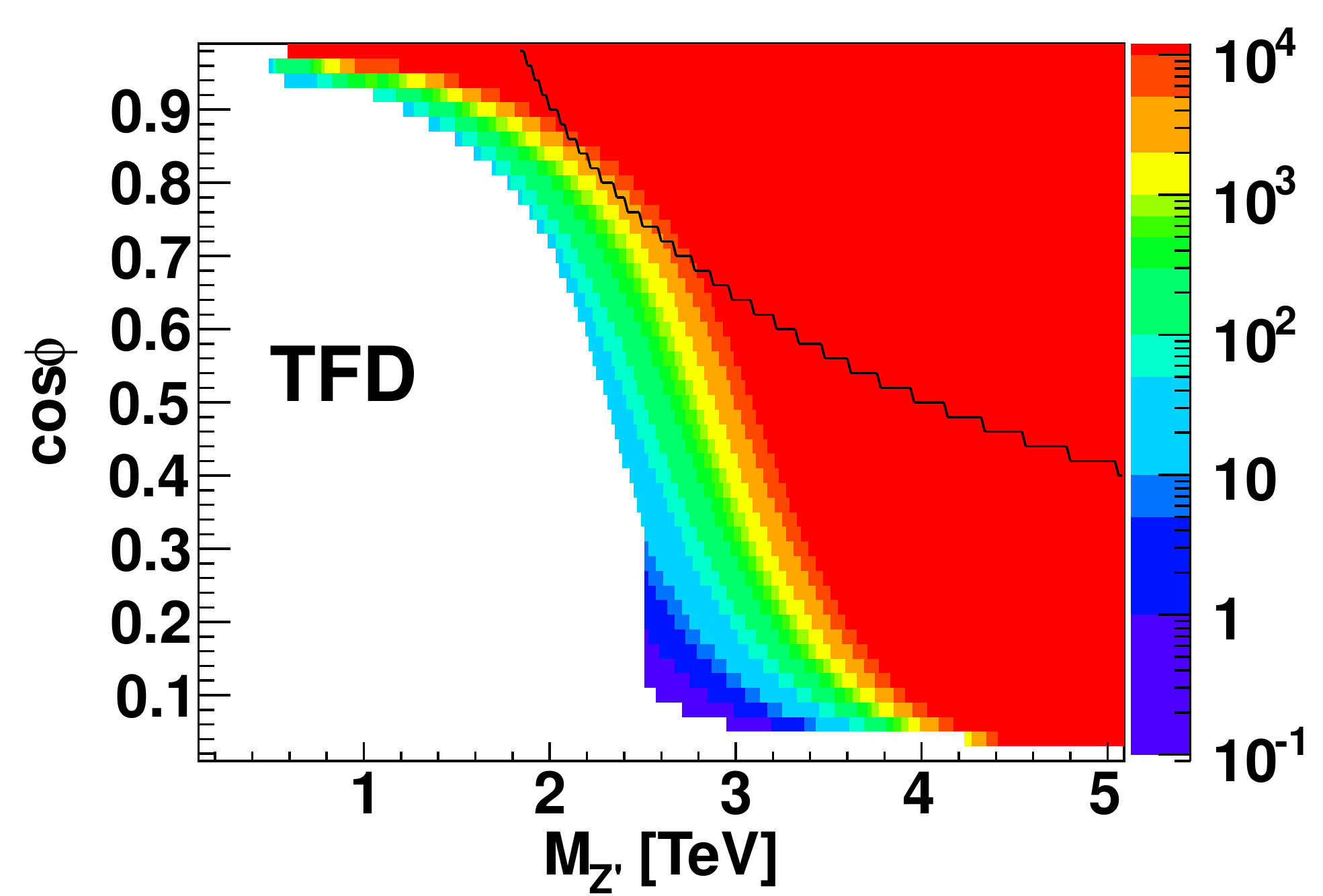}
	\includegraphics[width=0.32\textwidth]{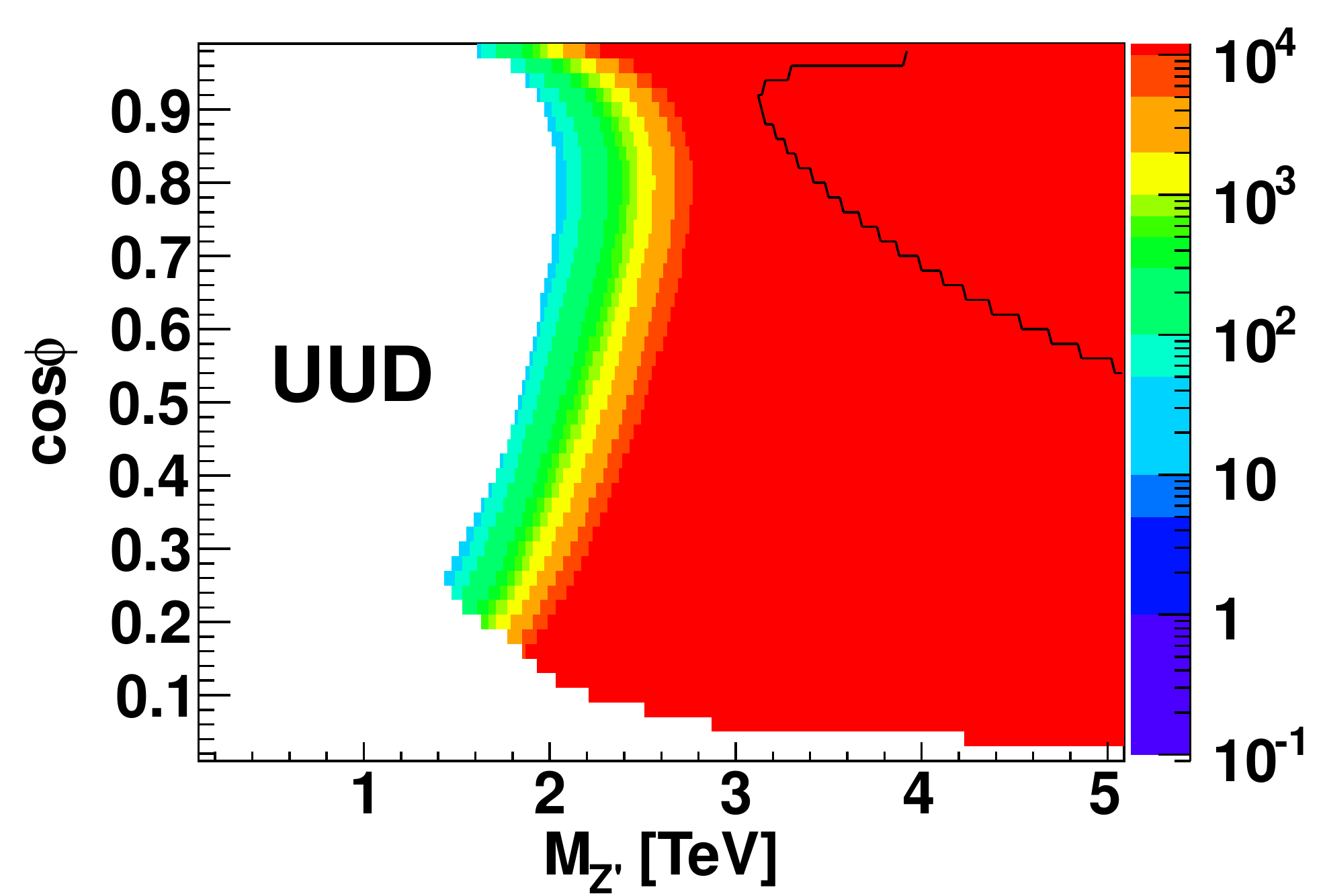}
\caption{$5\sigma$ Discovery potential (fb$^{-1}$) for different 
luminosity at LHC7 via $Z^\prime$ leptonic decay channel. 
The current combined constraints are within solid black contour. The color palette shows the integrated luminosity with unit fb$^{-1}$. 
The dashed black lines in LRD and SQD
represent MLR and MSQ models.}
\label{lumi7Z}
\end{figure}

Figure~\ref{lumi7Z} shows the $5\sigma$ discovery potential (fb$^{-1}$) 
at the  LHC7 via the $Z^\prime$ leptonic decay channel,
and the current combined constraints are within solid black contour.
We can see that for LRD(T), SQD, TFD, UUD, MLR and MSQ, 
further discovery via the $Z^\prime$ leptonic decay channel needs more than $100$ fb$^{-1}$, 
which is definitely far beyond the total integrated luminosity before LHC switches away from 7 TeV.
However, some corner of the parameter space of LPD(T), HPD(T) and FPD(T) 
can be further tested when LHC7 reaches $5.6$ fb$^{-1}$.
Especially, for HPD(T) and FPD(T), there are small regions where 
$Z^\prime$ can be discovered with a few fb$^{-1}$ luminosity 
or these parameters can be excluded with less than one fb$^{-1}$ luminosity.
%The reason why present LHC7 data has not excluded them
%is that  $Z^\prime$ masses larger than 2 TeV have not been reached yet.
%Therefore, further analysis based on present data collection 
%can make improvement on exclusion or find some little sign of new physics signal.
%This means, for LHC7 operation, 
At the LHC7, the $Z^\prime$ leptonic decay channel 
is more efficient than EWPTs on discovering LPD(T), HPD(T) and FPD(T).
For LRD(T), LPD(T) and HPD(T), EWPTs are more sensitive to the large $c_\phi$ region, where LHC7 cannot compete with EWPTs.
For SQD, TFD and UUD, both $W^\prime$ and $Z^\prime$ leptonic decay channel cannot make further test 
at LHC7, because the constraint from EWPTs for UUD is much stronger than Tevatron or LHC7 data,
as shown in Fig.~\ref{parameter_constraints}.  
\begin{figure}
	\includegraphics[width=0.32\textwidth]{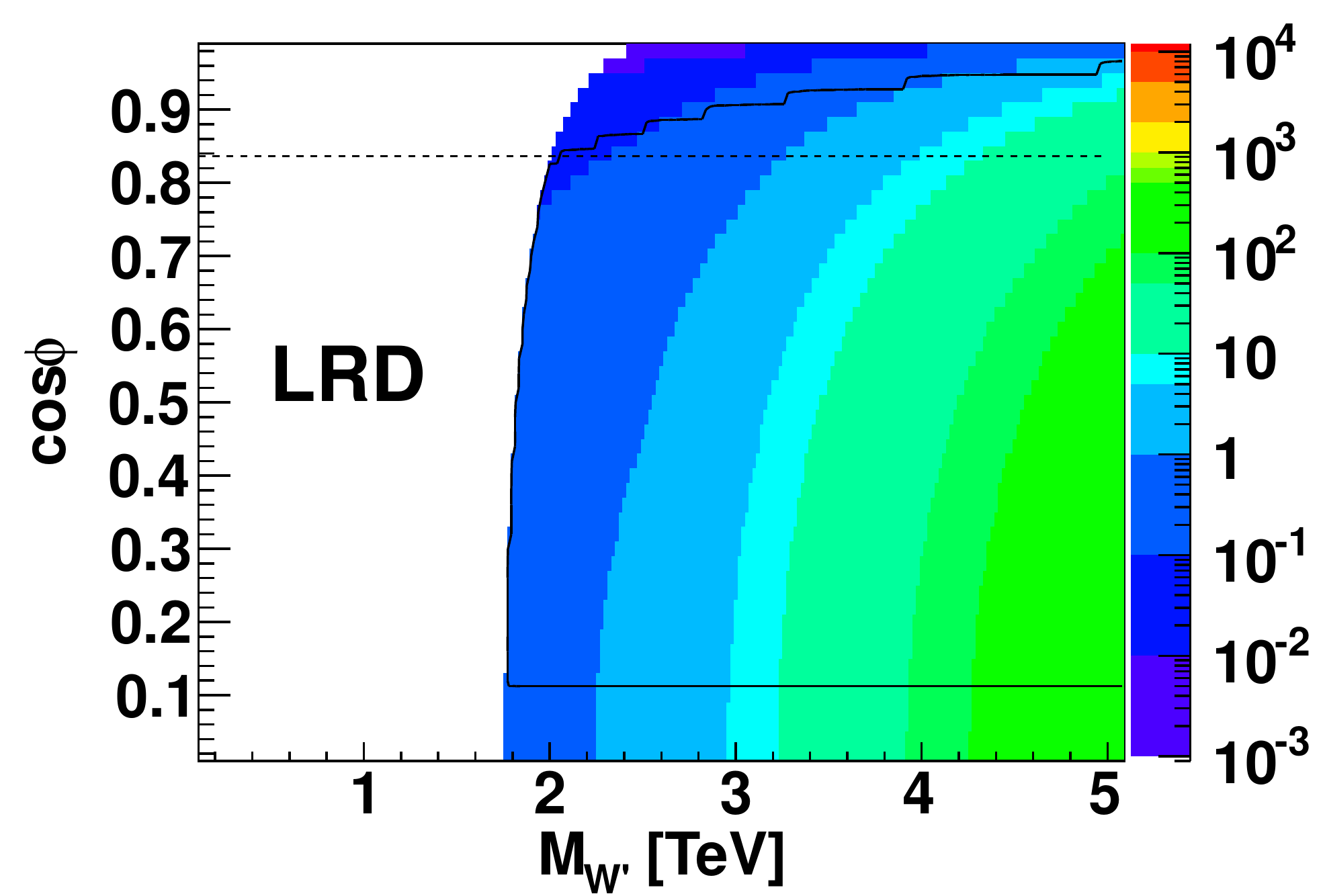}
	\includegraphics[width=0.32\textwidth]{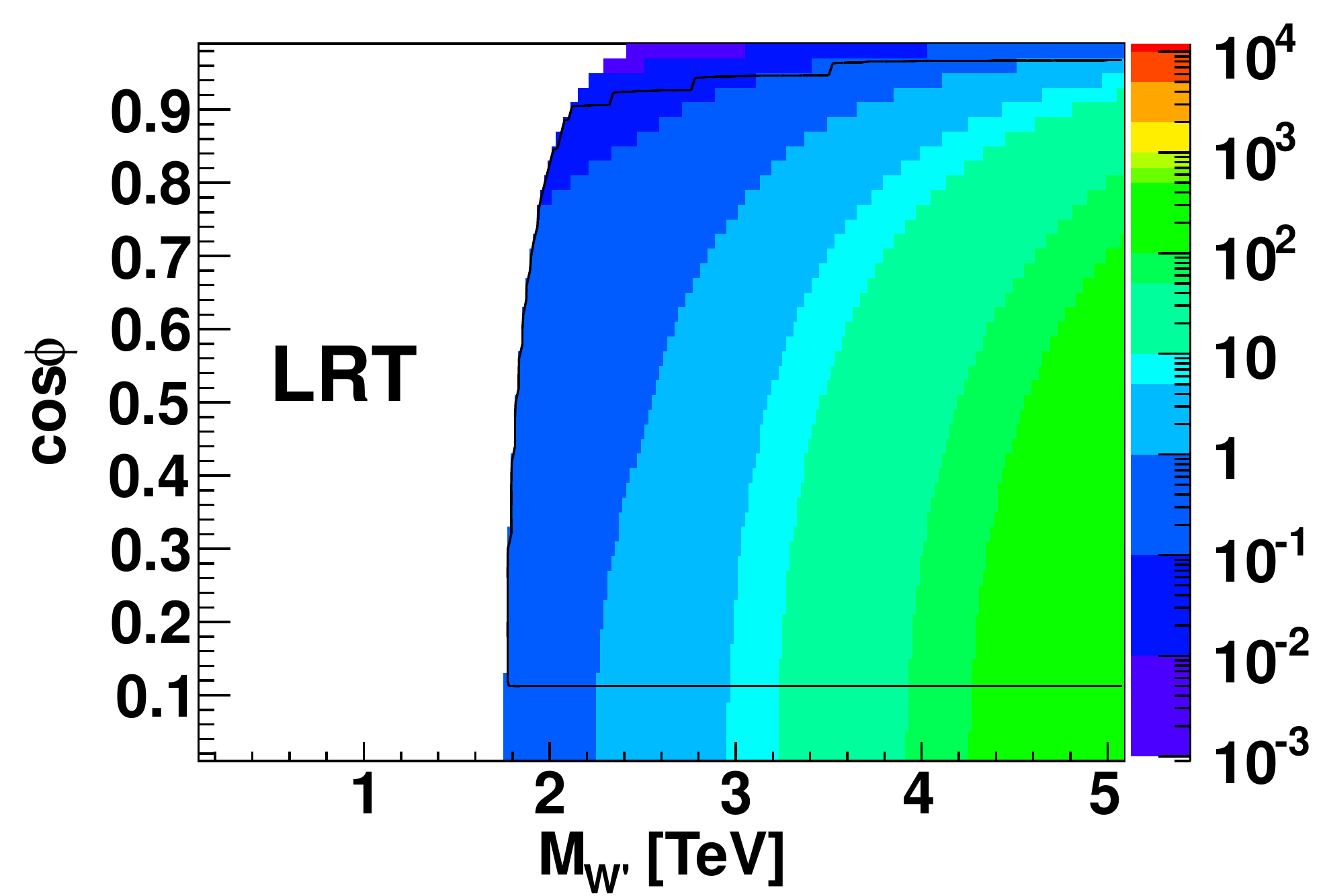}
	\includegraphics[width=0.32\textwidth]{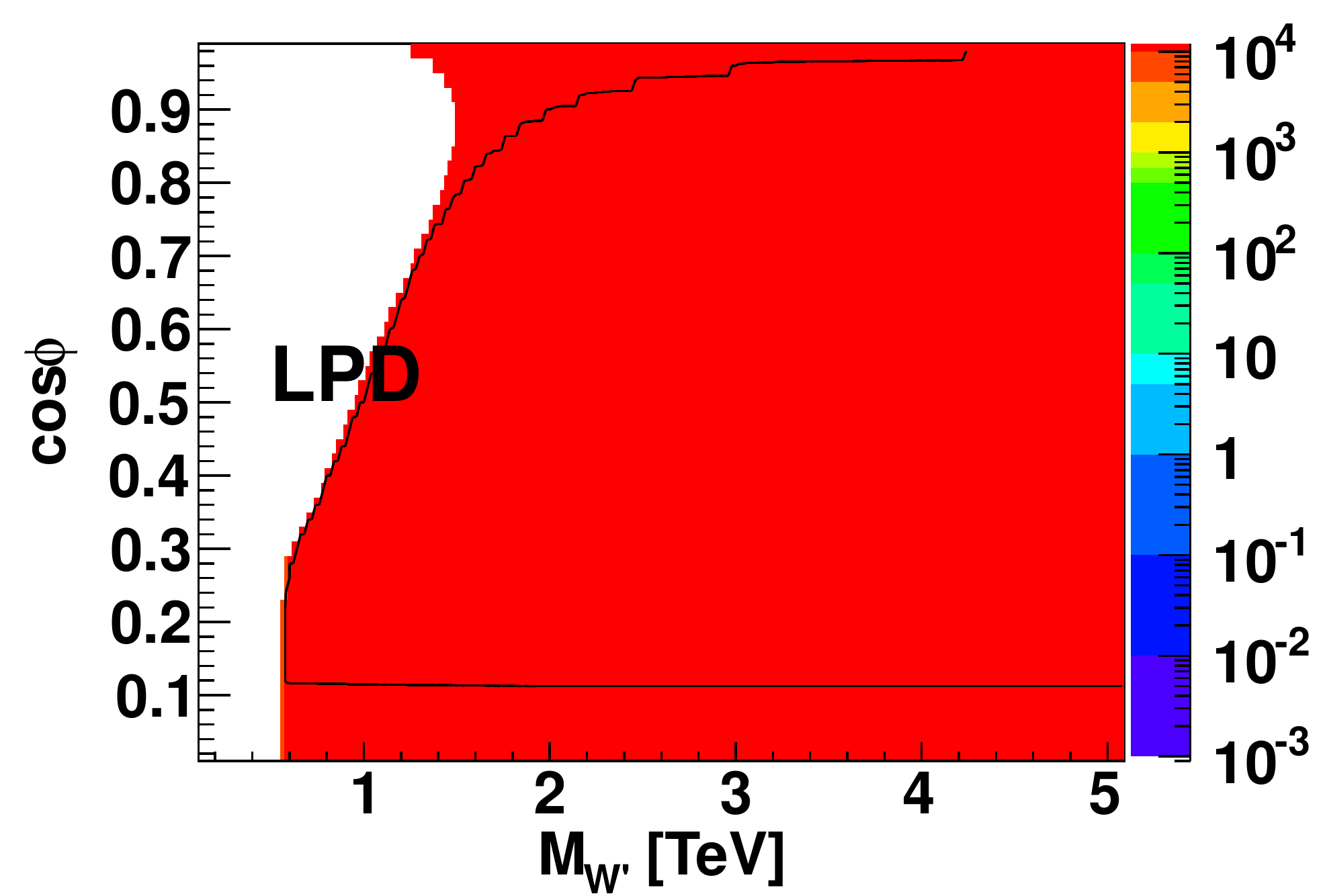}
	\includegraphics[width=0.32\textwidth]{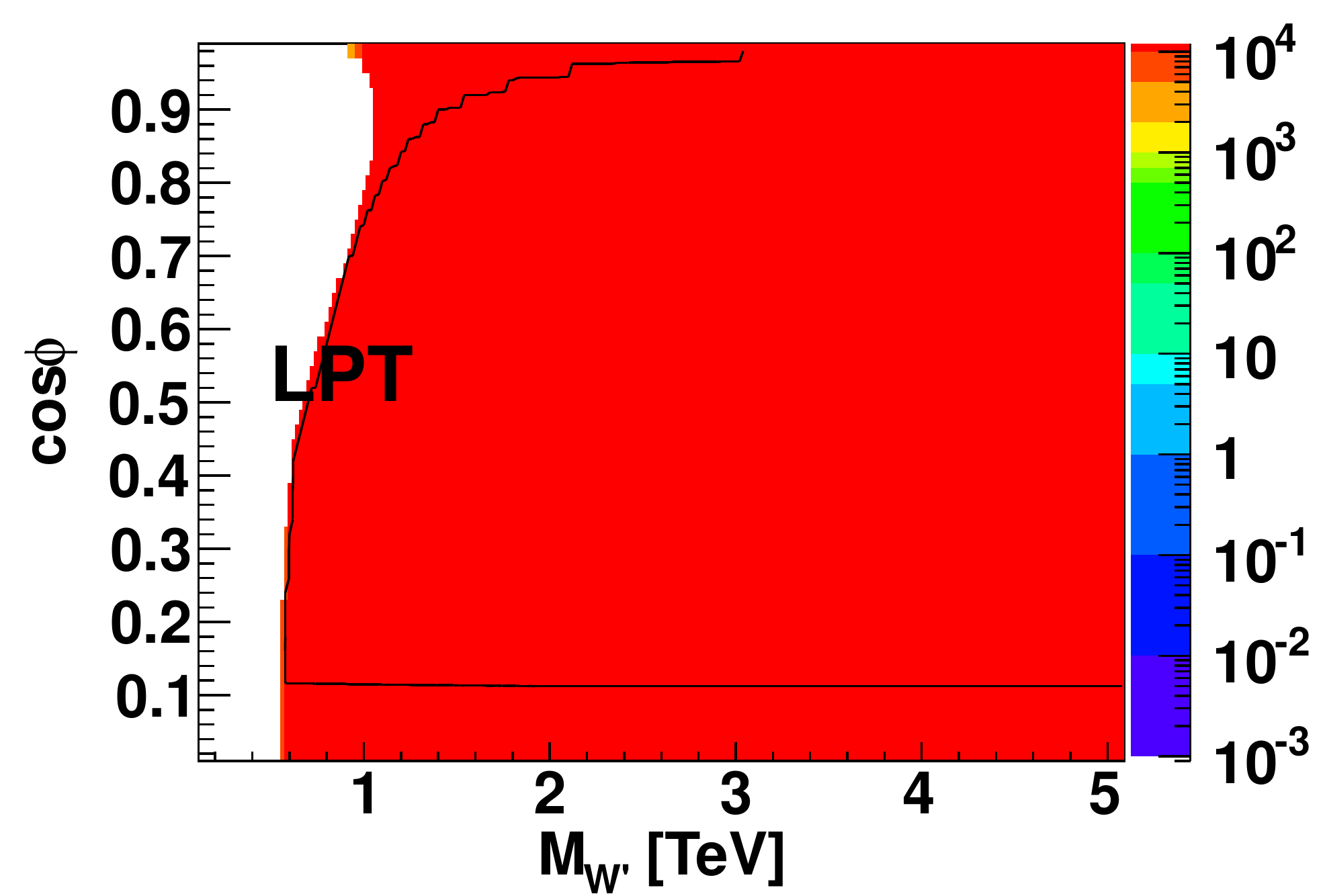}
	\includegraphics[width=0.32\textwidth]{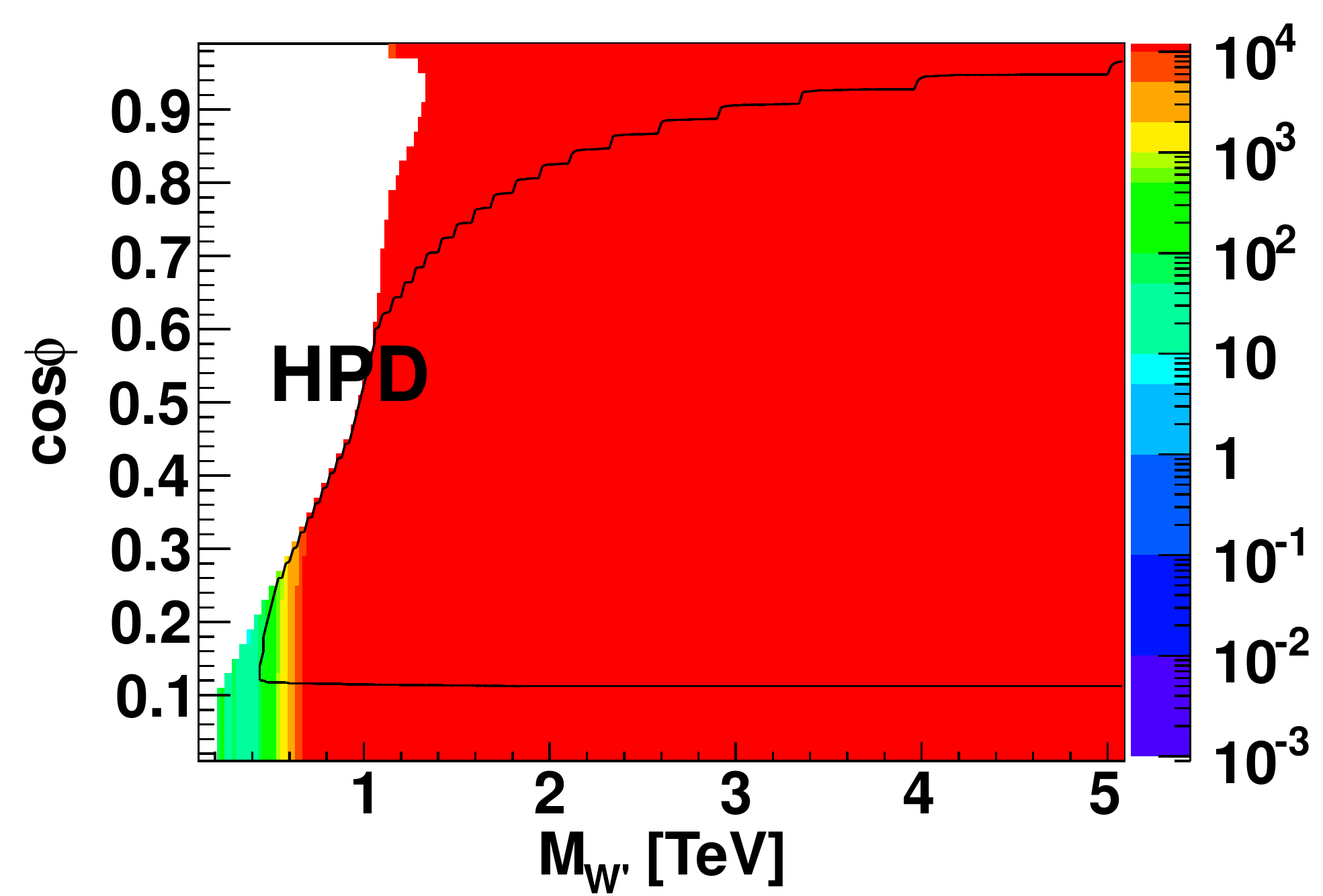}
	\includegraphics[width=0.32\textwidth]{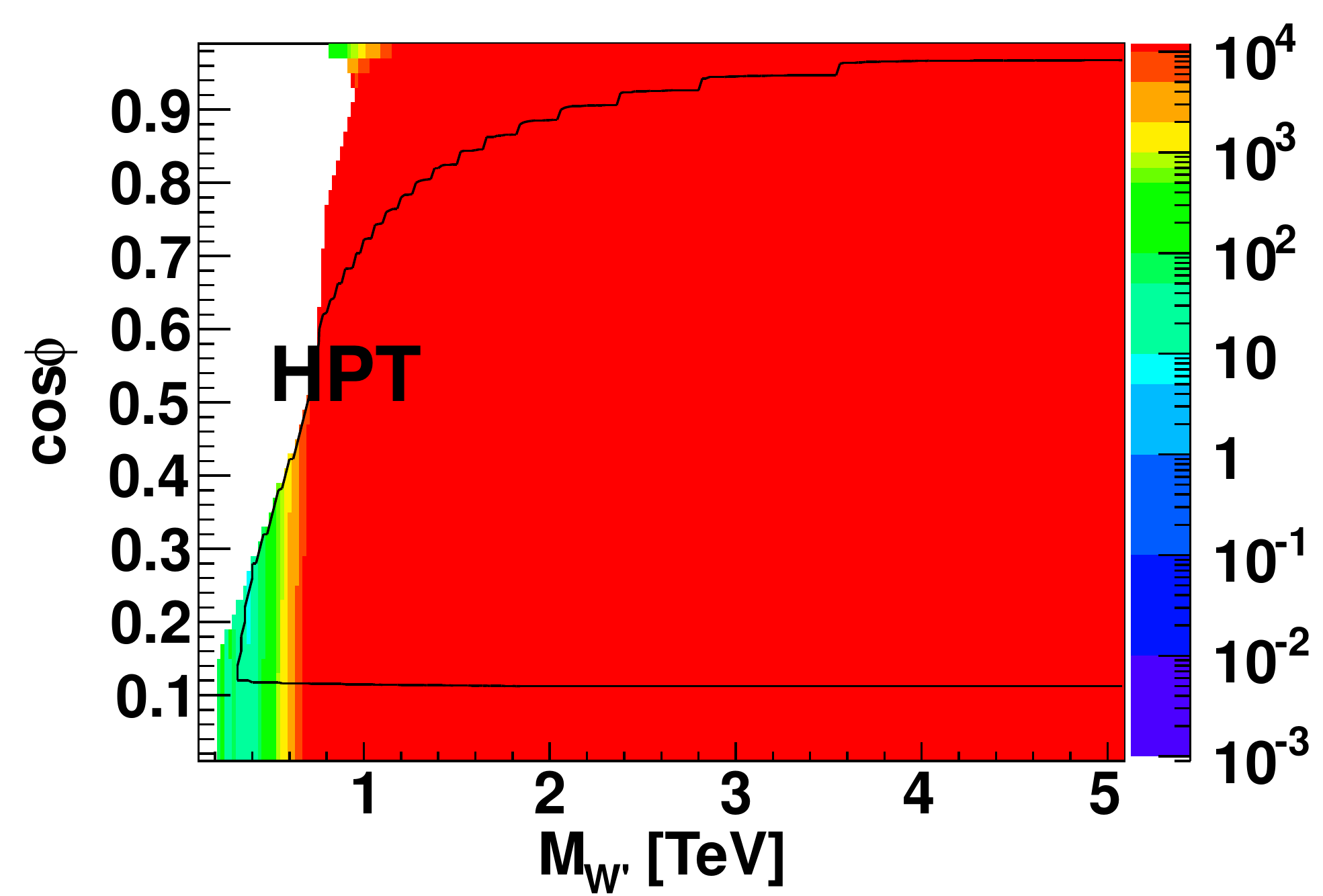}
	\includegraphics[width=0.32\textwidth]{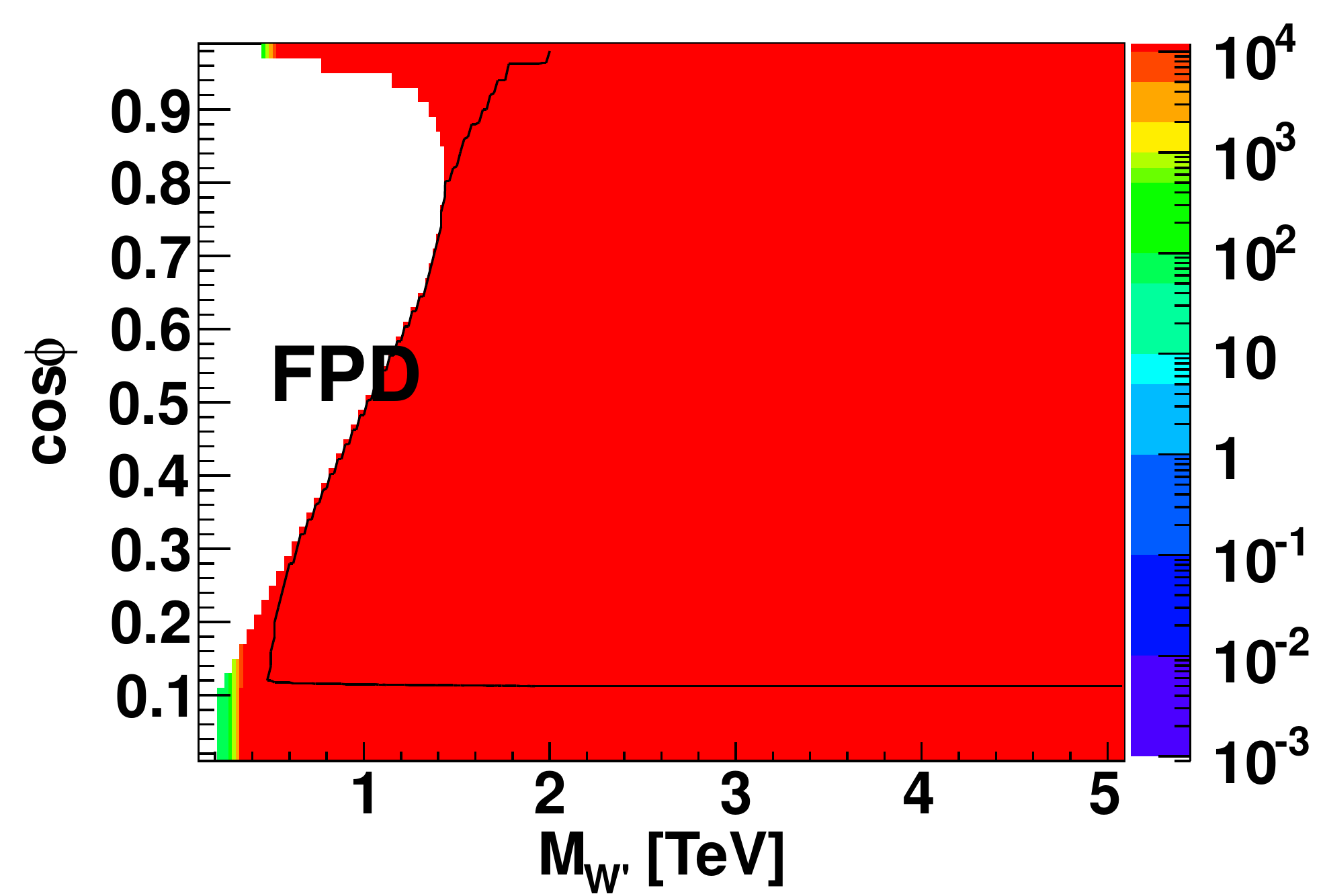}
	\includegraphics[width=0.32\textwidth]{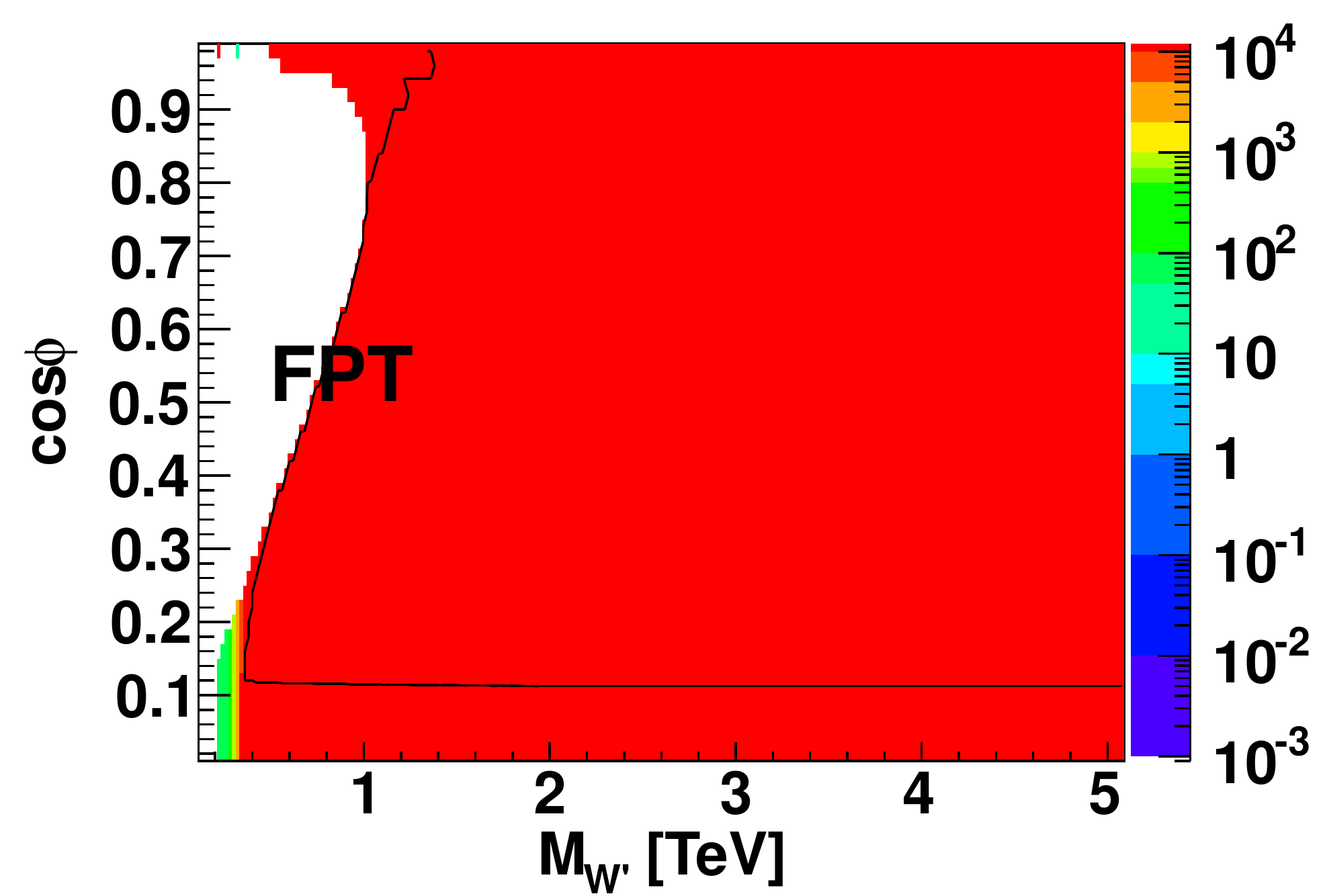}
	\includegraphics[width=0.32\textwidth]{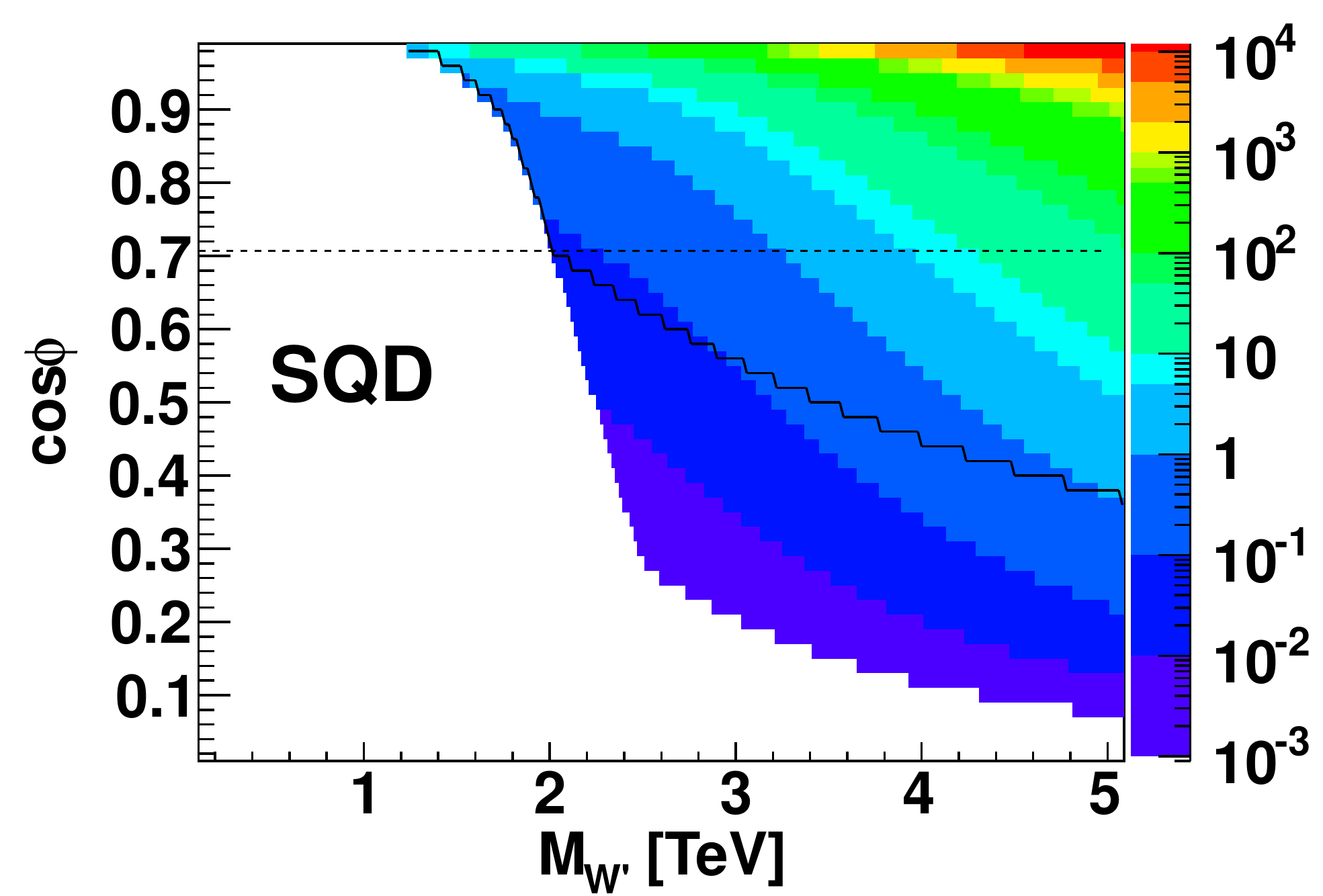}
	\includegraphics[width=0.32\textwidth]{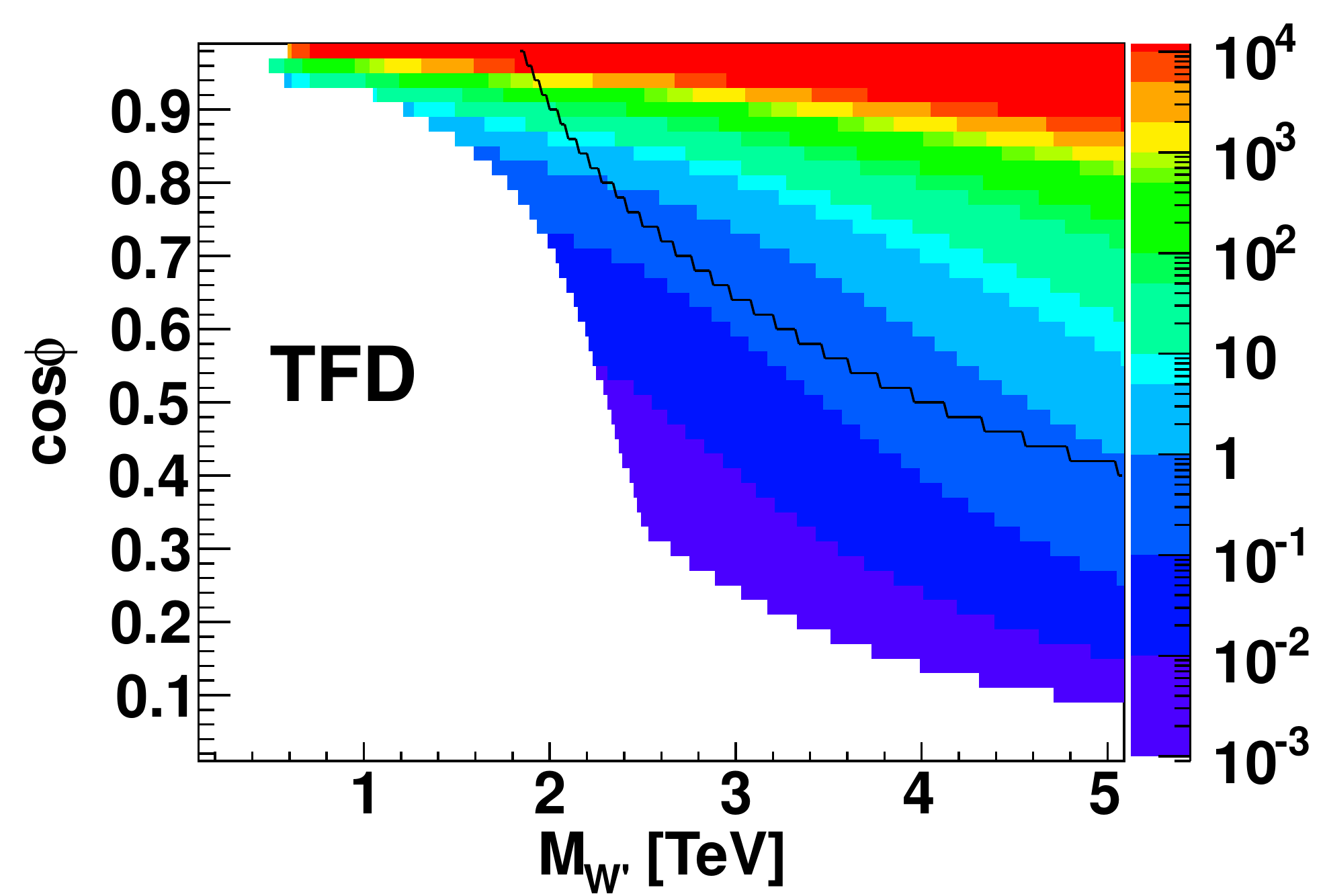}
	\includegraphics[width=0.32\textwidth]{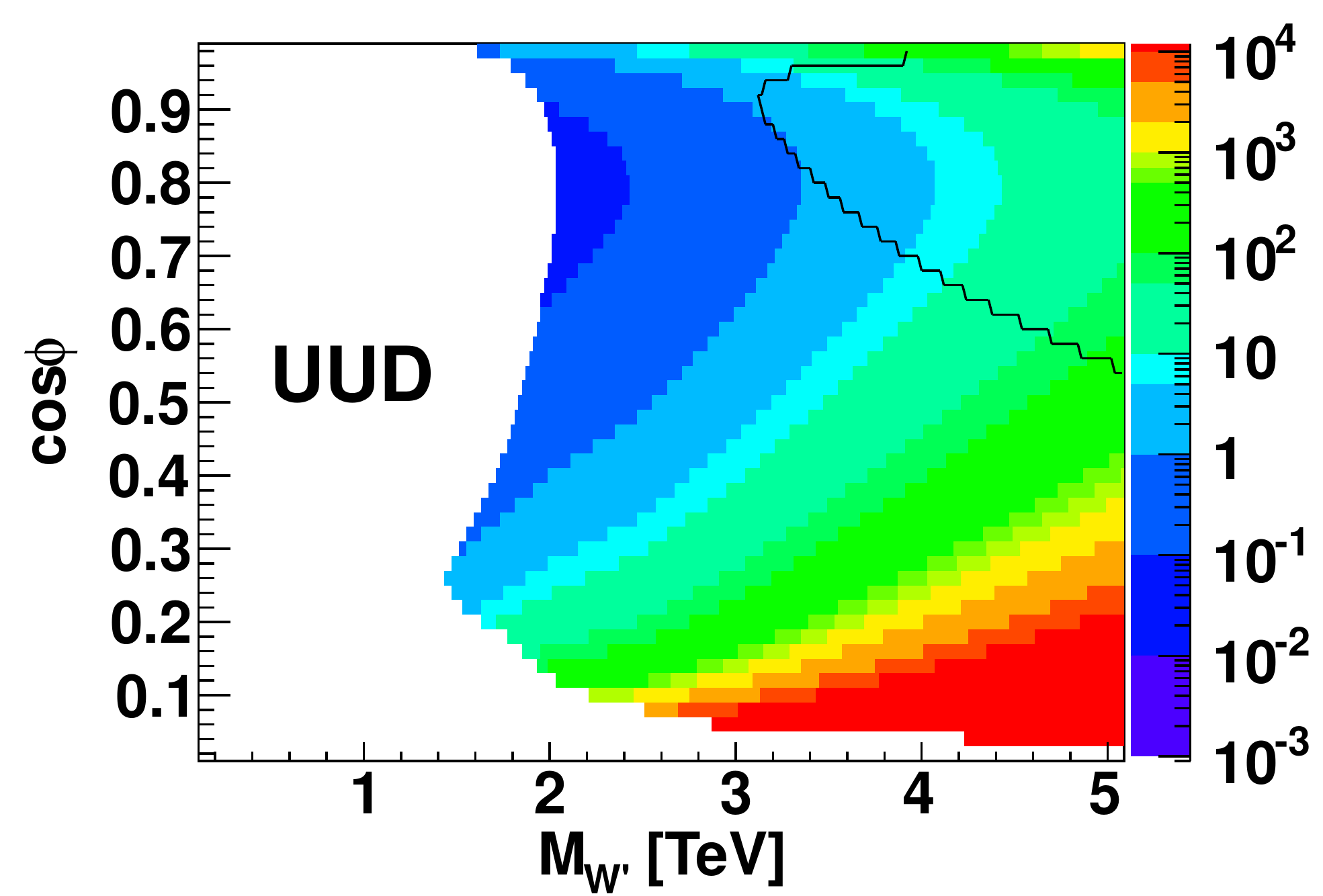}
\caption{$5\sigma$ Discovery potential (fb$^{-1}$) for different 
luminosity at LHC14 via $W^\prime$ leptonic decay channel. 
The color palette shows the integrated luminosity with unit fb$^{-1}$. 
The current constraints are within solid black contour. 
The dashed black lines in LRD and SQD
represent MLR and MSQ models.}
\label{lumi14W}
\end{figure}

Figure~\ref{lumi14W} presents the $5\sigma$ discovery potential (fb$^{-1}$) 
at the LHC14 via the $W^\prime$ leptonic decay channel,
and current constraints are within the solid black contour.
After the LHC14 collects 10 fb$^{-1}$, a sizable region of parameter space will be further tested, 
except all the phobic models, LPD(T), HPD(T) and FPD(T). 
For the phobic models, very large integrated luminosity is needed to have $5\sigma$ discovery 
because of the small total cross section in the $W^\prime$ leptonic decay channel,
which is either suppressed by the production rate of the $W^\prime$, such as HPD(T) and FPD(T), 
or suppressed by the decay branching ratio, such as LPD(T) and FPD(T).
With a 10 fb$^{-1}$ luminosity, 
for LRD(T), the discovery potential for $W^\prime$ mass can reach more than 3 TeV,
and the $W^\prime$ mass discovery for MLR can reach more than 4 TeV.
Furthermore, for the large $c_\phi$ region in LRD(T), 
LHC14 search via $W^\prime$ leptonic decay channel
can easily probe large $M_{W^\prime}$ region with several fb$^{-1}$.
In BP-II, the current constraints already pushed the $W^\prime$  
to the large mass region.
However, with a $10$ fb$^{-1}$ integrated luminosity, for SQD, TFD and UUD models, 
most of the allowed region below $5$ TeV $W^\prime$ mass can be further tested.
For relatively small $c_\phi$ in SQD and TFD, a few fb$^{-1}$ luminosity can even probe $W'$ boson beyond 5 TeV.
When the LHC is upgraded to 14 TeV, SQD, TFD and UUD can be further tested, 
exploring the region where current constraints cannot reach.
This shows that the capability of LHC14 is far beyond LHC7.
However, even LHC14 cannot tested all the phobic models, 
such as LPD(T), HPD(T) and FPD(T), via only $W^\prime$ leptonic decay channel.

\begin{figure}
\includegraphics[width=0.32\textwidth]{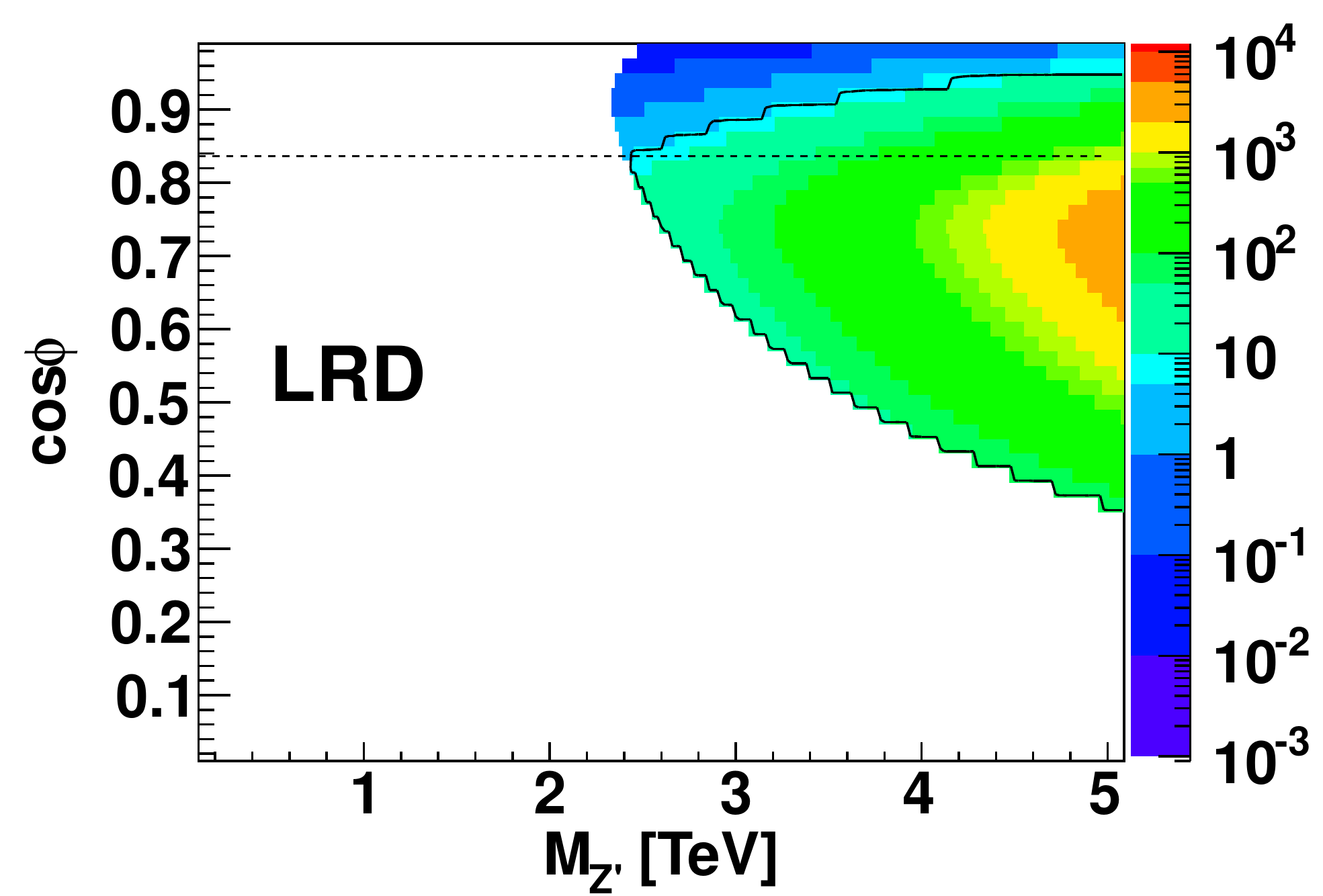}
\includegraphics[width=0.32\textwidth]{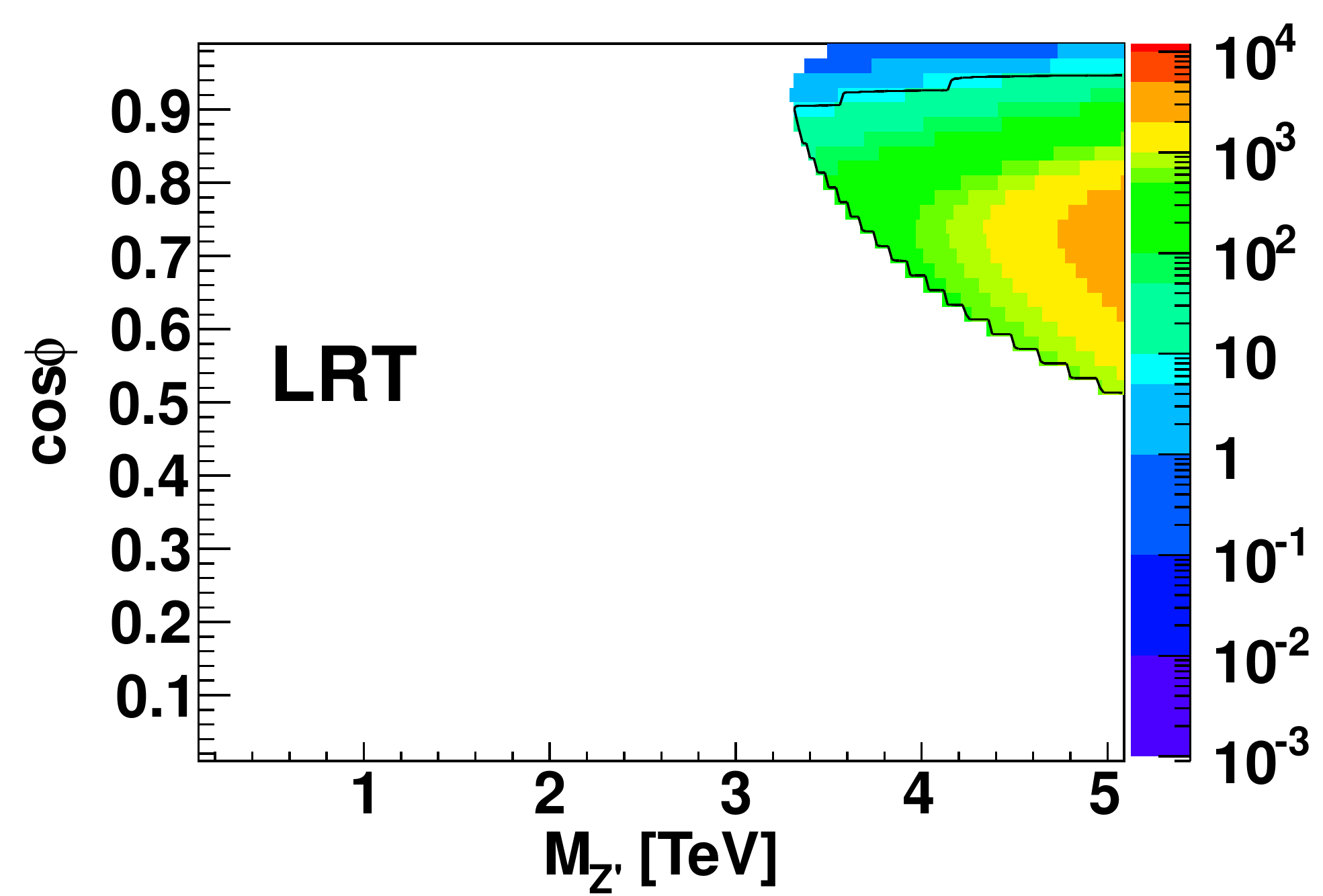}
\includegraphics[width=0.32\textwidth]{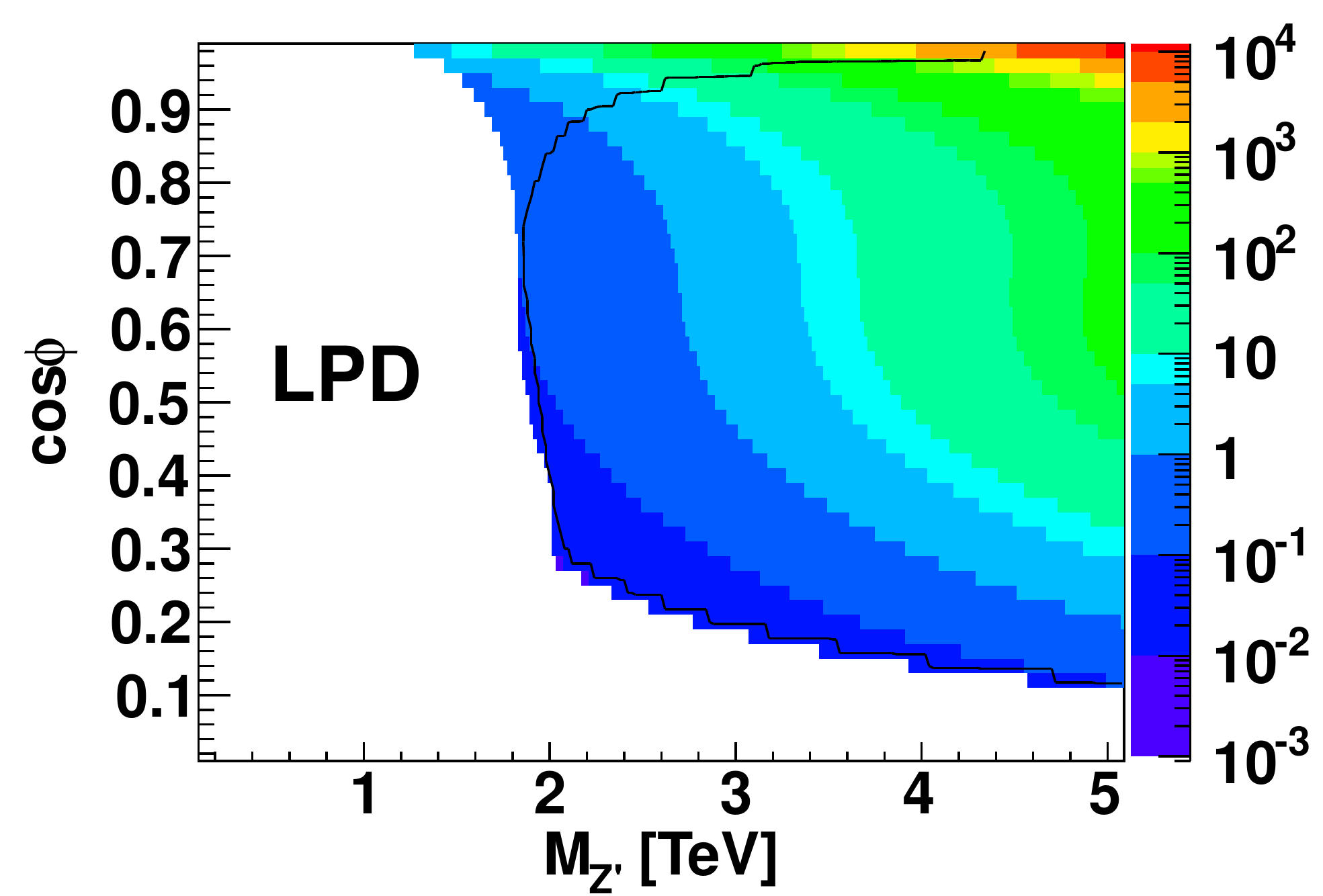}
\includegraphics[width=0.32\textwidth]{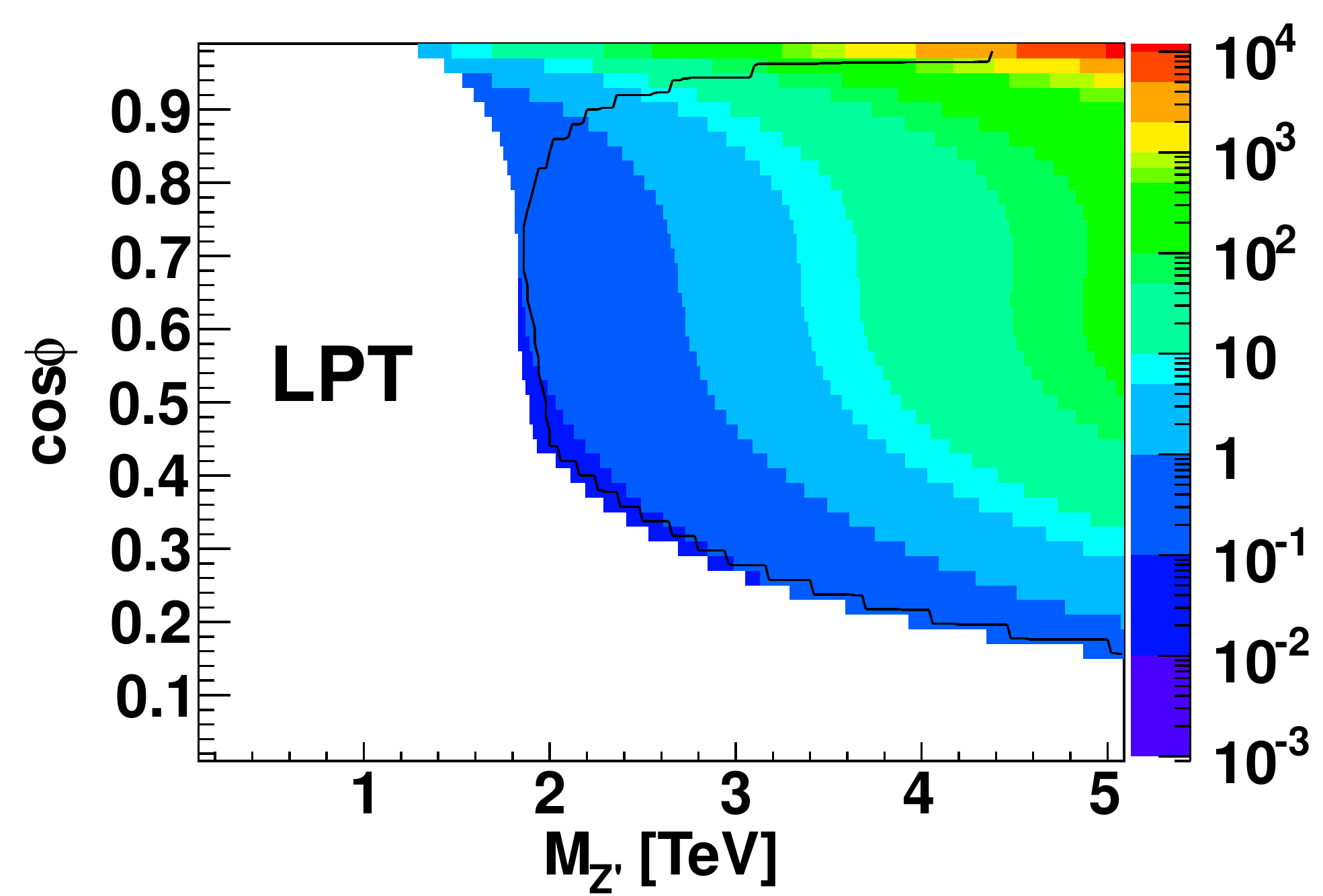}
\includegraphics[width=0.32\textwidth]{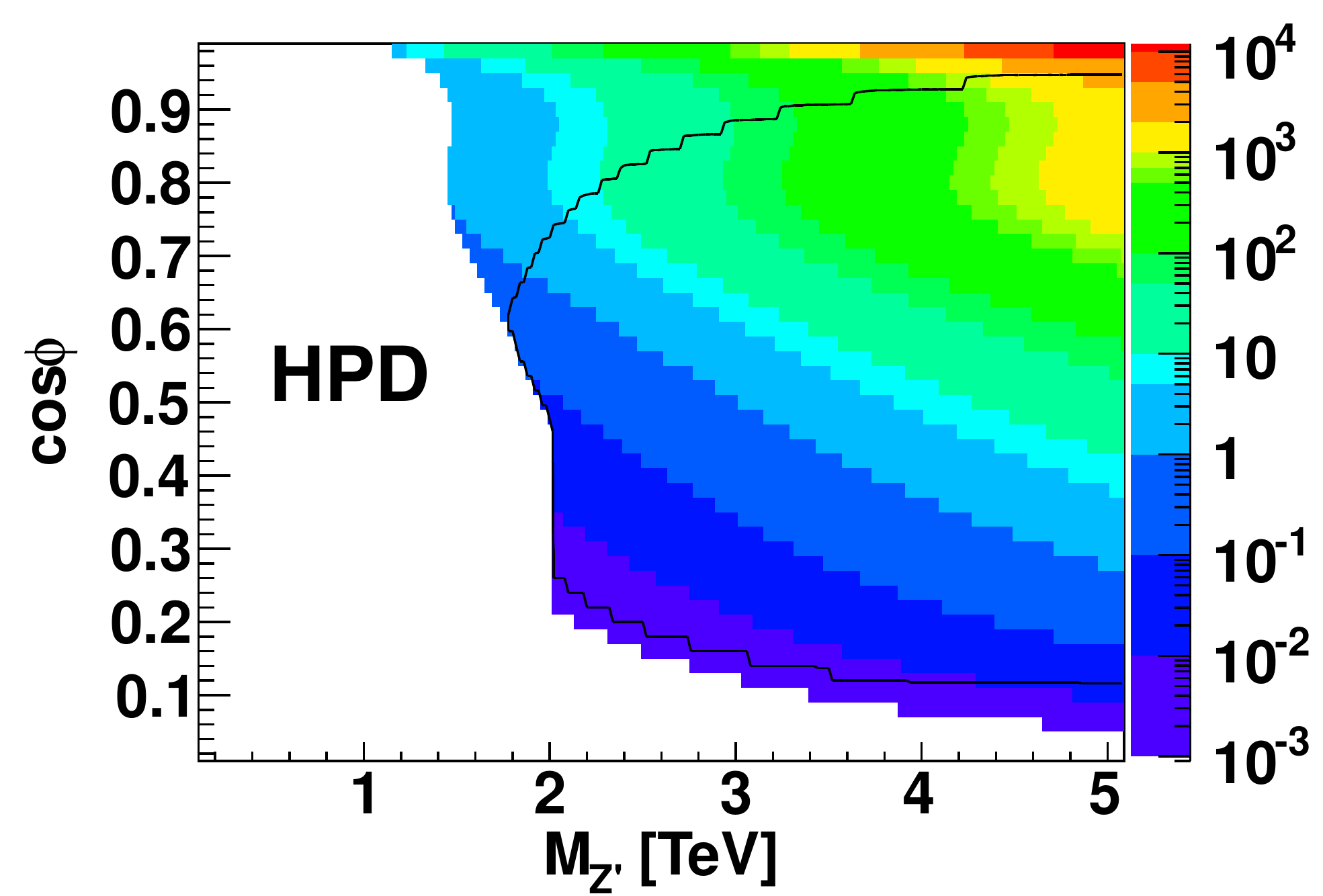}
\includegraphics[width=0.32\textwidth]{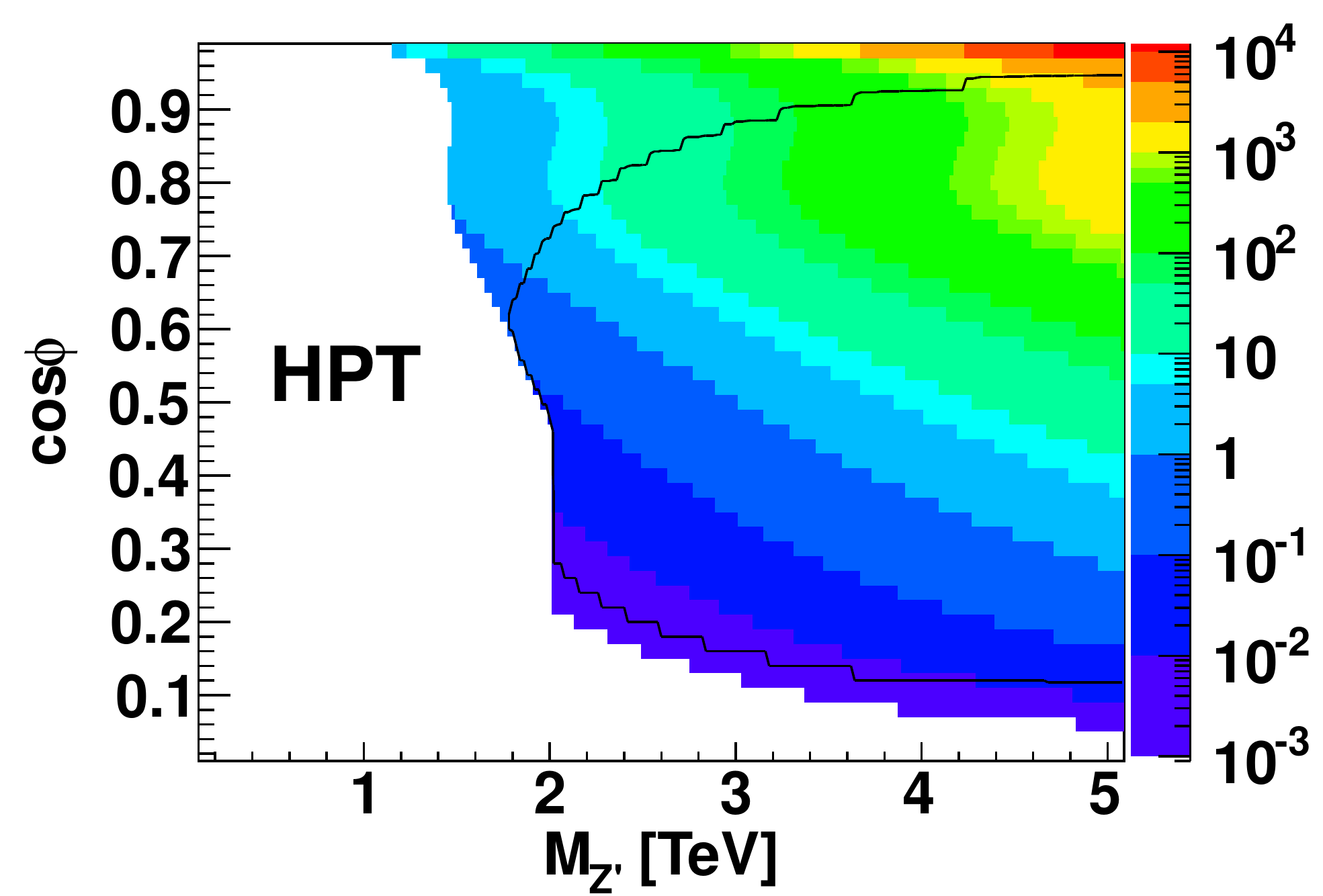}
\includegraphics[width=0.32\textwidth]{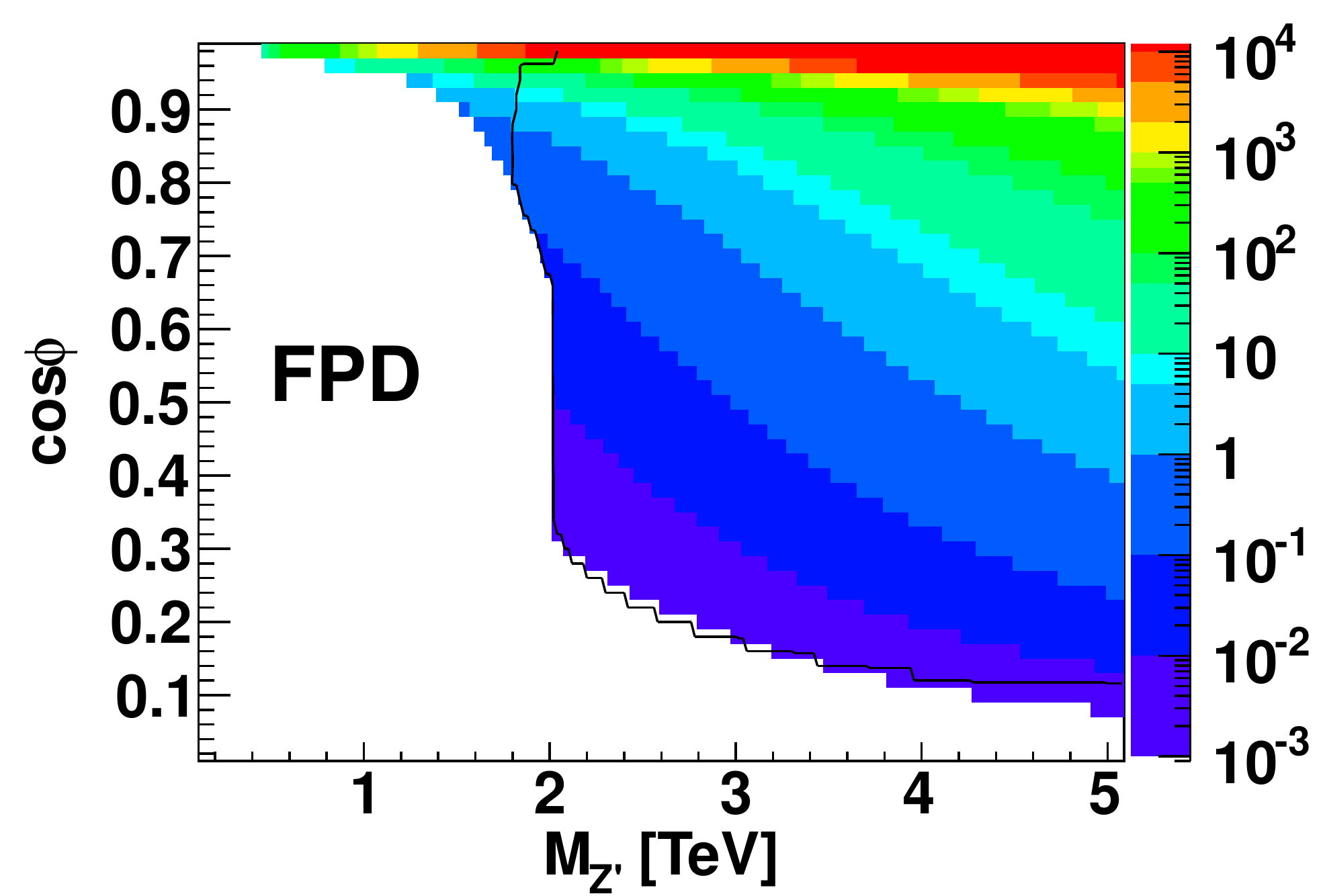}
\includegraphics[width=0.32\textwidth]{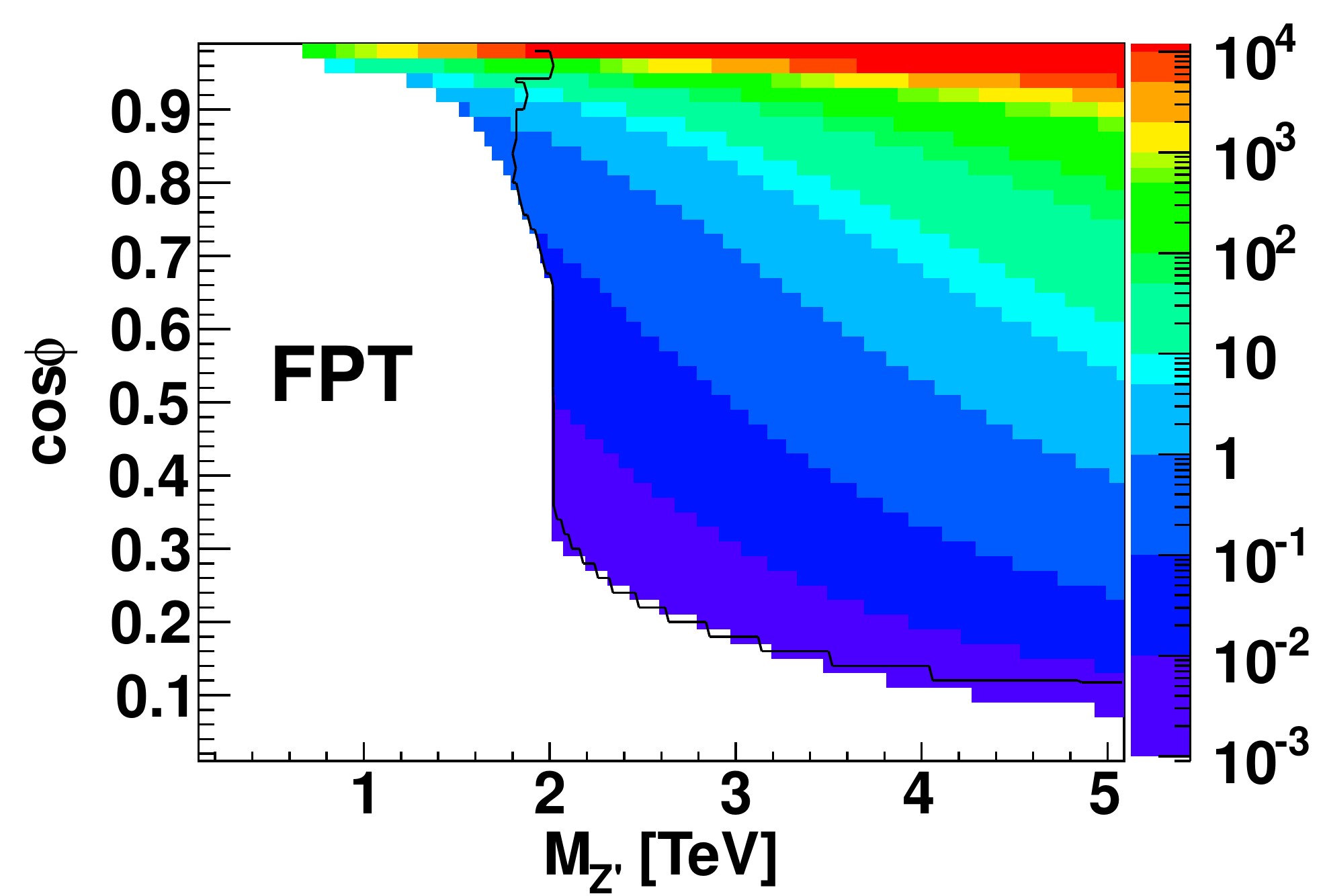}
\includegraphics[width=0.32\textwidth]{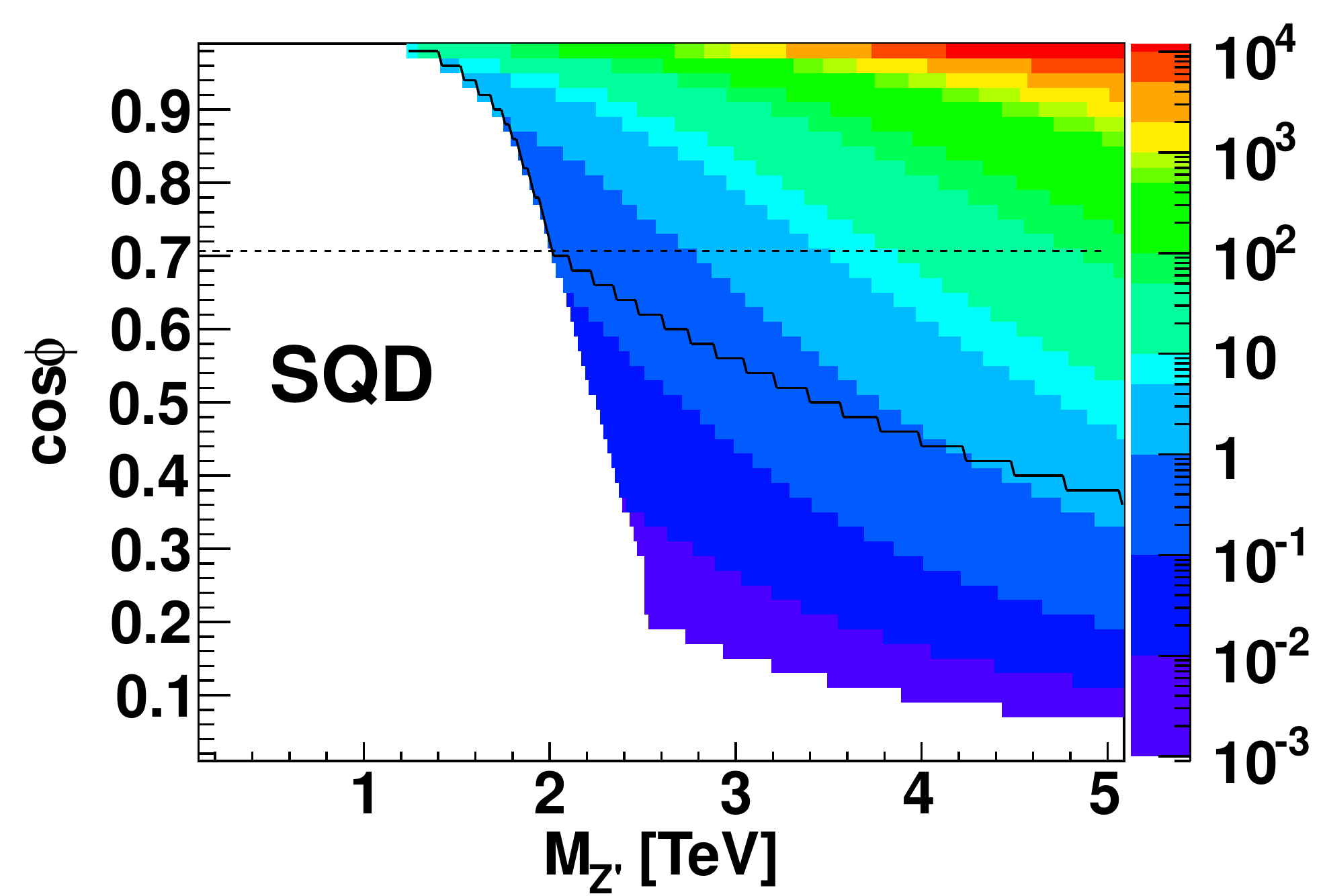}
\includegraphics[width=0.32\textwidth]{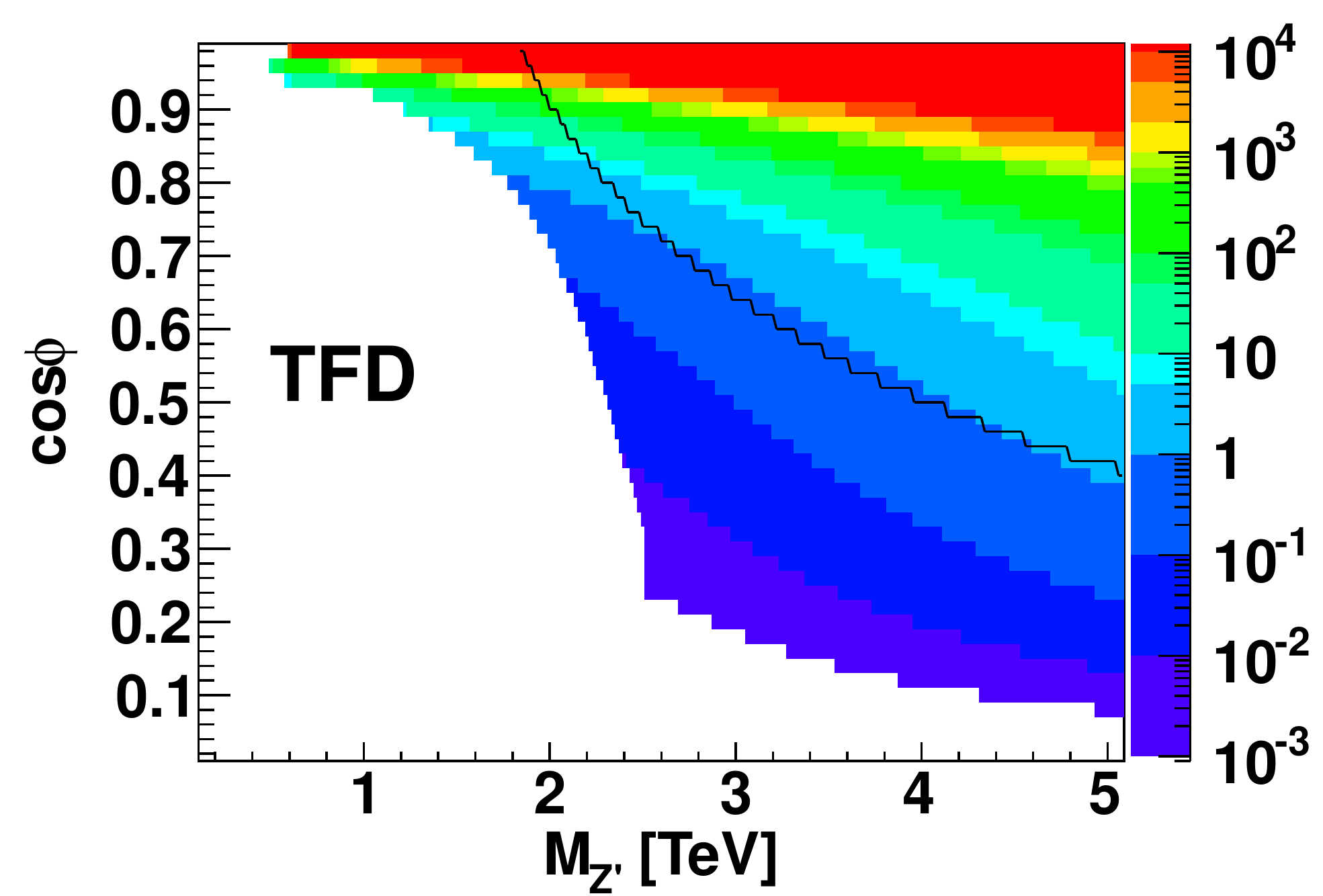}
\includegraphics[width=0.32\textwidth]{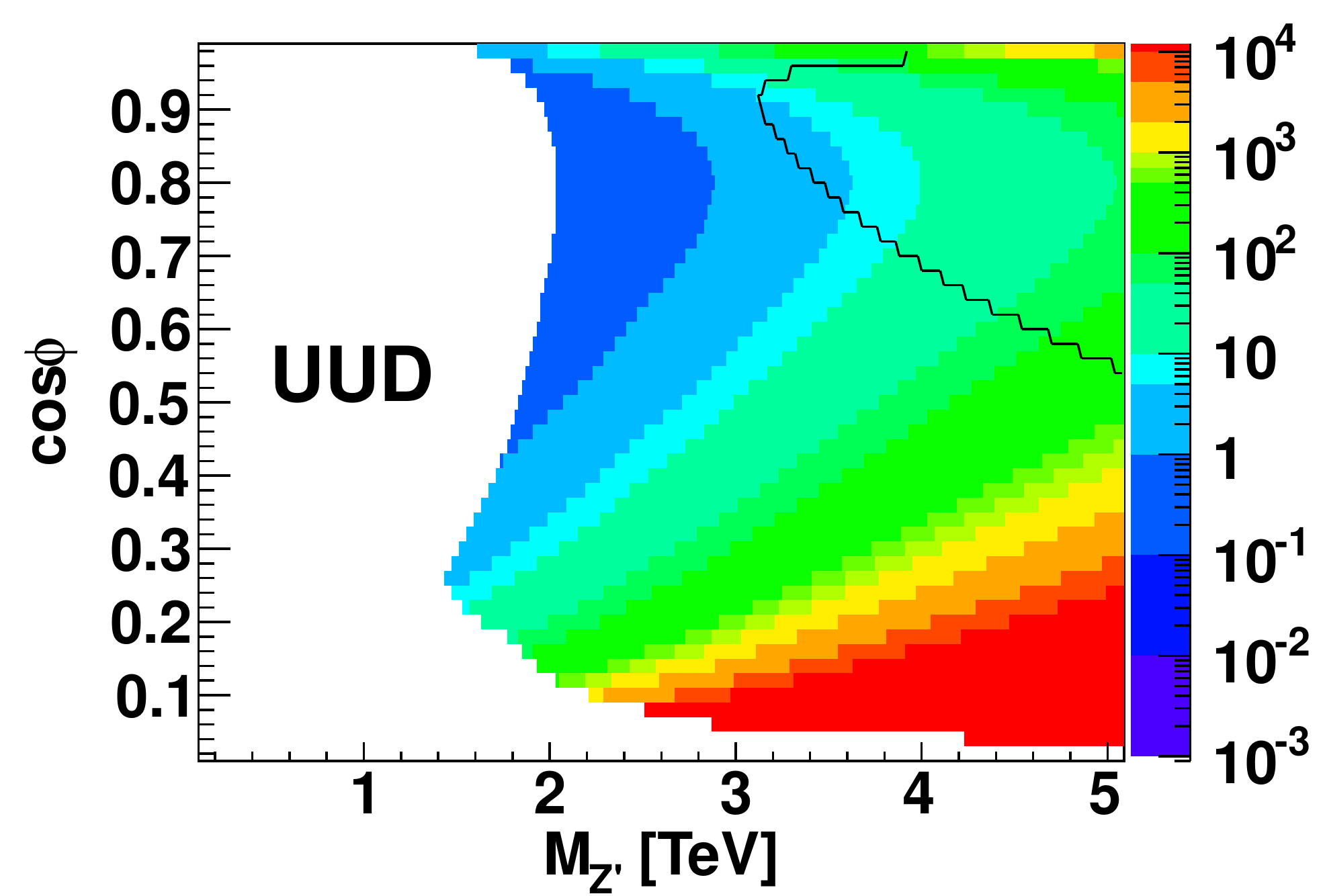}
\caption{$5\sigma$ Discovery potential  (fb$^{-1}$) for different 
luminosity  at LHC14 via $Z^\prime$ leptonic decay channel. 
The color palette shows the integrated luminosity with unit fb$^{-1}$. 
The current combined constraint are within solid black contour. 
The dashed black lines in LRD and SQD
represent MLR and MSQ models.}
\label{lumi14Z}
\end{figure}

Figure~\ref{lumi14Z} shows the $5~\sigma$ discovery potential (fb$^{-1}$) for the LHC14 
via the $Z^\prime$ leptonic decay channel,
and current combined constraints are within solid black contour.
For the models other than LRD(T), UUD and MLR, 
the LHC14 can already test the parameter space effectively
with the integrated luminosity  less than 1 fb$^{-1}$.
However, for the FPD(T), SQD and TFD models,  EWPTs are more sensitive to the large $c_\phi$ region.
Also if the luminosity can reach 10 fb$^{-1}$,  we can test a large parameter space region,
where we can either discover new physics based on these models or constrain the parameters in the relevant region.
For LRD(T), UUD and MLR, when integrated luminosity is accumulated to more than 100 fb$^{-1}$,
LHC14 data can have sizable parameter space further tested up to even beyond 5 TeV $M_{Z^\prime}$.
For LRD(T) $Z^\prime$ leptonic decay channel is less effective than $W^\prime$ channel.
However, for the phobic models, such as LPD(T), FPD(T) and HPD(T),
there is no ${\mathcal O}(1/x)$ suppression on the couplings of $Z'$ to fermions, unlike the couplings of $W'$ to fermions. 
So $Z^\prime$ leptonic decay channel is much more effective than $W^\prime$ for the investigation 
based on the LHC14 data. Especially, for the small $c_\phi$ region, a few pb$^{-1}$ luminosity can probe
very large $M_{Z^\prime}$. 
In the phobic models, observing a $Z^\prime$ alone
cannot rule out the possibility of non-Abelian gauge extension of new physics.

In BP-II, both $Z^\prime$ and $W^\prime$ leptonic decay channel 
are suitable to explore the allowed parameter space of the models. 
Since the mass of $W^\prime$ and $Z^\prime$ are degenerate in BP-II, 
discovering degenerated $W^\prime$ and $Z^\prime$ in the leptonic decay channels 
at the same time will be the distinct feature compared to the models in BP-I.
Compared to the LHC7 discovery potentials in Figs.~\ref{lumi7W} and~\ref{lumi7Z},
Figures~\ref{lumi14W} and~\ref{lumi14Z} show that for LHC the upgrade of the CM energy from 7 TeV to 14 TeV
is much more efficient than accumulation of luminosity.
For instance, for FPD(T) the $Z^\prime$ leptonic decay channel at LHC14 
with less than $1$ fb$^{-1}$ can explore 
some region of parameter space, while LHC7 needs more than $10^4$ fb$^{-1}$ luminosity to achieve the similar sensitivity.  
For all these $G(221)$ models, LHC14 can exceed 
the capability of current combined constraints and have promising discovery potential.

If the heavy gauge bosons $W^\prime$ and/or $Z^\prime$ are not discovered, the potential for discovery can be converted to the $95\%$ CL 
exclusion limits on the heavy gauge bosons $W^\prime$ and/or $Z^\prime$ using the relations  
$Z=S/3$ as discussed above. 
Equivalently, the luminosity for exclusion limits is about one order of magnitude lower 
than the discovery luminosity. 
Therefore, as shown in Fig.~\ref{lumi7W} supposing on signals found, 
via $W^\prime$ leptonic decay channel, $W^\prime$ mass in LRD(T) 
can be further excluded by about 100 GeV 
after the LHC7 collects $5.61$ fb$^{-1}$ luminosity.
Figure~\ref{lumi7Z} shows that via $Z^\prime$ leptonic decay channel, one can expect slightly 
further exclusion on LPD(T), HPD(T) and FPD(T) at LHC7.
At the LHC14, as show in Figs.~\ref{lumi14W} and~\ref{lumi14Z}, 
exclusion region can extend very fast when luminosity is accumulated.
For instance, via $W^\prime$ leptonic decay channel, 1 fb$^{-1}$ can exclude most of the parameter region 
for LRD(T), SQD, TFD and UUD, and 10 fb$^{-1}$ can 
completely remove the possibility of $M_{Z^\prime}$ less than 5 TeV in  these models 
if there is no any sign of $W^\prime$ production.
For LPD(T), HPD(T) and FPD(T), $Z^\prime$ leptonic decay channel at the LHC14 can 
be used to exclude most of the parameter space region
with only 1 fb$^{-1}$ luminosity. 
Then data with 10 fb$^{-1}$ luminosity at LHC14 may leave LPD(T) and HPD(T)
and FPD(T) only a corner of parameter space at large $c_\phi$ to survive.
The shapes of the exclusion contours are the same as these at the discovery contours.

\section{Conclusion}

\begin{table}[!htb]
\renewcommand{\arraystretch}{2.0}
 \begin{tabular}{|c||c|c||c|c|}
\hline 
Models  & Current $M_{W'}$ Limit   & $M_{W'}$ LHC14 Reach & Current $M_{Z'}$ Limit   & $M_{Z'}$ LHC14 Reach\tabularnewline
\hline
\hline 
LRD (LRT)
& 1.72 (1.76) \textrm{ TeV}& 3.2 - 5 \textrm{ TeV} & 2.25 (3.2) \textrm{ TeV}& 2.8 - 5 \textrm{ TeV} \tabularnewline
\hline 
LPD (LPT)
& 0.55 (0.55) \textrm{ TeV}& \textrm{No improvement} & 1.8 (1.8) \textrm{ TeV}& 3.5 - 5 \textrm{ TeV}\tabularnewline
\hline 
HPD (HPT)
& 0.46 (0.35) \textrm{ TeV}& 0.55 \textrm{ TeV} & 1.7 (1.7) \textrm{ TeV}& 3 - 5 \textrm{ TeV}\tabularnewline
\hline 
FPD (FPT)
& 0.5 (0.4) \textrm{ TeV}& \textrm{No improvement} & 1.75 (1.75) \textrm{ TeV}& 1.75 - 5 \textrm{ TeV}\tabularnewline
\hline 
SQD
& 1.25 \textrm{ TeV}&  3.5 - 5 \textrm{ TeV} & 1.25 \textrm{ TeV}& 1.5 - 5 \textrm{ TeV} \tabularnewline
\hline 
TFD
& 1.7 \textrm{ TeV}&  2 - 5 \textrm{ TeV} & 1.7 \textrm{ TeV}& 2 - 5 \textrm{ TeV} \tabularnewline
\hline 
UUD
& 3.1 \textrm{ TeV}&  4 - 5 \textrm{ TeV}& 3.1 \textrm{ TeV}& 3.3 - 5 \textrm{ TeV} \tabularnewline
\hline
\hline
\end{tabular}
\caption{The current lowest limits and discovery reaches at the LHC14 with 100 fb$^{-1}$ luminosity on $W'$ and $Z'$ masses. }
\label{tab:masslist}
\end{table}

In this paper we have discussed the potential for discovering, or
setting limits on, the extra heavy gauge bosons $W^\prime$ and/or $Z^\prime$ using
two different scenarios at the LHC: an early run with $\sqrt{s}=7$
TeV and total integrated luminosity of $5.61$ fb$^{-1}$; a long run with
$\sqrt{s}=14$ TeV and $100$ fb$^{-1}$ integrated luminosity. 
The EWPTs, Tevatron and LHC data have been used to set bounds on the
allowed parameter space. We showed that direct searches give tighter
bounds than EWPTs in BP-I. Although LHC data surpass the constraint from 
Tevatron and EWPTs constraints in LRD, LRT models, in other models
the parameter space depends non-trivially on the present bounds,
especially during the early LHC runs. The unexplored parameter space
%,we show that the parameter space
will become accessible for $5\sigma$ discovery at different time scales.  
In LRD(T) it is more efficient to use $W^\prime$ leptonic decay channel 
for discovery or exclusion than  $Z^\prime$ leptonic decay channel.
In the phobic models, it is challenging to discover a $W'$ decaying into leptonic mode. Hence, observing a $Z^\prime$ alone
cannot rule out the possibility of NP models with non-Abelian gauge extension of the standard model.
In BP-II models, both $Z^\prime$ and $W^\prime$ leptonic decay channel 
are suitable to explore the allowed parameter space. 
Discovering degenerate $W^\prime$ and $Z^\prime$ in the leptonic decay channels 
at the same time will be the distinct feature in BP-II.
In Table~\ref{tab:masslist}, we summarize the current constraints 
and LHC14 reaches with  100 fb$^{-1}$ luminosity on the $W'$ and $Z'$ masses in various models.
%Furthermore, the parameter contours
%enable limits to be set on models or a discovery to discriminate between
%particular models, although this discrimination requires the theoretical
%cross sections to be accurately calculated, and to be matched  well with
%the experiment data. 
If one needs to identify new physics
models more precisely, one has to %do the analysis by either combining
combine different discovery channels, such as top quark pair, single top quark production for the
heavy resonances, or study angular distributions, or other properly
defined asymmetries, in the most promising regions of parameter space of the models considered.
For example, the LPD(T), HPD(T), and FPD(T) models can be further explored by examining the single-top production,
the associate production of $W'$ and $W$(or $Z$) bosons, and the production of weak gauge boson pairs from electroweak gauge boson fusion processes.

\begin{acknowledgments}

We thanks Wade Fisher for discussing the LHC results on putting limits
on the $W^\prime$ mass. We thanks Reinhard Schwienhorst for his comments on part of this draft. 
The work of ZL, JHY and CPY was supported by the U.S. National Science Foundation under Grand No. PHY-0855561.
QHC was supported in part by the National Natural Science Foundation of China under Grants No. 11245003.
\end{acknowledgments}
	
\bibliography{wppaper}

\begin{thebibliography}{43}
\expandafter\ifx\csname natexlab\endcsname\relax\def\natexlab#1{#1}\fi
\expandafter\ifx\csname bibnamefont\endcsname\relax
  \def\bibnamefont#1{#1}\fi
\expandafter\ifx\csname bibfnamefont\endcsname\relax
  \def\bibfnamefont#1{#1}\fi
\expandafter\ifx\csname citenamefont\endcsname\relax
  \def\citenamefont#1{#1}\fi
\expandafter\ifx\csname url\endcsname\relax
  \def\url#1{\texttt{#1}}\fi
\expandafter\ifx\csname urlprefix\endcsname\relax\def\urlprefix{URL }\fi
\providecommand{\bibinfo}[2]{#2}
\providecommand{\eprint}[2][]{\url{#2}}

\bibitem[{\citenamefont{Mohapatra and
  Pati}(1975{\natexlab{a}})}]{Mohapatra:1974gc}
\bibinfo{author}{\bibfnamefont{R.}~\bibnamefont{Mohapatra}} \bibnamefont{and}
  \bibinfo{author}{\bibfnamefont{J.~C.} \bibnamefont{Pati}},
  \bibinfo{journal}{Phys.Rev.} \textbf{\bibinfo{volume}{D11}},
  \bibinfo{pages}{2558} (\bibinfo{year}{1975}{\natexlab{a}}).

\bibitem[{\citenamefont{Mohapatra and
  Pati}(1975{\natexlab{b}})}]{Mohapatra:1974hk}
\bibinfo{author}{\bibfnamefont{R.~N.} \bibnamefont{Mohapatra}}
  \bibnamefont{and} \bibinfo{author}{\bibfnamefont{J.~C.} \bibnamefont{Pati}},
  \bibinfo{journal}{Phys.Rev.} \textbf{\bibinfo{volume}{D11}},
  \bibinfo{pages}{566} (\bibinfo{year}{1975}{\natexlab{b}}).

\bibitem[{\citenamefont{Senjanovic and Mohapatra}(1975)}]{Senjanovic:1975rk}
\bibinfo{author}{\bibfnamefont{G.}~\bibnamefont{Senjanovic}} \bibnamefont{and}
  \bibinfo{author}{\bibfnamefont{R.~N.} \bibnamefont{Mohapatra}},
  \bibinfo{journal}{Phys.Rev.} \textbf{\bibinfo{volume}{D12}},
  \bibinfo{pages}{1502} (\bibinfo{year}{1975}).

\bibitem[{\citenamefont{Mohapatra and Senjanovic}(1981)}]{Mohapatra:1980yp}
\bibinfo{author}{\bibfnamefont{R.~N.} \bibnamefont{Mohapatra}}
  \bibnamefont{and}
  \bibinfo{author}{\bibfnamefont{G.}~\bibnamefont{Senjanovic}},
  \bibinfo{journal}{Phys.Rev.} \textbf{\bibinfo{volume}{D23}},
  \bibinfo{pages}{165} (\bibinfo{year}{1981}).

\bibitem[{\citenamefont{Chivukula et~al.}(2006)}]{Chivukula:2006cg}
\bibinfo{author}{\bibfnamefont{R.~S.} \bibnamefont{Chivukula}}
  \bibnamefont{et~al.}, \bibinfo{journal}{Phys. Rev.}
  \textbf{\bibinfo{volume}{D74}}, \bibinfo{pages}{075011}
  (\bibinfo{year}{2006}).

\bibitem[{\citenamefont{Barger et~al.}(1980{\natexlab{a}})\citenamefont{Barger,
  Keung, and Ma}}]{Barger:1980ix}
\bibinfo{author}{\bibfnamefont{V.~D.} \bibnamefont{Barger}},
  \bibinfo{author}{\bibfnamefont{W.-Y.} \bibnamefont{Keung}}, \bibnamefont{and}
  \bibinfo{author}{\bibfnamefont{E.}~\bibnamefont{Ma}},
  \bibinfo{journal}{Phys.Rev.} \textbf{\bibinfo{volume}{D22}},
  \bibinfo{pages}{727} (\bibinfo{year}{1980}{\natexlab{a}}).

\bibitem[{\citenamefont{Barger et~al.}(1980{\natexlab{b}})\citenamefont{Barger,
  Keung, and Ma}}]{Barger:1980ti}
\bibinfo{author}{\bibfnamefont{V.~D.} \bibnamefont{Barger}},
  \bibinfo{author}{\bibfnamefont{W.-Y.} \bibnamefont{Keung}}, \bibnamefont{and}
  \bibinfo{author}{\bibfnamefont{E.}~\bibnamefont{Ma}},
  \bibinfo{journal}{Phys.Rev.Lett.} \textbf{\bibinfo{volume}{44}},
  \bibinfo{pages}{1169} (\bibinfo{year}{1980}{\natexlab{b}}).

\bibitem[{\citenamefont{Georgi et~al.}(1989)\citenamefont{Georgi, Jenkins, and
  Simmons}}]{Georgi:1989ic}
\bibinfo{author}{\bibfnamefont{H.}~\bibnamefont{Georgi}},
  \bibinfo{author}{\bibfnamefont{E.~E.} \bibnamefont{Jenkins}},
  \bibnamefont{and} \bibinfo{author}{\bibfnamefont{E.~H.}
  \bibnamefont{Simmons}}, \bibinfo{journal}{Phys.Rev.Lett.}
  \textbf{\bibinfo{volume}{62}}, \bibinfo{pages}{2789} (\bibinfo{year}{1989}).

\bibitem[{\citenamefont{Georgi et~al.}(1990)\citenamefont{Georgi, Jenkins, and
  Simmons}}]{Georgi:1989xz}
\bibinfo{author}{\bibfnamefont{H.}~\bibnamefont{Georgi}},
  \bibinfo{author}{\bibfnamefont{E.~E.} \bibnamefont{Jenkins}},
  \bibnamefont{and} \bibinfo{author}{\bibfnamefont{E.~H.}
  \bibnamefont{Simmons}}, \bibinfo{journal}{Nucl.Phys.}
  \textbf{\bibinfo{volume}{B331}}, \bibinfo{pages}{541} (\bibinfo{year}{1990}).

\bibitem[{\citenamefont{Li and Ma}(1981)}]{Li:1981nk}
\bibinfo{author}{\bibfnamefont{X.}~\bibnamefont{Li}} \bibnamefont{and}
  \bibinfo{author}{\bibfnamefont{E.}~\bibnamefont{Ma}},
  \bibinfo{journal}{Phys.Rev.Lett.} \textbf{\bibinfo{volume}{47}},
  \bibinfo{pages}{1788} (\bibinfo{year}{1981}).

\bibitem[{\citenamefont{Malkawi et~al.}(1996)\citenamefont{Malkawi, Tait, and
  Yuan}}]{Malkawi:1996fs}
\bibinfo{author}{\bibfnamefont{E.}~\bibnamefont{Malkawi}},
  \bibinfo{author}{\bibfnamefont{T.~M.} \bibnamefont{Tait}}, \bibnamefont{and}
  \bibinfo{author}{\bibfnamefont{C.}~\bibnamefont{Yuan}},
  \bibinfo{journal}{Phys.Lett.} \textbf{\bibinfo{volume}{B385}},
  \bibinfo{pages}{304} (\bibinfo{year}{1996}).

\bibitem[{\citenamefont{He and Valencia}(2002)}]{He:2002ha}
\bibinfo{author}{\bibfnamefont{X.-G.} \bibnamefont{He}} \bibnamefont{and}
  \bibinfo{author}{\bibfnamefont{G.}~\bibnamefont{Valencia}},
  \bibinfo{journal}{Phys.Rev.} \textbf{\bibinfo{volume}{D66}},
  \bibinfo{pages}{013004} (\bibinfo{year}{2002}).

\bibitem[{\citenamefont{Hsieh et~al.}(2010)\citenamefont{Hsieh, Schmitz, Yu,
  and Yuan}}]{Hsieh:2010zr}
\bibinfo{author}{\bibfnamefont{K.}~\bibnamefont{Hsieh}},
  \bibinfo{author}{\bibfnamefont{K.}~\bibnamefont{Schmitz}},
  \bibinfo{author}{\bibfnamefont{J.-H.} \bibnamefont{Yu}}, \bibnamefont{and}
  \bibinfo{author}{\bibfnamefont{C.-P.} \bibnamefont{Yuan}},
  \bibinfo{journal}{Phys.Rev.} \textbf{\bibinfo{volume}{D82}},
  \bibinfo{pages}{035011} (\bibinfo{year}{2010}).

\bibitem[{\citenamefont{Rizzo}(2006)}]{Rizzo:2006nw}
\bibinfo{author}{\bibfnamefont{T.~G.} \bibnamefont{Rizzo}},
  \bibinfo{journal}{TASI06} pp. \bibinfo{pages}{537--575}
  (\bibinfo{year}{2006}), \eprint{hep-ph/0610104}.

\bibitem[{\citenamefont{Berger et~al.}(2011{\natexlab{a}})\citenamefont{Berger,
  Cao, Chen, and Zhang}}]{Berger:2011hn}
\bibinfo{author}{\bibfnamefont{E.~L.} \bibnamefont{Berger}},
  \bibinfo{author}{\bibfnamefont{Q.-H.} \bibnamefont{Cao}},
  \bibinfo{author}{\bibfnamefont{C.-R.} \bibnamefont{Chen}}, \bibnamefont{and}
  \bibinfo{author}{\bibfnamefont{H.}~\bibnamefont{Zhang}},
  \bibinfo{journal}{Phys. Rev.} \textbf{\bibinfo{volume}{D83}},
  \bibinfo{pages}{114026} (\bibinfo{year}{2011}{\natexlab{a}}),
  \eprint{1103.3274}.

\bibitem[{\citenamefont{Bauer et~al.}(2010)\citenamefont{Bauer, Ligeti,
  Schmaltz, Thaler, and Walker}}]{Bauer:2009cc}
\bibinfo{author}{\bibfnamefont{C.~W.} \bibnamefont{Bauer}},
  \bibinfo{author}{\bibfnamefont{Z.}~\bibnamefont{Ligeti}},
  \bibinfo{author}{\bibfnamefont{M.}~\bibnamefont{Schmaltz}},
  \bibinfo{author}{\bibfnamefont{J.}~\bibnamefont{Thaler}}, \bibnamefont{and}
  \bibinfo{author}{\bibfnamefont{D.~G.} \bibnamefont{Walker}},
  \bibinfo{journal}{Phys.Lett.} \textbf{\bibinfo{volume}{B690}},
  \bibinfo{pages}{280} (\bibinfo{year}{2010}).

\bibitem[{\citenamefont{Khachatryan et~al.}(2011)}]{Khachatryan:2010fa}
\bibinfo{author}{\bibfnamefont{V.}~\bibnamefont{Khachatryan}}
  \bibnamefont{et~al.} (\bibinfo{collaboration}{CMS Collaboration}),
  \bibinfo{journal}{Phys.Lett.} \textbf{\bibinfo{volume}{B698}},
  \bibinfo{pages}{21} (\bibinfo{year}{2011}).

\bibitem[{\citenamefont{Chatrchyan et~al.}(2011)}]{Chatrchyan:2011wq}
\bibinfo{author}{\bibfnamefont{S.}~\bibnamefont{Chatrchyan}}
  \bibnamefont{et~al.} (\bibinfo{collaboration}{CMS Collaboration}),
  \bibinfo{journal}{JHEP} \textbf{\bibinfo{volume}{1105}}, \bibinfo{pages}{093}
  (\bibinfo{year}{2011}).

\bibitem[{\citenamefont{Aad et~al.}(2011{\natexlab{a}})}]{Aad:2011yg}
\bibinfo{author}{\bibfnamefont{G.}~\bibnamefont{Aad}} \bibnamefont{et~al.}
  (\bibinfo{collaboration}{ATLAS Collaboration}), \bibinfo{journal}{Phys.Lett.}
  \textbf{\bibinfo{volume}{B705}}, \bibinfo{pages}{28}
  (\bibinfo{year}{2011}{\natexlab{a}}), \eprint{1108.1316}.

\bibitem[{\citenamefont{Aad
  et~al.}(2011{\natexlab{b}})}]{Collaboration:2011dca}
\bibinfo{author}{\bibfnamefont{G.}~\bibnamefont{Aad}} \bibnamefont{et~al.}
  (\bibinfo{collaboration}{ATLAS Collaboration}),
  \bibinfo{journal}{Phys.Rev.Lett.} \textbf{\bibinfo{volume}{107}},
  \bibinfo{pages}{272002} (\bibinfo{year}{2011}{\natexlab{b}}),
  \eprint{1108.1582}.

\bibitem[{\citenamefont{Langacker}(2009)}]{Langacker:2008yv}
\bibinfo{author}{\bibfnamefont{P.}~\bibnamefont{Langacker}},
  \bibinfo{journal}{Rev.Mod.Phys.} \textbf{\bibinfo{volume}{81}},
  \bibinfo{pages}{1199} (\bibinfo{year}{2009}).

\bibitem[{\citenamefont{Carena et~al.}(2004)\citenamefont{Carena, Daleo,
  Dobrescu, and Tait}}]{Carena:2004xs}
\bibinfo{author}{\bibfnamefont{M.~S.} \bibnamefont{Carena}},
  \bibinfo{author}{\bibfnamefont{A.}~\bibnamefont{Daleo}},
  \bibinfo{author}{\bibfnamefont{B.~A.} \bibnamefont{Dobrescu}},
  \bibnamefont{and} \bibinfo{author}{\bibfnamefont{T.~M.} \bibnamefont{Tait}},
  \bibinfo{journal}{Phys.Rev.} \textbf{\bibinfo{volume}{D70}},
  \bibinfo{pages}{093009} (\bibinfo{year}{2004}).

\bibitem[{\citenamefont{Salvioni et~al.}(2009)\citenamefont{Salvioni,
  Villadoro, and Zwirner}}]{Salvioni:2009mt}
\bibinfo{author}{\bibfnamefont{E.}~\bibnamefont{Salvioni}},
  \bibinfo{author}{\bibfnamefont{G.}~\bibnamefont{Villadoro}},
  \bibnamefont{and} \bibinfo{author}{\bibfnamefont{F.}~\bibnamefont{Zwirner}},
  \bibinfo{journal}{JHEP} \textbf{\bibinfo{volume}{0911}}, \bibinfo{pages}{068}
  (\bibinfo{year}{2009}).

\bibitem[{\citenamefont{Accomando et~al.}(2011)\citenamefont{Accomando,
  Belyaev, Fedeli, King, and Shepherd-Themistocleous}}]{Accomando:2010fz}
\bibinfo{author}{\bibfnamefont{E.}~\bibnamefont{Accomando}},
  \bibinfo{author}{\bibfnamefont{A.}~\bibnamefont{Belyaev}},
  \bibinfo{author}{\bibfnamefont{L.}~\bibnamefont{Fedeli}},
  \bibinfo{author}{\bibfnamefont{S.~F.} \bibnamefont{King}}, \bibnamefont{and}
  \bibinfo{author}{\bibfnamefont{C.}~\bibnamefont{Shepherd-Themistocleous}},
  \bibinfo{journal}{Phys.Rev.} \textbf{\bibinfo{volume}{D83}},
  \bibinfo{pages}{075012} (\bibinfo{year}{2011}).

\bibitem[{\citenamefont{Lynch et~al.}(2001)\citenamefont{Lynch, Simmons,
  Narain, and Mrenna}}]{Lynch:2000md}
\bibinfo{author}{\bibfnamefont{K.~R.} \bibnamefont{Lynch}},
  \bibinfo{author}{\bibfnamefont{E.~H.} \bibnamefont{Simmons}},
  \bibinfo{author}{\bibfnamefont{M.}~\bibnamefont{Narain}}, \bibnamefont{and}
  \bibinfo{author}{\bibfnamefont{S.}~\bibnamefont{Mrenna}},
  \bibinfo{journal}{Phys.Rev.} \textbf{\bibinfo{volume}{D63}},
  \bibinfo{pages}{035006} (\bibinfo{year}{2001}).

\bibitem[{\citenamefont{Schmaltz and Spethmann}(2011)}]{Schmaltz:2010xr}
\bibinfo{author}{\bibfnamefont{M.}~\bibnamefont{Schmaltz}} \bibnamefont{and}
  \bibinfo{author}{\bibfnamefont{C.}~\bibnamefont{Spethmann}},
  \bibinfo{journal}{JHEP} \textbf{\bibinfo{volume}{1107}}, \bibinfo{pages}{046}
  (\bibinfo{year}{2011}).

\bibitem[{\citenamefont{Maiezza et~al.}(2010)\citenamefont{Maiezza, Nemevsek,
  Nesti, and Senjanovic}}]{Maiezza:2010ic}
\bibinfo{author}{\bibfnamefont{A.}~\bibnamefont{Maiezza}},
  \bibinfo{author}{\bibfnamefont{M.}~\bibnamefont{Nemevsek}},
  \bibinfo{author}{\bibfnamefont{F.}~\bibnamefont{Nesti}}, \bibnamefont{and}
  \bibinfo{author}{\bibfnamefont{G.}~\bibnamefont{Senjanovic}},
  \bibinfo{journal}{Phys.Rev.} \textbf{\bibinfo{volume}{D82}},
  \bibinfo{pages}{055022} (\bibinfo{year}{2010}), \eprint{1005.5160}.

\bibitem[{\citenamefont{Grojean et~al.}(2011)\citenamefont{Grojean, Salvioni,
  and Torre}}]{Grojean:2011vu}
\bibinfo{author}{\bibfnamefont{C.}~\bibnamefont{Grojean}},
  \bibinfo{author}{\bibfnamefont{E.}~\bibnamefont{Salvioni}}, \bibnamefont{and}
  \bibinfo{author}{\bibfnamefont{R.}~\bibnamefont{Torre}},
  \bibinfo{journal}{JHEP} \textbf{\bibinfo{volume}{1107}}, \bibinfo{pages}{002}
  (\bibinfo{year}{2011}), \eprint{1103.2761}.

\bibitem[{\citenamefont{Nemevsek et~al.}(2011)\citenamefont{Nemevsek, Nesti,
  Senjanovic, and Zhang}}]{Nemevsek:2011hz}
\bibinfo{author}{\bibfnamefont{M.}~\bibnamefont{Nemevsek}},
  \bibinfo{author}{\bibfnamefont{F.}~\bibnamefont{Nesti}},
  \bibinfo{author}{\bibfnamefont{G.}~\bibnamefont{Senjanovic}},
  \bibnamefont{and} \bibinfo{author}{\bibfnamefont{Y.}~\bibnamefont{Zhang}},
  \bibinfo{journal}{Phys.Rev.} \textbf{\bibinfo{volume}{D83}},
  \bibinfo{pages}{115014} (\bibinfo{year}{2011}), \eprint{1103.1627}.

\bibitem[{\citenamefont{Torre}(2011)}]{Torre:2011wy}
\bibinfo{author}{\bibfnamefont{R.}~\bibnamefont{Torre}} (\bibinfo{year}{2011}),
  \eprint{1109.0890}.

\bibitem[{\citenamefont{Jezo et~al.}(2012)\citenamefont{Jezo, Klasen, and
  Schienbein}}]{Jezo:2012rm}
\bibinfo{author}{\bibfnamefont{T.}~\bibnamefont{Jezo}},
  \bibinfo{author}{\bibfnamefont{M.}~\bibnamefont{Klasen}}, \bibnamefont{and}
  \bibinfo{author}{\bibfnamefont{I.}~\bibnamefont{Schienbein}}
  (\bibinfo{year}{2012}), \bibinfo{note}{5 pages, 3 figures},
  \eprint{1203.5314}.

\bibitem[{\citenamefont{Keung and Senjanovic}(1983)}]{Keung:1983uu}
\bibinfo{author}{\bibfnamefont{W.-Y.} \bibnamefont{Keung}} \bibnamefont{and}
  \bibinfo{author}{\bibfnamefont{G.}~\bibnamefont{Senjanovic}},
  \bibinfo{journal}{Phys.Rev.Lett.} \textbf{\bibinfo{volume}{50}},
  \bibinfo{pages}{1427} (\bibinfo{year}{1983}).

\bibitem[{\citenamefont{Berger et~al.}(2011{\natexlab{b}})\citenamefont{Berger,
  Cao, Yu, and Yuan}}]{Berger:2011xk}
\bibinfo{author}{\bibfnamefont{E.~L.} \bibnamefont{Berger}},
  \bibinfo{author}{\bibfnamefont{Q.-H.} \bibnamefont{Cao}},
  \bibinfo{author}{\bibfnamefont{J.-H.} \bibnamefont{Yu}}, \bibnamefont{and}
  \bibinfo{author}{\bibfnamefont{C.-P.} \bibnamefont{Yuan}},
  \bibinfo{journal}{Phys.Rev.} \textbf{\bibinfo{volume}{D84}},
  \bibinfo{pages}{095026} (\bibinfo{year}{2011}{\natexlab{b}}),
  \eprint{1108.3613}.

\bibitem[{\citenamefont{Nadolsky et~al.}(2008)}]{Nadolsky:2008zw}
\bibinfo{author}{\bibfnamefont{P.~M.} \bibnamefont{Nadolsky}}
  \bibnamefont{et~al.}, \bibinfo{journal}{Phys. Rev.}
  \textbf{\bibinfo{volume}{D78}}, \bibinfo{pages}{013004}
  (\bibinfo{year}{2008}).

\bibitem[{\citenamefont{Amsler et~al.}(2008)}]{Amsler:2008zzb}
\bibinfo{author}{\bibfnamefont{C.}~\bibnamefont{Amsler}} \bibnamefont{et~al.}
  (\bibinfo{collaboration}{Particle Data Group}), \bibinfo{journal}{Phys.Lett.}
  \textbf{\bibinfo{volume}{B667}}, \bibinfo{pages}{1} (\bibinfo{year}{2008}).

\bibitem[{\citenamefont{Erler}(1999)}]{Erler:1999ug}
\bibinfo{author}{\bibfnamefont{J.}~\bibnamefont{Erler}} (\bibinfo{year}{1999}),
  \eprint{hep-ph/0005084}.

\bibitem[{\citenamefont{Abazov et~al.}(2011)}]{Abazov:2010ti}
\bibinfo{author}{\bibfnamefont{V.~M.} \bibnamefont{Abazov}}
  \bibnamefont{et~al.} (\bibinfo{collaboration}{D0 Collaboration}),
  \bibinfo{journal}{Phys.Lett.} \textbf{\bibinfo{volume}{B695}},
  \bibinfo{pages}{88} (\bibinfo{year}{2011}).

\bibitem[{\citenamefont{Aaltonen et~al.}(2011)}]{Aaltonen:2010jj}
\bibinfo{author}{\bibfnamefont{T.}~\bibnamefont{Aaltonen}} \bibnamefont{et~al.}
  (\bibinfo{collaboration}{CDF Collaboration}), \bibinfo{journal}{Phys.Rev.}
  \textbf{\bibinfo{volume}{D83}}, \bibinfo{pages}{031102}
  (\bibinfo{year}{2011}).

\bibitem[{\citenamefont{Aaltonen et~al.}(2009)}]{Aaltonen:2009qu}
\bibinfo{author}{\bibfnamefont{T.}~\bibnamefont{Aaltonen}} \bibnamefont{et~al.}
  (\bibinfo{collaboration}{CDF Collaboration}),
  \bibinfo{journal}{Phys.Rev.Lett.} \textbf{\bibinfo{volume}{103}},
  \bibinfo{pages}{041801} (\bibinfo{year}{2009}).

\bibitem[{\citenamefont{Aaltonen et~al.}(2008)}]{:2007dia}
\bibinfo{author}{\bibfnamefont{T.}~\bibnamefont{Aaltonen}} \bibnamefont{et~al.}
  (\bibinfo{collaboration}{CDF Collaboration}), \bibinfo{journal}{Phys.Rev.}
  \textbf{\bibinfo{volume}{D77}}, \bibinfo{pages}{051102}
  (\bibinfo{year}{2008}).

\bibitem[{\citenamefont{Collaboration}(2012)}]{CMS-PAS-EXO-11-092}
\bibinfo{author}{\bibfnamefont{C.}~\bibnamefont{Collaboration}}
  (\bibinfo{collaboration}{CMS Collaboration}),
  \bibinfo{journal}{CMS-PAS-EXO-11-092}  (\bibinfo{year}{2012}).

\bibitem[{\citenamefont{Aad et~al.}(2009)}]{Aad:2009wy}
\bibinfo{author}{\bibfnamefont{G.}~\bibnamefont{Aad}} \bibnamefont{et~al.}
  (\bibinfo{collaboration}{ATLAS Collaboration}) (\bibinfo{year}{2009}),
  \eprint{0901.0512}.

\bibitem[{\citenamefont{Adam~Bourdarios et~al.}(2009)}]{AdamBourdarios:2009zza}
\bibinfo{author}{\bibfnamefont{C.}~\bibnamefont{Adam~Bourdarios}}
  \bibnamefont{et~al.} (\bibinfo{collaboration}{ATLAS}) (\bibinfo{year}{2009}),
  \bibinfo{note}{aTL-PHYS-PUB-2009-063}.

\end{thebibliography}

\end{document}